\documentclass[aps,pra,reprint,superscriptaddress]{revtex4-2}

\PassOptionsToPackage{hyperfootnotes=false}{hyperref}

\usepackage{fontspec}

\usepackage{textcomp}
\usepackage{url}
\usepackage{booktabs}
\usepackage{array}[=2016-10-06]
\usepackage{amsfonts,amsmath,amssymb}
\usepackage{mathrsfs}
\usepackage{microtype}

\numberwithin{equation}{section}

\usepackage{float}
\usepackage{needspace}
\usepackage{xcolor}
\usepackage{hyperref}
\usepackage{bookmark}
\usepackage{orcidlink}
\hypersetup{
  bookmarksnumbered=true,
  bookmarksopen=true,
  bookmarksopenlevel=0,
  pdfpagemode=UseOutlines,
  colorlinks=true,
  citecolor=blue,
  linkcolor=red,
  urlcolor=blue,
  pdftitle={Structure-Fair Quantum Circuit Complexity: An Auditable Information-Theoretic Lower Bound},
  pdfauthor={HongZheng Liu, YiNuo Tian, Zhiyue Wu}
}
\newcommand{\RCCformaltext}[1]{\textbf{#1}}
\let\RCCclasssubsection\subsection
\makeatletter
\newcommand{\RCCcompactsubsections}{%
  \renewcommand{\subsection}{\@startsection{subsection}{2}{\z@}%
    {12pt plus 1pt minus 1pt}{7pt plus 0.5pt minus 1pt}%
    {\normalfont\small\bfseries\centering}}%
}
\makeatother
\makeatletter
\def\toclevel@section{0}
\def\toclevel@subsection{1}
\makeatother
\bookmarksetup{depth=-1}
\begin{document}

\title{\texorpdfstring{Structure-Fair Quantum Circuit Complexity:\\An Auditable Information-Theoretic Lower Bound}{Structure-Fair Quantum Circuit Complexity: An Auditable Information-Theoretic Lower Bound}}

\author{HongZheng Liu\,\orcidlink{0009-0002-2238-9187}}
\thanks{Corresponding author: \href{mailto:weiyouyeyu@foxmail.com}{\nolinkurl{weiyouyeyu@foxmail.com}} (H.L.)}
\affiliation{Independent, China}

\author{YiNuo Tian\,\orcidlink{0009-0005-8088-9894}}
\affiliation{Independent, China}

\author{Zhiyue Wu\,\orcidlink{0009-0003-4765-2049}}
\affiliation{Independent, China}

\begin{abstract}
Quantum circuit complexity is often used to characterize the physical cost of state preparation, but its physical meaning depends on the reference and counting rules; the entropy-removal costs of operations such as reset may be left out of resource accounting. We propose the principle of structural fairness and develop the Reference-Contingent Complexity (RCC) framework, jointly specifying the reference, generation capabilities, and atomic costs. We construct a model family that can approximate arbitrary finite-dimensional pure and mixed states. Within an admissible model fixed in advance, we prove a rigorous lower bound on universal optimal quantum circuit complexity. The target state's smooth one-shot information gap relative to the unbiased structured vacuum (the maximum-entropy state on the reference support) has an entropy-spectrum structure. Calibrated by the atomic control bandwidth and with finite-description corrections included, this gap sets a common cost floor for every admissible successful path. Predeclared final-state measurements and their finite-sample statistics thus yield independently verifiable one-sided complexity lower-bound certificates without reconstructing the generation history. Finally, exact structural allocation relations under changes of observation window and reference motivate a conjecture on the reference covariance of entropy and complexity: a reference can shift the complexity zero point, but cannot remove the burden of generating structure at no cost.
\end{abstract}

\maketitle
\bookmarksetup{depth=1}
\RCCcompactsubsections

\begingroup
\setlength{\parskip}{1.2pt plus 0.3pt minus 0.2pt}
\setlength{\abovedisplayskip}{8pt plus 2pt minus 2pt}
\setlength{\belowdisplayskip}{8pt plus 2pt minus 2pt}
\setlength{\abovedisplayshortskip}{4pt plus 1pt minus 1pt}
\setlength{\belowdisplayshortskip}{5pt plus 1pt minus 1pt}
\setlength{\jot}{2pt}
\section{Introduction}
\label{sec:1}

When quantum circuit complexity is used to quantify the physical cost of generation, a fundamental asymmetry in the generation background becomes apparent. Consider the product state $|0^n\rangle$ of $n$ qubits. In the standard universal quantum circuit model, this state is usually supplied as the initial state, from which complexity is counted \cite{PleschBrukner2011_StatePreparation}. Generating the same mathematical state from a maximally mixed or thermal background, or in an unpolarized physical system, still requires establishing a preferred direction, removing entropy, and applying control~\cite{SchulmanMorWeinstein2005_HBAC}. The state itself is unchanged, yet the shortest admissible generation path compatible with the background has changed.

Two aspects of the problem offer a starting point: the definition of complexity and the realizability of generation with finite resources. Standard quantum circuit complexity is always defined relative to a specified generation model \cite{BarencoEtAl1995_ElementaryGates,Knill1995_ApproximationQuantumCircuits,Aaronson2016_ComplexityStatesTransformations}. The initial state, allowed operations, and auxiliary resources specify the starting point and available capabilities; the error criterion determines the target domain; and the addressing, parallelism, and counting rules give concrete meaning to a single step \cite{ShendeEtAl2006_Synthesis,PleschBrukner2011_StatePreparation}. In simple backgrounds, these data often remain implicit in the notation. For code spaces, conserved sectors, or energy cutoffs, they enter directly into the definition of the generation task. Geometric approaches to complexity and their development in quantum field theory \cite{Nielsen2006_GeometricCircuitLowerBounds,NielsenEtAl2006_QuantumComputationGeometry,Brown2024_PolynomialEquivalenceComplexityGeometries,JeffersonMyers2017_CircuitComplexityQFT} express this dependence through the choice of reference point, allowed control directions, and cost geometry.

\Needspace{3\baselineskip}
The standard definition thus explains the origin of the opening disparity, but does not supply a common prescription for comparison across backgrounds. Once the background changes, structure previously assigned to the zero point may itself become a generation task. How this cost is accounted for determines whether the two complexity values are physically comparable.

Realizability with finite resources requires an admissible run to use finite control records and finite workspace, with precision and readout rules declared in advance. Yet a finite run can still encode generation capabilities that grow with system size. A global reset from an unpolarized background to a pure state can be written as a constant-length instruction, while the entropy it removes grows linearly with system size. The thermodynamics of information erasure has established that such information loss carries a physical cost under specified thermal and operational conditions \cite{Landauer1961_Irreversibility,DelRioEtAl2011_NegativeEntropy,ReebWolf2014_Landauer,FaistEtAl2015_MinimalWorkCost}. If these capabilities enter through atomic operations, auxiliary resources, or program priors without being charged, finite instruction length alone cannot ensure a fair comparison of generation costs.

Both considerations point to the provenance of structure and the responsibility for its cost. We therefore propose the principle of structural fairness as a physical postulate. Shared structure that the background supplies stably is absorbed into the complexity zero point. Any capability that produces target-directed structure relative to that background must appear explicitly in the predeclared dynamics, auxiliary resources, or program prior and be charged to the corresponding resource coordinates.

Beyond specifying how costs are counted, we need an independently testable physical basis for comparing them. Even with the reference and counting rules fixed, complexity is the infimum of the costs over all admissible generation histories. Both discrete circuit and continuous geometric formulations face this global optimization. Finding the infimum is generally difficult, whereas the statistical properties of the final state can be independently checked by repeated measurements \cite{FlammiaLiu2011_DirectFidelity,KlieschRoth2021_Certification}. Since every successful path must produce the target structure in its final state, can the structural difference between that state and the reference background yield a common cost lower bound for all admissible successful paths, without reconstructing the generation history?

Moving from control histories to final-state evidence requires a link between descriptions of the generation process and information in the state. Quantum algorithmic information theory has developed several frameworks for the algorithmic information of individual quantum states \cite{Gacs2001_QuantumAlgorithmicEntropy,Vitanyi2001_QuantumKolmogorovComplexity,BerthiaumeEtAl2001_QuantumKolmogorov}, providing a starting point for this connection. When the generation task permits finite error, smooth max-relative entropy \cite{Datta2009_MinMaxRelativeEntropies,TomamichelHayashi2013_Hierarchy,Tomamichel2016_FiniteResources} characterizes the one-shot reference gap that cannot be eliminated within the allowed error tolerance. The choice of reference, in turn, depends on which structures the background already supplies. Quantum resource theories characterize resources through free objects and free transformations \cite{HorodeckiEtAl2003_PureMixed,BartlettEtAl2007_ReferenceFrames,BrandaoGour2015_ReversibleResourceTheories,ChitambarGour2019_QRT}, providing a language for distinguishing background supply from newly generated structure.

We apply these tools to finite-dimensional target-state generation. Guided by the structural fairness postulate, we construct a class of physical models for Reference-Contingent Complexity (RCC), jointly specifying the reference background, admissible generation, and atomic costs. The readout rules are fixed before the target state is chosen. Control histories must be faithfully transcribed into programs whose lengths are bounded in terms of their path costs, and the program prior must satisfy reference admissibility to constrain any directional structure it already contains.

Explicit constructions provide models in this class that can approximate arbitrary finite-dimensional pure and mixed states, thereby establishing universal state-generation capability. Verifiable reference-qualification certificates are also available for a class of finite-control processes with measurement, feed-forward, and synchronous loops. The constructions and certification are developed in Appendices~\ref{app:F} and \ref{app:H}, respectively. The companion codebase rcc-refcert \cite{LiuTianWu2026_RCCRefcert} implements selected constructions and certificates and provides a reproducible example connecting reference-qualification checks to lower-bound computation within a fixed model.

For a member model $\mathfrak M_R$ fixed before the target state is chosen, universal optimal quantum circuit complexity $C_{\rm opt}^{(\epsilon)}(\rho;\mathfrak M_R)$ is the infimum of the costs, measured in atomic resource slots, over all admissible circuits that generate the target state $\rho$ within error $\epsilon$. Here, “universal” refers to state-generation capability, and “optimal” to all admissible paths in the fixed model.

To quantify structure in the final state beyond what the background supplies, we first fix the corresponding unbiased zero point. The reference set $R$ records the shared structure supplied by the background and determines the support projector $\Pi_R$, the support subspace $\mathcal H_R:=\operatorname{im}\Pi_R$, and its dimension $d_R$. Taking the maximum-entropy state on this support as the structured vacuum preserves the support specified by the reference without installing a preferred target direction within it. For a target state $\rho$ satisfying $\operatorname{supp}(\rho)\subseteq\mathcal H_R$, the structured vacuum and the state-side RCC value are
\begin{equation*}
\begin{aligned}
\sigma_R &:= \frac{\Pi_R}{d_R},
\\[2pt]
C_R(\rho) &:= \frac{D(\rho\Vert\sigma_R)}{\log_2\Gamma_R}
= \frac{\log_2 d_R-S(\rho)}{\log_2\Gamma_R}.
\end{aligned}
\end{equation*}

Here, the relative entropy $D(\rho\Vert\sigma_R)$ is the entropy deficit of the target state relative to the maximum entropy on the support. The number $\Gamma_R$ counts the action classes distinguishable in one atomic resource slot, and its logarithm gives the control bandwidth. The quantity $C_R$ converts this entropy deficit to the scale of atomic operations and forms the state-side linear leading term of the main lower bound.

We prove that the final state's one-shot reference gap gives a rigorous lower bound on the optimal generation cost within the model. The full nonasymptotic result retains finite error and transcription overhead. For $0<\epsilon\le\epsilon_0<1$, suppose that the smoothing loss satisfies the regularity condition $\delta_{\rm sm}(\rho,\epsilon)\le c_{\rm sm}\log_2(1/\epsilon)$ of Corollary~3.1 and that the relevant constants are uniformly bounded over the task family under consideration. The entropy–spectrum structure of the bound then takes the compact form
\begin{equation*}
\begin{aligned}
&C_{\rm opt}^{(\epsilon)}(\rho;\mathfrak M_R)
\ge C_R(\rho)+\frac{\Delta_{\rm spec}(\rho)}{\log_2\Gamma_R}
\\[1pt]
&\quad-O\!\Biggl(
\frac{\log_2\max\!\bigl\{1,C_{\rm opt}^{(\epsilon)}(\rho;\mathfrak M_R)\bigr\}
+\log_2(1/\epsilon)}{\log_2\Gamma_R}
\Biggr).
\end{aligned}
\end{equation*}

Here, $\Delta_{\rm spec}(\rho):=S(\rho)-H_{\min}(\rho)\ge0$, with $H_{\min}(\rho):=-\log_2\lVert\rho\rVert_\infty$. The entropy deficit measures the average information depth relative to the background, while $\Delta_{\rm spec}$ captures the additional one-shot contribution associated with the largest eigenvalue. This nonnegative spectral correction strengthens the lower bound. The $O(\cdot)$ term accounts for the smoothing loss allowed by the error tolerance and the finite-description correction. The nonrecursive inversion in Sec.~\ref{sec:3} gives a lower bound directly from final-state information.

The bound constrains all paths because every successful history must produce the target structure within the specified tolerance. Faithful transcription converts any admissible control history into a classical self-delimiting (prefix-free) program, with its description length controlled by the path cost. Reference admissibility then bounds the one-shot reference gap of a program's output by the program length plus the fixed reference-calibration constant. Together, these relations turn the final-state gap into a cost floor for every successful history. After cost inversion, the bound depends only on the final state and the fixed model, so it remains valid upon taking the infimum of the path costs. The final state can thus constrain the optimal generation cost within the model without retaining the path that actually occurred.

When the reference background changes, a state-side comparison of structure must also be coupled to a process-side conversion of paths. Under reference-state domination and resource-compilation conditions, we obtain cross-reference comparisons that account for compilation error and cost; the results are given in Appendix~\ref{app:A.10}.

Statistical tests can also certify these constraints on path costs. If a predeclared test effect has an acceptance probability on the target state reliably above the structured-vacuum baseline, the contrast provides evidence for the reference gap and rules out all admissible generation histories below the corresponding cost threshold. Quantum hypothesis testing \cite{WangRenner2012_HypothesisTesting} supplies a standard formulation for optimizing such effects. Combining these tests with error calibration and finite-sample control yields computable, independently verifiable one-sided complexity certificates based on measurement data. Statistical evidence from the final state thereby constrains the global optimum over paths, even when that optimum is difficult to determine directly. Our main contributions are as follows.

\begingroup
\setlength{\leftmargini}{1.5em}
\makeatletter
\def\@listi{\leftmargin\leftmargini
  \topsep 3pt minus 0.5pt
  \partopsep 0pt \itemsep 2pt plus 0.4pt minus 0.2pt \parsep 0pt}
\makeatother
\begin{enumerate}

\item \textbf{We propose the principle of structural fairness and construct a class of RCC physical models.} Taking realizability with finite resources as the operational premise, we place background structure, admissible generation, and atomic costs within a common scheme of resource accounting. Explicit constructions provide models with universal state-generation capabilities. Finite-control certificates and their companion implementations make reference qualification verifiable for the corresponding open processes.

\item \textbf{We prove a rigorous lower bound on universal optimal quantum circuit complexity.} Within an admissible model fixed in advance, the final state's one-shot reference gap constrains every admissible successful path and hence the optimal cost over all paths. Three reference-aligned constructions—classical probability concentration, nested uniform projector contraction, and coherent logical encoding—give explicit preparation upper bounds for the corresponding state families. Under their respective construction and asymptotic conditions, these upper bounds match the leading scaling of the lower bound.

\item \textbf{We construct one-sided complexity certificates.} Predeclared final-state tests, combined with quantum hypothesis testing and finite-sample certification, provide independently verifiable statistical evidence that excludes generation histories below the certified cost threshold.

\Needspace*{4\baselineskip}
\item \textbf{We develop windowed RCC and propose a conjecture on the reference covariance of entropy and complexity.} Under an unbiased reference, task-relevant windows resolve differences in internal structure among pure states that share the same global $C_R$ leading term, expanding a single global value into a structural profile. Exact finite-dimensional allocation relations further show how structure can be reassigned among effective entropy, explicit generation responsibility, and the reference background. A reference can shift the complexity zero point, but cannot eliminate generation responsibility at no cost.

\item \textbf{We establish quantitative interfaces connecting RCC to dynamics, energy–entanglement resources, and window thermodynamics.} These yield, respectively, model-consistent two-sided complexity intervals, conditional resource–output lower bounds, and window response relations, all with a common reference, error semantics, and atomic scale.

\end{enumerate}
\endgroup

Section~\ref{sec:2} introduces the physical models, structured vacuum, and state-side value. Section~\ref{sec:3} proves the main lower bound and discusses final-state certificates and near-tightness; Sec.~\ref{sec:4} develops the certificates into computable, composable audit protocols with finite-sample guarantees. Section~\ref{sec:5} studies properties at fixed reference and windowed RCC. Section~\ref{sec:6} establishes the dynamical, resource, and thermodynamic interfaces, Sec.~\ref{sec:7} discusses the physical meaning of complexity, related frameworks, and the entropy–complexity covariance conjecture, and Sec.~\ref{sec:8} concludes.

Appendices~\ref{app:F} and \ref{app:H} give the technical details of model construction and qualification certification. Appendix~\ref{app:A} proves the main lower bound, final-state certificates, and cross-reference comparisons through resource compilation; Appendix~\ref{app:G} gives the near-tightness constructions. Appendices~\ref{app:B}--\ref{app:D} treat support deviations, explicit certificate synthesis, and finite-sample statistics, respectively. Appendix~\ref{app:E} establishes the structural properties underlying fixed-reference and windowed RCC, and Appendix~\ref{app:I} develops the dynamical and thermodynamic interfaces.

\endgroup

\begingroup

\setlength{\parskip}{1.1pt plus 0.6pt minus 0.2pt}
\setlength{\abovedisplayskip}{8pt plus 0.8pt minus 3pt}
\setlength{\belowdisplayskip}{8pt plus 0.8pt minus 3pt}
\setlength{\abovedisplayshortskip}{4pt plus 0.4pt minus 2pt}
\setlength{\belowdisplayshortskip}{5pt plus 0.4pt minus 2pt}
\setlength{\jot}{2pt}
\newcommand{\IIstatement}[1]{%
  \par\addvspace{0.25\baselineskip}%
  \Needspace{3\baselineskip}%
  \noindent\hspace*{\parindent}\RCCformaltext{#1}\hspace{0.5em}\ignorespaces
}
\newcommand{\IIproofsketch}{%
  \par\addvspace{0.12\baselineskip}%
  \noindent\hspace*{\parindent}\RCCformaltext{Proof sketch.}\enspace\ignorespaces
}
\newcommand{\IIqed}{\nobreak\hspace{0.35em}\ensuremath{\square}%
  \par\addvspace{0.12\baselineskip}}

\section{Reference-Consistent Quantum Circuit Models and RCC}
\label{sec:2}

The background asymmetry in the Introduction calls for the reference dependence of complexity to be made explicit in the generation model. For finite-dimensional target-state generation, we use the principle of structural fairness to specify jointly what the background supplies, which dynamics are allowed, and how atomic resources are measured. These specifications define a reference-consistent quantum circuit model.

Admissible processes and the atomic scale determine generation costs; the reference set and structured vacuum determine the state-side zero point. Faithful transcription of control histories and reference constraints on the program prior connect the two. Native constructions and finite-control certification then show how such models can be realized and which states they can generate. With this structure in place, calibration by the atomic control bandwidth turns the reference information gap into the linear leading term of RCC. Section~\ref{sec:3} establishes the corresponding rigorous lower bound. All logarithms below are to base 2 unless stated otherwise.

\subsection{Structural Fairness and Universal Optimal Quantum Circuit Complexity}
\label{sec:2.1}

An admissible generation process may involve quantum channels, measurements, classical control, auxiliary systems, and coupling to an environment. It may also use randomization, discarding, or the preparation of a purification followed by a partial trace. The initial states, access rules, control precision, and resource scales of these components must all be fixed before the target state is chosen. Realizability with finite resources ensures that each run uses finite control records and finite workspace.

Yet the global-reset example shows that a finite device can still package directional generation capabilities that grow with system size into a single atomic action. Accounting for generation costs therefore also requires a distinction between structure already supplied by the background and structure added by the dynamics. We adopt structural fairness as a physical postulate governing how each enters the complexity accounting.

\IIstatement{Physical Principle 2.1 (Structural fairness).}
For a given statement about quantum circuit complexity, the reference background, preset resources, allowed operations, control interface, and program prior must be fixed jointly before the target state is chosen. Structure that the background can supply stably is absorbed into the reference zero point. Any directional initial state, auxiliary system, environment, macroinstruction, addressing capability, precision, postselection rule, or program bias that can shorten a target-generation path must either be explicitly supplied by a compatible reference background or appear as a physical input beyond that background in the allowed dynamics and be included in the resource coordinates. A change of reference, encoding, or interpretation interface alone must not introduce additional usable structure in the target direction.

This principle allows different physical backgrounds to define different complexity models, each with a coherent structural zero point and resource scale. The reference background, admissible generation, readout semantics, and atomic scale together constitute the model data
\begin{equation}
\label{eq:2.1}
\mathfrak M_R
=
(\mathcal H,\mathcal G,\mathcal F_R,R,\mathcal D_R,\Gamma_R).
\end{equation}

Here, $\mathcal H$ is the effective output space in which target states are compared; the states and operators $\rho,\Pi_R,\sigma_R$ below are all defined on this space. The set $\mathcal G$ specifies admissible gates, channels, and measurements. The data $\mathcal F_R$ collect the background initial states, preset resources, auxiliary systems, environment, and implementation data. They explicitly include the operationally reduced atomic physical interface $\mathfrak I_R^{(1)}$ defined below and, when needed, an extended implementation space, an extended reference state, and a fixed unconditional readout. The reference set $R$ records structure supplied before the target state is chosen; its state-side construction is developed in Subsec.~\ref{sec:2.2}. Finally, $\mathcal D_R$ specifies the effective output semantics and the associated distance $d_R^{\rm out}$.

The interface $\mathfrak I_R^{(1)}$ fixes the model's process-side scale. It specifies the serial, parallel, and resource-accounting rules for $L(\mathcal C)$ and determines $\mathcal A_R^{(1)}$, $\Gamma_R$, and $g_R$. We list $\Gamma_R$ separately in Eq.~\eqref{eq:2.1} because it provides the explicit calibration. For problems that scale with system size, the data are indexed as $\mathfrak M_{R,n}$ and organized, under the same interface rules, into a model family $\{\mathfrak M_{R,n}\}_{n\in\mathbb N}$. Qualification constants must be uniformly bounded across the family.

To give different implementations the same final-state semantics, admissible model outputs are always normalized unconditional states. For protocols with success flags or postselection, the success probability, repetitions, amplification, and associated resources must remain part of the complete physical process. A normalized conditional success branch considered without its realization probability and repetition process is outside the complexity comparison defined here.

Work registers, failure flags, and control records belong to the extended implementation. When such components are used, the model must fix in advance a completely positive and trace-preserving (CPTP) readout $\mathcal R_{R,n}$, independent of the target state and uniform across the family. This readout maps the extended output to the effective space and the extended reference state to $\sigma_{R,n}$, defined in Subsec.~\ref{sec:2.2}. Complete positivity then carries reference domination on the extended space through to the final output.

Once the model specifies admissible outputs, circuit length still needs a definite physical unit. The operationally reduced atomic physical interface $\mathfrak I_R^{(1)}$ specifies what constitutes one step. Each atomic resource slot has fixed spacetime support, serial and parallel rules, reliable control resolution, addressing, auxiliary resources, environment, and outcome records. Two single-slot commands are equivalent when both their dynamical semantics, defined on every allowed input (total semantics), and their static resource signatures agree. Taking this exact quotient of the raw commands gives a finite, alias-free set of action classes
\begin{equation}
\begin{aligned}
\label{eq:2.2}
&\mathcal A_R^{(1)},
\qquad
\Gamma_R:=|\mathcal A_R^{(1)}|,
\quad
2\le\Gamma_R<\infty,
\\
&g_R:=\log_2\Gamma_R.
\end{aligned}
\end{equation}

The quantity $g_R$ is the Hartley control bandwidth of one atomic slot \cite{Hartley1928_TransmissionInformation}, measuring distinguishable discrete action choices. Labels, Kraus representations, or global phases with identical full semantics do not increase $\Gamma_R$. Addresses, accessible outcomes, control memory, or resource side effects that can affect subsequent execution distinguish action classes. Continuous controls must first be represented by a finite codebook at the reliable resolution, or their required precision must enter an explicit resource coordinate.

Here, atomicity is always relative to the declared resource slot; it does not prescribe a unique minimal gate set across platforms. If a high-level action can inject directional structure that grows with system size within a single undecomposed slot, that capability must also be reflected in additional code length, a separate fuel coordinate, or a decomposition into lower-level actions. Appendix~\ref{app:F.6} develops the corresponding resource-recalibration interface.

In a native model admitting all finite action words, there are exactly $\Gamma_R^L$ atomic control histories of length $L$, so the raw action-choice information is exactly $L g_R$. General models constrained by syntax, dynamics, or control states realize only a subset of these histories.

A self-delimiting encoding must also specify the length of the finite history, adding $O(\log L)$ overhead. A rank block for the whole word, together with an Elias length header \cite{Elias1975_UniversalCodewordSets}, attains this asymptotic minimax rate for any finite $\Gamma_R$, as proved in Appendix~\ref{app:F.3}. The expression $L g_R+a\log\max\{1,L\}$ used in Sec.~\ref{sec:3} thus reflects the control capacity of the atomic interface.

With the output semantics and atomic scale fixed, target-state generation defines a global optimum over all admissible physical paths. We call the infimum of the number of atomic resource slots required to reach the target error domain universal optimal quantum circuit complexity.

\IIstatement{Definition 2.1 (Universal optimal quantum circuit complexity).}
Given the model data $\mathfrak M_R$, a target state $\rho$, and a tolerance $0\le\epsilon\le1$, let ${\rm Circ}(\mathfrak M_R)$ denote the domain of admissible circuits in the model. Define
\begin{equation}
\begin{aligned}
\label{eq:2.3}
C_{\rm opt}^{(\epsilon)}(\rho;\mathfrak M_R)
:=
\inf\bigl\{
L(\mathcal C)\,\bigm|\,
&\mathcal C\in{\rm Circ}(\mathfrak M_R),
\\
&d_R^{\rm out}(\omega_{\mathcal C},\rho)\le\epsilon
\bigr\}.
\end{aligned}
\end{equation}

Here, $\omega_{\mathcal C}$ is the circuit's normalized unconditional output, and $L(\mathcal C)$ is measured according to the serial, parallel, and resource rules of $\mathfrak I_R^{(1)}$. The value is $+\infty$ when the feasible set is empty. When the model is fixed and unambiguous, we write $C_{\rm opt}^{(\epsilon)}(\rho)$.

Equation~\eqref{eq:2.3} takes the infimum over all admissible circuits within a given generation background, defining the central process-side object of this work. Subsection~\ref{sec:2.3} establishes the qualifications that make these model data structure-fair. Within the resulting domain, Sec.~\ref{sec:3} derives an auditable final-state lower bound on this global optimum.

\subsection{Reference Background and Structured Vacuum}
\label{sec:2.2}

The preceding subsection fixed admissible generation paths and their atomic lengths. On the state side, we must also specify the physical background from which those paths begin. Conservation laws may confine evolution to a fixed-charge sector, stabilizer constraints may restrict accessible states to a code space, and a Hamiltonian together with an energy cutoff may delimit an effective Hilbert space in advance. These structures precede the choice of target state and determine which part of state space the generation problem can access. They should therefore be accounted for as background structure rather than charged again as structure produced by target generation.

Specifying the allowed support identifies the states that belong to the problem, but does not yet identify a state representing the absence of any selected target direction. The state-side reference therefore has two levels: the reference set determines the shared physical support, and the structured vacuum supplies an unbiased zero point within it. Together they provide the model's state-side baseline.

\IIstatement{Definition 2.2 (Reference background and structured vacuum).}
In the finite-dimensional unbiased setting used here, let $R=\{P_i\}$ be a set of compatible, mutually commuting constraint projectors fixed by independent physical priors. The reference set specifies the state domain in which the shared constraints hold simultaneously,
\begin{equation}
\label{eq:2.4}
\begin{gathered}
\Pi_R:=\prod_i P_i,
\qquad
\mathcal H_R:=\Pi_R\mathcal H,
\\
d_R:={\rm Tr}\,\Pi_R>0.
\end{gathered}
\end{equation}

In the absence of further statistical constraints, the associated unbiased structured vacuum is defined by
\begin{equation}
\label{eq:2.5}
\sigma_R:=\frac{\Pi_R}{d_R},
\qquad
S(\sigma_R)=\log_2 d_R.
\end{equation}

Equations~\eqref{eq:2.4} and~\eqref{eq:2.5} together specify the state-side reference background. The projector $\Pi_R$ absorbs shared structure already supplied by a sector, code space, or effective support. The state $\sigma_R$ is the unique maximum-entropy state on that support, with no preinstalled target-specific direction, phase, correlation, or low-entropy structure. Vacuum here denotes the model's structural zero point; it does not imply the absence of energy, dynamics, or environmental coupling. Proposition F.1 in Appendix~\ref{app:F} proves the uniqueness of this maximum-entropy zero point and gives the exact identity $D(\rho\Vert\sigma_R)=\log_2d_R-S(\rho)$, on which the state-side reference information gap in Subsec.~\ref{sec:2.4} is based.

The reference set must be fixed before target selection by conservation laws, an encoding structure, a Hamiltonian, the control architecture, or predeclared task conditions. An insufficient reference attributes stable background structure to the target; an excessively detailed reference absorbs into the zero point information that the generation process should produce. A minimally sufficient reference includes the shared structure that can be supplied stably and repeatedly, leaving target-specific structure to the actual generation dynamics.

For mutually exclusive superselection sectors, the task sector is selected first, and compatible constraints are then imposed within it. Mutually exclusive projectors cannot be multiplied as simultaneously satisfied conditions. In the non-Abelian case, one may first decompose the space into invariant blocks; Appendix~\ref{app:F.8} establishes the same maximum-entropy zero point on an unbiased invariant block.

We work below in the ideal reference domain ${\rm supp}(\rho)\subseteq\mathcal H_R$. Small support leakage can be included explicitly in certificates through projection conditioning, reference regularization, and conservative reductions, as developed in Appendix~\ref{app:B}. If a Hamiltonian, temperature, or chemical potential supplies an additional independent statistical background, a Gibbs-type weighted reference may be considered. Appendix~\ref{app:F.8} gives the corresponding free-energy identity and specifies the qualifications that must be checked anew for the program construction and one-shot relations in this extension.

\subsection{From Structural Fairness to a Class of Reference-Consistent Models}
\label{sec:2.3}

The preceding subsections specified the physical generation processes, atomic costs, and reference background. To constrain final-state information by process length, we must also encode control histories as programs of controlled length and examine how the interpreter distributes output weights across all short descriptions. The physical reference, history transcription, and program prior are thus the three objects that structural fairness must constrain jointly within the model.

Physical execution uses total CPTP channels or quantum instruments \cite{DaviesLewis1970_QuantumProbability}, the standard quantum-information description of processes with measurement, feedback, and environmental coupling that yield normalized unconditional outputs. At the description level, classical self-delimiting, prefix-free programs give finite control records unambiguous boundaries, and their Kraft weights define the full program prior. Within these two standard semantics, we implement Principle 2.1 through reference consistency, faithful transcription, and semantic reference admissibility, as follows.

Consider a model–interpreter family
\begin{equation*}
\widehat{\boldsymbol{\mathfrak M}}_R
:=
\left\{
\widehat{\mathfrak M}_{R,n}
=
(\mathfrak M_{R,n},U_{R,n})
\right\}_{n\in\mathbb N},
\end{equation*}

where $U_{R,n}$ depends only on the reference and model data fixed in advance. Let $\mathcal P_n$ be its full effective prefix-free program domain. Every $p\in\mathcal P_n$ outputs a normalized unconditional state $\rho_{p,n}$ in the effective space.

The same physical reference data must determine what the background supplies, which outputs are admissible, and how resources are counted.

\IIstatement{Definition 2.3 (Kinematic reference consistency).}
A physical model family $\{\mathfrak M_{R,n}\}_n$ satisfies kinematic reference consistency $\mathsf{RCon}$ if the structured vacuum, background initial states and free resources, input and output domains of allowed operations, unconditional readout, error semantics, and atomic accounting rules are all fixed by the same target-independent reference data. Initial states and repeatable structure supplied by the background enter $R$ or the corresponding extended reference state. Other auxiliary structure and control capabilities must remain explicit in the model and accounting rules. Directional reset, cooling, and measurement are allowed, with their inputs, outputs, and resource requirements specified relative to the same reference.

Appendix~\ref{app:F.1} establishes the corresponding fixed reference, operationally reduced atomic interface, and unconditional readout, and proves that reference domination on the extended space carries through to the effective output space.

Every admissible run has a finite control history, to which a self-delimiting program assigns a finite description. To turn history length into a constraint on final-state information, the compilation map must cover all admissible processes while preserving their output semantics and resource scale. Faithful transcription expresses this requirement as a correspondence between processes and programs with controlled description length.

\IIstatement{Definition 2.4 (Faithful transcription).}
The model must declare in advance whether it uses the exact or approximate branch; the approximate branch must also specify a nonempty certified precision domain $\Delta_{\rm TC}\subset(0,1)$. Suppose there exist a compilation map independent of the target state and uniform across the model family, and constants $a,b,\gamma\ge0$ independent of system size. The map must cover every admissible circuit and preserve its unconditional output semantics. For any admissible circuit $\mathcal C$ with output $\omega_{\mathcal C}$ and cost $L(\mathcal C)$, require the following relations, with the approximate branch holding for every $\delta_{\rm cmp}\in\Delta_{\rm TC}$:
\begin{equation}
\label{eq:2.6}
\begin{aligned}
&\text{exact transcription}
\\[-1pt]
&\rho_{p_{\mathcal C}}=\omega_{\mathcal C},
\\[-1pt]
&|p_{\mathcal C}|\le
g_{R,n}L(\mathcal C)
+a\log_2\max\{1,L(\mathcal C)\}
+\gamma;
\\[4pt]
&\text{approximate transcription}
\\[-1pt]
&d_{R,n}^{\rm out}\!\left(
\rho_{p_{\mathcal C,\delta_{\rm cmp}}},
\omega_{\mathcal C}
\right)
\le
\delta_{\rm cmp},
\\[-1pt]
&|p_{\mathcal C,\delta_{\rm cmp}}|\le
g_{R,n}L(\mathcal C)
+a\log_2\max\{1,L(\mathcal C)\}
\\[-1pt]
&\quad
+b\log_2(1/\delta_{\rm cmp})
+\gamma.
\end{aligned}
\end{equation}

Under these conditions, the model–interpreter family satisfies faithful transcription $\mathsf{TC}$. Here $g_{R,n}=\log_2\Gamma_{R,n}$. Native discrete models use the exact branch and may take $b=0$. In the approximate branch, the compilation error $\delta_{\rm cmp}$ and generation error $\epsilon_{\rm circ}$ are combined according to $\mathcal D_R$. For a metric error criterion, $\epsilon_{\rm circ}+\delta_{\rm cmp}\le\epsilon$ ensures that the program output remains within the total target tolerance. Addressing, precision, auxiliary resources, and failure records must enter $L$ or the transcription length. The same resource must not be counted more than once across $L$, $\Gamma_R$, and additional code length.

Faithful transcription constrains the description length and output semantics of each physical history. It does not determine how the full interpreter distributes all short descriptions. The full program domain induces the semidensity operator
\begin{equation}
\label{eq:2.7}
\begin{gathered}
M_n
:=
\sum_{p\in\mathcal P_n}2^{-|p|}\rho_{p,n},
\\
M_n\succeq0,
\qquad
{\rm Tr}\,M_n\le1.
\end{gathered}
\end{equation}

The operator $M_n$ aggregates the code-length weights and output directions of all effective programs, representing the overall structure assigned to the output space by the interpreter before target selection. Kraft's inequality controls the total weight, while the output directions depend on program semantics. For example, if every program in a domain of total Kraft weight 1 outputs $|0^n\rangle$, the minimal domination constant relative to $I/2^n$ is $2^n$. The asymmetry of global reset thus reappears at the program level.

For any final-state effect $0\le Q\le I$, the quantity ${\rm Tr}(QM_n)$ aggregates the program prior's response to that effect, while ${\rm Tr}(Q\sigma_{R,n})$ gives the background response to the same effect. Program length already enters $M_n$ through the Kraft weights. To implement structural fairness, we require the amplification of the prior response relative to the background response to be bounded, for all effects, by a single finite constant independent of the target state and uniform in system size. A zero background response must also imply a zero prior response. This applies Principle 2.1's constraint on uncharged directional structure to the full program prior.

\IIstatement{Proposition 2.1 (Operator characterization of structural fairness for the program prior).}
Under the prefix-free, unconditional program semantics above, the full program prior meets the effect-response requirement if and only if there is a finite constant $C_U\ge1$, independent of the target state and system size, such that
\begin{equation}
\label{eq:2.8}
\setlength{\abovedisplayskip}{10pt plus 0.5pt minus 2pt}
\setlength{\belowdisplayskip}{9pt plus 0.5pt minus 2pt}
\setlength{\abovedisplayshortskip}{6pt plus 0.5pt minus 2pt}
\setlength{\belowdisplayshortskip}{7pt plus 0.5pt minus 2pt}
M_n\preceq C_U\sigma_{R,n}
\qquad
\text{for all }n.
\end{equation}
We call the property expressed by Eq.~\eqref{eq:2.8} semantic reference admissibility $\mathsf{RA}$.

\IIproofsketch
If Eq.~\eqref{eq:2.8} holds, taking the trace against any effect $Q$ gives the uniform response bound. Conversely, if $M_n-C_U\sigma_{R,n}$ has a positive eigenvalue at some size, its positive spectral projection is an effect for which the prior response exceeds $C_U$ times the background response, contradicting the requirement. The minimal domination constant $C_n^\star$ therefore equals the maximal effect-response ratio at that size, and $\sup_n C_n^\star<\infty$ gives uniform domination across the model family. \mbox{Appendix}~\ref{app:F.2} gives the full variational characterization and support conditions.\IIqed

Proposition 2.1 converts a physical response requirement on the program prior into an operator condition. It can be checked using the interpreter, full program domain, and structured vacuum, before computing complexity for any particular target state. A uniform constant controls the entire size-indexed family, keeping the interpreter's directional bias on a common reference scale. Alongside this condition, $\mathsf{RCon}$ specifies the physical background and resources, and $\mathsf{TC}$ ensures that the program description covers every admissible generation history.

\begingroup
\setlength{\parskip}{1.1pt plus 0.6pt minus 1.1pt}
To quantify the information gap between an individual program output and the structured vacuum, we use max-relative entropy \cite{Datta2009_MinMaxRelativeEntropies}. For positive operators $\omega,\tau$, write $D_{\max}(\omega\Vert\tau):=\inf\{\lambda:\omega\preceq2^\lambda\tau\}$, with value $+\infty$ when the supports are incompatible. Every effective program satisfies $2^{-|p|}\rho_{p,n}\preceq M_n$. Writing $\chi_U:=\log_2C_U=O(1)$, Eq.~\eqref{eq:2.8} immediately gives the single-program constraint
\begin{equation}
\label{eq:2.9}
\setlength{\abovedisplayskip}{10pt plus 0.5pt minus 3pt}
\setlength{\belowdisplayskip}{10pt plus 0.5pt minus 3pt}
\setlength{\abovedisplayshortskip}{7pt plus 0.5pt minus 3pt}
\setlength{\belowdisplayshortskip}{8pt plus 0.5pt minus 3pt}
D_{\max}(\rho_{p,n}\Vert\sigma_{R,n})
\le
|p|+\chi_U.
\end{equation}

Equation~\eqref{eq:2.9} states that the one-shot reference gap of a program's output is at most its description length plus a constant independent of system size. Combined with $\mathsf{TC}$, this relates the cost of each physical history to its final-state information. The smooth form is given in Appendix~\ref{app:A.3}; reference anchoring of the program semidensity and the conversion between constants are treated in Appendix~\ref{app:F.2}.

A change of reference shifts the structural zero point and changes the generation capabilities available at no cost. Comparing complexity across backgrounds therefore requires joint control of reference-state domination and target-independent resource compilation, as developed in Appendix~\ref{app:A.10}. In the Introduction's comparison between $|0^n\rangle$ and a maximally mixed background, retaining a free global reset implemented by a constant-length program after changing the reference makes the required domination constant grow with size. Uniform $\mathsf{RA}$ then fails.

The requirements on the physical reference, process transcription, and program prior jointly determine RCC model qualification.

\IIstatement{Definition 2.5 (RCC-admissible reference-con\-sistent model family).}
For a model–interpreter family, RCC admissibility is defined by the conjunction of the three conditions above,
\begin{equation}
\label{eq:2.10}
\setlength{\abovedisplayskip}{11pt plus 0.5pt minus 5pt}
\setlength{\belowdisplayskip}{10pt plus 0.5pt minus 4pt}
\setlength{\abovedisplayshortskip}{7pt plus 0.5pt minus 2pt}
\setlength{\belowdisplayshortskip}{8pt plus 0.5pt minus 2pt}
\mathsf{Acc}_{\rm RCC}
\!\left(\widehat{\boldsymbol{\mathfrak M}}_R\right)
:=
\mathsf{RCon}
\wedge\mathsf{TC}
\wedge\mathsf{RA}.
\end{equation}

Equation~\eqref{eq:2.10} connects actual generation processes to their full program descriptions under a common background and resource scale. It defines the physical model class within which we study optimal complexity.

The reference qualification of a concrete model can be checked through a finite proof object. We say that a model family satisfies $\mathsf{RefCert}$ if target-independent data, fixed in advance and uniform across the size-indexed family, soundly imply Eq.~\eqref{eq:2.8}. Such a certificate provides a verifiable sufficient condition for reference domination of the full program prior.

\IIstatement{Proposition 2.2 (Reference-qualification certificate interface).}
For the same model family, if $\mathsf{RefCert}$ holds, then
\begin{equation}
\label{eq:2.11}
\mathsf{RCon}
\wedge\mathsf{TC}
\wedge\mathsf{RefCert}
\Longrightarrow
\mathsf{Acc}_{\rm RCC}
\!\left(\widehat{\boldsymbol{\mathfrak M}}_R\right).
\end{equation}

The certificate thus turns abstract reference domination into a verification task for a concrete model. The full quantifiers are given in Appendix~\ref{app:F.2}.
\par\endgroup

Explicit dynamical constructions establish realizability for this class of models. Consider a native model whose admissible circuits are all finite atomic words and whose initial state is the structured vacuum $\sigma_R$. Let $\Phi_a$ denote the total CPTP dynamical semantics of action class $a\in\mathcal A_R^{(1)}$. A sufficient construction makes the uniform action average preserve the structured vacuum. For models operating on an extended space, this condition is first established on the extended reference state and then carried to the effective space by the fixed unconditional readout. On the effective space, it reads
\begin{equation}
\label{eq:2.12}
\frac{1}{\Gamma_R}
\sum_{a\in\mathcal A_R^{(1)}}
\Phi_a(\sigma_R)
=
\sigma_R.
\end{equation}

The atomic alphabet is then said to be reference-balanced on $\sigma_R$. Individual actions may be directional; preservation of the reference under the uniform average makes the actual control labels carry the choice of direction. Paired reset channels provide a concrete example.

\IIstatement{Corollary 2.1 (Existence and nontriviality of native models).}
Suppose a full-word native model family satisfies $\mathsf{RCon}$, starts from $\sigma_{R,n}$ at each size $n$, and satisfies the size-indexed form of Eq.~\eqref{eq:2.12}, using its pre-readout form for extended outputs. Then it admits exact faithful transcription by whole-word rank blocks and Elias length headers. Its canonical interpreter satisfies $M_{{\rm rank},n}=K_{\Gamma_{R,n}}\sigma_{R,n}\preceq\sigma_{R,n}$, where $K_{\Gamma_{R,n}}\le1$ is the total Kraft weight of the canonical valid rank domain. These models therefore satisfy $\mathsf{Acc}_{\rm RCC}$. This class includes qubit models built from universal discrete unitary gates with explicit addressing, paired resets, and local dephasing channels, capable of approximately generating arbitrary pure and mixed states from a maximally mixed background.

Whole-word transcription matches the code rate of atomic control histories to the Hartley bandwidth and gives uniform domination under reference balance; see Appendix~\ref{app:F.3}. In the qubit model above, explicit resets first produce a pure initial state, and universal discrete gates then approximate the target pure state. For a mixed state, an amplitude state is first prepared approximately from its spectral weights, followed by local dephasing and a rotation to its eigenbasis. Appendix~\ref{app:F.5} gives the construction. Reference qualification and directional generation capability are thus realized within the same physical model.

Reference balance provides a concise sufficient route. A broader class of models can qualify through a stationary reference envelope, as developed in Appendix~\ref{app:F.4}. For actions with unequal single-step reference gains, the additional structure can also be assigned a resource cost in advance. Certified interpreters can be combined into a mixture interpreter that is universal relative to the specified certified class. These constructions are given in Appendices~\ref{app:F.6} and~\ref{app:F.7}, respectively.

Finite-control processes with measurement, feed-forward, and synchronous loops also admit certification. For models satisfying $\mathsf{RCon}$ and $\mathsf{TC}$ and equipped with a synchronous realization coupling prefix-code parsing to completely positive (CP) dynamics, terminating paths and the nonhalting residual can be treated within the same unconditional semantics. The contributions of paths terminating after finitely many steps form the program semidensity. Finite certificates verify its reference domination across the family, and certificate margins control stability under finite-precision perturbations.

These results are established in Appendices~\ref{app:H.1}--\ref{app:H.5}; Appendix~\ref{app:H.6} treats explicit costs for reference gain. The companion codebase rcc-refcert implements several of these finite models, allowing their program semidensities to be recomputed and their reference-domination certificates to be checked \cite{LiuTianWu2026_RCCRefcert}.

These constructions and certification methods establish a realizable class of structure-fair models. Within it, the native qubit model family is fixed before target selection. At every size, the same member model can approximate every pure and mixed state of that size to any positive tolerance, providing the universality required here. Optimality is considered within a fixed member model: Eq.~\eqref{eq:2.3} takes the infimum of the costs over all its successful paths. Section~\ref{sec:3} derives a rigorous lower bound with this order of quantifiers.

Once the background, dynamics, and program description are fixed, the structure remaining in the final state relative to that background can be examined independently. The next subsection uses relative entropy to quantify its average information content and converts it to the resource scale of atomic control.

\subsection{Reference Information Gap and the RCC Linear Leading Term}
\label{sec:2.4}

The structured vacuum represents a background in which no direction within the reference support has yet been selected. The average structure exhibited by a target state relative to this zero point can be quantified through the statistical distinguishability captured by quantum relative entropy, without prior knowledge of the actual generation path. For a target state satisfying ${\rm supp}(\rho)\subseteq\mathcal H_R$, define the reference information gap
\begin{equation}
\label{eq:2.13}
\setlength{\abovedisplayskip}{10pt plus 0.5pt minus 3pt}
\setlength{\belowdisplayskip}{10pt plus 0.5pt minus 3pt}
\setlength{\abovedisplayshortskip}{7pt plus 0.5pt minus 3pt}
\setlength{\belowdisplayshortskip}{8pt plus 0.5pt minus 3pt}
I_R(\rho)
:=
D(\rho\Vert\sigma_R)
=
\log_2 d_R-S(\rho),
\end{equation}
where $S(\rho):=-{\rm Tr}(\rho\log_2\rho)$. The right-hand side follows from the exact identity for $\sigma_R=\Pi_R/d_R$: $\log_2d_R$ is the maximum unbiased uncertainty allowed by the reference support, and $S(\rho)$ is the uncertainty remaining in the target state. Their difference is the average structure exhibited by the target relative to that background. The reference absorbs the shared support constraints, while the information gap within the support remains part of the target structure. Thus $I_R$ depends only on the final state and the reference data fixed in advance, giving a path-independent state-side value.

The quantity $I_R$ measures state-side structure in bits, whereas circuit length measures generation cost in atomic resource slots. Equation~\eqref{eq:2.2} and Appendix~\ref{app:F.3} identify $g_R=\log_2\Gamma_R$ as the control bandwidth of the atomic interface and characterize it as the asymptotic minimax code rate per step in the worst case. Converting the reference information gap by this bandwidth defines the state-side linear leading term of RCC,
\begin{equation}
\label{eq:2.14}
C_R(\rho)
:=
\frac{I_R(\rho)}{g_R}
=
\frac{D(\rho\Vert\sigma_R)}{\log_2\Gamma_R}
=
\frac{\log_2 d_R-S(\rho)}{\log_2\Gamma_R}.
\end{equation}

The numerator quantifies target structure not yet supplied by the background; the denominator calibrates the discrete choice capability of one control step according to the atomic interface. The condition $\mathsf{RCon}$ fixes both through the same reference and resource specification, giving the ratio a definite physical scale within the model. To report state-side structure on this scale, we specify a unit associated with the reference background and atomic interface.

\IIstatement{Definition 2.6 (The canonical R-structon unit).}
One R-structon, denoted ${\rm st}_R$, is defined as
\begin{equation*}
\setlength{\abovedisplayskip}{10pt plus 0.5pt minus 5pt}
\setlength{\belowdisplayskip}{9pt plus 0.5pt minus 4pt}
\setlength{\abovedisplayshortskip}{6pt plus 0.5pt minus 4pt}
\setlength{\belowdisplayshortskip}{7pt plus 0.5pt minus 4pt}
1\,{\rm st}_R
\equiv
\log_2\Gamma_R\ {\rm bits}
=
\ln\Gamma_R\ {\rm nats}.
\end{equation*}
One ${\rm st}_R$ corresponds to the Hartley action-choice bandwidth of a complete atomic control slot in the model; the value of $C_R(\rho)$ is reported in this unit. The unit is fixed by the reference background and atomic interface. Comparisons across models are established through resource compilation that respects reference compatibility and controls costs.

Turning the average reference gap into a finite-error cost bound for every admissible path also requires spectral skew, smoothing loss, self-delimiting transcription, and fixed model overheads. Section~\ref{sec:3} and Appendix~\ref{app:A} carry out this derivation and show how $C_R$ appears with unit coefficient as the leading term of the bound under regularity conditions. We next use encoding structures and conserved sectors to illustrate the physical backgrounds underlying these values.

\subsection{Physical References under Structural Fairness}
\label{sec:2.5}

Structural fairness requires reference structure to have an independent physical origin before target selection. Engineered encoding structures and natural conservation structures provide complementary examples. The former show how constraints stably maintained by a device become part of the background; the latter show how the dynamics themselves delimit the state domain of the generation problem.

Consider a stabilizer code encoding $k$ logical qubits into $N$ physical qubits. If $\{S_i\}_{i=1}^{N-k}$ are independent stabilizer generators, one may take
\begin{equation}
\label{eq:2.15}
\Pi_R
=
\prod_i\frac{I+S_i}{2},
\qquad
d_R=2^k,
\qquad
\sigma_R=\frac{\Pi_R}{2^k}.
\end{equation}

The encoding scheme stably supplies the stabilizer relations, coding redundancy, and code-space support, which therefore enter the reference background. Logical directions, phases, entanglement, and correlations within the code space remain the responsibility of target generation. Every logical pure state $\rho_L$ has $C_R(\rho_L)=k/\log_2\Gamma_R$. These pure states share the same global reference gap because the unbiased structured vacuum selects no one-dimensional direction in the code space. Their full process complexities can nevertheless differ substantially. Windowed RCC, developed in Sec.~\ref{sec:5}, resolves task-relevant differences in internal organization within the accessible object domain.

Now consider a system constrained by fixed particle number, charge, or total spin. If the task is restricted to the sector with charge $q$ and projector $P_q$, the natural choice is
\begin{equation}
\label{eq:2.16}
\Pi_R=P_q,
\qquad
\sigma_R=
\frac{P_q}{{\rm Tr}\,P_q}.
\end{equation}

The conservation law determines which states belong to the generation problem; occupation distributions, coherence, entanglement, and other correlations within the sector remain part of the target structure. If the device can exchange the conserved quantity with an external reservoir, both the allowed operations and resource channels change. The reservoir and its costs must then be included in a different reference-consistent model.

These examples show the complementary physical roles of the reference set and atomic interface. The set $R$ absorbs shared structure stably supplied by the device, $D(\rho\Vert\sigma_R)$ quantifies target structure not yet supplied within the allowed domain, and $\log_2\Gamma_R$ sets its operational scale according to the actual gate classes, addresses, parallel rules, and control resolution. Even with the same state-side reference, different platforms can have different atomic interfaces and hence different RCC scales. The quantity $\Gamma_R$ must therefore be derived from the actual control architecture, rather than assigned in advance from the Hilbert-space dimension or abstract gate names.

\subsection{Reference Selection, Upgrades, and Domain of Applicability}
\label{sec:2.6}

The examples also show that reference selection is part of the physical model. RCC's reference dependence anchors complexity to a generation background that can be independently identified. Structure included in the reference must have an independent origin, be repeatedly identifiable, and be fixed before target selection. Appendix~\ref{app:F.1} expresses this as $R=R(\mathcal P_{\rm phys})$, fixed by independent physical priors $\mathcal P_{\rm phys}$. Minimal sufficiency separates repeatedly available background structure from target-specific structure and features inferred from data after the fact. The corresponding unbiased structured vacuum is proved to be the unique maximum-entropy zero point on the reference support.

Choosing a zero point after observing the target, so as to make that target appear simple, redefines the generation problem. Such a choice cannot retrospectively supply the background structure fixed in advance for the current task.

New conserved quantities, stable logical constraints, or repeatable structural residuals can change our understanding of the background. Once such structure is established by independent data or mechanisms and incorporated into the reference specification before the next target is chosen, it defines a reference upgrade,
\begin{equation}
\label{eq:2.17}
\widehat{\mathfrak M}_R
=
(\mathfrak M_R,U_R)
\longrightarrow
\widehat{\mathfrak M}_{R'}
=
(\mathfrak M_{R'},U_{R'}).
\end{equation}

An upgrade jointly respecifies $R$, $\sigma_R$, free resources, allowed operations, the atomic interface, $\Gamma_R$, the program interpreter, error conventions, and audit protocols. It also requires renewed checks of $\mathsf{RCon}$, $\mathsf{TC}$, and either $\mathsf{RA}$ or $\mathsf{RefCert}$. Existing certificates continue to describe the generation problem in $\widehat{\mathfrak M}_R$. The new reference defines a new complexity object, leaving the physical background of existing results intact.

A reference upgrade may change both the support and the process model. For generalized reference changes that preserve a common support, two-sided operator bounds still provide quantitative control on the state side. Let $\sigma$ and $\tau$ be strictly positive reference states on the same finite-dimensional support, satisfying
\begin{equation}
\label{eq:2.18}
2^{-a_-}\sigma
\preceq
\tau
\preceq
2^{a_+}\sigma,
\qquad
a_-,a_+\ge0.
\end{equation}
Then, for every target state with compatible support,
\begin{equation}
\label{eq:2.19}
-a_+
\le
D(\rho\Vert\tau)-D(\rho\Vert\sigma)
\le
a_-.
\end{equation}

The same two-sided bound holds for $D_{\max}^{\epsilon}$ when a common trace-distance smoothing set is used. A reference change within a common support thus produces a state-side additive shift controlled by the operator comparison constants. If the change also modifies admissible operations and resource costs, the state-side comparison alone does not determine the relation between the two quantities $C_{\rm opt}^{(\epsilon)}$. A target-independent resource compilation, uniform across the model family and available in both directions, is then needed to control output errors, length rescaling, and additional costs. Appendices~\ref{app:F.8} and~\ref{app:A.10} give the state-side stability and process-side comparison interfaces, respectively.

The rigorous results of this work are currently established for finite-dimensional unbiased references. Non-Abelian unbiased blocks retain the maximum-entropy zero point. General weighted references, energetically asymmetric atomic costs, open systems with explicit bath memory or non-Markovian structure, expected costs at random stopping times, and cutoff-independent infinite-dimensional limits require their own program qualifications and dynamical resource specifications. These extensions change the state objects, cost semantics, and control constants on which the present theorems depend, and therefore require more than direct extrapolation from the finite-dimensional unbiased results.

Within the finite-dimensional unbiased reference domain, structural fairness has now been implemented as a class of RCC models with explicit backgrounds, allowed dynamics, and resource accounting. Native constructions and finite-control certification realize these requirements in concrete models. Section~\ref{sec:3} uses this foundation to relate the generation cost of every admissible history to the final-state reference gap, and thereby constrain universal optimal quantum circuit complexity within the model.

\endgroup

\begingroup

\frenchspacing
\newcommand{\IIIstatement}[1]{%
  \par\addvspace{0.25\baselineskip}%
  \Needspace{3\baselineskip}%
  \noindent\hspace*{\parindent}\RCCformaltext{#1}\hspace{0.5em}\ignorespaces
}

\setlength{\parskip}{1.1pt plus 0.2pt minus 0.2pt}
\setlength{\abovedisplayskip}{8pt plus 2pt minus 3pt}
\setlength{\belowdisplayskip}{8pt plus 2pt minus 3pt}
\setlength{\abovedisplayshortskip}{4pt plus 1pt minus 2pt}
\setlength{\belowdisplayshortskip}{5pt plus 1pt minus 2pt}
\setlength{\jot}{2pt}

\section{Rigorous Lower Bound on Universal Optimal Quantum Circuit Complexity}
\label{sec:3}

Section~\ref{sec:2} fixed the generation background, admissible processes, and atomic resource scale under the principle of structural fairness. Within such a model, how does the structure revealed by the final state relative to the background constrain the optimal cost over all admissible generation histories?

We derive a pathwise lower bound from the target state's smooth one-shot reference gap, then take the infimum over path costs to obtain a rigorous lower bound on universal optimal quantum circuit complexity. Final-state effects turn this result into testable complexity certificates. Explicit constructions for three reference-aligned subfamilies show when the bound can track the leading scaling of the actual generation cost.

“Universal” refers to the model's ability to generate every pure and mixed state to any positive tolerance. The model family constructed in Appendix~\ref{app:F.5} is fixed before the target is chosen, and its member at each system size has this ability. “Optimal” means the infimum of the costs of all admissible successful paths within a fixed member model, for a fixed target state and tolerance. This bound applies to any model family satisfying the qualification conditions of Sec.~\ref{sec:2}. We fix such a \mbox{family} and an arbitrary member
\begin{equation*}
\widehat{\mathfrak M}_R
=
(\mathfrak M_R,U_R),
\qquad
\sigma_R=\Pi_R/d_R,
\end{equation*}
and retain the uniform boundedness requirements on qualification constants from Sec.~\ref{sec:2}. We state the main theorem first for the exact branch of faithful transcription, then give the error corrections for approximate transcription. All logarithms are to base $2$.

\subsection{Main Theorem}
\label{sec:3.1}

To constrain path costs using the final state, we must connect the control record of each successful path to the structure it produces. The program description bounds the output state by a scalar multiple of the structured vacuum, with the factor determined by the program length and fixed reference calibration. Since the actual output may lie anywhere within the target state's tolerance neighborhood, the relevant quantity is the smallest gap that persists throughout that neighborhood. Smooth max-relative entropy characterizes this one-shot constraint \cite{Tomamichel2016_FiniteResources}.

Throughout this section, circuit output error, approximate transcription error, and program smoothing are measured by the normalized trace distance $T(\rho,\tau):=\frac12\lVert\rho-\tau\rVert_1$; thus $d_R^{\rm out}=T$. Other error conventions require converting tolerances using the corresponding bounds on $T$. With this convention, define
\begin{equation*}
D_{\max}^{\epsilon}(\rho\Vert\sigma_R)
:=
\inf_{\substack{
\tau\in\mathcal D(\mathcal H_R)\\
T(\tau,\rho)\le\epsilon
}}
\inf\left\{
\lambda:\tau\preceq2^{\lambda}\sigma_R
\right\}.
\end{equation*}

This is the smallest operator gap relative to the structured vacuum that persists within tolerance $\epsilon$ of the target state. The gap is measured in bits, whereas $C_{\rm opt}^{(\epsilon)}$ is measured in atomic resource slots. The control bandwidth and transcription rules fixed in Sec.~\ref{sec:2} provide the conversion between them.

In the canonical atomic-slot model, faithful transcription of an admissible path of cost $L$ yields a program whose length, apart from the fixed overhead $\gamma$, is bounded uniformly by
\begin{equation*}
\setlength{\abovedisplayskip}{10pt plus 0.5pt minus 5pt}
\setlength{\belowdisplayskip}{9pt plus 0.5pt minus 4pt}
\setlength{\abovedisplayshortskip}{6pt plus 0.5pt minus 4pt}
\setlength{\belowdisplayshortskip}{7pt plus 0.5pt minus 4pt}
\Phi_{g_R,a}(L)
:=
g_RL+a\log\max\{1,L\},
\qquad
g_R:=\log\Gamma_R,
\end{equation*}
where $a\ge0$ is a self-delimiting transcription constant uniform across the model family. Let $F_{g_R,a}$ be the inverse of $\Phi_{g_R,a}$ on $[0,\infty)$, extended to return zero for nonpositive inputs. It converts the information required at the program level into a cost measured in atomic resource slots. The transcription and logarithmic reference-calibration overheads combine as $\beta_U:=\gamma+\chi_U=O(1)$. For a general nondecreasing code-length envelope, inversion uses the left generalized inverse defined in Appendix~\ref{app:A.7}.

The following theorem gives the rigorous lower bound on universal optimal quantum circuit complexity.

\IIIstatement{Theorem 3.1 (Lower bound on universal optimal quantum circuit complexity).}
Let a canonical reference-consistent model family satisfy $\mathsf{RCon}\wedge\mathsf{TC}\wedge\mathsf{RA}$, with the exact transcription branch of $\mathsf{TC}$, and let $\widehat{\mathfrak M}_R$ be any member of this family. For every target state satisfying ${\rm supp}(\rho)\subseteq\mathcal H_R$ and every $0\le\epsilon\le1$,
\begin{equation}
\label{eq:3.1}
\setlength{\abovedisplayskip}{10pt plus 0.5pt minus 2pt}
\setlength{\belowdisplayskip}{10pt plus 0.5pt minus 2pt}
\setlength{\abovedisplayshortskip}{7pt plus 0.5pt minus 1pt}
\setlength{\belowdisplayshortskip}{8pt plus 0.5pt minus 2pt}
\boxed{
C_{\rm opt}^{(\epsilon)}(\rho;\mathfrak M_R)
\ge
F_{g_R,a}\!\left[
D_{\max}^{\epsilon}(\rho\Vert\sigma_R)-\beta_U
\right].
}
\end{equation}

The right-hand side of Eq.~\eqref{eq:3.1} depends only on the target state and the model data fixed in advance. The effective outputs of all admissible successful histories lie in the same $\epsilon$-neighborhood of the target, so the reference gap that persists at this tolerance constrains the final states of all these histories.

Applying this constraint to each path and then taking the infimum of the costs over all admissible paths gives a common lower bound on the optimal quantum circuit complexity within the model. The bound holds even when the optimal circuit is unknown or the infimum is not attained. Subsection~\ref{sec:3.2} explains how the model conditions establish this pathwise connection.

For the unbiased structured vacuum $\sigma_R=\Pi_R/d_R$, we have $D_{\max}(\rho\Vert\sigma_R)=\log d_R-H_{\min}(\rho)$ and $D(\rho\Vert\sigma_R)=\log d_R-S(\rho)=g_RC_R(\rho)$. The exact identities relating max-relative entropy, relative entropy, and min-entropy therefore give the decomposition
\begin{equation*}
D_{\max}^{\epsilon}(\rho\Vert\sigma_R)
=
g_RC_R(\rho)
+\Delta_{\rm spec}(\rho)
-\delta_{\rm sm}(\rho,\epsilon),
\end{equation*}
where
\begin{equation*}
\begin{aligned}
\Delta_{\rm spec}(\rho)
&:=
S(\rho)-H_{\min}(\rho)
\\
&=
D_{\max}(\rho\Vert\sigma_R)
-D(\rho\Vert\sigma_R),
\\[2pt]
H_{\min}(\rho)
&:=
-\log\lVert\rho\rVert_\infty ,
\end{aligned}
\end{equation*}
and
\begin{equation*}
\delta_{\rm sm}(\rho,\epsilon)
:=
D_{\max}(\rho\Vert\sigma_R)
-D_{\max}^{\epsilon}(\rho\Vert\sigma_R)
\ge0.
\end{equation*}

Equation~\eqref{eq:3.1} can thus be written equivalently as
\begin{equation}
\label{eq:3.2}
\begin{aligned}
C_{\rm opt}^{(\epsilon)}(\rho;\mathfrak M_R)
&\ge
F_{g_R,a}\!\Bigl[
g_RC_R(\rho)
+\Delta_{\rm spec}(\rho)
\\[-1pt]
&\quad
-\delta_{\rm sm}(\rho,\epsilon)
-\beta_U
\Bigr].
\end{aligned}
\end{equation}

Equation~\eqref{eq:3.2} displays the full physical structure of the main lower bound. The term $g_RC_R(\rho)$ is the target state's average reference gap relative to the structured vacuum; $\Delta_{\rm spec}(\rho)$ accounts for the additional one-shot contribution associated with the largest eigenvalue; $\delta_{\rm sm}(\rho,\epsilon)$ records the structure absorbed at the current resolution; and $\beta_U$ collects the transcription and reference-calibration overheads, uniformly bounded with respect to the target state and system size. The inverse $F_{g_R,a}$ converts these information quantities into a complexity lower bound measured in atomic resource slots.

Here $C_R(\rho)$ is the state-side linear leading term, normalized by the control bandwidth. Under the regularity condition below, it enters the complexity lower bound with unit coefficient; the full bound retains the corrections for spectral skew, finite resolution, and self-delimiting transcription.

\IIIstatement{Corollary 3.1 (Compact form under smoothing regularity).}
Suppose the target-state family under consideration satisfies, throughout a fixed precision range $0<\epsilon\le\epsilon_0<1$,
\begin{equation*}
\setlength{\abovedisplayskip}{10pt plus 0.5pt minus 5pt}
\setlength{\belowdisplayskip}{9pt plus 0.5pt minus 4pt}
\setlength{\abovedisplayshortskip}{6pt plus 0.5pt minus 4pt}
\setlength{\belowdisplayshortskip}{7pt plus 0.5pt minus 4pt}
\delta_{\rm sm}(\rho,\epsilon)
\le
c_{\rm sm}\log(1/\epsilon),
\end{equation*}
with $c_{\rm sm}=O(1)$ uniformly bounded across that family. Then there is a likewise uniformly bounded $\widetilde c_1=O(1)$ such that
\begin{equation}
\label{eq:3.3}
\begin{aligned}
&
C_{\rm opt}^{(\epsilon)}(\rho;\mathfrak M_R)
\ge
C_R(\rho)
+\frac{\Delta_{\rm spec}(\rho)}{\log\Gamma_R}
\\
&\quad-
\frac{\widetilde c_1}{\log\Gamma_R}
\Bigl[
\log\max\!\left\{
1,C_{\rm opt}^{(\epsilon)}(\rho;\mathfrak M_R)
\right\}
\\[-1pt]
&\qquad\qquad
+\log(1/\epsilon)
\Bigr].
\end{aligned}
\end{equation}

Equation~\eqref{eq:3.3} makes the unit-coefficient leading term $C_R(\rho)$ explicit and thus identifies the leading scale of the lower bound. Since its correction still contains $C_{\rm opt}^{(\epsilon)}$, numerical inversion and the subsequent audit procedures use the nonrecursive forms in Eq.~\eqref{eq:3.1} or Eq.~\eqref{eq:3.2}.

For a model with approximate transcription, compilation error must enter both the smoothing radius of the program output and the transcription code length. Let $\epsilon_{\rm circ}$ be the circuit output tolerance, $\Delta_{\rm TC}$ the domain of compilation precisions actually certified for the model, and $b\ge0$ the corresponding precision-encoding coefficient. Assume
\begin{equation*}
\left\{
\delta_{\rm cmp}\in\Delta_{\rm TC}:
\epsilon_{\rm circ}+\delta_{\rm cmp}\le1
\right\}
\neq\varnothing .
\end{equation*}
Then
\begin{equation}
\label{eq:3.4}
\begin{aligned}
&
C_{\rm opt}^{(\epsilon_{\rm circ})}(\rho;\mathfrak M_R)
\ge
\sup_{\substack{
\delta_{\rm cmp}\in\Delta_{\rm TC}\\
\epsilon_{\rm circ}+\delta_{\rm cmp}\le1
}}
\\[1pt]
&\quad
F_{g_R,a}\!\left[
\begin{aligned}
&
D_{\max}^{\epsilon_{\rm circ}+\delta_{\rm cmp}}
(\rho\Vert\sigma_R)
-\beta_U
\\[-1pt]
&
-b\log(1/\delta_{\rm cmp})
\end{aligned}
\right].
\end{aligned}
\end{equation}

Equation~\eqref{eq:3.4} is the approximate-transcription version of the main result. Compilation error enlarges the total error radius of the program output relative to the target and adds the description overhead required to achieve that precision. The left-hand side remains the universal optimal quantum circuit complexity at circuit tolerance $\epsilon_{\rm circ}$. Compilation precisions are compared within the nonempty domain $\Delta_{\rm TC}$ actually certified for the model, preserving the distinct parameter meanings of circuit error, program error, and the complexity being bounded.

These results use $\mathsf{RCon}\wedge\mathsf{TC}\wedge\mathsf{RA}$, with the formula appropriate to the exact or approximate branch of $\mathsf{TC}$. Proposition 2.1 gives the operator characterization of $\mathsf{RA}$; a concrete model can also obtain this qualification through $\mathsf{RefCert}$ and Proposition 2.2 of Sec.~\ref{sec:2}.

The loss due to finite resolution can be determined from the target spectrum. Appendix~\ref{app:A.6} reduces normalized trace-distance smoothing to a spectral clipping problem, and Appendix~\ref{app:A.8} uses this reduction to obtain the regularity estimate in Eq.~\eqref{eq:3.3}. Appendix~\ref{app:A} gives the full treatment of general envelopes, support, and endpoints.

We now start from an arbitrary admissible path and derive the final-state constraint it must satisfy.

\subsection{Proof Mechanism}
\label{sec:3.2}

The proof converts the control record of an actual generation path into an information constraint on the final state. Kinematic reference consistency $\mathsf{RCon}$ ensures that the reference state, effective output, and atomic accounting are specified by the same model data fixed in advance. The path cost and final-state gap being compared therefore belong to the same physical generation problem.

Let $\mathcal C$ be an arbitrary admissible circuit in the fixed model, with cost $L=L(\mathcal C)$ measured in atomic resource slots and effective output $\omega_{\mathcal C}$ satisfying
\begin{equation*}
T(\omega_{\mathcal C},\rho)\le\epsilon .
\end{equation*}

Faithful transcription $\mathsf{TC}$ assigns to this physical path a prefix-free program $p_{\mathcal C}$ that records the admissible control history and reproduces its effective output exactly, with code length at most $\Phi_{g_R,a}(L)+\gamma$. To extract from such path descriptions the minimum code length required by the target state, define the machine-relative quantum-state generation complexity
\begin{equation*}
K_{\epsilon}^{U_R}(\rho\mid\mathfrak M_R)
:=
\min\left\{
|p|:
p\in\mathcal P,\ 
T(\rho_p,\rho)\le\epsilon
\right\}.
\end{equation*}

We set $K_{\epsilon}^{U_R}=+\infty$ if no program meets the error requirement. When a feasible program exists, the nonnegative integer code lengths ensure that the minimum is attained. This is prefix Kolmogorov complexity conditioned on the fixed physical model and interpreter, with an allowed output tolerance; it retains the meaning of a shortest self-delimiting description. Faithful transcription supplies a candidate program for every actual path, whose length is at least the corresponding $K_{\epsilon}^{U_R}$.

To turn code length into an operator constraint on the output state, we use the program semidensity operator from Sec.~\ref{sec:2} and omit the size index:
\begin{equation*}
M_U
:=
\sum_{p\in\mathcal P}2^{-|p|}\rho_p.
\end{equation*}

The sum runs over the full effective program domain of the fixed interpreter. Each program contributes its output state $\rho_p$ to $M_U$ with Kraft weight $2^{-|p|}$, so shorter descriptions contribute greater operator weight in their output directions. Below, we abbreviate $K_{\epsilon}^{U_R}(\rho\mid\mathfrak M_R)$ as $K_{\epsilon}^{U_R}(\rho)$. Path transcription, the shortest program, the program semidensity operator, and the reference state are related by
\begin{equation}
\label{eq:3.5}
\setlength{\jot}{0pt}
\begin{aligned}
\Phi_{g_R,a}(L)+\gamma
&\ge
|p_{\mathcal C}|
\ge
K_{\epsilon}^{U_R}(\rho),
\\
K_{\epsilon}^{U_R}(\rho)
&\ge
D_{\max}^{\epsilon}(\rho\Vert M_U)
\\[-1pt]
&\ge
D_{\max}^{\epsilon}(\rho\Vert\sigma_R)-\chi_U .
\end{aligned}
\end{equation}

The first step of Eq.~\eqref{eq:3.5} is the code-length bound from $\mathsf{TC}$, and the second follows from the definition of the shortest program. The third uses the positive term $2^{-|p|}\rho_p\preceq M_U$ contributed by each program to convert description length into an operator constraint in the output direction. The prefix-free property ensures, through Kraft's inequality, that $M_U$ has finite total weight. This step uses a one-way relation between program length and algorithmic probability.

The final step invokes reference admissibility $\mathsf{RA}$. Proposition 2.1 characterizes it as $M_U\preceq2^{\chi_U}\sigma_R$, calibrating the program prior to the physical structured vacuum. Kraft's inequality controls total weight, while this domination relation constrains how that weight is distributed across directions in Hilbert space. Since $\chi_U$ is uniformly bounded across the model family, the path's actual description must account for any final-state gap beyond the fixed calibration allowance.

Combining the fixed overheads as $\beta_U=\gamma+\chi_U$ and using the monotonicity of $\Phi_{g_R,a}$ shows that every admissible successful path satisfies
\begin{equation*}
\begin{aligned}
\Phi_{g_R,a}(L)
&\ge
D_{\max}^{\epsilon}(\rho\Vert\sigma_R)-\beta_U
\\
\Longrightarrow\qquad
L
&\ge
F_{g_R,a}\!\left[
D_{\max}^{\epsilon}(\rho\Vert\sigma_R)-\beta_U
\right].
\end{aligned}
\end{equation*}

The right-hand side contains only the target state and fixed model data and is independent of the path. Taking the infimum of the costs over all admissible successful paths gives Eq.~\eqref{eq:3.1}.

Approximate transcription preserves this proof mechanism, adding compilation error to the total smoothing radius and the precision description to the program code length. For every certified compilation precision $\delta_{\rm cmp}\in\Delta_{\rm TC}$ satisfying $\epsilon_{\rm circ}+\delta_{\rm cmp}\le1$, the same admissible path satisfies
\begin{equation}
\label{eq:3.6}
\begin{aligned}
\Phi_{g_R,a}(L)
&\ge
D_{\max}^{\epsilon_{\rm circ}+\delta_{\rm cmp}}
(\rho\Vert\sigma_R)
-\beta_U
\\
&\quad
-b\log(1/\delta_{\rm cmp}) .
\end{aligned}
\end{equation}

Compilation error enters the total smoothing radius of the program output through the triangle inequality, while the precision description enters the code length. The same physical path satisfies Eq.~\eqref{eq:3.6} for every certified compilation precision. We may therefore first take the supremum of the inverted bounds over the certified precision domain, and then take the infimum over all path costs, yielding Eq.~\eqref{eq:3.4}.

The detailed derivations of the program bridge, path transcription, and Eq.~\eqref{eq:3.5} are given in Appendices~\ref{app:A.2}--\ref{app:A.5}. Appendix~\ref{app:A.7} treats general monotone inversion and the infimum over paths.

The same proof also allows final-state effects to provide conservative lower estimates of the reference gap. \mbox{Applying} the same cost inversion then yields testable complexity certificates.

\subsection{Final-State Effects and Hypothesis-Testing Certificates}
\label{sec:3.3}

To turn the main lower bound into a testable physical statement, we compare the response of the same final-state effect on the target state and on the structured vacuum. A single effect fixed in advance already gives a certificate; optimizing over effects leads to the standard hypothesis-testing relative entropy.

Take any predeclared final-state effect $0\le Q\le I$. It may represent a discrimination event, occupation of a projected subspace, or an acceptance outcome in an experimental protocol: a binary question posed to the final state and fixed in advance. If the output $\omega_{\mathcal C}$ of an admissible circuit $\mathcal C$ satisfies $T(\omega_{\mathcal C},\rho)\le r$, the variational characterization of trace distance and program operator domination under exact transcription give, respectively,
\begin{equation*}
\begin{aligned}
{\rm Tr}(Q\omega_{\mathcal C})
&\ge
{\rm Tr}(Q\rho)-r,
\\
{\rm Tr}(Q\omega_{\mathcal C})
&\le
2^m{\rm Tr}(Q\sigma_R),
\\
m
&=
|p_{\mathcal C}|+\chi_U .
\end{aligned}
\end{equation*}

The first inequality guarantees an acceptance margin robust to generation error. The second bounds the same acceptance probability in terms of the baseline response of the structured vacuum and the actual program length. Together, they turn the effect's target acceptance margin into a lower bound on program length, and hence, through faithful transcription and inversion, into a complexity certificate.

\IIIstatement{Theorem 3.2 (Complexity certificate from an arbitrary final-state effect).}
Under the exact-transcription conditions of Theorem 3.1, let $0\le r<1$ and take any predeclared effect $0\le Q\le I$ satisfying ${\rm Tr}(Q\rho)>r$. Then
\begin{equation}
\label{eq:3.7}
\boxed{
C_{\rm opt}^{(r)}(\rho;\mathfrak M_R)
\ge
F_{g_R,a}\!\left[
\log
\frac{{\rm Tr}(Q\rho)-r}
{{\rm Tr}(Q\sigma_R)}
-\beta_U
\right].
}
\end{equation}

Physically, the higher the reliable acceptance rate of $Q$ on the target and the lower its baseline response on the structured vacuum, the stronger the target-specific structure it certifies. Equation~\eqref{eq:3.7} converts this final-state distinguishability directly into a minimum complexity threshold shared by all admissible generation histories. Every feasible effect gives a conservative certificate; the choice of effect determines its strength.

Low-rank projections give an intuitive special case. If $Q=P\le\Pi_R$ is a rank-$k$ projection, then ${\rm Tr}(P\sigma_R)=k/d_R$, and the input to the inverse becomes
\begin{equation*}
\log
\frac{d_R[{\rm Tr}(P\rho)-r]}{k}
-\beta_U .
\end{equation*}
Thus, reliable occupation of a low-rank structured subspace by the target state can itself certify a corresponding minimum generation cost. Section~\ref{sec:4} develops one-sided confidence bounds on the acceptance rate and reference response from finite samples, turning this state-level result into a verifiable statistical certificate.

Optimizing over all effects at a specified target acceptance rate leads to the standard hypothesis-testing relative entropy \cite{WangRenner2012_HypothesisTesting}. For $0\le\eta<1$, define
\begin{equation*}
D_H^\eta(\rho\Vert\sigma_R)
:=
-\log
\inf_{0\le Q\le I}
\left\{
{\rm Tr}(Q\sigma_R):
{\rm Tr}(Q\rho)\ge1-\eta
\right\}.
\end{equation*}

Here $2^{-D_H^\eta}$ is the minimum acceptance probability on the structured vacuum among tests whose target acceptance rate is at least $1-\eta$. The parameter $\eta$ is the allowed probability of rejecting the target state, while $r$ is the trace-distance tolerance for generating it. The quantity $1-\eta-r$ is precisely the acceptance margin guaranteed after generation error.

Appendix~\ref{app:A.9} further proves that, under the finite-dimensional support-compatibility conditions used here, robust optimization over all effects has the exact representation
\begin{equation*}
\begin{aligned}
&
\sup_{\substack{
0\le Q\le I\\
{\rm Tr}(Q\rho)>r
}}
\log
\frac{{\rm Tr}(Q\rho)-r}
{{\rm Tr}(Q\sigma_R)}
\\[-1pt]
&\qquad
=
\sup_{0\le\eta<1-r}
\left[
D_H^\eta(\rho\Vert\sigma_R)
+\log(1-\eta-r)
\right].
\end{aligned}
\end{equation*}

This identity shows that the full optimization of arbitrary-effect certificates is characterized exactly by the hypothesis-testing relative entropy together with the robust acceptance margin. For the unbiased reference and normalized trace-distance smoothing used here, the positive part of this optimized value is exactly $D_{\max}^{r}(\rho\Vert\sigma_R)$. Applying the same cost inversion therefore recovers the complexity lower bound in Eq.~\eqref{eq:3.1}; see Appendix~\ref{app:A.9} for the proof.

For any fixed testing tolerance satisfying $\eta+r<1$, Theorem 3.2 immediately gives the following canonical form.

\IIIstatement{Corollary 3.2 (Hypothesis-testing complexity certificate).}
Under the model conditions of Theorem 3.2, if $\eta+r<1$, then
\begin{equation}
\label{eq:3.8}
\begin{aligned}
C_{\rm opt}^{(r)}(\rho;\mathfrak M_R)
&\ge
F_{g_R,a}\!\Bigl[
D_H^\eta(\rho\Vert\sigma_R)
-\beta_U
\\[-1pt]
&\qquad
-\log\frac{1}{1-\eta-r}
\Bigr].
\end{aligned}
\end{equation}

Equation~\eqref{eq:3.8} formulates effect selection as a standard final-state discrimination task and retains the one-shot loss due jointly to generation and testing tolerances. An actual protocol may use a feasible effect fixed in advance or, when the available measurements permit, approach the hypothesis-testing optimum. Both constrain the same $C_{\rm opt}^{(r)}$.

Under approximate transcription, compilation error both reduces the final-state acceptance margin and adds to the precision code length. Fix any $0\le\eta<1-\epsilon_{\rm circ}$. If~the certified feasible set specified below is nonempty, the certificate optimized over hypothesis tests is
\begin{equation}
\label{eq:3.9}
\begin{aligned}
&
C_{\rm opt}^{(\epsilon_{\rm circ})}(\rho;\mathfrak M_R)
\ge
\sup_{\substack{
\delta_{\rm cmp}\in\Delta_{\rm TC}\\
\eta+\epsilon_{\rm circ}+\delta_{\rm cmp}<1
}}
\\[1pt]
&\quad
F_{g_R,a}\!\left[
\begin{aligned}
&
D_H^\eta(\rho\Vert\sigma_R)
-\beta_U
\\[-1pt]
&
-b\log(1/\delta_{\rm cmp})
\\[-1pt]
&
-\log\frac{1}{
1-\eta-\epsilon_{\rm circ}-\delta_{\rm cmp}
}
\end{aligned}
\right].
\end{aligned}
\end{equation}

The supremum is taken over the nonempty feasible set of precisions actually certified for the model and satisfying the acceptance-margin condition. The left-hand side remains the complexity at circuit tolerance $\epsilon_{\rm circ}$. Compilation error enters the program radius and description overhead, while testing error controls the final-state acceptance condition. Effect selection, testing tolerance, generation tolerance, compilation error, and the complexity being bounded thus retain distinct operational meanings.

The proof of Eq.~\eqref{eq:3.7} and the arbitrary-effect version for approximate transcription are given in Appendix~\ref{app:A.9}.

Theorem 3.2 establishes auditable lower-bound certificates for the global extremal quantity of universal optimal quantum circuit complexity. The target and reference responses of a final-state effect yield complexity guarantees valid for every admissible generation path. These responses can be computed in an ideal model or conservatively estimated from finite experimental samples. Each nontrivial certificate simultaneously excludes all admissible generation histories below its complexity threshold.

We must also determine when this final-state lower bound approaches the true optimal quantum circuit complexity. We next use explicit constructions to examine its near-tightness and resolution limits.

\setlength{\parskip}{3.2pt plus 1pt minus 0.3pt}
\setlength{\abovedisplayskip}{12pt plus 0.5pt minus 3pt}
\setlength{\belowdisplayskip}{12pt plus 0.5pt minus 3pt}

\subsection{Near-Tightness across State Families and Resolution Limits}
\label{sec:3.4}

For which physical state families does the RCC bound in the main theorem capture the leading growth of universal optimal quantum circuit complexity? The answer determines its quantitative explanatory power in concrete tasks.

Near-tightness compares two bounds on the same complexity: RCC supplies the lower bound, and an explicit admissible circuit supplies an upper-bound witness. When they agree at the level of leading scaling within the same reference model, error convention, and cost scale, we call the bound near-tight.

Consider a family of reference-consistent models with exact transcription and uniformly bounded qualification and transcription constants, together with their target states. Write
\begin{equation}
\begin{aligned}
\label{eq:3.10}
C_n
&:=
C_{\rm opt}^{(\epsilon_n)}
(\rho_n;\mathfrak M_{R_n}),
\\
Q_n^{(\epsilon_n)}
&:=
\frac{
D_{\max}^{\epsilon_n}
(\rho_n\Vert\sigma_{R_n})
}{g_n},
\\
g_n
&:=\log\Gamma_{R_n},
\qquad
0<\epsilon_n\le\frac12 .
\end{aligned}
\end{equation}

The quantity $Q_n^{(\epsilon_n)}$ converts the final state's smooth one-shot information gap relative to the structured vacuum into a resource-slot scale by dividing by the atomic control bandwidth. Near-tightness means that it and $C_n$ differ in their leading growth by at most constant factors. Convergence of the ratio of the upper and lower bounds to one is the stronger property of asymptotic saturation.

\IIIstatement{Open Problem 3.1 (Structural classification of near-tight state families).}
For which reference-aligned structured-state families does the condition $Q_n^{(\epsilon_n)}\longrightarrow\infty$ imply
\begin{equation}
\label{eq:3.11}
C_n
=
\Theta\!\left(Q_n^{(\epsilon_n)}\right),
\end{equation}
and under what perturbations, mixing operations, and compilation conditions does this scaling relation persist? A complete classification remains open. The three proven subfamilies below show that distinct structural mechanisms can achieve such agreement in scaling. For any given state family, the general criterion in Appendix~\ref{app:G} reduces the question to constructing admissible generation paths of cost $O(Q_n^{(\epsilon_n)})$ within the same reference model and on the same atomic scale.

\IIIstatement{Proposition 3.1 (Criterion for matching scaling across a state family).}
Suppose $Q_n^{(\epsilon_n)}\to\infty$, and there exist an explicit family of admissible circuits generating $\rho_n$ to within error $\epsilon_n$ and a constant $A<\infty$ independent of $n$, such that their costs in atomic resource slots satisfy
\begin{equation*}
L_n
\le
A Q_n^{(\epsilon_n)}
+o\!\left(Q_n^{(\epsilon_n)}\right).
\end{equation*}
Then
\begin{equation}
\setlength{\abovedisplayskip}{9pt plus 0.5pt minus 2pt}
\setlength{\belowdisplayskip}{9pt plus 0.5pt minus 2pt}
\setlength{\abovedisplayshortskip}{6pt plus 0.5pt minus 1pt}
\setlength{\belowdisplayshortskip}{9pt plus 0.5pt minus 2pt}
\label{eq:3.12}
\begin{aligned}
Q_n^{(\epsilon_n)}
-o\!\left(Q_n^{(\epsilon_n)}\right)
&\le
C_n
\\
&\le
A Q_n^{(\epsilon_n)}
+o\!\left(Q_n^{(\epsilon_n)}\right).
\end{aligned}
\end{equation}
Consequently, $C_n=\Theta(Q_n^{(\epsilon_n)})$. The left-hand side of Eq.~\eqref{eq:3.12} follows from canonical inversion of the main theorem: fixed transcription and reference calibration contribute only a subleading loss of $O(1+\log Q_n^{(\epsilon_n)})$. The right-hand side follows from the feasibility of the explicit circuits. When $A=1$ and $L_n=(1+o(1))Q_n^{(\epsilon_n)}$, the two bounds further establish $C_n=(1+o(1))Q_n^{(\epsilon_n)}$, so the construction asymptotically saturates the main RCC lower bound.

Approximate transcription preserves the structure of this criterion, provided the total loss in scale from the radius shift and precision code length is also $o(Q_n^{(\epsilon_n)})$. If $g_n$ grows with system size, the upper-bound construction must retain the actual bandwidth factor or achieve bandwidth-saturating compilation on the atomic-slot scale.

This criterion reduces near-tightness to a concrete physical question: can a uniform generation mechanism realize the reference structure revealed by the final state at a cost of the same order? Appendix~\ref{app:G} gives explicit constructions for three subfamilies satisfying the criterion under their stated conditions on compilation, precision, stable occupation margins, and bandwidth.

\begin{enumerate}
\setlength{\topsep}{0.35\baselineskip}
\setlength{\partopsep}{0pt}
\setlength{\itemsep}{0.4\baselineskip}
\setlength{\parsep}{0pt}

\item \textbf{Classically concentrated states.}
Probability mass remains concentrated on a set of labels spanning a subspace of reduced dimension relative to the reference space, and low-rank occupation supplies the final-state one-shot gap. If the corresponding amplitudes admit uniformly efficient preparation, reference-compatible dephasing completes a generation path with matching scaling under the cost and precision conditions of Appendix~\ref{app:G.3}.

\item \textbf{Nested projector states and background mixing.}
A predeclared chain of projections contracts the support successively at uniform cost. The total rank contraction determines both the final-state information gap and the construction length. The low-cost background mixing in Appendix~\ref{app:G.4} preserves this generation mechanism, while a stable occupation margin on the final projector preserves the leading scale of the gap under smoothing.

\item \textbf{Logical pure states.}
The reference subspace consists of $k_n\ge1$ logical qubits, and the smooth one-shot information gap of a pure state is $k_n+\log(1-\epsilon_n)$. Under the uniform encoding and precision-overhead conditions of Appendix~\ref{app:G.5}, the corresponding subfamily admits reference-consistent generation paths of cost $O(k_n)$. With uniformly bounded control bandwidth, this cost has the same order as the bandwidth-normalized lower bound; growing bandwidth requires corresponding bandwidth-saturating compilation.

\end{enumerate}

These three constructions prove near-tightness at the level of leading scaling across the respective families under the stated conditions. They share a physical feature: the structural compression revealed by the final state is also the structure actually used in generation. The necessary cost certified by RCC can therefore track the leading cost of the explicit construction on the same parameter scale.

Whether these mechanisms retain near-tightness under more general perturbations, mixing operations, and uniform generation structures remains part of the classification sought in Open Problem 3.1.

\raggedbottom
\setlength{\parskip}{2pt plus 0.2pt minus 0.2pt}
\setlength{\abovedisplayskip}{10pt plus 1pt minus 3pt}
\setlength{\belowdisplayskip}{10pt plus 1pt minus 3pt}
Near-tightness also delineates the resolution limits of global RCC. For a fixed unbiased structured vacuum with $d_n\ge2$, all pure states within the support share the smooth one-shot floor $Q_n^{(\epsilon_n)}=[\log d_n+\log(1-\epsilon_n)]/g_n$. Their specific amplitudes, phases, long-range arrangements, and algorithmic incompressibility may nevertheless drive the true process complexity far above this scale.

\looseness=-1
Knill's counting bound gives a concrete example. In a unitary circuit model where each gate acts on at most a fixed number of qubits, fix any state-vector two-norm error $0<\varepsilon_{\rm K}<\sqrt{2}$. If the gate count $b_n$ satisfies $b_n\log b_n=o(2^n)$, the Haar measure of the set of $n$-qubit pure states approximable by such circuits tends to zero~\cite{Knill1995_ApproximationQuantumCircuits}.

The nonnegative gap between universal optimal quantum circuit complexity and the corresponding RCC lower bound accommodates residual process complexity not certified by the current final-state object domain. This may arise from path-specific organization, microscopic random detail, addressing and architectural constraints, or the conservatism of the certificate itself.

This separation characterizes the level of detail resolved by the certificate. Global RCC certifies the common necessary cost of reaching the target's structural class from the structured vacuum. For tasks that must distinguish internal organization beyond the common pure-state floor, Sec.~\ref{sec:5} introduces predeclared reference-compatible windows. In structured-state families satisfying Eq.~\eqref{eq:3.12}, residual process cost introduces no new parameter scale above $Q_n^{(\epsilon_n)}$; under asymptotic saturation, it becomes subleading.

For general target states, feasible paths can also supply two-sided information. Section~\ref{sec:6} combines the final-state certificates of Sec.~\ref{sec:4} with dynamical paths compiled admissibly into the model. When the target domain, reference model, error metric, and cost scale agree, and $0\le\epsilon_U\le\epsilon_*\le\epsilon_L\le1/2$,
\begin{equation*}
L_{\rm RCC}^{(\epsilon_L)}
\le
C_{\rm opt}^{(\epsilon_*)}
\le
C_{\gamma,R}^{(\epsilon_U)},
\end{equation*}
where the right-hand side is the cost of an actual feasible path measured in atomic resource slots. Near-tight constructions bring the two bounds together at the leading scale across the family. A general dynamical path gives a complexity interval that can narrow as final-state certificates strengthen and implementation protocols improve.

This section thus establishes a rigorous connection between final-state information and universal optimal quantum circuit complexity. Final-state certificates constrain every admissible generation path, while matching constructions determine the leading scale of optimal complexity in the corresponding structured-state families. The next section applies these theoretical certificates to finite data, incorporating readout error and joint confidence control.

\Needspace*{9\baselineskip}

\endgroup

\begingroup

\newcommand{\IVVstatement}[1]{%
  \par\addvspace{0.14\baselineskip}%
  \Needspace{4\baselineskip}%
  \noindent\hspace*{\parindent}\RCCformaltext{#1}\enspace\ignorespaces
}
\newcommand{\IVVqed}{\nobreak\hspace{0.35em}\ensuremath{\square}%
  \par\addvspace{0.14\baselineskip}}

\begingroup
\frenchspacing
\renewcommand{\IVVstatement}[1]{%
  \par\addvspace{0.25\baselineskip}%
  \Needspace{3\baselineskip}%
  \noindent\hspace*{\parindent}\RCCformaltext{#1}\hspace{0.5em}\ignorespaces
}
\renewcommand{\IVVqed}{\nobreak\hspace{0.35em}\ensuremath{\square}%
  \par\addvspace{0.25\baselineskip}}
\setlength{\parskip}{1.1pt plus 0.2pt minus 0.2pt}
\setlength{\abovedisplayskip}{8pt plus 2pt minus 3pt}
\setlength{\belowdisplayskip}{8pt plus 2pt minus 3pt}
\setlength{\abovedisplayshortskip}{4pt plus 1pt minus 2pt}
\setlength{\belowdisplayshortskip}{5pt plus 1pt minus 2pt}
\setlength{\jot}{2pt}

\section{From Final-State Evidence to Operational Auditing}
\label{sec:4}

Under the unbiased-reference and smoothing conventions of Sec.~\ref{sec:3}, ideal optimization over effects recovers the main theorem's lower bound; see Corollary A.4 in Appendix~\ref{app:A.9}. In an actual audit, final-state responses are estimated from finite counts, calibration records, and accessible observations, with readout error and support leakage also taken into account. This section develops one-sided lower confidence bounds on universal optimal quantum circuit complexity from these data.

Final-state effects declared in advance yield direct certificates by comparing target and structured-vacuum responses, while hypothesis testing organizes the optimization over effects. Low-rank projections reveal concentrated occupation, and reference-compatible measurements extract information from accessible statistics.

Each audit route uses observational primitives compatible with the platform. One-sided confidence estimation and cost inversion then combine these routes, on a common confidence event, into an independently verifiable RCC lower-bound certificate.

\subsection{Audit Task and One-Sided Certificates}
\label{sec:4.1}

An RCC audit constructs a one-sided lower confidence bound on universal optimal quantum circuit complexity from finite samples. To specify the generation task being certified, fix an RCC-admissible model $\widehat{\mathfrak M}_R=(\mathfrak M_R,U_R)$, a target state $\rho$, and a generation tolerance $\epsilon$.

\IVVstatement{Definition 4.1 (RCC one-sided complexity lower-bound certificate).}
Suppose an audit protocol declared in advance produces a nonnegative value $L_{\rm RCC}$ from finite data, and let $0<\delta_{\rm stat}<1$. If, under the sampling model declared for the protocol,
\begin{equation}
\label{eq:4.1}
{\rm Pr}_{\rm data}\!\left[
C_{\rm opt}^{(\epsilon)}(\rho;\mathfrak M_R)
\ge
L_{\rm RCC}
\right]
\ge
1-\delta_{\rm stat},
\end{equation}
then $L_{\rm RCC}$ is an RCC one-sided complexity lower-bound certificate at confidence level $1-\delta_{\rm stat}$, abbreviated below as an RCC lower-bound certificate.

Equation~\eqref{eq:4.1} turns the pathwise result of Sec.~\ref{sec:3} into an exclusion statement based on finite samples: on the corresponding coverage event, all admissible generation paths shorter than $L_{\rm RCC}$ are incompatible with the certified final-state evidence. The certificate's strength depends on the final-state structure that the protocol can reliably resolve; its validity rests on the direction of the bound and its statistical coverage.

The certificate refers to a fixed physical generation task. Changes to the reference set, structured vacuum, admissible operations, program interface, resource scale, or error convention generally change the complexity being audited. A verifiable report therefore specifies the model, generation tolerance, audit route, and confidence level, together with the necessary reference treatment and calibration information.

\subsection{From Finite Data to a Complexity Lower Bound}
\label{sec:4.2}

Constructing a certificate satisfying Eq.~\eqref{eq:4.1} requires propagating finite-sample uncertainty through to the final complexity lower bound. In the exact transcription branch, the conservative conversion proceeds as follows:
\begin{equation}
\label{eq:4.2}
\begin{aligned}
&\text{Reference domain and finite data}
\longrightarrow
\mathcal F_{\rm stat},
\\
&\mathcal F_{\rm stat}
\longrightarrow
\underline{\mathcal I}_j(\epsilon)
\le
D_{\max}^{\epsilon}(\rho\Vert\sigma_R),
\\
&\underline{\mathcal I}_j(\epsilon)
\longrightarrow
L_j^{\rm final}
\le
C_{\rm opt}^{(\epsilon)}(\rho;\mathfrak M_R).
\end{aligned}
\end{equation}
The first step fixes the reference domain and the treatment of support. In the ideal case, the target state is supported on $\mathcal H_R$. When support leakage is present, using a normalized projected conditional state, a support-regularized reference state, or leakage discounting addresses different physical questions. For leakage discounting, let $q_R={\rm Tr}(\Pi_R\rho_{\rm phys})$ and, when $q_R>0$, define $\rho_R=\Pi_R\rho_{\rm phys}\Pi_R/q_R$. If a joint confidence event provides conservative endpoints $q_{R,L}$ and $C_{R,L}(\rho_R)$, then
\begin{equation}
\label{eq:4.3}
C_{{\rm leak},L}
:=
q_{R,L}\,[C_{R,L}(\rho_R)]_+
\end{equation}
gives a conservative lower bound on the average information attributable to the reference subspace, expressed in RCC units. To certify generation complexity from this bound, one must establish the one-shot lower bound in Eq.~\eqref{eq:4.2} for the state–reference pair after treatment, then apply the inversion in Eq.~\eqref{eq:4.4}. The corresponding generation model accounts for the success probability and implementation cost of conditioning. Appendix~\ref{app:B} establishes the deterministic relations for the three reference treatments; Appendix~\ref{app:D.5} handles finite-sample propagation for leakage and readout calibration.

The second step constructs a confidence set $\mathcal F_{\rm stat}$ covering the true physical parameters from finite samples. We use the fixed measurements and independent, identically distributed repeated preparations of Appendix~\ref{app:D}; other sampling models require corresponding coverage guarantees. The target acceptance probability enters through a lower endpoint, and the structured-vacuum response through an upper endpoint. Projection occupations, measurement distributions, and readout matrices enter the appropriate joint feasible sets, over which the certified input is optimized conservatively.

Witnesses, measurement settings, and error parameters are declared before the final audit samples are generated. If the data are used to select a route, simultaneous confidence guarantees or a training–validation split must cover that selection. Appendix~\ref{app:D} proves coverage for the statistical endpoints and feasible sets.

The third step converts one-sided statistics into a certified information input $\underline{\mathcal I}_j(\epsilon)$. In the exact transcription branch, Appendix~\ref{app:C} proves that a general nondecreasing transcription envelope and its canonical atomic-slot special case give, respectively,
\begin{equation}
\label{eq:4.4}
\begin{gathered}
C_{\rm opt}^{(\epsilon)}(\rho;\mathfrak M_R)
\ge
L_j^{\rm final},
\\
L_j^{\rm final}
:=
\begin{cases}
\Lambda_R^{\leftarrow}\!\left[
\underline{\mathcal I}_j(\epsilon)-\chi_U
\right],
& \begin{gathered}\text{general}\\[-2pt]\text{envelope}\end{gathered},\\[2pt]
F_{g_R,a}\!\left[
\underline{\mathcal I}_j(\epsilon)-\beta_U
\right],
& \begin{gathered}\text{canonical}\\[-2pt]\text{model}\end{gathered},
\end{cases}
\end{gathered}
\end{equation}
where $\beta_U=\gamma+\chi_U$. Equation~\eqref{eq:4.4} converts certified information into a complexity lower bound measured in atomic resource slots; the right-hand side is determined by the certified input and model constants. Appendices~\ref{app:C.1}--\ref{app:C.2} give the left generalized inverse for a general envelope, the exact expression for $F_{g_R,a}$, and conservative ways to evaluate it. Under approximate transcription, the quantity being audited remains $C_{\rm opt}^{(\epsilon_{\rm circ})}$. Appendix~\ref{app:C.5} handles the total program radius $\epsilon_{\rm circ}+\delta_{\rm cmp}$, the compilation code-length correction, and optimization over compilation precision.

\subsection{Three Types of Final-State Evidence}
\label{sec:4.3}

The three audit routes resolve different observable aspects of the same reference information gap. Final-state effects probe operational distinguishability, low-rank projections probe stable concentration along structural directions, and reference-compatible measurements probe the average information accessible within the current object domain.

\IVVstatement{Audit route 1 (Final-state effects and hypothesis testing).}
Comparing the response of the same effect on the target state and on the structured vacuum yields a complexity lower bound from their distinguishability. Take an effect $0\le Q\le I$ declared in advance, and write the target acceptance probability and structured-vacuum response as
\begin{equation*}
a={\rm Tr}(Q\rho),
\qquad
b={\rm Tr}(Q\sigma_R).
\end{equation*}
Finite samples provide a conservative lower endpoint $a_L$ for $a$ and a conservative upper endpoint $b_U$ for $b$. When $a_L>\epsilon$, the direct finite-sample input to Theorem 3.2 is
\begin{equation}
\label{eq:4.5}
\underline{\mathcal I}_{Q}(\epsilon)
:=
\log\frac{a_L-\epsilon}{b_U}
\le
D_{\max}^{\epsilon}(\rho\Vert\sigma_R).
\end{equation}

A higher guaranteed target acceptance probability and a lower structured-vacuum baseline response certify stronger structure specific to the target through Eq.~\eqref{eq:4.5}. When $b$ is computed exactly from the fixed model, it uses no statistical failure budget. When estimated from reference samples, it enters through the corresponding upper endpoint.

If the test's target acceptance probability is further certified by $a_L\ge1-\eta$, with $\eta+\epsilon<1$, then
\begin{equation}
\label{eq:4.6}
\begin{aligned}
D_H^\eta(\rho\Vert\sigma_R)
&\ge
D_{H,L}^\eta
:=
-\log b_U,
\\
\underline{\mathcal I}_{H}(\epsilon,\eta)
&:=
D_{H,L}^\eta
-\log\frac{1}{1-\eta-\epsilon}
\\
&\le
D_{\max}^{\epsilon}(\rho\Vert\sigma_R).
\end{aligned}
\end{equation}
Equation~\eqref{eq:4.6} gives a hypothesis-testing lower-bound certificate through Corollary 3.2. The fixed-effect route retains the acceptance margin actually certified, while the hypothesis-testing route casts optimization over effects in a standard information-theoretic form. Both probe the same final-state–reference contrast. Appendix~\ref{app:D.2} gives the finite-sample intervals, conditions that may remain uncertified, and rules for a route that fails certification.

\IVVstatement{Audit route 2 (Low-rank projection witnesses).}
A low-rank projection certifies stable concentration of the target state in a specified subspace, providing a structured special case of the fixed-effect route. Let $P\le\Pi_R$ be a nonzero projection declared in advance, with $k={\rm Tr}(P)$ and $p={\rm Tr}(P\rho)$, and let $p_L$ denote a finite-sample lower endpoint for the occupation probability. The structured-vacuum response in this direction is exactly $k/d_R$. For any $\epsilon_{\rm sm}\ge\epsilon$, define
\begin{equation}
\label{eq:4.7}
\begin{aligned}
\underline{\mathcal I}_{P}(\epsilon)
&:=
\Psi_{k,d_R}^{\epsilon_{\rm sm}}(p_L)
\le
D_{\max}^{\epsilon}(\rho\Vert\sigma_R),
\\
\Psi_{k,d}^{s}(x)
&=
\begin{cases}
\displaystyle
\left[\log\frac{(x-s)d}{k}\right]_+,
&x>s,\\[2mm]
0,
&x\le s.
\end{cases}
\end{aligned}
\end{equation}
The standard choice is $\epsilon_{\rm sm}=\epsilon$. Equation~\eqref{eq:4.7} yields a final certificate through Theorem 3.1 and cost inversion for the appropriate transcription branch. The one-shot input is positive when the guaranteed occupation, after subtracting the smoothing margin, remains above the structured-vacuum baseline. This route is therefore particularly suited to logical subspaces, stabilizer code spaces, approximate code spaces, and low-dimensional structural blocks specified in advance. If audit data are used to select the projection, simultaneous confidence bounds must cover the candidate family, or independent training and validation samples must be used. Appendix~\ref{app:D.3} gives the full construction.

\IVVstatement{Audit route 3 (Reference-compatible measurements).}
This route extracts the state-side structural gap directly from accessible distributions, without global state tomography. Let $\mathcal M$ be a fixed reference-compatible measurement or dephasing channel, and write
\begin{equation*}
p=\mathcal M(\rho),
\qquad
u=\mathcal M(\sigma_R).
\end{equation*}
The data-processing inequality for quantum relative entropy, together with a finite-sample confidence set, gives~\cite{Lindblad1975_CPMapsEntropy}
\begin{equation}
\label{eq:4.8}
D(\rho\Vert\sigma_R)
\ge
D_{\rm cl}(p\Vert u)
\ge
D_{{\rm cl},L}.
\end{equation}
Further coarse-graining of the same measurement output cannot increase the ideal relative entropy. With finite samples, the certified lower bound also depends on the precision of statistical estimation. When the reference output is uniform over $m$ outcomes, $D_{\rm cl}(p\Vert u)=\log m-H(p)$; physically, this is the information that remains visible relative to the unbiased reference within the current object domain.

Dividing Eq.~\eqref{eq:4.8} by $g_R=\log_2\Gamma_R$ gives the lower bound $D_{{\rm cl},L}/g_R$ on the state-side linear leading term $C_R$. As shown in Appendix~\ref{app:C.3}, we obtain a one-shot input by adding a lower bound on spectral skew to $D_{{\rm cl},L}$, measured in bits, and subtracting an upper bound on smoothing loss. Cost inversion then converts this input into a complexity lower bound.

Without independent information on spectral skew, its lower bound can conservatively be set to zero. Until the smoothing loss is controlled, $D_{{\rm cl},L}/g_R$ remains a lower bound on the state-side RCC leading term. Appendices~\ref{app:D.4}--\ref{app:D.5} treat classical relative entropy, entropy confidence bounds, and readout-error propagation.

\subsection{Joint Certificates and the Roles of Error}
\label{sec:4.4}

To combine evidence from different observations under a common confidence guarantee, let $\mathcal J_{\rm cert}$ index the audit routes that have yielded complexity lower bounds through the appropriate transcription branch. Suppose the statistical events used by these routes jointly form
\begin{equation*}
\mathcal E_{\rm stat}
=
\bigcap_{k=1}^{K}\mathcal E_k,
\qquad
{\rm Pr}(\mathcal E_{\rm stat})
\ge
1-\delta_{\rm stat}.
\end{equation*}
Equation~\eqref{eq:4.4} gives the route-level lower bounds for exact transcription; Appendix~\ref{app:C.5} gives those for approximate transcription, where $\epsilon=\epsilon_{\rm circ}$.

\IVVstatement{Proposition 4.1 (Joint finite-sample certificate from multiple audit routes).}
Suppose that, on $\mathcal E_{\rm stat}$, every $j\in\mathcal J_{\rm cert}$ completes cost inversion for the appropriate transcription branch and yields a lower bound $L_j^{\rm final}$ on the same target quantity $C_{\rm opt}^{(\epsilon)}(\rho;\mathfrak M_R)$. Then
\begin{equation}
\label{eq:4.9}
\begin{aligned}
&L_{\rm RCC}
:=
\max\!\left(
\{0\}
\cup
\{L_j^{\rm final}:j\in\mathcal J_{\rm cert}\}
\right),
\\
&{\rm Pr}_{\rm data}\!\left[
C_{\rm opt}^{(\epsilon)}(\rho;\mathfrak M_R)
\ge
L_{\rm RCC}
\right]
\ge
1-\delta_{\rm stat}.
\end{aligned}
\end{equation}
\IVVstatement{Proof.}
On $\mathcal E_{\rm stat}$, all valid route-level lower bounds hold simultaneously, so their maximum is still bounded above by the same $C_{\rm opt}^{(\epsilon)}$. Applying the coverage probability of this event gives Eq.~\eqref{eq:4.9}.\IVVqed

Equation~\eqref{eq:4.9} allows statistical dependence between audit routes. On the common coverage event, one can select the strongest certified lower bound from complementary final-state evidence. If $\mathcal J_{\rm cert}$ is empty, $L_{\rm RCC}=0$ indicates that the current protocol has yet to yield a nontrivial complexity lower bound. The raw one-shot inputs, state-side values, and conditions still to be met by each route are retained separately.

Errors enter the certificate according to their physical roles. The generation tolerance $\epsilon$, smoothing radius $\epsilon_{\rm sm}$, and testing parameter $\eta$ enter the one-shot relations. Support leakage determines how the state–reference pair is treated, while readout error enters the feasible set compatible with calibration data. The parameter $\delta_{\rm stat}$ controls joint statistical coverage alone. Approximate transcription additionally retains compilation error and the precision code-length correction.

These parameters cannot be reduced to a single additive total error. If the feasible set for readout calibration is empty, the audit reports that the data are incompatible with the declared model, rather than issuing a lower-bound certificate. Appendix~\ref{app:D.6} establishes the joint confidence budget, route statuses, and complete reporting format.

For canonical models with a uniform reference, ideal support and readout, and exact transcription, the companion software rcc-refcert \cite{LiuTianWu2026_RCCRefcert} computes cost lower bounds under the declared model assumptions, using a complete exact spectrum or counts for a projection fixed in advance. The counting route uses independent, identically distributed sampling at a fixed sample size and one-sided Hoeffding endpoints, then shares canonical cost inversion with the exact-spectrum route. The records retain the inputs, evidence provenance, and information needed for scalar replay with the same software version.

This section turns the theoretical final-state certificates of Sec.~\ref{sec:3} into operational audits that can be reported, combined, and independently verified with finite samples. Finite data determine how much target structure the current protocol can resolve, while the audited quantity $C_{\rm opt}^{(\epsilon)}$ remains fixed. Larger samples, improved readout calibration, or access to a broader object domain can increase the certificate's resolution and certify a higher complexity floor. Section~\ref{sec:5} turns to the state-side source of these certificates, examining RCC structure at fixed reference and how different accessible object domains resolve the internal organization of pure states.

\endgroup

\frenchspacing
\renewcommand{\IVVstatement}[1]{%
  \par\addvspace{0.25\baselineskip}%
  \Needspace{3\baselineskip}%
  \noindent\hspace*{\parindent}\RCCformaltext{#1}\hspace{0.5em}\ignorespaces
}
\setlength{\parskip}{1.1pt plus 0.2pt minus 0.2pt}
\setlength{\abovedisplayskip}{6pt plus 0.6pt minus 2pt}
\setlength{\belowdisplayskip}{6pt plus 0.6pt minus 2pt}
\setlength{\abovedisplayshortskip}{4pt plus 1pt minus 2pt}
\setlength{\belowdisplayshortskip}{5pt plus 1pt minus 2pt}
\setlength{\jot}{2pt}

\section{Fixed Reference Structure, Object Domains, and Windowed RCC}
\label{sec:5}

Sections~\ref{sec:3} and~\ref{sec:4} connected the target final state to optimal generation complexity and developed corresponding lower-bound certificates. This section examines how the state-side linear leading term $C_R$ resolves internal structure. Under an unbiased reference, all pure states share the same global value, leaving differences in local, correlational, and logical organization to be resolved.

To make these differences visible in the complexity value, the task must also specify which subsystems, correlations, or observable directions it can resolve. We represent this object domain, chosen in advance, by a reference-compatible window and construct windowed RCC while keeping the generation model, structural zero point, and atomic bandwidth fixed. Different windows thus resolve the global leading term into a complexity profile with physical labels.

\subsection{Unbiased Zero Point and Global Leading Term}
\label{sec:5.1}

To compare complexity values under mixing, coarse-graining, and symmetry transformations, we retain the unbiased zero point and resource scale used above. Fix~an~RCC-admissible reference-consistent quantum circuit–description model $\widehat{\mathfrak M}_R=(\mathfrak M_R,U_R),$ and fix the reference set $R$, reference support $\mathcal H_R$, dimension $d_R$, and atomic control bandwidth $g_R=\log\Gamma_R>0$. The unbiased zero point absorbs only the support constraints specified by $R$. The normalized state that favors no direction within this support is precisely the maximum-entropy state. The structured vacuum and state-side linear leading term are therefore
\begin{equation}
\label{eq:5.1}
\setlength{\jot}{1pt}
\begin{aligned}
\sigma_R
&=
\operatorname*{arg\,max}_{\omega:\,{\rm supp}(\omega)\subseteq\mathcal H_R}
S(\omega)
=
\frac{\Pi_R}{d_R},
\\[1mm]
C_R(\rho)
&=
\frac{D(\rho\Vert\sigma_R)}{\log\Gamma_R}
=
\frac{\log d_R-S(\rho)}{\log\Gamma_R}.
\end{aligned}
\end{equation}

These expressions hold for every state supported on $\mathcal H_R$. Equation~\eqref{eq:5.1} converts the target state's entropy gap relative to the unbiased background into a complexity value on the scale of atomic resource slots. The range of this value and the equality conditions are
\begin{equation}
\label{eq:5.2}
\begin{aligned}
0
&\le
C_R(\rho)
\le
\frac{\log d_R}{\log\Gamma_R},
\\
C_R(\rho)=0
&\Longleftrightarrow
\rho=\sigma_R,
\\
C_R(\rho)=\frac{\log d_R}{\log\Gamma_R}
&\Longleftrightarrow
{\rm rank}(\rho)=1.
\end{aligned}
\end{equation}
The structured vacuum is the unique zero point at fixed reference, and all pure states attain the upper boundary of the global leading term.

At fixed background, introducing classical uncertainty cannot create additional structure visible on average. If $\rho=\sum_i p_i\rho_i$, convexity of quantum relative entropy gives
\begin{equation}
\label{eq:5.3}
C_R(\rho)
\le
\sum_i p_i C_R(\rho_i).
\end{equation}
The complexity value of a mixture is no greater than the average of its components' values.

Coarse-graining and symmetry transformations at fixed background must respect the same structural zero point. Let $\mathcal E$ be a CPTP map and $U$ a unitary transformation, satisfying $\mathcal E(\sigma_R)=\sigma_R$ and $U\sigma_RU^\dagger=\sigma_R$, respectively. Data processing and unitary invariance give
\begin{equation}
\label{eq:5.4}
\setlength{\jot}{0pt}
\begin{aligned}
\mathcal E(\sigma_R)=\sigma_R
&\Longrightarrow
C_R(\mathcal E(\rho))\le C_R(\rho),
\\
U\sigma_RU^\dagger=\sigma_R
&\Longrightarrow
C_R(U\rho U^\dagger)=C_R(\rho).
\end{aligned}
\end{equation}
Coarse-graining that preserves the reference zero point lowers or preserves the global value; a unitary transformation that preserves the reference leaves it unchanged. Appendices~\ref{app:E.1}--\ref{app:E.3} give the full proofs.

\subsection{Common Pure-State Floor and Object Domains}
\label{sec:5.2}

The pure-state equality condition in Eq.~\eqref{eq:5.2} also leaves a question of resolution. Every pure state $\rho_\psi=|\psi\rangle\langle\psi|$ supported on $\mathcal H_R$ has zero entropy, so Eq.~\eqref{eq:5.1} gives
\begin{equation}
\label{eq:5.5}
\setlength{\abovedisplayskip}{8pt plus 0.5pt minus 2pt}
\setlength{\belowdisplayskip}{8pt plus 0.5pt minus 2pt}
\setlength{\abovedisplayshortskip}{6pt plus 0.5pt minus 2pt}
\setlength{\belowdisplayshortskip}{7pt plus 0.5pt minus 2pt}
C_R(\rho_\psi)
=
\frac{D(\rho_\psi\Vert\sigma_R)}{\log\Gamma_R}
=
\frac{\log d_R}{\log\Gamma_R}.
\end{equation}
This common value supplies a uniform leading term in the lower bound for pure-state generation. Stabilizer relations, phase structure, and entanglement patterns can nevertheless correspond to very different optimal generation histories.

Resolving these differences requires specifying the object domain accessible to the current task under the same background. The reference background determines where the comparison starts, admissible processes determine which generation histories enter the optimization, and the object domain determines which relations within the target state can be identified as structure. Once subsystem partitions, local correlation functions, stabilizer generators, logical operators, or low-energy modes are fixed, different pure states can exhibit different spectra, correlations, and accessible expectation values through the corresponding effective objects.

Entanglement cuts provide a direct physical example. In a study of SO(5) deconfined criticality in the $J$--$Q_3$ model, different bipartitions induce different boundary conditions. The additional gapless edge modes introduced by a nonordinary cut contribute to both the entanglement entropy and the entanglement spectrum, revealing different boundary contributions from the same bulk critical background in different object domains \cite{ZhuEtAl2026_BipartiteEntanglementSurfaceCriticality}. The cut thus helps specify the physical object actually accessed by an entanglement probe.

The window construction makes the object domain declared in advance explicit as observable structure. It thereby gives physical resolution to pure-state directions while keeping the structural zero point free of target-specific bias.

\subsection{From Object Domains to Windowed RCC}
\label{sec:5.3}

Operational indistinguishability is the starting point for incorporating object domains into the theory. For a task, experiment, or readout protocol fixed in advance, two global states that yield the same statistics for all allowed observations represent the same effective object. The corresponding structural value reflects the state differences that the protocol can distinguish.

To make this distinction a stable theoretical object, one must fix both the observations that can be invoked and the composition rules that the protocol actually supports. A list of operators alone does not ensure that the object domain is closed under protocol composition. In finite dimensions, a unital von Neumann subalgebra $\mathcal A_\Xi$, validated against the resource rules, provides a canonical representation. Algebraic closure here organizes only the products, adjoints, and compositions already supported by the protocol; it does not automatically admit composite observations whose resource costs have yet to be accounted for.

Each indistinguishability class also needs an effective state on which entropy and relative entropy can be evaluated. This representative should preserve all accessible expectation values in $\mathcal A_\Xi$, remove directions invisible to the object domain, and preserve the same structured vacuum. The trace-preserving conditional expectation $E_\Xi$ onto $\mathcal A_\Xi$ meets these requirements, giving the window defined below \cite{Umegaki1954_ConditionalExpectation}.

\IVVstatement{Definition 5.1 (Reference-compatible window).}
In a fixed reference-consistent model, a reference-compatible window declared by the task in advance and fixed by the corresponding resource rules is denoted by
\begin{equation*}
\Xi=(\mathcal A_\Xi,E_\Xi),
\end{equation*}
where $\mathcal A_\Xi\subseteq B(\mathcal H_R)$ is a finite-dimensional von Neumann subalgebra containing the identity $\Pi_R$, and $E_\Xi:B(\mathcal H_R)\to\mathcal A_\Xi$ is the trace-preserving conditional expectation onto $\mathcal A_\Xi$. For all $A,B\in\mathcal A_\Xi$ and $X\in B(\mathcal H_R)$, it satisfies
\begin{equation}
\label{eq:5.6}
\setlength{\jot}{1pt}
\setlength{\abovedisplayskip}{8pt plus 0.5pt minus 2pt}
\setlength{\belowdisplayskip}{8pt plus 0.5pt minus 2pt}
\setlength{\abovedisplayshortskip}{6pt plus 0.5pt minus 2pt}
\setlength{\belowdisplayshortskip}{7pt plus 0.5pt minus 2pt}
\begin{aligned}
E_\Xi^2&=E_\Xi,
\quad
E_\Xi\ {\rm CPTP},
\quad
{\rm Ran}(E_\Xi)=\mathcal A_\Xi,
\\
E_\Xi(AXB)&=AE_\Xi(X)B,
\qquad
E_\Xi(\sigma_R)=\sigma_R.
\end{aligned}
\end{equation}

Idempotence and the range condition make $E_\Xi$ a projection onto the current observable algebra. The bimodule condition preserves the observational structure within the algebra, and reference preservation gives different windows the same complexity zero point. Together, these properties yield a direct physical equivalence: global states that the current task cannot distinguish have exactly the same effective state in the window. For any $\rho$ and $\tau$,
\begin{equation}
\label{eq:5.7}
\setlength{\jot}{0pt}
\setlength{\abovedisplayskip}{6pt plus 0.5pt minus 2pt}
\setlength{\belowdisplayskip}{6pt plus 0.5pt minus 2pt}
\begin{aligned}
\rho\sim_\Xi\tau
&\Longleftrightarrow
{\rm Tr}(A\rho)={\rm Tr}(A\tau)
\quad\forall A\in\mathcal A_\Xi
\\
&\Longleftrightarrow
E_\Xi(\rho)=E_\Xi(\tau).
\end{aligned}
\end{equation}

Thus, $E_\Xi(\rho)$ is the canonical representative of $\rho$ in the current object domain: it preserves all accessible expectation values and identifies exactly those global states that the protocol cannot distinguish. The reference set $R$ determines which states enter the comparison, while the window algebra $\mathcal A_\Xi$ determines which differences between these states remain identifiable as physical structure. Appendix~\ref{app:E.4} proves this equivalence.

Once a canonical effective state is available, complexity values across object domains must remain comparable. A window therefore changes only the target state's effective representative for the current task; the structured vacuum $\sigma_R$ and atomic control bandwidth $\log\Gamma_R$ remain fixed. The comparison rule for global RCC then restricts naturally to the window's object domain.

\IVVstatement{Definition 5.2 (Windowed RCC).}
For a fixed reference-consistent model and a reference-compatible window $\Xi=(\mathcal A_\Xi,E_\Xi)$, the windowed RCC of a target state $\rho$ is defined as
\begin{equation}
\label{eq:5.8}
C_{R,\Xi}(\rho)
=
\frac{D(E_\Xi(\rho)\Vert\sigma_R)}{\log\Gamma_R}
=
\frac{\log d_R-S(E_\Xi(\rho))}{\log\Gamma_R}.
\end{equation}

Using the same structured vacuum and atomic control bandwidth, Eq.~\eqref{eq:5.8} combines structural fairness and directional resolution in a single value. A coarser window absorbs invisible differences into the effective state's mixing and uncertainty. As the object domain is refined, some of these differences become distinguishable structure and reappear as a reference gap relative to the same zero point. Here $E_\Xi$ specifies the effective description. If an experimental protocol actually implements this channel, its control, ancillary resources, and dissipation must still be included in the process accounting of $\mathfrak M_R$.

\subsection{Local–Correlation Decomposition and Window Profiles}
\label{sec:5.4}

A subsystem partition provides the most direct physical setting for this construction. A global pure state always has zero entropy, whereas the entropy of a reduced state varies with the partition and entanglement structure. Directional differences unresolved by the global $C_R$ can therefore reappear in local object domains. To see this, let
\begin{equation*}
\mathcal H_R
=
\mathcal H_A\otimes\mathcal H_{\bar A},
\qquad
\sigma_R
=
\frac{I_A}{d_A}
\otimes
\frac{I_{\bar A}}{d_{\bar A}},
\end{equation*}

and let the current task access all observations on subsystem $A$ alone, taking $\mathcal A_A=B(\mathcal H_A)\otimes I_{\bar A}$. The window map must preserve the reduced state on $A$ while restoring the inaccessible complementary subsystem to the unbiased background. The corresponding conditional expectation and windowed RCC are therefore
\begin{equation}
\label{eq:5.9}
\begin{aligned}
E_A(\rho)
&=
{\rm Tr}_{\bar A}(\rho)
\otimes
\frac{I_{\bar A}}{d_{\bar A}}
=
\rho_A
\otimes
\frac{I_{\bar A}}{d_{\bar A}},
\\[1mm]
C_{R,A}(\rho)
&=
\frac{\log d_A-S(\rho_A)}
{\log\Gamma_R}.
\end{aligned}
\end{equation}

The complementary subsystem's maximally mixed entropy cancels exactly in Eq.~\eqref{eq:5.9}, leaving the state-side gap that subsystem $A$ can reveal independently relative to its unbiased local background. If $\rho_{A\bar A}$ is globally pure, $S(\rho_A)$ is the bipartite entanglement entropy. As entanglement increases, the gap in each independent local window decreases, and more structure is carried by correlations in the joint object domain. This change describes the distribution of state-side structure; it does not assume the same additive decomposition for the full process complexity.

The local windows show how much structure each side can reveal independently, leaving a natural question: where is the part of the global leading term assigned to neither local object domain? Under a product-form unbiased reference, this remainder is precisely the total correlation between the two regions.

\IVVstatement{Proposition 5.1 (Local–correlation decomposition).}
Under the bipartite product-form unbiased reference above, let $C_{R,\bar A}(\rho)=[\log d_{\bar A}-S(\rho_{\bar A})]/\log\Gamma_R$. Then
\begin{equation}
\label{eq:5.10}
\begin{aligned}
C_R(\rho_{A\bar A})
&=
C_{R,A}(\rho)
+
C_{R,\bar A}(\rho)
\\
&\quad+
\frac{I(A:\bar A)_\rho}{\log\Gamma_R}.
\end{aligned}
\end{equation}

Equation~\eqref{eq:5.10} decomposes the global state-side leading term exactly into the two local-window leading terms and the reference gap carried by correlations, $I(A:\bar A)_\rho/\log\Gamma_R$. For a general mixed state, quantum mutual information measures the total correlation between the two regions; for a bipartite pure state, it reduces to $2S(\rho_A)$. Structure that neither local object domain can reveal independently thus acquires an exact correlation component in the joint object domain. More general correlation algebras can likewise be chosen to address many-body correlations, logical entanglement, and task-specific nonlocal structure. Appendix~\ref{app:E.5} gives the full derivation.

As a task gains access to more observations, the window value should retain the structure already resolved. If a finer window contains all observables of a coarser window, the trace-preserving conditional expectations satisfy the tower property because the algebras are nested. Data processing then gives a definite ordering of the values as resolution changes.

\IVVstatement{Proposition 5.2 (Monotonicity under nested windows).}
Under the same reference, suppose the observable algebras of two windows are nested, so that their trace-preserving conditional expectations satisfy the tower property. Write
\begin{equation*}
\mathcal A_{\Xi_1}
\subseteq
\mathcal A_{\Xi_2},
\qquad
E_{\Xi_1}
=
E_{\Xi_1}\circ E_{\Xi_2},
\qquad
\Xi_1\preceq\Xi_2.
\end{equation*}

\Needspace{6\baselineskip}
Then
\begin{equation}
\label{eq:5.11}
\setlength{\abovedisplayskip}{8pt plus 0.5pt minus 1pt}
\setlength{\belowdisplayskip}{8pt plus 0.5pt minus 1pt}
\setlength{\abovedisplayshortskip}{6pt plus 0.5pt minus 1pt}
\setlength{\belowdisplayshortskip}{8pt plus 0.5pt minus 1pt}
0
\le
C_{R,\Xi_1}(\rho)
\le
C_{R,\Xi_2}(\rho)
\le
C_R(\rho).
\end{equation}

The finer the object domain, the more distinguishable structure can contribute to the value; a coarser window identifies more internal directions within the same operational equivalence class. Incomparable windows resolve structure along different observational directions and generally have no common ordering. Appendix~\ref{app:E.6} proves Eq.~\eqref{eq:5.11} using the tower property of conditional expectations and the data-processing inequality, and gives complete reference coarse-graining and the identity window as the two endpoints.

This monotonicity allows values from multiple windows to form a structural profile. A single window gives a value for a particular partition or observational direction, while a family of windows declared in advance records how the same target state's visible structure changes with the object domain. Given a family $\mathcal W$ of reference-compatible windows, define
\begin{equation}
\label{eq:5.12}
\setlength{\jot}{0pt}
\begin{aligned}
{\rm Prof}_{R,\mathcal W}(\rho)
&=
\bigl(
C_{R,\Xi}(\rho)
\bigr)_{\Xi\in\mathcal W},
\\
\Xi
&\longmapsto
C_{R,\Xi}(\rho).
\end{aligned}
\end{equation}

A nested family of windows gives a monotone curve as resolution varies; a general family gives a multicomponent picture from different observational directions. The family of effective states $\bigl(E_\Xi(\rho)\bigr)_{\Xi\in\mathcal W}$ retains all distinguishable information in the respective object domains. The scalar profile extracts each window's complexity leading-term value, allowing comparisons by partition or observational direction.

A four-qubit example illustrates this resolution directly. Consider $|\Phi^+\rangle_{12}\otimes|\Phi^+\rangle_{34}$ and the Greenberger–Horne–Zeilinger (GHZ) state $|{\rm GHZ}_4\rangle$. Under the unbiased reference on the full space, these pure states have the same $C_R$. For two-qubit subsystem windows, the former gives $C_{R,A}=2/\log\Gamma_R$ for $A=\{1,2\}$ and zero for $A=\{1,3\}$; the latter gives $C_{R,A}=1/\log\Gamma_R$ for every two-qubit subsystem. The same global leading term thus resolves into distinct window profiles, displaying the different distributions of window values for the locally paired state and the GHZ state across two-body object domains.

Actual readout turns the window profile from a descriptive decomposition into auditable structure. Let $\{P_i\}\subseteq\mathcal A_\Xi$ be a complete family of orthogonal projections declared in advance, and write
\begin{equation*}
p_i
=
{\rm Tr}(P_i\rho)
=
{\rm Tr}(P_iE_\Xi(\rho)),
\qquad
u_i
=
{\rm Tr}(P_i\sigma_R).
\end{equation*}
Pinching and the data-processing inequality give
\begin{equation}
\label{eq:5.13}
C_{R,\Xi}(\rho)
\ge
\frac{D_{\rm cl}(p\Vert u)}{\log\Gamma_R}.
\end{equation}

Windows and projection readouts declared in advance can thus translate the internal organization of pure states into measurable deviations of classical distributions. Probes aligned with stabilizers, graph structure, local correlations, or low-energy modes can make the corresponding structure particularly visible in specific object domains.

Selection among multiple windows and finite-sample confidence control continue to follow the joint statistical rules of Sec.~\ref{sec:4} and Appendix~\ref{app:D}. Appendices~\ref{app:E.7}--\ref{app:E.8} establish the pinching decomposition, classical lower bounds, and properties of window profiles.

\subsection{Relational Complexity and the Process Interface}
\label{sec:5.5}

The complexity characterized by RCC is relational. The cost of generating a target state depends on the reference background and on which generation processes are admissible. Structural fairness incorporates structure already supplied by the background into the zero point and accounts for the capacity to generate additional structure in the resource scale, giving this relation a definite physical measure.

The object domain gives this relation a specific observational resolution. Under the same background and generation conventions, local, correlation, and other task-specific windows each access the corresponding organization within the target state, forming a complexity profile. Changing windows thus resolves structural differences under the same background. A reference upgrade changes the structure already supplied by the background and requires a corresponding update of the entire reference-consistent model. Section~\ref{sec:2} and Appendix~\ref{app:F} specify reference selection, upgrades, and mismatch control; Proposition 5.2 gives the monotone ordering of nested windows at fixed reference.

Window values also turn visible local and correlational structure into evidence about generation complexity. By Eq.~\eqref{eq:5.11}, windowed RCC provides a conservative lower bound on the global leading term $C_R$. Converting this bound to the information scale and incorporating a lower bound on spectral skew and an upper bound on smoothing loss for the original target state $\rho$ gives the one-shot input to the main theorem in Sec.~\ref{sec:3}. Cost inversion then yields a lower-bound certificate for $C_{\rm opt}^{(\epsilon)}(\rho;\mathfrak M_R)$. The structure resolved by a window can therefore impose a common minimum cost on all admissible paths that generate the original target state.

For pure states, $\Delta_{\rm spec}(\rho_\psi)=0$, and Corollary A.2 in Appendix~\ref{app:A} gives the smoothing loss. For general mixed states, an independent lower bound on spectral skew can be included, or set to zero when no additional spectral information is available. Appendix~\ref{app:E.9} proves this conversion, Appendix~\ref{app:C.3} gives the corresponding certified information input, and finite-sample endpoints and joint coverage follow Sec.~\ref{sec:4} and Appendix~\ref{app:D}.

Windowed RCC thus connects the resolution of internal organization to the certification of generation complexity on a common resource scale. Section~\ref{sec:6} will relate these values to actual generation paths, dynamical resources, and thermodynamic responses; changes in the reference background itself are taken up in Sec.~\ref{sec:7}.

\endgroup

\begingroup

\frenchspacing
\newcommand{\VIstatement}[1]{%
  \par\addvspace{0.25\baselineskip}%
  \Needspace{3\baselineskip}%
  \noindent\hspace*{\parindent}\RCCformaltext{#1}\hspace{0.5em}\ignorespaces
}
\setlength{\parskip}{1.1pt plus 0.2pt minus 0.2pt}
\setlength{\abovedisplayskip}{6pt plus 0.6pt minus 2pt}
\setlength{\belowdisplayskip}{6pt plus 0.6pt minus 2pt}
\setlength{\abovedisplayshortskip}{4pt plus 1pt minus 2pt}
\setlength{\belowdisplayshortskip}{5pt plus 1pt minus 2pt}
\setlength{\jot}{2pt}

\section{Dynamical and Thermodynamic Interfaces of RCC}
\label{sec:6}

The preceding sections established complexity lower bounds from final-state evidence and resolved the state-side linear leading term into window values. This section examines how these results constrain physical implementations. Compilable paths and final-state certificates jointly bound optimal complexity; the energy–entanglement performance relation turns complexity requirements into lower bounds on dynamical resources; and the object domains opened by a finite budget bring windowed complexity into entropy and thermodynamic response relations.

\subsection{Dynamical Paths and Two-Sided Complexity Intervals}
\label{sec:6.1}

Section~\ref{sec:3} used matching constructions to establish near-tightness at the level of leading scaling for three state families under their respective assumptions. For a general task without such a construction, an actually executed or explicitly compilable generation path can help determine how far optimal complexity lies above the RCC lower endpoint. If the path realizes the target within the same model and error semantics, its cost gives an upper bound within that model. Establishing this bound does not require a prior proof that the path is optimal.

Let $\mathfrak r_\gamma$ be a record of a dynamical implementation, including continuous controls or a gate sequence, initial resources, locality and addressing, ancillary systems, measurement branches, precision requirements, and success conditions. If a model-consistent compiler converts this record into a finite admissible circuit in $\mathfrak M_R$ that generates the target state $\rho$ within tolerance $\epsilon_U$, denote the compiled circuit's cost in atomic resource slots by $C_{\gamma,R}^{(\epsilon_U)}(\rho)$.

This definition brings Hamiltonian simulation, gate synthesis, control precision, ancillary resources, branch handling, and target-specific data accessed at runtime into the single cost function fixed in Sec.~\ref{sec:2}. If compilation fails, or if the compiled object fails to satisfy path–model compatibility, target accuracy, and finite-cost requirements jointly, the upper endpoint is set to $+\infty$.

\VIstatement{Proposition 6.1 (Model-consistent two-sided complexity interval).}
If the lower and upper endpoints use the same target domain, reference model, normalized unconditional output, error metric, and cost function in atomic resource slots, and
\begin{equation*}
0
\le
\epsilon_U
\le
\epsilon_*
\le
\epsilon_L
\le
\frac12,
\end{equation*}
then, on the event where both the lower-bound certificate and the upper-bound certification hold,\nopagebreak[4]
\begin{equation}
\label{eq:6.1}
\setlength{\abovedisplayskip}{10pt plus 0.5pt minus 2pt}
\setlength{\belowdisplayskip}{10pt plus 0.5pt minus 2pt}
\setlength{\abovedisplayshortskip}{8pt plus 0.5pt minus 2pt}
\setlength{\belowdisplayshortskip}{10pt plus 0.5pt minus 2pt}
L_{\rm RCC}^{(\epsilon_L)}
(\rho;\widehat{\mathfrak M}_R)
\le
C_{\rm opt}^{(\epsilon_*)}(\rho;\mathfrak M_R)
\le
C_{\gamma,R}^{(\epsilon_U)}(\rho).
\end{equation}

The left inequality follows from the main theorem in Sec.~\ref{sec:3} and certificate synthesis in Sec.~\ref{sec:4}; the right follows because the compiled circuit belongs to the feasible set defining optimal complexity. Monotonicity of complexity in the allowed error connects both endpoints to the intermediate tolerance.

Equation~\eqref{eq:6.1} brackets the unknown global optimum between auditable operational endpoints: stronger final-state evidence raises the lower endpoint, while better implementations and their compilations lower the upper endpoint. The endpoint difference bounds the compiled implementation's excess cost above the optimum at the intermediate tolerance; when the lower endpoint is positive, their ratio bounds the corresponding multiplicative overhead. Appendix~\ref{app:I.1} treats different target states, compilation across models, postselection, random stopping times, and finite calibration errors.

When both endpoints are obtained from finite data, $L_{\rm RCC}^{(\epsilon_L)}$ and the dynamical upper envelope must be reported on the same joint confidence event. If the actual compiled circuit, its complete output, target error, and length have all been certified deterministically, its discrete cost is itself a deterministic upper endpoint. An empty interval serves as a model diagnostic: the corresponding confidence events have failed to cover the true objects jointly, or the target, reference, error, support treatment, compilation rules, and counting units have not yet been aligned.

Admissible compilation supplies the upper endpoint in discrete cost. When an implementation originates in continuous control, one must also determine whether the Hamiltonian, state trajectory, and final-state overlap describe a single consistent physical path. Exact quantum speed limits (QSLs) provide geometric quantities for this check.

Consider an absolutely continuous finite-dimensional pure-state unitary path $\gamma:t\mapsto|\psi(t)\rangle$ on $[0,t_{\rm f}]$, and fix a time-independent orthonormal basis $\mathcal B$. Hall's exact uncertainty relation and the exact QSLs developed by Pati et al. resolve energy fluctuations into contributions from probability redistribution within the fixed basis and relative-phase motion \cite{Hall2001_ExactUncertaintyRelations,PatiEtAl2025_ExactQuantumSpeedLimits}. The corresponding Wootters probability-path length and Fubini–Study projective length satisfy \cite{Wootters1981_StatisticalDistance,AnandanAharonov1990_GeometryQuantumEvolution}
\begin{equation}
\label{eq:6.2}
\begin{aligned}
\ell_{W,\mathcal B}[\gamma]
&=
\frac{1}{\hbar}
\int_0^{t_{\rm f}}
\Delta H_{{\rm nc},\mathcal B}(t)\,dt
\\
&\le
\ell_{\rm FS}[\gamma]
=
\frac{1}{\hbar}
\int_0^{t_{\rm f}}
\Delta H(t)\,dt.
\end{aligned}
\end{equation}

\begingroup
\widowpenalty=10000
Here $\Delta H_{{\rm nc},\mathcal B}$ is the nonclassical energy-fluctuation component in Hall's decomposition that drives changes in the occupation probabilities of the fixed basis. The length $\ell_{W,\mathcal B}$ captures probability motion, while $\ell_{\rm FS}$ also includes changes in relative phase. If the probability trajectory, the two energy-fluctuation quantities, and the final-state overlap jointly satisfy Eq.~\eqref{eq:6.2} and its endpoint conditions, they support the geometric consistency of the same continuous record. Appendix~\ref{app:I.1} gives the exact identities and equality conditions. With its discrete cost supplied by admissible compilation, the path can be paired with final-state evidence to bound its excess cost above the optimum.
\par\endgroup

\Needspace{6\baselineskip}
\subsection{Dynamical Resource Consequences of the Complexity Lower Bound}
\label{sec:6.2}

A complexity lower bound also imposes dynamical requirements on generation. Under local interactions, state evolution depends on energy fluctuations, while the formation of entanglement across a partition is constrained by propagation speed. To turn the RCC lower endpoint into a joint constraint on energy fluctuations and entanglement output, these two restrictions must be connected to task complexity.

Our previously proposed Energy–Entanglement Performance Equation (RECT-$\eta$) relates energy fluctuations, local interactions, entanglement output, and task complexity within the same process window \cite{LiuTianWu2025_CostNonlocality}. We use its original dynamical model: a finite-dimensional closed quantum system evolves unitarily from a pure state under a bounded local Hamiltonian compatible with the Lieb–Robinson-type entanglement-growth bound being used. The initial state is a product state across the target bipartition, and $S_E$ is the entanglement quantity across that partition controlled by the growth bound. The link between task complexity and evolution time uses the RECT-$\eta$ postulate of orthogonal advancement.

In this model, the speed-limit and entanglement-growth bounds constrain the same evolution time. The postulate of orthogonal advancement decomposes task complexity into effective steps that are distinguishable in sequence, and the Mandelstam–Tamm QSL limits how quickly those steps can be completed at a given energy fluctuation \cite{MandelstamTamm1945_EnergyTime}. A Lieb–Robinson-type entanglement-growth bound limits how quickly local interactions can spread entanglement across the target partition \cite{LiebRobinson1972_FiniteGroupVelocity,BravyiHastingsVerstraete2006_LRGeneration}. Two efficiency coefficients record how closely the actual process approaches these limits. Eliminating the common evolution time gives the exact performance identity within this domain:
\begin{equation*}
\begin{aligned}
\sigma_{\rm avail}S_E
&=
\kappa_{\rm RECT}
C_{\rm opt}^{\rm RECT}(\mathcal T),
\\[2pt]
\kappa_{\rm RECT}
&=
\frac{\eta_{\rm LR}}{\eta_{\rm QSL}}
\frac{\pi\gamma_{\rm LR}J}{2},
\\[2pt]
0<\eta_{\rm QSL}
&\le 1,
\qquad
0\le\eta_{\rm LR}\le 1.
\end{aligned}
\end{equation*}

\begingroup
\widowpenalty=10000
Here $\Delta t$ is the duration of the process window, and $\sigma_{\rm avail}:=(\Delta t)^{-1}\int_0^{\Delta t}\Delta H(t)\,dt$ is the time average of the energy standard deviation over that window. The quantity $S_E$ is the entanglement output across the target partition, $J$ is the local interaction scale, and $\gamma_{\rm LR}$ is the entanglement-growth constant denoted by $\gamma$ in the original RECT paper. The quantity $C_{\rm opt}^{\rm RECT}(\mathcal T)$ is the minimum logical depth required to implement the target unitary task $\mathcal T$; the target unitary, implementation tolerance, and error convention are all part of the task's defining data.
\par\endgroup

\begingroup
\setlength{\parskip}{1.1pt plus 0.2pt minus 1.1pt}
Applying this unitary task to a fixed initial state gives the corresponding target-state generation task. Suppose the initial state, target output, gate set, and error semantics are compatible, and every RECT implementation corresponds, under the chosen serial and parallel counting rules, to an admissible RCC path whose cost in atomic resource slots equals its logical depth. The two feasible sets then give
\begin{equation*}
\setlength{\abovedisplayskip}{8pt plus 0.3pt minus 3pt}
\setlength{\belowdisplayskip}{8pt plus 0.3pt minus 3pt}
\setlength{\abovedisplayshortskip}{6pt plus 0.3pt minus 3pt}
\setlength{\belowdisplayshortskip}{7pt plus 0.3pt minus 3pt}
C_{\rm opt}^{\rm RECT}(\mathcal T)
\ge
C_{\rm opt}^{(\epsilon)}(\rho;\mathfrak M_R)
\ge
L_{\rm RCC}^{(\epsilon)}
(\rho;\widehat{\mathfrak M}_R).
\end{equation*}

Since $\eta_{\rm LR}$ can vanish, a resource-feasibility criterion additionally requires a strictly positive lower bound that holds uniformly over the current task domain:
\begin{equation*}
\setlength{\abovedisplayskip}{8pt plus 0.3pt minus 3pt}
\setlength{\belowdisplayskip}{8pt plus 0.3pt minus 3pt}
\setlength{\abovedisplayshortskip}{6pt plus 0.3pt minus 3pt}
\setlength{\belowdisplayshortskip}{7pt plus 0.3pt minus 3pt}
\kappa_{\rm RECT}
\ge
\underline{\kappa}_{\rm RECT}>0.
\end{equation*}
The quantities $S_E$, $C_{\rm opt}^{\rm RECT}$, $\gamma_{\rm LR}$, and both efficiencies are dimensionless; $\sigma_{\rm avail}$, $J$, $\kappa_{\rm RECT}$, and $\underline{\kappa}_{\rm RECT}$ all have units of energy.

This positive lower bound must be calibrated independently from external information declared in advance. It must not be calculated from the same unknown $C_{\rm opt}^{\rm RECT}(\mathcal T)$, the product $\sigma_{\rm avail}S_E$ for the process being certified, or an algebraically equivalent quantity.

\VIstatement{Corollary 6.1 (RCC-induced lower bound on the product of energy fluctuations and entanglement output).}
In the RECT-$\eta$ dynamical model above, suppose the RECT task implementation and the RCC target-state generation task are aligned as specified, and independent calibration provides $\underline{\kappa}_{\rm RECT}>0$. Then
\begin{equation}
\label{eq:6.3}
\setlength{\abovedisplayskip}{8pt plus 0.5pt minus 1pt}
\setlength{\belowdisplayskip}{8pt plus 0.5pt minus 1pt}
\setlength{\abovedisplayshortskip}{6pt plus 0.5pt minus 1pt}
\setlength{\belowdisplayshortskip}{8pt plus 0.5pt minus 1pt}
\sigma_{\rm avail}S_E
\ge
\underline{\kappa}_{\rm RECT}
L_{\rm RCC}^{(\epsilon)}.
\end{equation}

Equation~\eqref{eq:6.3} gives a resource–output threshold for the target task. For a fixed target entanglement output $S_E>0$, the required energy-fluctuation scale is bounded below by $\underline{\kappa}_{\rm RECT}L_{\rm RCC}^{(\epsilon)}/S_E$; implementations below this threshold are incompatible with the target task. Different implementations within the same reference model can also be compared by their margins above the common lower bound.

If $\eta_{\rm QSL}$ and $\eta_{\rm LR}$ can be calibrated separately, the full RECT-$\eta$ identity can further distinguish whether the principal performance bottleneck lies in state advancement or local entanglement propagation. Comparing protocols in these two efficiency coordinates then characterizes the engineering performance frontier within the same task domain.

If both the performance coefficient and the RCC certificate are obtained from finite data, they must be combined on a joint confidence event. The dynamical derivation of RECT-$\eta$ is given in the paper that introduced it \cite{LiuTianWu2025_CostNonlocality}. The comparison chain between complexity objects is established here; Appendix~\ref{app:I.1} gives the general treatment of resource compilation across models and joint confidence with finite data. Within this domain, final-state evidence therefore rules out implementations with an insufficient product of energy fluctuations and entanglement output.

\par\endgroup
\subsection{Thermodynamic Consequences of Windowed RCC}
\label{sec:6.3}

Finite generation capacity also affects thermodynamic accessibility. Under given macroscopic constraints, which states can be generated and which structures can be controlled and resolved still depend on the physical resources available. Expressing this limitation through a complexity budget brings generation cost into the definition of the thermodynamic object domain.

Our previously proposed Complexity Window Thermodynamics (CWT) uses finite generation capacity to delimit accessibility and builds a statistical description within the resulting object domain. With the reference state, elementary operations, and precision convention fixed, the conditions $C^*(\rho_{\rm tar})\le\xi$ or $C^*(U)\le\xi$ specify the targets that can be generated or implemented within budget. The task protocol then determines the effective object domain that can be controlled and resolved \cite{LiuTianWu2025_CWT}. Window entropy and the complexity-generation potential are defined on this effective state space.

The optimal quantum circuit complexity $C^*$ required by CWT can be taken to be $C_{\rm opt}^{(\epsilon)}(\rho;\mathfrak M_R)$ in an RCC reference-consistent model. With the same target, reference zero point, admissible operations, error semantics, and atomic-resource-slot scale, the event on which the RCC certificate holds also gives
\begin{equation*}
L_{\rm RCC}^{(\epsilon)}(\rho)>\xi
\quad\Longrightarrow\quad
C_{\rm opt}^{(\epsilon)}(\rho;\mathfrak M_R)>\xi.
\end{equation*}
The final-state certificate thus directly rules out generating the target state within the current budget. Which additional observables the budget makes accessible is determined by the chosen control and measurement protocol.

Windowed RCC from Sec.~\ref{sec:5} gives a complexity leading-term value for the same object domain. To compare it quantitatively with CWT window entropy, take the shared window constructed in Appendix~\ref{app:I}, defined by a trace-preserving conditional expectation in finite dimensions:
\begin{equation*}
\Xi_\xi=(\mathcal A_\xi,E_\xi),
\qquad
\tau_\xi=E_\xi(\rho).
\end{equation*}

Here $\mathcal A_\xi$ is the effective observable algebra supported by budget $\xi$, and $E_\xi$ gives the canonical representative in that object domain. Within this shared window family, increasing the budget refines the object domain. Tower compatibility of nested windows ensures that the effective state in a coarse window can be recovered by further coarse-graining the effective state in a finer window.

The budget $\xi$ limits available resources, the window $\Xi_\xi$ represents the object domain supported by the protocol, and $C_{R,\Xi_\xi}(\rho)$ measures the target state's complexity leading term revealed within it. The first two specify what can be resolved; the third gives the corresponding state-side value. A common cost function or explicit resource calibration places budget and value on the same scale.

To compare window entropy with complexity values, we now express information in natural logarithmic units. On the shared support, define $D_{\rm nat}(\tau\Vert\sigma):={\rm Tr}[\tau(\ln\tau-\ln\sigma)]$, and write
\begin{equation*}
s_\xi
=
-{\rm Tr}(\tau_\xi\ln\tau_\xi),
\qquad
\mathscr S_\xi
=
k_Bs_\xi,
\qquad
r_R
=
\ln d_R.
\end{equation*}
The atomic control bandwidth introduced in Sec.~\ref{sec:2} is $g_R=\log_2\Gamma_R$; in natural information units it is $\ln\Gamma_R=(\ln2)g_R$. Thus
\begin{equation*}
C_{R,\Xi_\xi}(\rho)
=
\frac{
D_{\rm nat}(\tau_\xi\Vert\sigma_R)
}{
\ln\Gamma_R
}
=
\frac{
r_R-s_\xi
}{
\ln\Gamma_R
}.
\end{equation*}

\VIstatement{Proposition 6.2 (Window capacity relation at fixed reference).}
For a fixed finite-dimensional unbiased reference $\sigma_R=\Pi_R/d_R$ and a reference-compatible shared window,
\begin{equation}
\label{eq:6.4}
\setlength{\abovedisplayskip}{9pt plus 0.5pt minus 1pt}
\setlength{\belowdisplayskip}{8pt plus 0.5pt minus 1pt}
\setlength{\abovedisplayshortskip}{7pt plus 0.5pt minus 1pt}
\setlength{\belowdisplayshortskip}{8pt plus 0.5pt minus 1pt}
\mathscr S_\xi
+
k_B\ln\Gamma_R\,
C_{R,\Xi_\xi}(\rho)
=
k_Br_R
=
k_B\ln d_R.
\end{equation}

Equation~\eqref{eq:6.4} follows exactly from the unbiased-reference identity $D_{\rm nat}(\tau_\xi\Vert\sigma_R)=r_R-s_\xi$. At fixed reference, a decrease in window entropy is exactly matched by an increase in the complexity leading term, with $k_B\ln\Gamma_R$ converting the state-side complexity value to thermodynamic entropy units. The fixed reference capacity is thus shared between unresolved structure, counted as window entropy, and revealed structure, counted as windowed RCC.

For a fixed input state and finite-dimensional reference, exact nested windows form a finite staircase. Each strict refinement increases the dimension of the effective algebra. The corresponding threshold marks the entry of a new observation channel, control protocol, or structural sector into the object domain. Along this staircase, windowed RCC is nondecreasing and window entropy is nonincreasing; finite differences of Eq.~\eqref{eq:6.4} give the basic response.

A differentiable response can be established through calibrated effective protocols, ensemble interpolation, or controlled passage to a continuum description. Such interpolations need not themselves be conditional expectations. Consider a region where the chosen branch forms a differentiable, locally reversible CWT state manifold, preserves the required window monotonicity, and keeps $R$, $d_R$, and $\Gamma_R$ fixed. The complexity-generation potential is then
\begin{equation*}
\Pi_\xi
=
-T_\xi
\left(
\frac{\partial\mathscr S_\xi}{\partial\xi}
\right)_{E,V,N,\ldots}.
\end{equation*}

Here $T_\xi$ is the CWT window temperature on this effective branch, and $\Pi_\xi$ is the generalized force conjugate to the complexity budget $\xi$. It enters the reversible extended first law of CWT as $+\Pi_\xi d\xi$. Differentiating Eq.~\eqref{eq:6.4} with respect to $\xi$ along this branch gives a quantitative expression for this thermodynamic potential in terms of windowed RCC.

\VIstatement{Corollary 6.2 (Conditional thermodynamic response of windowed RCC).}
On the effective smooth branch specified above,
\begin{equation}
\label{eq:6.5}
\Pi_\xi
=
k_BT_\xi\ln\Gamma_R
\left(
\frac{\partial C_{R,\Xi_\xi}}{\partial\xi}
\right)_{E,V,N,\ldots}.
\end{equation}

Equation~\eqref{eq:6.5} expresses the complexity-generation potential as the marginal response of windowed RCC to budget. The derivative $\partial_\xi C_{R,\Xi_\xi}$ measures the complexity leading term revealed per unit increase in budget, and $k_BT_\xi\ln\Gamma_R$ converts this response to the thermodynamic scale. Under a common reference, macroscopic constraints, and budget scale, it allows comparisons of the marginal efficiency with which different windows or calibrated control protocols resolve structure, distinguishing regions of rapid structural resolution from those approaching saturation. Window monotonicity gives $\partial_\xi C_{R,\Xi_\xi}\ge0$, so $\Pi_\xi\ge0$ on a positive-temperature branch.

This response describes how the complexity leading term of a fixed target changes within the object domains opened by the budget. Total apparatus work, dissipation, and control energy consumption are determined by a separate hardware model. When the reference background, support, or atomic bandwidth varies, the response acquires the corresponding corrections from Appendix~\ref{app:I.6}. Appendices~\ref{app:I.2}--\ref{app:I.6} give the full treatment of shared windows, path integrals, and nonsmooth response.

Under the stated model and calibration conditions, we obtain a certified bound on an implementation's excess cost above the unknown optimum and use RECT-$\eta$ to convert final-state evidence into a dynamical resource–output threshold. For CWT shared windows at fixed reference, we establish an exact balance between window entropy and the complexity leading term. On the specified effective smooth branch, the budget response of windowed RCC gives the complexity-generation potential. Section~\ref{sec:7} returns to the reference dependence of complexity, examining related complexity objects and possible covariant structures relating entropy and complexity.

\endgroup

\let\subsection\RCCclasssubsection
\begingroup
\RCCcompactsubsections
\widowpenalty=10000
\setlength{\parskip}{1.1pt plus 0.2pt minus 0.2pt}
\setlength{\abovedisplayskip}{8pt plus 1pt minus 2pt}
\setlength{\belowdisplayskip}{8pt plus 1pt minus 2pt}
\setlength{\abovedisplayshortskip}{5pt plus 0.5pt minus 2pt}
\setlength{\belowdisplayshortskip}{6pt plus 0.5pt minus 2pt}
\setlength{\jot}{2pt}
\frenchspacing
\newcommand{\VIIstatement}[1]{%
  \par\addvspace{0.25\baselineskip}%
  \Needspace{3\baselineskip}%
  \noindent\hspace*{\parindent}\RCCformaltext{#1}\hspace{0.5em}\ignorespaces
}

\section{Discussion and Outlook}
\label{sec:7}

The pathwise results above show that final-state structure imposes a common lower bound on the cost of every admissible generation history. Complexity thus connects auditable final-state evidence to a path space that cannot be exhaustively searched. This connection holds with the reference background and generation rules fixed. When these conditions change, a new background may already supply structure that the process previously had to generate. Complexity changes accordingly. To understand that change, we must ask more closely what physical content it measures.

\subsection{From Algorithmic Description to Physical Generation Depth}
\label{sec:7.1}

Kolmogorov complexity characterizes the algorithmic information in an individual object by the length of the shortest program that generates it on a fixed universal machine. Conditional complexity places the given background explicitly in the condition. Quantum algorithmic information theory extends this idea to quantum objects. Gács constructs quantum algorithmic entropy from a universal semicomputable semidensity operator, while Vitányi describes individual pure quantum states using classical programs and incorporates approximation error into the complexity definition \cite{Gacs2001_QuantumAlgorithmicEntropy,Vitanyi2001_QuantumKolmogorovComplexity}. In these constructions, complexity retains the part of an object's description that admits no further compression.

A program can be short even when the generation capabilities it invokes are costly. Low-entropy initial states, reset and cooling, target-specific macros, or high-precision continuous parameters may be hidden in the universal machine, interpreter, or runtime environment. The $|0^n\rangle$ example in Sec.~\ref{sec:1} shows how the same mathematical state entails different generation responsibilities when starting from a free pure state or an unpolarized background. Inferring physical generation cost from a shortest description therefore requires the resource specification to include the capabilities supplied by these conditions.

The Principle of Structural Fairness makes this requirement a physical condition of an RCC model. The reference fixes the generation zero point, admissible processes and program qualification specify the available capabilities, and the atomic scale supplies a common unit. Under this specification, the pathwise result in Sec.~\ref{sec:3} converts the final state's one-shot reference gap relative to the structured vacuum into a complexity lower bound for every admissible generation history. This constraint covers both known implementations and undiscovered paths: however far path search proceeds, the target structure imposes the same minimum generation requirement.

Complexity thereby acquires the meaning of physical generation depth. It measures the irreducible structural burden of the target relative to its generation background, with its value given by the model's atomic cost function. Here, depth is distinct from circuit depth defined by the number of parallel layers. It is also relational. The background determines which structures are already available, the allowed processes determine how the remaining structure can form, and the target determines when the generation task is complete. Once these conditions are fixed, a final-state certificate constrains this depth across all possible histories from the output side.

RCC gives this physical interpretation a concrete form: final-state structure supplies a common lower bound on generation cost. In this reference-consistent account of generation, if information measures uncertainty, complexity measures residual structure.

\subsection{Different Physical Projections of Complexity}
\label{sec:7.2}

Physical generation depth takes different forms across theories. Shortest descriptions ask how much information must be written down to generate an object; control geometry asks how to realize it along a shortest path; dynamical spreading asks how structure unfolds during the actual evolution. These quantities measure different objects directly. Aligning the initial state, target, reference, and resource scale reveals the complementary constraints they impose on the same generation task. Table~\ref{tab:1} lists the respective objects; we next discuss the physical meaning of these differences.

\begin{table*}[t]
\caption{Related complexity objects: direct quantities and interfaces with RCC}
\label{tab:1}
\centering
\small
\renewcommand{\arraystretch}{1.20}

\begin{tabular}{
@{}
>{\raggedright\arraybackslash}p{0.20\textwidth}
@{\hspace{1.5em}}
>{\raggedright\arraybackslash}p{0.74\textwidth}
@{}
}
\toprule
Object & Direct quantity and interface with RCC \\
\midrule

Quantum algorithmic complexity
&
Measures the shortest program under fixed description rules; RCC subjects the program background, physical initial state, allowed resources, and atomic scale to structural fairness together.
\\[2pt]

Nielsen geometric complexity
&
Measures the shortest path on a control manifold; with task and resource scales aligned, it complements RCC's final-state lower bound on generation cost.
\\[2pt]

\mbox{Krylov/spread}\newline\mbox{complexity}
&
Measures dynamical spreading for a specified Hamiltonian, seed object, and inner product; RCC derives from final-state structure a cost lower bound for every admissible generation history.
\\[2pt]

Quantum resource theory
&
Measures conversion capabilities, distillation, or resource consumption under free operations; RCC supplies a state monotone under reference-preserving operations and an interface from it to universal target-state circuit lower bounds.
\\[2pt]

Field-theoretic and high-energy complexity
&
Measures paths in field space or gravitational candidates such as volume and action; RCC can provide reference calibration, a cost scale in atomic resource slots, and an auditable structural lower bound in a boundary generation model.
\\

\bottomrule
\end{tabular}
\end{table*}

Rigorous lower bounds have also been obtained from distinguishability, random circuits, and entanglement generation. Brandão et al. define strong complexity through the size of a measurement circuit needed to distinguish a target state from the maximally mixed state with near-optimal bias, and study complexity growth using unitary designs \cite{BrandaoEtAl2021_ModelsComplexityGrowth}. For prescribed random-circuit architectures satisfying the required connectivity condition, Haferkamp et al. prove that exact circuit complexity generated by Haar-random two-qubit gates grows linearly with probability one, up to an exponential scale \cite{HaferkampEtAl2022_LinearComplexityGrowth}. Eisert uses entangling power to bound the circuit cost of local control \cite{Eisert2021_EntanglingPowerComplexity}.

RCC uses the final-state reference gap as evidence and, through faithful transcription, reference qualification, and atomic bandwidth, constrains every admissible generation path within the same model. This interface also brings measurement, feedback, and mixed-state generation into a common resource accounting.

Nielsen's geometric approach expresses the difficulty of generation as path length, replacing discrete circuits with a continuous shortest-path problem on a control manifold. Suitable Finsler metrics can lower-bound circuit size, and a class of complexity geometries is polynomially equivalent to quantum circuits under controlled penalty and precision conditions \cite{Nielsen2006_GeometricCircuitLowerBounds,NielsenEtAl2006_QuantumComputationGeometry,Brown2024_PolynomialEquivalenceComplexityGeometries}. The original geometric object is a unitary transformation. For target-state generation, the geometric quantity should take the infimum over all allowed unitaries that map the chosen initial state of the geometric task into the target error domain, or be formulated on the corresponding orbit of that initial state. Incorporating mixed-state tasks through unitary dilation also requires a common specification of ancillary systems, the environment, and their costs \cite{Stinespring1955_PositiveFunctions}.

Krylov and spread complexity concern the evolution that actually occurs. Starting from a given Hamiltonian and initial object, they measure the spreading of an operator or state in a basis generated by a recurrence \cite{ParkerEtAl2019_UniversalOperatorGrowth,RabinoviciEtAl2021_OperatorComplexity,BalasubramanianEtAl2022_SpreadComplexity}. Nielsen geometry seeks the shortest path among allowed controls; Krylov spreading describes how specified dynamics unfold; RCC supplies an endpoint lower bound on generation cost that every admissible path must satisfy. Once the initial state, target error domain, control resources, and object window are aligned, these quantities constrain the same physical task through the optimal path, actual evolution, and final-state structure, respectively.

Whether a path is costly also depends on which capabilities are freely available. Quantum resource theories organize constrained tasks through free states, free operations, and the preorder of state conversion. Monotones, conversion rates, and resource consumption characterize operational capabilities \cite{ChitambarGour2019_QRT}. In the resource theory of informational nonequilibrium with a trivial Hamiltonian, the maximally mixed state supplies the free background, and $\log d-S(\rho)$ quantifies the corresponding nonuniformity \cite{GourEtAl2015_InformationalNonequilibrium}. Work on quantum reference frames shows that establishing and sharing a phase or orientation reference can itself be an operational resource \cite{BartlettEtAl2007_ReferenceFrames}.

At fixed structured vacuum, Sec.~\ref{sec:5} establishes that the RCC linear leading term is nonincreasing under reference-preserving channels. Program qualification and atomic bandwidth further connect it to a rigorous lower bound on universal optimal quantum circuit complexity. A resource theory of generation complexity built on this connection must therefore account for the physical origin of the free background. How it is established, how it is compiled across models, and which resources enable its generation capabilities all determine the physical content of complexity comparisons.

Generation depth may also take a geometric form in high-energy physics. The complexity=volume (CV) and complexity=action (CA) conjectures proposed by Leonard Susskind and collaborators relate boundary-state complexity to the volume of an extremal spatial slice and to the gravitational action of a Wheeler–DeWitt region, respectively \cite{StanfordSusskind2014_ComplexityShockWaves,BrownEtAl2016_HolographicComplexityBulkAction,BrownEtAl2016_ComplexityActionBlackHoles}. In this picture, the irreducible generation history behind a boundary state might appear in the bulk as spatial depth or accumulated action. The generation burden measured by complexity could thereby acquire an expression in spacetime geometry.

Understanding such a correspondence requires a physical calibration of boundary complexity. Circuit complexity in field theory depends on the reference state, allowed gates, cost function, and ultraviolet cutoff. Even for free Gaussian states, its value can change substantially with these choices \cite{JeffersonMyers2017_CircuitComplexityQFT}. RCC places these dependencies within the generation task: the reference state and code subspace specify background structure, the gate set and cost function specify allowed processes, the cutoff fixes the current object domain, and atomic bandwidth fixes the unit of measurement.

Once these choices are aligned within a finite code subspace, fixed sector, or energy cutoff, an RCC certificate for the boundary final state supplies an independent lower bound on generation cost. If CV or CA gives the correct leading scaling of the corresponding boundary complexity, the volume or action value under a common calibration should accommodate this boundary structural requirement. Systematic deviations in state-family ordering, window profiles, or reference changes can then be used to test the compatibility of the geometric candidate, boundary model, and resource compilation.

Reference dependence thus matters beyond numerical calibration. Changing the reference may absorb some boundary structure into the background, reducing the depth that must be generated explicitly. Comparing geometric values across cutoffs, code subspaces, or reference states requires tracking where that structure has gone and which resources establish the new background. Comparisons of paths, resources, and geometry converge here on how generation responsibility is transferred when the reference changes.

\subsection{Reference Covariance of Entropy and Complexity}
\label{sec:7.3}

The physical structure of a target state remains unchanged, yet the complexity required to generate it can vary with the background. When a new background already supplies part of the target structure, that structure becomes an available condition for generation. Requirements previously borne by target preparation shift to establishing and supplying the background. How should this transfer be measured? The pathwise lower bound in Sec.~\ref{sec:3} already connects generation requirements to final-state evidence. We can therefore examine the structural gaps and complexity guarantees at different references to track the relation between explicit generation responsibility and background supply.

Transformation laws can preserve a common physical meaning across the values of a quantity at different references. The relational character of complexity suggests a similar possibility: if the structure taken up by the background can be calibrated independently, do generation requirements at different references obey compatible covariance laws?

Observation windows bring entropy into this picture. For the same physical state, structure unresolved by a coarse window appears as effective entropy; window refinement brings more structure into the explicit complexity value. Windows change the visibility of structure, while references change where responsibility for generating it lies. Together, these changes suggest that entropy, complexity, and the structure supplied by the background may be linked by a common set of compensation laws.

Concrete RCC relations provide a starting point for examining this covariance proposal. The shared windows in Sec.~\ref{sec:6} give an exact benchmark at fixed background. To compare reference capacity with complexity converted through the bandwidth, we use bits throughout this section and write
\begin{equation*}
\setlength{\jot}{0pt}
\begin{aligned}
S_{\rm bit}(\tau)
&:=
-{\rm Tr}(\tau\log_2\tau),
\\[-1pt]
D_{\rm bit}(\tau\Vert\sigma)
&:=
\frac{D_{\rm nat}(\tau\Vert\sigma)}{\ln2},
\qquad
g_R=\log_2\Gamma_R.
\end{aligned}
\end{equation*}

Fix a finite-dimensional unbiased reference $\sigma_R=\Pi_R/d_R$, and let $\tau_\Xi=E_\Xi(\rho)$ be the effective state in a reference-compatible window. Then
\begin{equation}
\label{eq:7.1}
S_{\rm bit}(\tau_\Xi)
+
g_R C_{R,\Xi}(\rho)
=
\log_2 d_R.
\end{equation}

Equation~\eqref{eq:7.1} allocates the fixed reference capacity between two quantities. Structure unresolved by the current window contributes to effective entropy; structure revealed relative to the structured vacuum contributes to the linear leading term of windowed complexity. Along nested-window refinement, the decrease in window entropy exactly matches the increase in the complexity leading term under the same bandwidth conversion. The exact staircase and effective smooth branch in Sec.~\ref{sec:6} express this structural transfer through finite differences and response, respectively. Its complexity guarantee is supplied by the pathwise lower bound in Sec.~\ref{sec:3}.

The window relation describes how structure is allocated at fixed background. We now allow the background itself to change and ask whether structure previously assigned to the target can be compensated exactly in a new reference. To do so, we fix the same effective window and compare the same effective state on a common support. Let $\mathcal H_B\subseteq\mathcal H_A$, and take nested unbiased references
\begin{equation*}
\sigma_A
=
\frac{\Pi_A}{d_A},
\qquad
\sigma_B
=
\frac{\Pi_B}{d_B}.
\end{equation*}

To measure the structure that the new reference takes up relative to the original background, define the directed state-side reference-structure charge
\begin{equation*}
\mathcal Q^{\rm ref}_{A\to B}
:=
D_{\rm bit}(\sigma_B\Vert\sigma_A).
\end{equation*}

For every effective state $\tau_\Xi$ supported on $\mathcal H_B$, the following identity holds exactly:
\begin{equation}
\label{eq:7.2}
S_{\rm bit}(\tau_\Xi)
+
D_{\rm bit}(\tau_\Xi\Vert\sigma_B)
+
\mathcal Q^{\rm ref}_{A\to B}
=
\log_2 d_A.
\end{equation}

Equation~\eqref{eq:7.2} divides the original reference capacity into unresolved structure, structure that still requires explicit generation relative to the new reference, and the state-side structure charge taken up by the new background. If model $B$ has atomic bandwidth $g_B=\log_2\Gamma_B$, the second term is $D_{\rm bit}(\tau_\Xi\Vert\sigma_B)=g_BC_{B,\Xi}$. The bandwidth converts the complexity leading term $C_{B,\Xi}$, measured in atomic resource slots, to bits, placing all three terms on the same scale. The structure removed from the target term is taken up in equal measure by the charge of the new background, making the compensation under reference change traceable. Whether the new background can physically supply this structure depends on reference preparation, resource origins, and model compilation.

Weighted references further distinguish directions within the support, so the structure taken up under a reference change may depend on the target state. To track this dependence, suppose
\begin{equation*}
{\rm supp}(\tau)
\subseteq
{\rm supp}(\sigma_B)
\subseteq
{\rm supp}(\sigma_A),
\end{equation*}

and take operator logarithms on the respective supports. Define the reference-transformation defect as the part of the reference change not absorbed by $\mathcal Q^{\rm ref}_{A\to B}$:
\begin{equation}
\label{eq:7.3}
\begin{aligned}
&\Delta_{A\to B}(\tau)
:=
D_{\rm bit}(\tau\Vert\sigma_A)
\\
&\quad
-D_{\rm bit}(\tau\Vert\sigma_B)
-\mathcal Q^{\rm ref}_{A\to B}
\\
&\quad=
{\rm Tr}\!\left[
(\tau-\sigma_B)
(\log_2\sigma_B-\log_2\sigma_A)
\right].
\end{aligned}
\end{equation}

Including this defect gives an exact identity for structural allocation at general references:
\begin{equation}
\label{eq:7.4}
\begin{aligned}
&S_{\rm bit}(\tau)
+D_{\rm bit}(\tau\Vert\sigma_B)
\\
&\quad
+\mathcal Q^{\rm ref}_{A\to B}
+\Delta_{A\to B}(\tau)
=
-{\rm Tr}\!\left(\tau\log_2\sigma_A\right).
\end{aligned}
\end{equation}

The right-hand side can also be written as $S_{\rm bit}(\tau)+D_{\rm bit}(\tau\Vert\sigma_A)$, the cross entropy of the effective state relative to the original reference. For nested unbiased references, it reduces to the state-independent capacity $\log_2d_A$, while $\Delta_{A\to B}(\tau)=0$, recovering Eq.~\eqref{eq:7.2}. A covariance relation must also remain consistent across successive reference changes. If the supports of three references are nested along $A\to B\to C$, and ${\rm supp}(\tau)\subseteq{\rm supp}(\sigma_C)$, then
\begin{equation}
\label{eq:7.5}
\mathcal Q^{\rm ref}_{A\to C}
=
\mathcal Q^{\rm ref}_{A\to B}
+
\mathcal Q^{\rm ref}_{B\to C}
+
\Delta_{A\to B}(\sigma_C).
\end{equation}

The same definitions give an exact composition relation for the defect itself:
\begin{equation*}
\Delta_{A\to C}(\tau)
=
\Delta_{A\to B}(\tau)
+
\Delta_{B\to C}(\tau)
-
\Delta_{A\to B}(\sigma_C).
\end{equation*}

The defect corrections in the two composition relations cancel exactly. Writing the full state-side structural transfer as $\widetilde{\mathcal Q}_{A\to B}(\tau):=\mathcal Q^{\rm ref}_{A\to B}+\Delta_{A\to B}(\tau)$, we obtain
\begin{equation*}
\widetilde{\mathcal Q}_{A\to C}(\tau)
=
\widetilde{\mathcal Q}_{A\to B}(\tau)
+
\widetilde{\mathcal Q}_{B\to C}(\tau).
\end{equation*}

The full state-side structural transfer therefore obeys an exact additive composition law. For fixed finite-dimensional references, set $K_{AB}:=\log_2\sigma_B-\log_2\sigma_A$ on ${\rm supp}(\sigma_B)$, with the second logarithm compressed to this support. Then $|\Delta_{A\to B}(\tau)|\le\lambda_{\max}(K_{AB})-\lambda_{\min}(K_{AB})$, so its departure from the zero-defect benchmark also has a state-independent bound given by the spectral width. These relations hold rigorously under the fixed support conditions. Comparisons across models, reference-upgrade specifications, and finite reference jump terms provide local ingredients for a further physical interpretation; see Appendices~\ref{app:A.10}, \ref{app:F}, and \ref{app:I.6}.

These exact relations support the physical intuition with which we began. Window changes redistribute structure between effective entropy and the complexity leading term. Reference changes introduce a structure charge and its target-dependent correction, while the full state-side transfer composes along reference transformations. Together, these results suggest a common physical mechanism by which structure is taken up when the reference changes. If the associated charge can be calibrated independently through background establishment and resource supply, this mechanism could extend beyond the present state-side relations to a broader class of physically realizable reference and object-domain transformations. We therefore propose the following conjecture.

\VIIstatement{Conjecture 7.1 (Reference covariance of entropy and complexity).}
There exists a nontrivial class of allowed transformations between reference-consistent descriptions, determined in advance by the physical conditions of the model. Its reference transformations and compatible object-domain transformations are physically realizable, satisfy structural fairness, and have a common cost calibration; the class is closed under composition. For every $A\to B$ in this class, a common embedding or compatible window transport must ensure that descriptions $A$ and $B$ refer to the same physical output and share the external generation target, output semantics, error specification, and success criterion. There then exists a directed physical reference-structure charge $\mathscr Q^{\rm phys}_{A\to B}$, calibrated independently for that transformation, such that
\begin{equation}
\label{eq:7.6}
\boxed{
\begin{aligned}
S_A^{\rm eff,bit}+\mathscr C_A^{\rm bit}
&=
S_B^{\rm eff,bit}+\mathscr C_B^{\rm bit}
\\
&\quad+
\mathscr Q^{\rm phys}_{A\to B}[\tau].
\end{aligned}
}
\end{equation}
The associated structure charges must be compatible with composition of the physical transformations.

A transformation $A\to B$ carries information about reference establishment, resource supply, and compatible transport of the object domain. The allowed class required by the conjecture extends beyond fixed windows and the zero-defect case of nested unbiased references, incorporating the state-side relations already established into broader physical transformations.

Here, $S_X^{\rm eff,bit}:=S_{\rm bit}(\tau_X)$ is the effective entropy in description $X$, measured in bits. The quantity $\mathscr C_X^{\rm gen}[\tau_X]$ denotes the explicit physical generation responsibility assigned by that description, measured in atomic resource slots. The conversion $\mathscr C_X^{\rm bit}:=g_X\mathscr C_X^{\rm gen}[\tau_X]$ places it on a common information scale, where $g_X=\log_2\Gamma_{R_X}$. The directed structure charge $\mathscr Q^{\rm phys}_{A\to B}[\tau]$ is also measured in bits.

In the state-side sector established here, if $\tau_X=E_{\Xi_X}(\rho_X)$, then
\begin{equation*}
\left.
\mathscr C_X^{\rm gen}[\tau_X]
\right|_{\rm state}
=
C_{R_X,\Xi_X}(\rho_X),
\quad
\mathscr C_X^{\rm bit}
=
D_{\rm bit}(\tau_X\Vert\sigma_{R_X}).
\end{equation*}
This projection is the bandwidth-calibrated linear leading term introduced above. The full RCC certificate connects it to an operational lower bound on $C_{\rm opt}^{(\epsilon)}$. The conjecture thus links to the complexity guarantees of this work and asks further how this generation responsibility changes with the reference.

For state-side window refinement at fixed reference, Eq.~\eqref{eq:7.1} ensures that $S_X^{\rm eff,bit}+\mathscr C_X^{\rm bit}$ remains equal to the same reference capacity $\log_2d_R$. Equation~\eqref{eq:7.6} therefore holds exactly with $\mathscr Q^{\rm phys}=0$. This zero structure charge describes state-side structural allocation alone; establishing, resolving, and physically implementing an object window can still carry separate costs. Support-compatible reference transformations within a fixed window correspond to
\begin{equation*}
\mathscr Q^{\rm phys}_{A\to B}[\tau]
=
\mathcal Q^{\rm ref}_{A\to B}
+
\Delta_{A\to B}(\tau).
\end{equation*}
Equation~\eqref{eq:7.4} gives their exact state-side realization.

The generalized structure charge may depend on $\tau$ as a transformation functional. Its declared domain, functional form, and resource origins must be fixed in advance by the allowed transformation, reference establishment, or resource supply; it cannot be defined retrospectively as the difference between the two sides of Eq.~\eqref{eq:7.6}. This independent calibration makes Eq.~\eqref{eq:7.6} a physical claim and reconnects generation capabilities absorbed into the background to actual resources. The conjecture fails if the allowed class contains transformations that systematically eliminate explicit generation responsibility without a corresponding structure charge taking it up, or if structural transfer admits no law compatible with composition of the physical transformations.

Testing this claim requires connecting the state-side defect to actual resource budgets. For a predeclared task family of size $n$, let $\mathfrak S_n$ be a comparison domain containing the target effective states and the intermediate reference states along the composition chain. Let $\mathscr I_n^{\rm str}\to\infty$ be the family's predeclared leading structural information scale, measured in bits. One quantitative route is to ask whether resource budgets yield the target-independent bound
\begin{equation}
\label{eq:7.7}
\sup_{\tau\in\mathfrak S_n}
\left|
\Delta_{A_n\to B_n}(\tau)
\right|
\le
\mathfrak b_{A\to B}(n).
\end{equation}

Finite-dimensional spectral width already guarantees the existence of some finite upper bound. The physical test concerns the scaling, resource origins, and stability under composition of $\mathfrak b_{A\to B}(n)$. When $\mathfrak b_{A\to B}(n)=o(\mathscr I_n^{\rm str})$, the zero-defect relation holds at the leading scale. If the defect grows at the same order, its contribution at that order should be accounted for by a predeclared, physically calibrated, composable budget for reference establishment, resource supply, or model compilation. Equation~\eqref{eq:7.7} provides a candidate quantitative test of Eq.~\eqref{eq:7.6}; more general covariance laws may take other mathematical forms.

RCC supplies the complexity coordinate and process guarantees at fixed reference on which this picture builds. A general covariance theory would connect coordinates at different references and explain how their values change with background structure and the cost of establishing it. Entropy characterizes unresolved structure in a given object domain; complexity characterizes the generation history required for structure to become physically realized. If the covariance expressed by Eq.~\eqref{eq:7.6} holds, the physical meaning of complexity would extend from generation depth within a fixed model to generation responsibility carried consistently across references. This could also provide the geometric or action-based pictures suggested by CV and CA with a reference-consistent foundation for complexity, calibrated by boundary final-state evidence.

\subsection{Domain of Applicability and Theoretical Extensions}
\label{sec:7.4}

Moving from the present results to this picture requires extending generation guarantees at fixed reference to changes in the background itself. Equations~\eqref{eq:7.2}--\eqref{eq:7.5} compare the same effective state in the same window. When both window and reference change, a common embedding or an explicit window transport map must ensure that both sides continue to describe the same effective object. Classifying allowed transformations and their structure charges must therefore proceed together with reference preparation, bandwidth conversion, and model compilation, so that the compensation relations and their composition laws acquire an operational physical meaning.

Weighted or Gibbs-type references bring temperature, the Hamiltonian, bath contact, and energy exchange into the structured vacuum, extending the complexity zero point from an unbiased structural background to a physical environment with thermodynamic content. The one-shot gap, smoothing, coding, statistical, and compilation corrections in the full RCC certificate must be tracked through this change as well.

Suppose that, after bandwidth conversion, the gaps between the full certificates and optimal complexity at both ends of a transformation are subleading relative to the scale of structural transfer being compared, and that the remaining corrections are also subleading on that scale. The covariance relation for the certificates then transfers, at that scale, to the within-model optimal complexity $C_{\rm opt}^{(\epsilon)}$. Whether reference cycles produce path dependence, and which combinations remain invariant under transformation, will further determine the global structure of a covariance theory.

Extending the objects to be generated brings these questions to broader physical processes. Unitary transformations, quantum channels, and open dynamics require explicit input families, task distances, and costs for the environment and branches. Continuous-variable systems and quantum field theory require energy or mode cutoffs, reference limits consistent with the cutoffs, and resource compilation across scales. Finite code subspaces and boundary generation models provide controlled starting points. Tracking RCC certificates as the cutoff changes can test the stability of complexity ordering and leading scaling. This permits an audit of the physical content of boundary complexity across scales and supplies a basis for comparison with volume, action, or other geometric candidates in concrete models.

These extensions also depend on how tightly RCC constrains generation cost. Appendix~\ref{app:G} gives rigorous constructions with matching leading scaling for three structured-state families. Near-tightness for broader families will determine how widely RCC can both rule out histories that are too short and track the scaling of optimal complexity. Window implementation costs and selection principles determine which structures can be revealed in a given experimental or theoretical task. The complexity budget and conjugate response in CWT give this selection thermodynamic content. If general reference covariance holds, responses across windows, control protocols, and reference backgrounds can be compared under a common generation scale and resource specification.

Along these directions, the question of the generation length required by a task broadens into questions about how physical structure forms, how it is taken up by the background, and whether that transfer obeys a unified transformation law. Rigorous results at fixed reference already provide a testable starting point. The prospect of reference covariance is to make complexity a basic coordinate of physical generation, connecting local depths in different backgrounds into a unified description of structural formation and transfer.

\section{Conclusions}
\label{sec:8}

In what sense can the minimum length of a quantum circuit be understood as a physical cost? The generation responsibility represented by this length depends on where generation begins, what resources the process can draw on, and how much physical capability is accounted for at each step. When low-entropy references, macro-instructions, or program priors carry unaccounted directional structure, equal circuit lengths can conceal different physical histories.

Taking realizability with finite resources as an operational premise, we have proposed structural fairness and built the Reference-Contingent Complexity (RCC) framework upon it. The reference background fixes the structural zero point, the allowed dynamics fix the available generation capabilities, and program qualification and the atomic scale place physical histories, description interfaces, and cost units under a common resource specification.

Native discrete constructions provide a family of admissible models fixed before target selection, with each size-indexed member able to approximate every pure or mixed target state. These constructions establish that the model class is nonempty and nontrivial and contains universal models. Finite-control certificates provide verifiable reference qualification for the corresponding open processes. Reference dependence thus enters the physical definition of complexity, identifying the structure already supplied by the background and the point from which the remaining structure is counted.

Within a fixed RCC model, final-state structure provides a testable constraint on the optimal generation cost. The final state does not specify the path the system actually followed, but it retains the target structure that every successful path must produce. We have proved that the target's one-shot reference gap relative to the structured vacuum, converted by the atomic control bandwidth and with finite-description corrections included, gives a rigorous lower bound on universal optimal quantum circuit complexity $C_{\rm opt}^{(\epsilon)}$ within the model. The constraint applies to every admissible successful history and then passes to the infimum of all implementation costs, covering paths both known and yet to be discovered.

This pathwise property also makes the bound available for operational auditing. Predeclared final-state effects and their finite-sample statistics can yield independently verifiable one-sided certificates that directly exclude all admissible generation histories below the certificate threshold.

The windowed complexity value also depends on what the task can resolve. Under an unbiased reference, pure states within the support share the same global RCC leading term. As local, correlation, or logical object domains become accessible to the task, windowed RCC progressively reveals structure previously assigned to effective entropy as a complexity profile and relates the resolved structure to the required generation length. From the process side, explicit generation constructions for three classes of reference-aligned structured-state families show that, under their respective construction assumptions and asymptotic conditions, the RCC lower bound and feasible upper bounds match at the leading scale. The bound can therefore track the growth of the optimal complexity within the model.

Once the structural zero point and atomic cost unit are fixed, final-state certificates can be connected to actual dynamics. Admissible compiled paths supply process-side witnesses for the upper end of the complexity interval, while exact quantum speed limits check the geometric compatibility of the corresponding continuous dynamical records. With aligned tasks and an independently calibrated uniform positive lower bound on the performance coefficient, RCC certificates also give a lower bound on the product of energy fluctuations and entanglement output in RECT-$\eta$. A finite generation budget brings complexity into thermodynamics. Within the shared windows and effective smooth branches of CWT, windowed RCC further gives quantitative relations between the complexity value, window entropy, thermodynamic state functions, and responses.

This raises a further question: when the generation zero point itself changes, how does the physical content of complexity carry over? The exact state-side relations established here already show that structure can be reassigned among effective entropy, explicit generation responsibility, and the reference background. Conjecture 7.1 makes a broader physical claim: responsibility for structure can be transferred, but cannot simply disappear. Generation capabilities absorbed into the background must be taken up by an independently calibrated reference-structure charge compatible with composition of the physical transformations. RCC supplies this picture with physical generation models at fixed reference and complexity coordinates calibrated by bandwidth. The full certificate connects final-state structure to pathwise constraints and thereby bounds the optimal generation cost within the model.

The rigorous results of this work rest on finite-dimensional target-state generation, an unbiased structured vacuum, and a fixed RCC-admissible model. Full extensions of RCC to weighted references, the generation of quantum processes, and continuous-variable systems remain to be established, along with a more general theory of reference covariance. Within this domain, the opening question has a definite answer: the minimum quantum circuit length measures the minimum generation responsibility required by the target structure relative to the specified generation background. The reference determines where generation begins and which structure the background already supplies; admissible generation processes account for the rest. Quantum circuit complexity thus gives quantitative expression to the physical fact that structure cannot arise for free.

\makeatletter
\def\section{\@startsection{section}{1}{\z@}%
  {12pt plus 1pt minus 1pt}{6pt plus 0.5pt minus 0.5pt}%
  {\normalfont\small\bfseries\centering}}
\makeatother
\section*{Code Availability}
\newcommand{\RCCcode}[1]{\texttt{\spaceskip=\fontdimen2\font\relax#1}}
\urlstyle{same}
\setlength{\emergencystretch}{1em}

\looseness=-1
The rcc-refcert 0.1.3 software \cite{LiuTianWu2026_RCCRefcert} (\href{https://doi.org/10.5281/zenodo.22768920}{archived release}; \href{https://github.com/yeruciwenrou-dotcom/rcc-refcert}{source code}) computes program semidensities, numerically verifies reference certificates for selected finite constructions (Appendices~\ref{app:F} and \ref{app:H}), and provides input templates and examples for final-state calculations (Sec.~\ref{sec:4}). With~Python~3.10 or later, \RCCcode{python -m pip install rcc-refcert==0.1.3} installs the package; \RCCcode{python -m rcc\_refcert reproduce --check} reproduces and checks the release results. Records distinguish scalar enclosures, numerical certificates, and sampling coverage; full model qualification and family-level uniform guarantees rest on this paper's analytical conditions and proofs.

\par\endgroup
\onecolumngrid
\raggedbottom

\begin{acknowledgments}
The authors are especially grateful to OpenAI’s ChatGPT (GPT-5.6 Sol) for its substantial assistance with proof development, formalization, cross-checking, and manuscript revision, and warmly thank Google’s Gemini 3.1 Pro for its valuable suggestions on physical intuition, theoretical aesthetics, and scholarly presentation. All AI-assisted derivations and arguments were critically examined and independently verified by the authors, who take full responsibility for the scientific content and conclusions of this work.
\end{acknowledgments}

\begingroup

\setlength{\parindent}{2em}
\setlength{\parskip}{0pt}

\setlength{\abovedisplayskip}{7pt plus 2pt minus 2pt}
\setlength{\belowdisplayskip}{7pt plus 2pt minus 2pt}
\setlength{\abovedisplayshortskip}{4pt plus 2pt}
\setlength{\belowdisplayshortskip}{5pt plus 2pt minus 2pt}

\setlength{\jot}{2pt}

\appendix

\begingroup

\setlength{\parindent}{1.5em}
\setlength{\parskip}{1pt plus 0.3pt minus 0.2pt}
\frenchspacing
\linespread{1.035}\selectfont
\setlength{\abovedisplayskip}{8.5pt plus 2pt minus 2pt}
\setlength{\belowdisplayskip}{8.5pt plus 2pt minus 2pt}
\setlength{\abovedisplayshortskip}{5.5pt plus 2pt minus 1pt}
\setlength{\belowdisplayshortskip}{6.5pt plus 2pt minus 1pt}
\setlength{\jot}{2.5pt}
\setlength{\emergencystretch}{1.25em}
\clubpenalty=10000
\widowpenalty=10000
\displaywidowpenalty=10000
\brokenpenalty=5000
\raggedbottom

\newcommand{\Astatement}[1]{%
  \par\addvspace{0.4\baselineskip}%
  \Needspace{4\baselineskip}%
  \indent\textbf{#1}\hspace{0.5em}\ignorespaces
}
\newcommand{\Aproof}{%
  \par\addvspace{0.25\baselineskip}%
  \Needspace{3\baselineskip}%
  \indent\textbf{Proof.}\hspace{0.5em}\ignorespaces
}
\newcommand{\Aqed}{%
  \unskip\nobreak\hspace{0.35em}\ensuremath{\square}%
  \par\addvspace{0.3\baselineskip}%
}
\makeatletter
\renewcommand{\subsection}{\@startsection{subsection}{2}{\z@}%
  {14pt plus 1pt minus 1pt}{8pt plus 0.5pt minus 1pt}%
  {\normalfont\small\bfseries\centering}}
\makeatother
\let\AOriginalSubsection\subsection
\renewcommand{\subsection}[1]{%
  \par\Needspace{3\baselineskip}%
  \AOriginalSubsection{#1}%
}

\section{Proof of the RCC Main Theorem}
\label{app:A}

\indent
Section~\ref{sec:2} specifies RCC-admissible models and the general domain of validity. Appendix~\ref{app:F} constructs canonical native models, and Appendix~\ref{app:H} provides a method for certifying reference domination in finite-control models. Together, they show that this domain contains constructible, nontrivial, and auditable quantum circuit models. Fix any model family satisfying the RCC admissibility condition of Sec.~\ref{sec:2},
\(
\mathsf{Acc}_{\rm RCC}
\!\left(\widehat{\boldsymbol{\mathfrak M}}_R\right)
=
\mathsf{RCon}\wedge\mathsf{TC}\wedge\mathsf{RA}
\),
and write it as
\begin{equation*}
\widehat{\boldsymbol{\mathfrak M}}_R
=
\left\{
\widehat{\mathfrak M}_{R,n}
=
(\mathfrak M_{R,n},U_{R,n})
\right\}_{n\in\mathbb N}.
\end{equation*}
Within this family, we prove the main RCC lower bound of Sec.~\ref{sec:3} and its hypothesis-testing certificates. Fix an arbitrary system size $n$ and suppress the size index whenever this causes no ambiguity. The transcription overheads and semantic reference-domination constants used in the main proof are uniformly bounded across the declared model family.

At fixed $n$, $\mathfrak M_R$ defines the universal optimal quantum circuit complexity $C_{\rm opt}^{(\epsilon)}(\rho;\mathfrak M_R)$, while $U_R$ defines the program complexity $K_\epsilon^{U_R}(\rho\mid\mathfrak M_R)$ conditional on the full model. Once the full model is fixed, we abbreviate the latter as $K_\epsilon^{U_R}(\rho)$. The conditioning data $\mathfrak M_R$ specify the admissible atomic actions, available structure, and resource and error conventions; the superscript $U_R$ identifies the interpreter and its encoding conventions.

The proof transcribes each admissible circuit into a program, then uses the program semidensity and reference admissibility to connect description length to the final-state information gap. Monotone inversion of the transcription envelope, followed by the infimum over path costs, yields the lower bound on optimal complexity. Spectral clipping, final-state effects, and resource compilation then provide computable expressions, audit certificates, and comparisons across models.

All logarithms in this appendix are base $2$. We specialize the output distance of Sec.~\ref{sec:2} to trace distance. At each size, the atomic interface may have any finite integer $2\le\Gamma_{R,n}<\infty$, and the conditioning background contains no target-specific information. Continuous parameters, precision, measurement, postselection, and the costs of invoking preset or free resources are included in the physical resource accounting of admissible discrete circuits. Both admissible circuits and programs output normalized unconditional states. The error ranges for the exact proof chain, regularity corollaries, hypothesis testing, and approximate compilation are specified in the corresponding subsections.

If a model retains failure flags, control records, or resource registers in an extended output space, $\omega_{\mathcal C}$, $\rho_p$, and $\sigma_R$ here refer to the effective outputs after a fixed unconditional CPTP readout declared in advance. If reference domination is established in the extended space as $\widehat M_n\preceq C\widehat\sigma_{R,n}$, and the readout $\mathcal R_n$ satisfies $\mathcal R_n(\widehat\sigma_{R,n})=\sigma_{R,n}$, complete positivity immediately gives
\begin{equation*}
M_n
=
\mathcal R_n(\widehat M_n)
\preceq
C\sigma_{R,n}.
\end{equation*}
Certificates on the extended space therefore pass to the effective output space used by the main theorem. Faithful transcription and error comparisons are likewise stated with the same readout semantics.

Model-side $O(1)$ constants may depend on the fixed interface rules, their encoding, compilation and smoothing conventions, and fixed mixing parameters. They are independent of the target state, $d_R$, and system size $n$, and are uniformly bounded across the model family. The constant $c_{\rm sm}$ in A.8 is a regularity constant for the declared target-state family. The main proof assumes ideal support, ${\rm supp}(\rho)\subseteq\mathcal H_R$. Support leakage is treated in Appendix~\ref{app:B}; comparisons of reference states and quantum circuit models are treated in Appendix~\ref{app:A.10}.

\subsection{Proof Target and Basic Conventions}
\label{app:A.1}

\indent
The main proof establishes a one-shot lower bound on the cost of every admissible circuit, with a right-hand side containing only final-state information and model constants fixed in advance. Let $\Lambda_R$ be the target-independent monotone transcription envelope supplied by $\mathsf{TC}$, let $\chi_U$ be the semantic reference-domination constant on the full program domain, and write $\Lambda_R^{\leftarrow}$ for the left generalized inverse on the allowed cost set. In the exact-transcription branch, the claim to be proved is
\begin{equation}
\label{eq:A.1}
C_{\rm opt}^{(\epsilon)}(\rho;\mathfrak M_R)
\ge
\Lambda_R^{\leftarrow}\!\left[
D_{\max}^{\epsilon}(\rho\Vert\sigma_R)-\chi_U
\right].
\end{equation}

\indent
This form requires only that $\Lambda_R$ be nondecreasing. It therefore covers continuous and discrete costs, as well as atomic models with equal or reassigned costs. For the canonical atomic-slot model of Sec.~\ref{sec:2}, set
\begin{equation*}
g_R:=\log_2\Gamma_R,
\qquad
\Phi_{g_R,a}(L)
:=
g_RL+a\log\max\{1,L\},
\qquad
\beta_U:=\gamma+\chi_U .
\end{equation*}
Then Eq.~\eqref{eq:A.1} becomes a nonrecursive lower bound expressed through $\Phi_{g_R,a}^{-1}$. Appendix~\ref{app:A.7} gives its exact Lambert-$W$ form and the variational form for approximate transcription over the certified precision domain. The spectral-skew term, error distance, and ideal-support condition of the main proof are, respectively,
\begin{align}
\Delta_{\rm spec}(\rho)
&=
S(\rho)-H_{\min}(\rho),
\qquad
H_{\min}(\rho)=-\log\|\rho\|_\infty
\label{eq:A.2}
\\
T(\rho,\tau)
&=
\frac12\|\rho-\tau\|_1
\label{eq:A.3}
\\
{\rm supp}(\rho)
&\subseteq
\mathcal H_R
\label{eq:A.4}
.
\end{align}

\indent
With the reference set defined in Sec.~\ref{sec:2},
\begin{equation}
\label{eq:A.5}
\begin{aligned}
\Pi_R
&=
\prod_i P_i,
&
\mathcal H_R
&=
\Pi_R\mathcal H,
\\[2pt]
d_R
&=
{\rm Tr}\,\Pi_R>0,
&
\sigma_R
&=
\frac{\Pi_R}{d_R}.
\end{aligned}
\end{equation}

\indent
For positive semidefinite operators $X,Y$, we use the standard extended definition of max-relative entropy,
\begin{equation}
\label{eq:A.6}
D_{\max}(X\Vert Y)
=
\inf\{\lambda:X\le 2^\lambda Y\}.
\end{equation}

\indent
Under normalized trace-distance smoothing,
\begin{equation}
\label{eq:A.7}
D_{\max}^{\epsilon}(\rho\Vert\sigma_R)
=
\inf_{\substack{
\tau\in\mathcal D(\mathcal H_R)\\
T(\tau,\rho)\le\epsilon
}}
D_{\max}(\tau\Vert\sigma_R).
\end{equation}
The same normalized smoothing convention applies to any positive semidefinite reference operator $Y$, including the subnormalized program semidensity $M_U$: replace the second argument in Eq.~\eqref{eq:A.7} by $Y$. For a candidate state with nonzero support on $\mathcal H_R^\perp$, $D_{\max}$ relative to $\sigma_R$ is $+\infty$. Define the smoothing loss by
\begin{equation}
\label{eq:A.8}
\delta_{\rm sm}(\rho,\epsilon)
:=
D_{\max}(\rho\Vert\sigma_R)
-
D_{\max}^{\epsilon}(\rho\Vert\sigma_R)
\ge0.
\end{equation}
The exact proof chain allows $0\le\epsilon\le1$, with $\epsilon=0$ recovering the unsmoothed result. The pure-state regularity corollary is stated on the usual range $0<\epsilon\le1/2$. Hypothesis-testing certificates additionally require $\eta+\epsilon<1$, and approximate transcription requires $\epsilon_{\rm circ}+\delta_{\rm cmp}$ to lie in the relevant smoothing range. Each result below uses the error domain stated in its subsection.

\subsection{Machine-Relative Quantum-State Generation Complexity from Classical Prefix Programs}
\label{app:A.2}

\indent
To compare the description costs of admissible generation histories, define the shortest self-delimiting program length needed to generate a target state within a given tolerance.

\indent
Let $U_R$ be the program interpreter of the fixed model $\widehat{\mathfrak M}_R=(\mathfrak M_R,U_R)$ and let its domain $\mathcal P$ of valid output programs be prefix-free. Each $p\in\mathcal P$ outputs a normalized effective state $\rho_p\in\mathcal D(\mathcal H)$.

\indent
The machine-relative quantum-state generation complexity induced by classical prefix-free programs, conditional on the physical model $\mathfrak M_R$, is
\begin{equation}
\label{eq:A.9}
K_\epsilon^{U_R}
\bigl(\rho\mid\mathfrak M_R\bigr)
:=
\min\left\{
|p|:\,
T(\rho_p,\rho)\le\epsilon
\right\}.
\end{equation}

\indent
If no program meets the error requirement, set $K_\epsilon^{U_R}=+\infty$. Otherwise, the minimum is attained because program lengths are nonnegative integers. With $\widehat{\mathfrak M}_R$ fixed, we write $K_\epsilon^{U_R}(\rho)$.

Here the algorithmic description is a classical self-delimiting bit string, and the quantum state is the interpreter's output. This follows the Kolmogorov approach in quantum algorithmic information theory that describes quantum objects by classical programs~\cite{Vitanyi2001_QuantumKolmogorovComplexity,LiVitanyi2019_KolmogorovComplexity}; definitions taking quantum programs as inputs form a different class~\cite{BerthiaumeEtAl2001_QuantumKolmogorov}. The quantity $K_\epsilon^{U_R}$ inherits the coding structure of standard conditional prefix Kolmogorov complexity. The interpreter, admissible atomic actions, available structure, readout, and error and encoding conventions together constitute the fixed condition; target-specific information is charged only through program length. This supplies the algorithmic-information bridge required by the main theorem.

The main proof uses two logically independent properties of $U_R$. Appendix~\ref{app:A.3} invokes semantic reference admissibility $\mathsf{RA}$, characterized by Proposition 2.1, while Appendix~\ref{app:A.4} uses faithful transcription $\mathsf{TC}$ covering all admissible physical circuits.

\subsection{Program Semidensity, the Algorithmic Bridge, and Reference Calibration}
\label{app:A.3}

\indent
To turn description length into a final-state constraint, collect all valid programs into an operator with Kraft weights and compare this program prior with the structured vacuum. Define the program semidensity operator induced by $U_R$ on its full effective program domain as
\begin{equation}
\label{eq:A.10}
M_U
:=
\sum_{p\in\mathcal P}2^{-|p|}\rho_p .
\end{equation}

\Astatement{Lemma A.1 (Well-definedness of the program semidensity).}
$M_U$ is a well-defined positive semidefinite semidensity operator satisfying
\begin{equation}
\label{eq:A.11}
M_U\ge0,
\qquad
{\rm Tr}\,M_U\le1.
\end{equation}

\Aproof
Every finite partial sum is positive semidefinite. Since $\mathcal P$ is prefix-free, Kraft's inequality gives
\begin{equation}
\label{eq:A.12}
\sum_{p\in\mathcal P}2^{-|p|}\le1.
\end{equation}

\indent
For finite subsets $F\subseteq G\subset\mathcal P$,
\begin{equation*}
\left\|
\sum_{p\in G\setminus F}2^{-|p|}\rho_p
\right\|_1
=
\sum_{p\in G\setminus F}2^{-|p|}.
\end{equation*}

\indent
The finite partial sums thus form a Cauchy net in trace norm and converge. Positivity, continuity of the trace, and Eq.~\eqref{eq:A.12} give Eq.~\eqref{eq:A.11}.\Aqed

\indent
In the main text, semantic reference admissibility constrains the directional bias of the full program prior relative to the structured vacuum; Proposition 2.1 gives its operator characterization. At fixed size, this takes the following form: there is a finite constant $C_U$, uniform across the model family and independent of the target state and system size, such that
\begin{equation}
\label{eq:A.13}
M_U
\le
C_U\sigma_R
=
2^{\chi_U}\sigma_R,
\qquad
\chi_U:=\log C_U.
\end{equation}

\indent
For the main theorem, any valid $C_U$ can be enlarged to $\max\{1,C_U\}$ without loss, so that $\chi_U\ge0$. If a canonical coding convention yields $C_U<1$, the corresponding fixed additive gain may be retained separately; it does not affect the general result.

\indent
To match the certificate parametrization in Appendices~\ref{app:F} and \ref{app:H}, define the anchored qualification operator
\begin{equation}
\label{eq:A.14}
M_\alpha
:=
(1-\alpha)M_U+\alpha\sigma_R,
\qquad
0<\alpha<1,
\end{equation}
and use
\begin{equation}
\label{eq:A.15}
M_\alpha
\le
2^{\kappa_U}\sigma_R
\end{equation}
as the parametrization to be certified. Equation~\eqref{eq:A.15} implies Eq.~\eqref{eq:A.13}, with the choice
\begin{equation}
\label{eq:A.16}
C_U
=
\frac{2^{\kappa_U}-\alpha}{1-\alpha},
\qquad
\chi_U
=
\log\frac{2^{\kappa_U}-\alpha}{1-\alpha}.
\end{equation}
Thus $\kappa_U$ and $\alpha$ convert the qualification certificate into the semantic RA constant. The main proof chain below uses $\chi_U$ directly.

\Astatement{Lemma A.2 (Support consistency).}
If Eq.~\eqref{eq:A.13} holds and $\sigma_R=\Pi_R/d_R$, then
\begin{equation}
\label{eq:A.17}
{\rm supp}(M_U)\subseteq\mathcal H_R,
\qquad
{\rm supp}(\rho_p)\subseteq\mathcal H_R
\quad
\text{for every }p\in\mathcal P.
\end{equation}

\Aproof
For positive semidefinite operators $0\le A\le B$, ${\rm supp}(A)\subseteq{\rm supp}(B)$. Equation~\eqref{eq:A.13} gives ${\rm supp}(M_U)\subseteq{\rm supp}(\sigma_R)=\mathcal H_R$. Termwise positivity of the program semidensity also gives
\begin{equation}
\label{eq:A.18}
M_U
\ge
2^{-|p|}\rho_p.
\end{equation}

\indent
Hence ${\rm supp}(\rho_p)\subseteq{\rm supp}(M_U)$.\Aqed

\Astatement{Lemma A.3 (The algorithmic semidensity bridge).}
For every valid program $p$,
\begin{equation}
\label{eq:A.19}
D_{\max}(\rho_p\Vert M_U)
\le
|p|.
\end{equation}

\indent
Moreover, for any target state and smoothing radius,
\begin{equation}
\label{eq:A.20}
K_\epsilon^{U_R}(\rho)
\ge
D_{\max}^{\epsilon}(\rho\Vert M_U).
\end{equation}

\Aproof
Equation~\eqref{eq:A.18} is equivalent to $\rho_p\le2^{|p|}M_U$, which gives Eq.~\eqref{eq:A.19} directly. If $K_\epsilon^{U_R}=\nobreak+\infty$, Eq.~\eqref{eq:A.20} is trivial. Otherwise, choose a shortest program $p^\ast$. Its output satisfies $T(\rho_{p^\ast},\rho)\le\epsilon$, so $D_{\max}^{\epsilon}(\rho\Vert M_U)\allowbreak\le\nobreak D_{\max}(\rho_{p^\ast}\Vert M_U)\allowbreak\le\nobreak |p^\ast|=K_\epsilon^{U_R}(\rho)$.\Aqed

\Astatement{Lemma A.4 (Reference calibration by RA).}
Under Eq.~\eqref{eq:A.13}, every normalized state $\tau$ satisfies
\begin{equation}
\label{eq:A.21}
D_{\max}(\tau\Vert\sigma_R)
\le
D_{\max}(\tau\Vert M_U)+\chi_U.
\end{equation}
Consequently,
\begin{equation}
\label{eq:A.22}
D_{\max}^{\epsilon}(\rho\Vert M_U)
\ge
D_{\max}^{\epsilon}(\rho\Vert\sigma_R)-\chi_U.
\end{equation}

\Aproof
If $D_{\max}(\tau\Vert M_U)=+\infty$, Eq.~\eqref{eq:A.21} is trivial. Otherwise, for every $\lambda>D_{\max}(\tau\Vert M_U)$,
\begin{equation*}
\tau
\le
2^\lambda M_U
\le
2^{\lambda+\chi_U}\sigma_R.
\end{equation*}

\indent
Letting $\lambda\downarrow D_{\max}(\tau\Vert M_U)$ gives Eq.~\eqref{eq:A.21}. Taking the infimum over the same trace-distance smoothing ball yields
\begin{equation*}
D_{\max}^{\epsilon}(\rho\Vert\sigma_R)
\le
D_{\max}^{\epsilon}(\rho\Vert M_U)+\chi_U.
\end{equation*}

\indent
Rearranging gives Eq.~\eqref{eq:A.22}.\Aqed

\indent
Combining Lemmas A.3 and A.4 gives the central algorithmic--physical bridge of this appendix:
\begin{equation}
\label{eq:A.23}
\boxed{
K_\epsilon^{U_R}(\rho)
\ge
D_{\max}^{\epsilon}(\rho\Vert M_U)
\ge
D_{\max}^{\epsilon}(\rho\Vert\sigma_R)-\chi_U .
}
\end{equation}

\indent
Equation~\eqref{eq:A.23} says that if a program of length $k$ generates the target state within error $\epsilon$, its smooth max-relative entropy gap relative to the structured vacuum is at most $k+\chi_U$. Prefix-free coding and the program semidensity turn a short description into an operator bound; reference admissibility then calibrates this algorithmic prior against the physical structured vacuum. Kraft's inequality controls only the total trace of $M_U$, while RA controls its direction in Hilbert space. Together they connect algorithmic description to the physical information gap.

\indent
Equations~\eqref{eq:A.18} and \eqref{eq:A.13} also give a direct physical information-gap bound for each program:
\begin{equation}
\label{eq:A.24}
D_{\max}(\rho_p\Vert\sigma_R)
\le
|p|+\chi_U.
\end{equation}

\Astatement{Remark A.1 (General conditions and certification routes).}
The main proof chain takes $\mathsf{TC}$ and $\mathsf{RA}$ from Sec.~\ref{sec:2} as general inputs. Appendix~\ref{app:F.3} supplies both exact $\mathsf{TC}$ and $M_{\rm rank}\preceq\sigma_R$ for canonical whole-word models that start from $\sigma_R$ and satisfy reference balance. Appendix~\ref{app:F.5} establishes their nontrivial dynamics for generating low-entropy states, and Appendix~\ref{app:F.7} establishes relative universal closure within the class of soundly certified interfaces. Appendix~\ref{app:H.1} uses synchronized realization and coverage of the halting domain to reorganize finite-control terminating paths into the program semidensity. Appendices~\ref{app:H.3}--\ref{app:H.5} provide reference certificates at fixed size, at the model-family level, and under perturbations. Full $\mathsf{TC}$ for a finite-control model is established by a separate compilation certificate for that model. These two routes give constructible, certifiable sources of Eq.~\eqref{eq:A.13}; the main theorem applies to all models satisfying the general qualification conditions.

\Astatement{Corollary A.1 (Reference compatibility of program-representable resources).}
Suppose a normalized resource state $\omega_j$ can be output by an admissible program $p_j$ within trace distance $\epsilon_j$, and write $\ell_j=|p_j|$. If $U_R$ is reference-admissible, then
\begin{equation}
\label{eq:A.25}
D_{\max}^{\epsilon_j}(\omega_j\Vert\sigma_R)
\le
\ell_j+\chi_U.
\end{equation}

\Aproof
By Eq.~\eqref{eq:A.24}, the program output $\rho_{p_j}$ satisfies
\begin{equation*}
D_{\max}(\rho_{p_j}\Vert\sigma_R)
\le
\ell_j+\chi_U.
\end{equation*}

\indent
Since $T(\rho_{p_j},\omega_j)\le\epsilon_j$, this output is a feasible smoothing candidate for $D_{\max}^{\epsilon_j}(\omega_j\Vert\sigma_R)$. Taking the infimum over candidates gives the result.\Aqed
In particular, if a family of resource states can be invoked as free or constant-cost resources with uniformly $O(1)$ program lengths, their smooth max-relative entropy gaps relative to the structured vacuum must also be uniformly $O(1)$ across the model family. External resources outside the output family of $U_R$ enter the model through the separate reference-compatibility calibration of Appendix~\ref{app:F}.

\subsection{Faithful Transcription and Monotone Code-Length Envelopes}
\label{app:A.4}

\indent
Faithful transcription records admissible control histories as programs and bounds program length by physical path cost. With $\mathsf{TC}$ as in Sec.~\ref{sec:2}, let $\mathcal L_R\subseteq[0,\infty)$ be the admissible cost set. Exact transcription requires a nondecreasing, target-independent envelope $\Lambda_R$, fixed in advance, such that every admissible circuit $\mathcal C$ has a program $p_{\mathcal C}$ in the prefix-free domain satisfying
\begin{equation}
\label{eq:A.26}
\rho_{p_{\mathcal C}}
=
\omega_{\mathcal C},
\qquad
|p_{\mathcal C}|
\le
\Lambda_R\!\bigl(L(\mathcal C)\bigr).
\end{equation}

\indent
Approximate transcription requires that, for every $\delta_{\rm cmp}$ in a nonempty certified precision domain $\Delta_{\rm TC}\subset(0,1)$ common to the model family, there be a program $p_{\mathcal C,\delta_{\rm cmp}}$ satisfying
\begin{equation}
\label{eq:A.27}
\begin{aligned}
T\!\left(
\rho_{p_{\mathcal C,\delta_{\rm cmp}}},
\omega_{\mathcal C}
\right)
&\le\delta_{\rm cmp},
\\
|p_{\mathcal C,\delta_{\rm cmp}}|
&\le
\Lambda_R\!\bigl(L(\mathcal C)\bigr)
+b\log(1/\delta_{\rm cmp}),
\end{aligned}
\end{equation}
where $b\ge0$ is uniformly bounded across the model family. If $T(\omega_{\mathcal C},\rho)\le\epsilon_{\rm circ}$, the triangle inequality for trace distance gives
\begin{equation}
\label{eq:A.28}
\epsilon_{\rm tot}
:=
\epsilon_{\rm circ}+\delta_{\rm cmp},
\qquad
T\!\left(
\rho_{p_{\mathcal C,\delta_{\rm cmp}}},
\rho
\right)
\le
\epsilon_{\rm tot}.
\end{equation}

\indent
The suprema over $\delta_{\rm cmp}$ below are used only when $\Delta_{\rm TC}$ has a nonempty intersection with the relevant total-error constraint. This intersection is the set of certifiable compilation precisions at the given target tolerance.
In the canonical atomic-slot model of Sec.~\ref{sec:2}, take $a,b,\gamma\ge0$ without loss and define
\begin{equation}
\label{eq:A.29}
\Phi_{g_R,a}(L)
:=
g_RL+a\log\max\{1,L\},
\qquad
g_R=\log_2\Gamma_R>0,
\qquad
\Lambda_R(L)
=
\Phi_{g_R,a}(L)+\gamma.
\end{equation}

\indent
In the exact branch, the output of $p_{\mathcal C}$ is the effective circuit output, so the program is a feasible candidate for $K_{\epsilon_{\rm circ}}^{U_R}$. In the approximate branch, Eq.~\eqref{eq:A.28} supplies a feasible candidate at total radius $\epsilon_{\rm tot}$. Equations~\eqref{eq:A.26} and \eqref{eq:A.27} then give, respectively,
\begin{equation}
\label{eq:A.30}
\begin{aligned}
K_{\epsilon_{\rm circ}}^{U_R}(\rho)
&\le
\Phi_{g_R,a}(L)+\gamma,
&&\text{exact transcription},
\\[2pt]
K_{\epsilon_{\rm tot}}^{U_R}(\rho)
&\le
\Phi_{g_R,a}(L)
+b\log(1/\delta_{\rm cmp})
+\gamma,
&&\text{approximate transcription}.
\end{aligned}
\end{equation}

\indent
The length header, halting, addressing, and compilation data not already included in the atomic interface contribute the subleading overhead in $\Phi$. Continuous parameters, precision, measurement branches, auxiliary resources, and failure records already included in $L$, $\Gamma_R$, or the physical resource account are counted only once. Compilation precision that must still be specified at the program level contributes $b\log(1/\delta_{\rm cmp})$ and is recorded separately from the target-state tolerance $\epsilon_{\rm circ}$. Under exact transcription, the target error enters entirely through the feasible circuit set and the smoothed quantity at the same radius; the program envelope remains $\Phi_{g_R,a}(L)+\gamma$.

\Astatement{Remark A.2 (An exact canonical check for arbitrary finite $\Gamma_R$).}
For a reference-balanced atomic alphabet with initial state $\sigma_R$, Appendix~\ref{app:F.3} uses an Elias-gamma length header
\begin{equation*}
s(L)
=
2\lfloor\log(L+1)\rfloor+1
\end{equation*}
and a fixed-length whole-word rank block for every finite atomic word, obtaining
\begin{equation}
\label{eq:A.31}
|p_w|
=
s(L)+\left\lceil L\log\Gamma_R\right\rceil
\end{equation}

\noindent
and
\begin{equation}
\label{eq:A.32}
M_{\rm rank}
=
K_{\Gamma_R}\sigma_R
\preceq
\sigma_R.
\end{equation}

\indent
The canonical whole-word model can therefore take $\chi_U=0$. If the word output satisfies $T(\rho_w,\rho)\le\epsilon_{\rm circ}$, then
\begin{equation}
\label{eq:A.33}
D_{\max}^{\epsilon_{\rm circ}}(\rho\Vert\sigma_R)
\le
\left\lceil L\log\Gamma_R\right\rceil+s(L).
\end{equation}

\indent
This cross-check covers every finite integer $\Gamma_R\ge2$. In the nondyadic case, unused binary ranks only make $K_{\Gamma_R}<1$; the physical action set and operator domination remain unchanged.

The certification $M_{\rm rank}\preceq\sigma_R$ here applies to native models whose full effective program domain consists of canonical whole-word codes. An interpreter with additional program branches must establish semantic RA on their full union, after which Eq.~\eqref{eq:A.23} holds on the full program domain.

Models in Appendix~\ref{app:F.6} or Appendix~\ref{app:H.6} with variable code lengths, path fuel, or reassigned costs must establish the corresponding $\Lambda_R$ before applying the general proof below. This envelope gives the appropriate conversion from reassigned cost to program length.

\Astatement{Remark A.3 (Support leakage).}
If the actual output fails the ideal-support condition, the lower bounds below use the leakage corrections and conditioning conventions of Appendix~\ref{app:B}.

\subsection{The Basic One-Shot Inequality for Every Admissible Circuit}
\label{app:A.5}

\indent
Combining reference calibration at the program level with faithful transcription of physical paths establishes the same final-state information constraint for every admissible circuit.

\Astatement{Proposition A.1 (The circuitwise basic inequality).}
Let $\mathcal C$ be any admissible quantum circuit in the fixed model, with cost $L=L(\mathcal C)$ and effective output $\omega_{\mathcal C}$.

\indent
If the model uses exact transcription and $T(\omega_{\mathcal C},\rho)\le\epsilon$, then
\begin{equation}
\label{eq:A.34}
\boxed{
\Lambda_R(L)
\ge
|p_{\mathcal C}|
\ge
K_\epsilon^{U_R}(\rho)
\ge
D_{\max}^{\epsilon}(\rho\Vert M_U)
\ge
D_{\max}^{\epsilon}(\rho\Vert\sigma_R)-\chi_U .
}
\end{equation}

\indent
In the canonical atomic-slot model, this is equivalent to
\begin{equation}
\label{eq:A.35}
\boxed{
\Phi_{g_R,a}(L)
\ge
D_{\max}^{\epsilon}(\rho\Vert\sigma_R)-\beta_U,
\qquad
\beta_U:=\gamma+\chi_U .
}
\end{equation}

\indent
If the model uses approximate transcription and the circuit satisfies $T(\omega_{\mathcal C},\rho)\le\epsilon_{\rm circ}$, then for every $\delta_{\rm cmp}\in\Delta_{\rm TC}$ such that $\epsilon_{\rm tot}=\epsilon_{\rm circ}+\delta_{\rm cmp}\le1$, we have
\begin{equation}
\label{eq:A.36}
\boxed{
\Phi_{g_R,a}(L)
\ge
D_{\max}^{\epsilon_{\rm tot}}(\rho\Vert\sigma_R)
-\beta_U
-b\log(1/\delta_{\rm cmp}) .
}
\end{equation}

\Aproof
In the exact branch, Eq.~\eqref{eq:A.26}, the definition of the shortest program, and Eq.~\eqref{eq:A.23} give Eq.~\eqref{eq:A.34} in sequence. Substituting $\Lambda_R=\Phi_{g_R,a}+\gamma$ and rearranging gives Eq.~\eqref{eq:A.35}. In the approximate branch, the second line of Eq.~\eqref{eq:A.30} and the bound in Eq.~\eqref{eq:A.23} at radius $\epsilon_{\rm tot}$ give Eq.~\eqref{eq:A.36}.\Aqed

\indent
Proposition A.1 shows that, under exact transcription, the target state's smooth one-shot information gap relative to the structured vacuum, less the reference-calibration constant, is a common lower bound on the transcription budget of every admissible circuit. This constraint depends only on the final state and fixed model data. The approximate branch includes the corresponding corrections for total error and additional code length.

\indent
Here $\log\Gamma_R$ is the Hartley control bandwidth of one atomic resource slot. In the exact branch, target tolerance enters Eq.~\eqref{eq:A.35} through the feasible circuit set and $D_{\max}^{\epsilon}$, preserving the error accounting at the same radius fixed in Appendix~\ref{app:A.4}.

\subsection{The Spectral-Skew Identity and Exact Spectral Clipping}
\label{app:A.6}

\indent
This subsection connects the one-shot quantity $D_{\max}^{\epsilon}$ to the relative-entropy leading term in the main text and solves normalized trace-distance smoothing exactly for the structured vacuum.

\Astatement{Lemma A.5 (The spectral-skew identity).}
If
\begin{equation*}
{\rm supp}(\rho)\subseteq\mathcal H_R,
\qquad
\sigma_R=\frac{\Pi_R}{d_R},
\end{equation*}
then
\begin{equation}
\label{eq:A.37}
D_{\max}(\rho\Vert\sigma_R)
=
D(\rho\Vert\sigma_R)
+
\Delta_{\rm spec}(\rho).
\end{equation}

\Aproof
On $\mathcal H_R$, $\log\sigma_R=-\log d_R$, so
\begin{equation}
\label{eq:A.38}
D(\rho\Vert\sigma_R)
=
\log d_R-S(\rho),
\qquad
D_{\max}(\rho\Vert\sigma_R)
=
\log d_R-H_{\min}(\rho).
\end{equation}

\indent
Subtracting the two identities proves the claim.\Aqed

\indent
Thus $\Delta_{\rm spec}$ is exactly the spectral difference between $D_{\max}$ and the average relative entropy. It enters the lower bound as a correction determined by the target spectrum itself. Together with Eq.~\eqref{eq:A.8}, this gives
\begin{equation}
\label{eq:A.39}
D_{\max}^{\epsilon}(\rho\Vert\sigma_R)
=
D(\rho\Vert\sigma_R)
+
\Delta_{\rm spec}(\rho)
-
\delta_{\rm sm}(\rho,\epsilon).
\end{equation}

\Astatement{Theorem A.1 (Spectral clipping for normalized trace-distance smoothing).}
Write the spectral decomposition of $\rho$ on $\mathcal H_R$ as
\begin{equation*}
\rho
=
\sum_{i=1}^{d_R}\lambda_i|i\rangle\langle i|,
\qquad
\lambda_1\ge\lambda_2\ge\cdots\ge\lambda_{d_R}\ge0.
\end{equation*}

\indent
Define
\begin{equation}
\label{eq:A.40}
\begin{aligned}
R_\rho(t)
&:=
\sum_{i=1}^{d_R}(\lambda_i-t)_+,
\\
t_\epsilon
&:=
\inf\left\{
t\in[1/d_R,\lambda_1]:
R_\rho(t)\le\epsilon
\right\}.
\end{aligned}
\end{equation}
Then
\begin{equation}
\label{eq:A.41}
\boxed{
D_{\max}^{\epsilon}(\rho\Vert\sigma_R)
=
\log(d_Rt_\epsilon).
}
\end{equation}

\indent
If $\epsilon\ge T(\rho,\sigma_R)=R_\rho(1/d_R)$, then $t_\epsilon=1/d_R$ and hence $D_{\max}^{\epsilon}=0$. If $0\le\epsilon<T(\rho,\sigma_R)$, then $t_\epsilon$ is the unique threshold satisfying $R_\rho(t_\epsilon)=\epsilon$.

\Aproof
For every normalized state $\tau$ supported on $\mathcal H_R$,
\begin{equation*}
D_{\max}(\tau\Vert\sigma_R)
=
\log\!\left(d_R\|\tau\|_\infty\right).
\end{equation*}

\indent
It therefore suffices to minimize $\|\tau\|_\infty$ over candidates satisfying $T(\tau,\rho)\le\epsilon$. Let $\mathcal P_\rho$ denote pinching in the spectral projections of $\rho$. This map fixes $\rho$, and data processing for trace distance gives
\begin{equation*}
T\!\left(\mathcal P_\rho(\tau),\rho\right)
\le
T(\tau,\rho).
\end{equation*}

\indent
If $\tau\le t\Pi_R$, then $\mathcal P_\rho(\tau)\le t\Pi_R$. Pinching thus preserves feasibility and does not increase the operator norm, so the optimum can be attained among candidates commuting with $\rho$. The problem reduces to clipping the classical probability vector $\lambda=(\lambda_i)$:
\begin{equation*}
\inf_{\substack{
q_i\ge0,\ \sum_iq_i=1\\
q_i\le t
}}
\frac12\sum_i|q_i-\lambda_i|.
\end{equation*}

\indent
For any probability vector satisfying $q_i\le t$, set $A_t=\{i:\lambda_i>t\}$. Its total variation distance obeys
\begin{equation*}
\frac12\sum_i|q_i-\lambda_i|
\ge
\sum_{i\in A_t}(\lambda_i-q_i)
\ge
\sum_{i\in A_t}(\lambda_i-t)
=
R_\rho(t).
\end{equation*}

\indent
Conversely, clip every component with $\lambda_i>t$ to $t$, removing total mass $R_\rho(t)$. The total capacity available below the threshold, without exceeding $t$, is
\begin{equation*}
\sum_i\left[t-\min\{\lambda_i,t\}\right]
=
d_Rt-1+R_\rho(t)
\ge
R_\rho(t),
\end{equation*}
where the last inequality uses $t\ge1/d_R$. All removed mass can therefore be redistributed among the lower spectral components. This constructs a normalized probability vector whose largest component is at most $t$ and whose total variation distance from $\lambda$ is exactly $R_\rho(t)$. Thus the minimum attainable distance at a prescribed spectral cap $t$ is precisely $R_\rho(t)$. Minimizing the threshold subject to $R_\rho(t)\le\epsilon$ gives Eq.~\eqref{eq:A.41}.\Aqed

\indent
Equations~\eqref{eq:A.38} and \eqref{eq:A.41} give the exact spectral expression for the smoothing loss:
\begin{equation}
\label{eq:A.42}
\boxed{
\delta_{\rm sm}(\rho,\epsilon)
=
\log\frac{\lambda_1}{t_\epsilon}.
}
\end{equation}

\indent
In the unsaturated range $0\le\epsilon<T(\rho,\sigma_R)$, suppose the unique threshold satisfies
\begin{equation*}
\lambda_k\ge t_\epsilon\ge\lambda_{k+1}.
\end{equation*}

\indent
Writing $S_k=\sum_{i=1}^k\lambda_i$, we obtain
\begin{equation}
\label{eq:A.43}
t_\epsilon
=
\frac{S_k-\epsilon}{k}.
\end{equation}

\indent
The smoothed value for a general mixed state can therefore be computed directly over finitely many spectral intervals. Theorem A.1 gives the exact optimum under the smoothing convention used here, from which the usual regularity estimates follow. Equation~\eqref{eq:A.39} becomes the fully explicit one-shot decomposition
\begin{equation}
\label{eq:A.44}
D_{\max}^{\epsilon}(\rho\Vert\sigma_R)
=
D(\rho\Vert\sigma_R)
+
\Delta_{\rm spec}(\rho)
-
\log\frac{\lambda_1}{t_\epsilon}.
\end{equation}

\indent
Theorem A.1 uses normalized candidate states and trace-distance smoothing. Subnormalized states or purified-distance smoothing impose different constraints on the redistribution of spectral mass and have their own smoothing formulas.

\subsection{Monotone Inversion and the Main Lower Bound on Optimal Complexity}
\label{app:A.7}

\indent
The preceding inequalities give the minimum description budget required by the final state. We now invert them into path-cost bounds and pass the common lower bound to the infimum over all paths.

\indent
For any nondecreasing envelope $\Lambda_R$, define its left generalized inverse on the admissible cost set $\mathcal L_R$ by
\begin{equation*}
\Lambda_R^{\leftarrow}(y)
:=
\inf\left\{
\ell\in\mathcal L_R:
\Lambda_R(\ell)\ge y
\right\},
\qquad
\inf\varnothing:=+\infty.
\end{equation*}

\Astatement{Theorem A.2 (The one-shot inversion bound on universal optimal quantum circuit complexity).}
Under exact $\mathsf{TC}$ and semantic $\mathsf{RA}$,
\begin{equation}
\label{eq:A.45}
\boxed{
C_{\rm opt}^{(\epsilon)}(\rho;\mathfrak M_R)
\ge
\Lambda_R^{\leftarrow}\!\left[
D_{\max}^{\epsilon}(\rho\Vert\sigma_R)-\chi_U
\right].
}
\end{equation}

\indent
In the canonical atomic-slot model, let $F_{g,a}$ be the lower inverse of $\Phi_{g,a}$ on $[0,\infty)$, with nonpositive inputs mapped to zero. Then
\begin{equation*}
C_{\rm opt}^{(\epsilon)}(\rho;\mathfrak M_R)
\ge
F_{g_R,a}\!\left[
D_{\max}^{\epsilon}(\rho\Vert\sigma_R)-\beta_U
\right].
\end{equation*}

\Aproof
If no admissible circuit of finite cost meets the target error requirement, the left-hand side is $+\infty$ and the claim holds. Otherwise, Eq.~\eqref{eq:A.34} shows that the cost $L$ of every admissible circuit meeting the target error requirement lies in
\begin{equation*}
\left\{
\ell\in\mathcal L_R:
\Lambda_R(\ell)
\ge
D_{\max}^{\epsilon}(\rho\Vert\sigma_R)-\chi_U
\right\}.
\end{equation*}

\indent
Each circuit therefore obeys the lower bound on the right-hand side of Eq.~\eqref{eq:A.45}. Since the bound is independent of the particular circuit, taking the infimum over feasible costs proves the claim, including when the optimum is not attained.\Aqed

\indent
Because $g>0$ and $a\ge0$, $\Phi_{g,a}$ is continuous and strictly increasing. Its inverse is
\begin{equation}
\label{eq:A.46}
F_{g,a}(y)
=
\begin{cases}
0,
&y\le0,
\\[3pt]
y/g,
&0<y\le g,
\\[7pt]
\dfrac{a}{g\ln2}
W_0\!\left[
\dfrac{g\ln2}{a}
\exp\!\left(\dfrac{y\ln2}{a}\right)
\right],
&y>g,\ a>0,
\\[11pt]
y/g,
&y>g,\ a=0.
\end{cases}
\end{equation}

\indent
Here $W_0$ is the principal branch of the Lambert-$W$ function~\cite{CorlessEtAl1996_LambertW}. For $y>g$, the inversion equation is
\begin{equation*}
y\ln2
=
gL\ln2+a\ln L.
\end{equation*}

\indent
For the $a>0$ branch, setting $z=(g\ln2/a)L$ gives $ze^z=(g\ln2/a)e^{y\ln2/a}$ and hence Eq.~\eqref{eq:A.46}. At the breakpoint $y=g$, both branches give $L=1$. If admissible costs are integers, the continuous inverse $F_{g_R,a}$ can be rounded up. More general discrete models use the generalized inverse in Eq.~\eqref{eq:A.45} directly.

\indent
This inversion also passes the robustness of qualification in Appendix~\ref{app:H.5} directly to the complexity certificate. Suppose the logarithmic reference-calibration constant increases from $\chi_U$ to $\chi_U+\Delta\chi$, with $\Delta\chi\ge0$. For $0\le L_1\le L_2$,
\begin{equation*}
\Phi_{g_R,a}(L_2)-\Phi_{g_R,a}(L_1)
\ge
g_R(L_2-L_1).
\end{equation*}

\indent
Hence $F_{g_R,a}$ is $1/g_R$-Lipschitz, and the continuous lower-bound value decreases by at most $\Delta\chi/g_R$. Certificate perturbations thus enter through a controlled conversion by the bandwidth.

\indent
Substituting Eq.~\eqref{eq:A.39} into the canonical-model bound gives the nonrecursive one-shot form for universal optimal quantum circuit complexity:
\begin{equation}
\label{eq:A.47}
\boxed{
C_{\rm opt}^{(\epsilon)}(\rho;\mathfrak M_R)
\ge
F_{g_R,a}\!\left[
D(\rho\Vert\sigma_R)
+
\Delta_{\rm spec}(\rho)
-
\delta_{\rm sm}(\rho,\epsilon)
-
\beta_U
\right].
}
\end{equation}

\indent
In particular, since $C_R(\rho)=\frac{D(\rho\Vert\sigma_R)}{g_R}$, the input to the inverse in Eq.~\eqref{eq:A.47} can be written as
\begin{equation*}
g_RC_R(\rho)+\Delta_{\rm spec}(\rho)-\delta_{\rm sm}(\rho,\epsilon)-\beta_U.
\end{equation*}

\indent
For approximate $\mathsf{TC}$, if Eq.~\eqref{eq:A.27} holds for every $\delta_{\rm cmp}$ in the certified domain $\Delta_{\rm TC}$, the same physical circuit obeys all the fixed-precision bounds simultaneously. Hence
\begin{equation}
\label{eq:A.48}
\boxed{
C_{\rm opt}^{(\epsilon_{\rm circ})}(\rho;\mathfrak M_R)
\ge
\sup_{\substack{
\delta_{\rm cmp}\in\Delta_{\rm TC}\\
\epsilon_{\rm circ}+\delta_{\rm cmp}\le1
}}
F_{g_R,a}\!\Bigl[
D_{\max}^{\epsilon_{\rm circ}+\delta_{\rm cmp}}
(\rho\Vert\sigma_R)
-
\beta_U
-
b\log(1/\delta_{\rm cmp})
\Bigr].
}
\end{equation}

\indent
If TC certifies only one fixed compilation precision, only the bound at that $\delta_{\rm cmp}$ is retained. If instead the total program radius $\bar\epsilon$ is fixed in advance and $\epsilon_{\rm circ}=\bar\epsilon-\delta_{\rm cmp}$, the left-hand side becomes $C_{\rm opt}^{(\bar\epsilon-\delta_{\rm cmp})}$. Different values of $\delta_{\rm cmp}$ then refer to different complexity quantities. Monotonicity in tolerance alone does not justify replacing them by a formula of the same form for $C_{\rm opt}^{(\bar\epsilon)}$.

\indent
This completes the general lower bound. Equation~\eqref{eq:A.45} applies to any nondecreasing transcription envelope, Eq.~\eqref{eq:A.47} gives the nonrecursive bound for the canonical atomic-slot model, and Eq.~\eqref{eq:A.48} retains all error and compilation costs of approximate transcription. These results convert the final-state information gap into a bound on optimal complexity for general envelopes, exact canonical models, and approximate transcription.

\subsection{Explicit Estimates, Smoothing Regularity, and Spectral Examples}
\label{app:A.8}

\indent
The inverse form retains all exact constants. For direct estimation, an elementary lower bound is
\begin{equation}
\label{eq:A.49}
F_{g,a}(y)
\ge
\begin{cases}
0,
&y\le0,
\\[2pt]
y/g,
&0<y\le g,
\\[5pt]
\displaystyle
\max\left\{
1,\,
\frac{y-a\log(y/g)}{g}
\right\},
&y>g.
\end{cases}
\end{equation}

\indent
For $y>g$, set $L=F_{g,a}(y)>1$. Since $y=gL+a\log L\ge gL$, we have $L\le y/g$ and therefore
\begin{equation*}
y
=
gL+a\log L
\le
gL+a\log(y/g).
\end{equation*}

\indent
Rearranging gives the third branch, with the term $1$ supplied by $L>1$. The other two branches are exact.

\indent
For fixed $g>0$ and $a\ge0$, as $y$ becomes large,
\begin{equation}
\label{eq:A.50}
F_{g,a}(y)
=
\frac{y}{g}
-
\frac{a}{g}\log\frac{y}{g}
+
O\!\left(\frac{\log y}{y}\right).
\end{equation}
Thus, when the one-shot input diverges, self-delimiting overhead contributes only a subleading logarithmic correction; the RCC leading term enters the complexity bound with unit cost coefficient.

\indent
To connect with the state-family formulation in Sec.~\ref{sec:3}, let $\boldsymbol\rho=\{\rho_n\}_{n\in\mathbb N}$ be a target-state family paired with the model family fixed at the start of this appendix, and continue to suppress the size index. Suppose there is a constant $c_{\rm sm}=O(1)$, independent of $n$ and $\epsilon$, such that every member of the family satisfies
\begin{equation}
\label{eq:A.51}
\delta_{\rm sm}(\rho,\epsilon)
\le
c_{\rm sm}\log(1/\epsilon),
\end{equation}
throughout a fixed precision interval $0<\epsilon\le\epsilon_0<1$. We then say that the target-state family satisfies the smoothing regularity condition. This provides an explicit, uniform upper bound across the family on the exact spectral loss of Theorem A.1, and Eq.~\eqref{eq:A.47} gives
\begin{equation}
\label{eq:A.52}
C_{\rm opt}^{(\epsilon)}
\ge
F_{g_R,a}\!\left[
g_RC_R(\rho)
+
\Delta_{\rm spec}(\rho)
-
c_{\rm sm}\log(1/\epsilon)
-
\beta_U
\right].
\end{equation}

\indent
Equation~\eqref{eq:A.52} retains the full nonrecursive inversion in the canonical model. To display the linear leading term with unit coefficient used in Sec.~\ref{sec:3}, write
\begin{equation*}
Y
:=
g_RC_R(\rho)
+\Delta_{\rm spec}(\rho)
-c_{\rm sm}\log(1/\epsilon)
-\beta_U,
\qquad
C
:=
C_{\rm opt}^{(\epsilon)}(\rho;\mathfrak M_R).
\end{equation*}

\indent
Set $y_0:=\log(1/\epsilon_0)>0$ and choose uniformly
\begin{equation*}
\widetilde c_1
\ge
\max\{a,c_{\rm sm}\}
+
\frac{\max\{\beta_U,0\}}{y_0}.
\end{equation*}

\indent
If $C=+\infty$, the compact inequality below holds in the extended-real sense. Assume henceforth that $C<\infty$.

\indent
If $Y\le0$, then $g_RC_R(\rho)+\Delta_{\rm spec}(\rho)\le c_{\rm sm}\log(1/\epsilon)+\beta_U$. By the choice of $\widetilde c_1$ and $\log(1/\epsilon)\ge y_0$, the right-hand side of the compact inequality below is nonpositive, so the claim follows from $C\ge0$. Now suppose $Y>0$ and set $L_*=F_{g_R,a}(Y)$. Equation~\eqref{eq:A.52} gives $L_*\le C$, while the inverse relation yields
\begin{equation*}
Y
=
g_RL_*
+a\log\max\{1,L_*\}
\le
g_RC
+a\log\max\{1,C\}.
\end{equation*}
Therefore,
\begin{equation*}
C
\ge
C_R(\rho)
+\frac{\Delta_{\rm spec}(\rho)}{g_R}
-\frac{a}{g_R}\log\max\{1,C\}
-\frac{c_{\rm sm}}{g_R}\log(1/\epsilon)
-\frac{\beta_U}{g_R}.
\end{equation*}

\indent
The chosen $\widetilde c_1$ depends only on the declared model family, target-state family, and precision interval, and is uniformly $O(1)$ across the family. We obtain the compact regular form of Corollary 3.1 in the main text:
\begin{equation*}
\boxed{
\begin{aligned}
C_{\rm opt}^{(\epsilon)}
(\rho;\mathfrak M_R)
\ge{}&
\frac{D(\rho\Vert\sigma_R)}{\log\Gamma_R}
\\
&-
\frac{\widetilde c_1}{\log\Gamma_R}
\left[
\log\max\!\left\{
1,
C_{\rm opt}^{(\epsilon)}(\rho;\mathfrak M_R)
\right\}
+\log(1/\epsilon)
\right]
+
\frac{\Delta_{\rm spec}(\rho)}{\log\Gamma_R}.
\end{aligned}
}
\end{equation*}

\indent
These expressions expose different features of the bound. Equation~\eqref{eq:A.47} retains the actual smoothing loss and full inversion; Eq.~\eqref{eq:A.52} gives a nonrecursive bound under smoothing regularity. The compact form above displays $C_R(\rho)=D(\rho\Vert\sigma_R)/\log\Gamma_R$ as the linear leading term with unit coefficient, together with its logarithmic deficits. For a general mixed-state family without verified Eq.~\eqref{eq:A.51}, use Eq.~\eqref{eq:A.47} with the exact $\delta_{\rm sm}$.

\Astatement{Corollary A.2 (The full smoothing formula for pure states).}
Let $d_R\ge2$, and let $P=|\psi\rangle\langle\psi|$ be supported on $\mathcal H_R$. Then
\begin{equation}
\label{eq:A.53}
D_{\max}^{\epsilon}(P\Vert\sigma_R)
=
\begin{cases}
\log[d_R(1-\epsilon)],
&0\le\epsilon\le1-\dfrac1{d_R},
\\[5pt]
0,
&\epsilon\ge1-\dfrac1{d_R}.
\end{cases}
\end{equation}
Correspondingly,
\begin{equation}
\label{eq:A.54}
\delta_{\rm sm}(P,\epsilon)
=
\begin{cases}
-\log(1-\epsilon),
&0\le\epsilon\le1-\dfrac1{d_R},
\\[5pt]
\log d_R,
&\epsilon\ge1-\dfrac1{d_R}.
\end{cases}
\end{equation}

\Aproof
The pure-state spectrum is $(1,0,\ldots,0)$, so Theorem A.1 gives
\begin{equation*}
t_\epsilon
=
\max\left\{
\frac1{d_R},1-\epsilon
\right\}.
\end{equation*}

\indent
Substitution into Eqs.~\eqref{eq:A.41} and \eqref{eq:A.42} proves the result.\Aqed

\indent
On the usual range $0<\epsilon\le1/2$, Eq.~\eqref{eq:A.54} gives
\begin{equation*}
\delta_{\rm sm}(P,\epsilon)
=
-\log(1-\epsilon)
\le
\log(1/\epsilon).
\end{equation*}

\indent
Pure-state families therefore satisfy Eq.~\eqref{eq:A.51} automatically, with $c_{\rm sm}=1$.

\indent
A simple dimension-independent sufficient condition is also available for general mixed states. By the definition of spectral clipping,
\begin{equation*}
t_\epsilon
\ge
\max\left\{
\frac1{d_R},\,
\lambda_1-\epsilon
\right\}.
\end{equation*}

\indent
If $\epsilon\le q\lambda_1$ for some fixed $q<1$, then
\begin{equation}
\label{eq:A.55}
\delta_{\rm sm}(\rho,\epsilon)
\le
-\log(1-q)
=
O(1).
\end{equation}

\indent
The same smoothing regularity thus covers general mixed-state families satisfying this spectral condition.

\Astatement{Remark A.4 (The limits of smoothing regularity for mixed-state families).}
Let $d\ge2$, $0<\theta\le1/2$, and $\sigma=I/d$, and define
\begin{equation}
\label{eq:A.56}
\rho_{d,\theta}
=
\left(\frac1d+\theta\right)|1\rangle\langle1|
+
\sum_{j=2}^{d}
\left(
\frac1d-\frac{\theta}{d-1}
\right)
|j\rangle\langle j|.
\end{equation}
Then $T(\rho_{d,\theta},\sigma)=\theta$, so $t_\theta=1/d$ and
\begin{equation}
\label{eq:A.57}
D_{\max}^{\theta}(\rho_{d,\theta}\Vert\sigma)
=
0,
\qquad
\delta_{\rm sm}(\rho_{d,\theta},\theta)
=
\log(1+d\theta).
\end{equation}

\indent
For fixed $\theta$, this loss can grow with $d$ as $\log d$. The general main theorem therefore retains the exact $\delta_{\rm sm}$ or $t_\epsilon$; the regular form applies to state families for which Eq.~\eqref{eq:A.51} has been verified.

\subsection{Arbitrary-Effect Certificates and Hypothesis-Testing Bounds}
\label{app:A.9}

\indent
We derive an auditable lower bound directly from an arbitrary final-state effect, then obtain the hypothesis-testing relative entropy form by optimization.

\Astatement{Lemma A.6 (Auditing with an arbitrary effect).}
Suppose a normalized state $\omega$ satisfies
\begin{equation*}
T(\omega,\rho)\le r,
\qquad
\omega\le2^m\sigma_R.
\end{equation*}

\indent
For any effect $0\le Q\le I$, the variational characterization of trace distance and operator domination give, respectively,
\begin{equation*}
{\rm Tr}(Q\omega)
\ge
{\rm Tr}(Q\rho)-r,
\qquad
{\rm Tr}(Q\omega)
\le
2^m{\rm Tr}(Q\sigma_R).
\end{equation*}
Thus, whenever ${\rm Tr}(Q\rho)>r$,
\begin{equation}
\label{eq:A.58}
\boxed{
m
\ge
\log
\frac{{\rm Tr}(Q\rho)-r}
{{\rm Tr}(Q\sigma_R)}.
}
\end{equation}

If the denominator is zero and the numerator is positive, the right-hand side is $+\infty$ in the extended-real convention: no finite reference-domination exponent exists. For a reference-admissible program, one may take $m=|p|+\chi_U$.\Aqed

\indent
We use the following hypothesis-testing relative entropy:
\begin{equation*}
D_H^\eta(\rho\Vert\sigma_R)
=
-\log
\inf_{0\le Q\le I}
\left\{
{\rm Tr}(Q\sigma_R):
{\rm Tr}(Q\rho)\ge1-\eta
\right\}.
\end{equation*}

\indent
If $0\le r,\eta$ and $r+\eta<1$, then
\begin{equation}
\label{eq:A.59}
\boxed{
D_H^\eta(\rho\Vert\sigma_R)
\le
D_{\max}^{r}(\rho\Vert\sigma_R)
+
\log\frac1{1-\eta-r}.
}
\end{equation}

\Aproof
Take any $T(\tau,\rho)\le r$ and $\tau\le2^\lambda\sigma_R$. Every test effect satisfying ${\rm Tr}(Q\rho)\ge1-\eta$ obeys
\begin{equation*}
{\rm Tr}(Q\tau)
\ge
1-\eta-r
\end{equation*}
and
\begin{equation*}
{\rm Tr}(Q\tau)
\le
2^\lambda{\rm Tr}(Q\sigma_R).
\end{equation*}

\indent
Hence
\begin{equation*}
{\rm Tr}(Q\sigma_R)
\ge
(1-\eta-r)2^{-\lambda}.
\end{equation*}

\indent
Optimizing successively over $Q$, $\lambda$, and the smoothing candidate $\tau$ gives Eq.~\eqref{eq:A.59}.\Aqed

\indent
Any effect satisfying Eq.~\eqref{eq:A.58} also gives a circuit complexity certificate directly. Under exact transcription,
\begin{equation}
\label{eq:A.60}
\boxed{
C_{\rm opt}^{(r)}(\rho;\mathfrak M_R)
\ge
F_{g_R,a}\!\left[
\log
\frac{{\rm Tr}(Q\rho)-r}
{{\rm Tr}(Q\sigma_R)}
-
\beta_U
\right].
}
\end{equation}

\indent
If $Q=P\le\Pi_R$ is a rank-$k$ projection, then
\begin{equation*}
{\rm Tr}(P\sigma_R)
=
\frac{k}{d_R},
\end{equation*}
so the input to the inverse becomes
\begin{equation*}
\log
\frac{d_R[{\rm Tr}(P\rho)-r]}{k}
-
\beta_U.
\end{equation*}

\indent
A reliable occupation probability in a low-rank structured subspace therefore directly bounds the corresponding description and execution costs from below.

\indent
Under approximate transcription, fix any $\delta_{\rm cmp}$ in the certified domain. If a circuit generates $\rho$ within error $\epsilon_{\rm circ}$, its program output approximates $\rho$ within total radius $\epsilon_{\rm circ}+\delta_{\rm cmp}$, and the code-length envelope acquires the additional term $b\log(1/\delta_{\rm cmp})$. For any predeclared effect $Q$, Eq.~\eqref{eq:A.58} and canonical inversion therefore give, over the nonempty certified feasible set,
\begin{equation*}
C_{\rm opt}^{(\epsilon_{\rm circ})}
(\rho;\mathfrak M_R)
\ge
\sup_{\substack{
\delta_{\rm cmp}\in\Delta_{\rm TC}\\
{\rm Tr}(Q\rho)>\epsilon_{\rm circ}+\delta_{\rm cmp}
}}
F_{g_R,a}\!\Bigg[
\log
\frac{
{\rm Tr}(Q\rho)-\epsilon_{\rm circ}-\delta_{\rm cmp}
}{
{\rm Tr}(Q\sigma_R)
}
-\beta_U
-b\log(1/\delta_{\rm cmp})
\Bigg].
\end{equation*}

\indent
The same physical circuit obeys the bound at every fixed compilation precision in the certified domain, so the supremum is valid. Equation~\eqref{eq:3.9} in the main text gives the hypothesis-testing form obtained by further optimization.

\indent
Define the robust effect-audit quantity
\begin{equation*}
\mathcal A_r(\rho\Vert\sigma_R)
:=
\sup_{\substack{
0\le Q\le I\\
{\rm Tr}(Q\rho)>r
}}
\log
\frac{{\rm Tr}(Q\rho)-r}
{{\rm Tr}(Q\sigma_R)}.
\end{equation*}

\Astatement{Corollary A.3 (Hypothesis-testing representation of the robust effect-audit quantity).}
In finite dimensions with compatible support,
\begin{equation}
\label{eq:A.61}
\boxed{
\mathcal A_r(\rho\Vert\sigma_R)
=
\sup_{0\le\eta<1-r}
\left[
D_H^\eta(\rho\Vert\sigma_R)
+
\log(1-\eta-r)
\right].
}
\end{equation}

\Aproof
For any effect $Q$, set $\eta=1-{\rm Tr}(Q\rho)$. Since $Q$ is a feasible test at this $\eta$,
\begin{equation*}
D_H^\eta(\rho\Vert\sigma_R)
\ge
-\log{\rm Tr}(Q\sigma_R).
\end{equation*}
This shows that the right-hand side of Eq.~\eqref{eq:A.61} is at least the left-hand side. Conversely, for any $\eta<1-r$, choose an effect $Q_\eta$ attaining the hypothesis-testing optimum. Its target acceptance probability is at least $1-\eta$; substitution into the definition of $\mathcal A_r$ gives the reverse inequality.\Aqed

\indent
To determine whether effect certification loses any lower-bound strength relative to the main theorem, we must compare $\mathcal A_r$ directly with the smoothed information input.

\Astatement{Corollary A.4 (Ideal effect optimization and the smoothed information bound).}
Let $\rho\in\mathcal D(\mathcal H_R)$, $\sigma_R=\Pi_R/d_R$, and $0\le r<1$. Under the normalized trace-distance smoothing of Theorem A.1,
\begin{equation}
D_{\max}^{r}(\rho\Vert\sigma_R)
=
\max\{0,\mathcal A_r(\rho\Vert\sigma_R)\}.
\label{eq:A.61a}
\end{equation}
With the conventions $\beta_U\ge0$ and $F_{g_R,a}$ mapping nonpositive inputs to zero, the same complexity inversion gives
\begin{equation}
\sup_{\substack{0\le Q\le I\\
{\rm Tr}(Q\rho)>r}}
F_{g_R,a}\!\left[
\log\frac{{\rm Tr}(Q\rho)-r}{{\rm Tr}(Q\sigma_R)}
-\beta_U
\right]
=
F_{g_R,a}\!\left[
D_{\max}^{r}(\rho\Vert\sigma_R)-\beta_U
\right].
\label{eq:A.61b}
\end{equation}

\Aproof
Apply Lemma A.6 to any normalized smoothing candidate, then take the infimum over the reference-domination exponent and the candidate state. This gives $\mathcal A_r\le D_{\max}^{r}$. If $0<r<T(\rho,\sigma_R)$, the spectral cap $t_r>1/d_R$ of Theorem A.1 satisfies $\sum_i(\lambda_i-t_r)_+=r$. Let $P$ project onto all eigendirections with $\lambda_i\ge t_r$, and denote its rank by $k$. Then
\begin{equation*}
{\rm Tr}(P\rho)-r=kt_r,
\qquad
{\rm Tr}(P\sigma_R)=\frac{k}{d_R}.
\end{equation*}
Thus $P$ is a feasible effect whose logarithmic response ratio is $\log(d_Rt_r)=D_{\max}^{r}$. When $r=0$, the projection onto the maximal eigenspace likewise attains $D_{\max}^{0}$. If $r\ge T(\rho,\sigma_R)$, the smoothing neighborhood already contains $\sigma_R$, so $D_{\max}^{r}=0$. Together with $\mathcal A_r\le D_{\max}^{r}$, this gives the zero branch of Eq.~\eqref{eq:A.61a}. When $D_{\max}^{r}>0$, the projection above attains the audit supremum, and monotonicity of $F_{g_R,a}$ gives Eq.~\eqref{eq:A.61b}. When $D_{\max}^{r}=0$, both sides vanish because $\beta_U\ge0$. Since $r<1$, the identity effect is always feasible, so the optimization domain is nonempty.\Aqed

\indent
Ideal optimization over all effects therefore recovers the full complexity lower bound in Eq.~\eqref{eq:3.1}. With the same reference, smoothing radius, and transcription conditions, any additional loss in an actual certificate comes from effect selection, the accessible measurement range, readout restrictions, and statistical estimation. The distance between this bound and the true optimal complexity is addressed by the near-tightness analysis in Subsec.~\ref{sec:3.4}.

\indent
Equation~\eqref{eq:A.59} and Theorem A.2 give the nonrecursive form of the hypothesis-testing certificate in Sec.~\ref{sec:3} for exact transcription:
\begin{equation}
\label{eq:A.62}
\boxed{
C_{\rm opt}^{(r)}(\rho;\mathfrak M_R)
\ge
F_{g_R,a}\!\left[
D_H^\eta(\rho\Vert\sigma_R)
-
\beta_U
-
\log\frac1{1-\eta-r}
\right],
\qquad
\eta+r<1.
}
\end{equation}

\indent
For approximate transcription, the corresponding form is
\begin{equation}
\label{eq:A.63}
\boxed{
C_{\rm opt}^{(\epsilon_{\rm circ})}(\rho;\mathfrak M_R)
\ge
\sup_{\substack{
\delta_{\rm cmp}\in\Delta_{\rm TC}\\
\eta+\epsilon_{\rm circ}+\delta_{\rm cmp}<1
}}
F_{g_R,a}\!\Bigg[
D_H^\eta(\rho\Vert\sigma_R)
-
\beta_U
-
b\log(1/\delta_{\rm cmp})
-
\log\frac1{
1-\eta-\epsilon_{\rm circ}-\delta_{\rm cmp}
}
\Bigg].
}
\end{equation}

\indent
If approximate TC covers the full interval $0<\delta_{\rm cmp}<A$, where $A=1-\eta-\epsilon_{\rm circ}>0$, then for $b>0$ the sum of the two explicit error penalties in Eq.~\eqref{eq:A.63} is strictly convex. Its stationary point satisfies $\delta_{\rm cmp}/(A-\delta_{\rm cmp})=b$, so the total penalty is minimized at
\begin{equation*}
\delta_{\rm cmp}^\ast
=
\frac{b}{b+1}A
\end{equation*}
with minimum value
\begin{equation*}
(b+1)\log\frac{b+1}{A}
-
b\log b.
\end{equation*}

\indent
For $b=0$, use the continuous-limit convention $0\log0=0$. The supremum is approached as $\delta_{\rm cmp}\downarrow0$, recovering $\log(1/A)$. If the certified precision domain is restricted, use the supremum in Eq.~\eqref{eq:A.63} over the actual certified domain.

\indent
This subsection turns final-state evidence into lower-bound certificates for universal optimal quantum circuit complexity. Every feasible effect gives a guarantee valid for each path, while full effect optimization recovers the corresponding smoothed information bound for the unbiased reference used here. Equations~\eqref{eq:A.62} and \eqref{eq:A.63} give the hypothesis-testing forms for exact and approximate transcription, respectively.

\subsection{Reference-State Comparison and Compilation between Quantum Circuit Models}
\label{app:A.10}

\indent
The $|0^n\rangle$ example from the Introduction tests reference admissibility across a model family by exposing the low-entropy structure that a constant-length program may carry. We then establish quantitative comparisons across models when both reference-state domination and resource compilation are controlled. The example tests structural fairness; the comparison bounds show how a reference change can become an auditable transformation between models.

\indent
Let
\begin{equation*}
P_n=|0^n\rangle\langle0^n|,
\qquad
\sigma_n=\frac{I}{2^n}.
\end{equation*}
Then
\begin{equation*}
D_{\max}(P_n\Vert\sigma_n)=n.
\end{equation*}

\indent
Suppose the output $\rho_{p_n}$ of program $p_n$ satisfies $T(\rho_{p_n},P_n)\le\epsilon_n$. The variational characterization of trace distance gives
\begin{equation*}
{\rm Tr}(P_n\rho_{p_n})
\ge
1-\epsilon_n.
\end{equation*}

\indent
Evaluating $M_U\ge2^{-|p_n|}\rho_{p_n}$ together with the semantic domination relation
\begin{equation*}
M_U\le2^{\chi_{U,n}}\sigma_n
\end{equation*}
in the $P_n$ direction gives
\begin{equation*}
\chi_{U,n}
\ge
n-|p_n|+\log(1-\epsilon_n).
\end{equation*}

\indent
In the anchored parametrization, the same calculation yields the exact necessary condition
\begin{equation}
\label{eq:A.64}
\kappa_{U,n}
\ge
\log\!\left[
\alpha
+
(1-\alpha)(1-\epsilon_n)2^n2^{-|p_n|}
\right].
\end{equation}

\indent
If $-\log(1-\epsilon_n)=O(1)$ and $|p_n|=o(n)$, Eq.~\eqref{eq:A.64} forces $\kappa_{U,n}$ to grow with $n$. In particular, if $P_n$ is a zero-cost preset initial state, or more generally its invocation satisfies $\Lambda_{R,n}(L_n)=O(1)$, one can take $|p_n|=O(1)$. Reference admissibility across the family relative to the maximally mixed structured vacuum then fails as size grows. A constant number of atomic steps need not have a constant transcription budget if $g_{R,n}=\log_2\Gamma_{R,n}$ grows. That control bandwidth already accounts for part of the structural cost and must remain in the comparison. Taking $P_n$ as a preset initial state and taking $\sigma_n$ as the generation zero point therefore define different models. Proposition F.8 gives the full incompatibility result.

\indent
This test identifies how changing the reference without a corresponding resource adjustment can break model qualification. For a controlled reference transformation, two comparisons are needed: reference states determine state-side information, while quantum circuit models determine admissible processes and their costs.

\Astatement{Lemma A.7 (One-shot and average reference-state comparisons on a common domain).}
Let $\sigma_A$ and $\sigma_B$ be normalized reference states on the same declared finite-dimensional domain $\mathcal H_\ast$, and let the target state $\rho$ be supported on $\mathcal H_\ast$. Both references use the same reference-independent trace-distance smoothing set
\begin{equation}
\label{eq:A.65}
\mathcal B_\epsilon(\rho)
=
\left\{
\tau\in\mathcal D(\mathcal H_\ast):
T(\tau,\rho)\le\epsilon
\right\}.
\end{equation}

\indent
If there is a finite constant $a_{A\to B}\ge0$ such that
\begin{equation}
\label{eq:A.66}
\sigma_A
\le
2^{a_{A\to B}}\sigma_B,
\end{equation}
then
\begin{equation}
\label{eq:A.67}
D_{\max}^{\epsilon}(\rho\Vert\sigma_B)
\le
D_{\max}^{\epsilon}(\rho\Vert\sigma_A)
+
a_{A\to B}.
\end{equation}

\indent
If ${\rm supp}(\rho)\subseteq{\rm supp}(\sigma_A)$, the same domination relation also gives the average relative-entropy comparison
\begin{equation}
\label{eq:A.68}
D(\rho\Vert\sigma_B)
\le
D(\rho\Vert\sigma_A)
+
a_{A\to B}.
\end{equation}

\indent
If there is also a finite constant $a_{B\to A}\ge0$ in the reverse direction such that
\begin{equation*}
\sigma_B
\le
2^{a_{B\to A}}\sigma_A,
\end{equation*}
then the two-sided bound is
\begin{equation}
\label{eq:A.69}
D_{\max}^{\epsilon}(\rho\Vert\sigma_A)-a_{B\to A}
\le
D_{\max}^{\epsilon}(\rho\Vert\sigma_B)
\le
D_{\max}^{\epsilon}(\rho\Vert\sigma_A)+a_{A\to B}.
\end{equation}

\indent
When the corresponding relative entropies are finite, we also have
\begin{equation}
\label{eq:A.70}
D(\rho\Vert\sigma_A)-a_{B\to A}
\le
D(\rho\Vert\sigma_B)
\le
D(\rho\Vert\sigma_A)+a_{A\to B}.
\end{equation}

\Aproof
Take any $\tau\in\mathcal B_\epsilon(\rho)$. If $D_{\max}(\tau\Vert\sigma_A)=+\infty$, the required one-sided inequality is trivial. Consider the case $D_{\max}(\tau\Vert\sigma_A)<+\infty$. For every $\lambda>D_{\max}(\tau\Vert\sigma_A)$, the definition of $D_{\max}$ gives
\begin{equation*}
\tau
\le
2^\lambda\sigma_A.
\end{equation*}

\indent
Equation~\eqref{eq:A.66} then yields
\begin{equation*}
\tau
\le
2^{\lambda+a_{A\to B}}\sigma_B.
\end{equation*}

\indent
Hence
\begin{equation*}
D_{\max}(\tau\Vert\sigma_B)
\le
D_{\max}(\tau\Vert\sigma_A)+a_{A\to B}.
\end{equation*}

\indent
Taking the infimum over the common smoothing set $\mathcal B_\epsilon(\rho)$ gives Eq.~\eqref{eq:A.67}. Interchanging $A$ and $B$ gives the reverse bound, and combining the two gives Eq.~\eqref{eq:A.69}.

\indent
For relative entropy, set $c=2^{a_{A\to B}}\ge1$. For every $\delta>0$, Eq.~\eqref{eq:A.66} gives
\begin{equation*}
\sigma_A+\delta I
\le
c(\sigma_B+\delta I).
\end{equation*}

\indent
Operator monotonicity of the logarithm gives
\begin{equation*}
\log(\sigma_A+\delta I)
\le
a_{A\to B}I+\log(\sigma_B+\delta I).
\end{equation*}

\indent
Taking the expectation in $\rho$ and letting $\delta\downarrow0$ yields Eq.~\eqref{eq:A.68}; the support condition ensures that the limits are finite. Interchanging $A$ and $B$ gives Eq.~\eqref{eq:A.70}.\Aqed
Thus, when the domination constants in both directions are uniformly bounded, the two reference readings differ by at most controlled constants. Any size dependence of the domination constants is retained in the comparison bounds.

Beyond state-side comparison, comparing universal optimal quantum circuit complexities requires a compilation map between admissible circuits.

\Astatement{Proposition A.2 (Quantum circuit complexity comparison under uniform resource compilation).}
Let $\mathfrak M_A$ and $\mathfrak M_B$ be reference-consistent quantum circuit models whose normalized unconditional output states are identified in the same finite-dimensional target domain $\mathcal H_\ast$. Write ${\rm Out}_A$ and ${\rm Out}_B$ for their respective normalized unconditional outputs. Suppose a resource compilation map ${\rm Comp}_{A\to B}$ maps every admissible circuit $\mathcal C$ in $\mathfrak M_A$ to an admissible circuit in $\mathfrak M_B$. For some fixed $\eta\ge0$, assume there are constants $s_{A\to B}>0$ and $b_{A\to B}\ge0$, independent of the target state and the particular circuit, such that
\begin{align}
T\!\left(
{\rm Out}_B({\rm Comp}_{A\to B}(\mathcal C)),
{\rm Out}_A(\mathcal C)
\right)
&\le\eta,
\label{eq:A.71}
\\
L_B({\rm Comp}_{A\to B}(\mathcal C))
&\le
s_{A\to B}L_A(\mathcal C)+b_{A\to B}.
\label{eq:A.72}
\end{align}
Then, for every target tolerance satisfying $\epsilon\ge0$ and $\epsilon+\eta\le1$,
\begin{equation}
\label{eq:A.73}
C_{\rm opt}^{(\epsilon+\eta)}(\rho;\mathfrak M_B)
\le
s_{A\to B}
C_{\rm opt}^{(\epsilon)}(\rho;\mathfrak M_A)
+
b_{A\to B}.
\end{equation}

\Aproof
If $C_{\rm opt}^{(\epsilon)}(\rho;\mathfrak M_A)=+\infty$, the claim is trivial. Otherwise, for every $\xi>0$ there is an admissible circuit $\mathcal C_\xi$ in $\mathfrak M_A$ such that
\begin{equation*}
T({\rm Out}_A(\mathcal C_\xi),\rho)
\le
\epsilon
\end{equation*}
and
\begin{equation*}
L_A(\mathcal C_\xi)
\le
C_{\rm opt}^{(\epsilon)}(\rho;\mathfrak M_A)+\xi.
\end{equation*}

\indent
Equation~\eqref{eq:A.71} and the triangle inequality for trace distance give
\begin{equation*}
T\!\left(
{\rm Out}_B({\rm Comp}_{A\to B}(\mathcal C_\xi)),
\rho
\right)
\le
\epsilon+\eta.
\end{equation*}

\indent
The compiled circuit is therefore a feasible candidate for $C_{\rm opt}^{(\epsilon+\eta)}(\rho;\mathfrak M_B)$. Together with Eq.~\eqref{eq:A.72}, this gives
\begin{equation*}
C_{\rm opt}^{(\epsilon+\eta)}(\rho;\mathfrak M_B)
\le
s_{A\to B}
\left[
C_{\rm opt}^{(\epsilon)}(\rho;\mathfrak M_A)+\xi
\right]
+
b_{A\to B}.
\end{equation*}

\indent
Letting $\xi\downarrow0$ proves Eq.~\eqref{eq:A.73}.\Aqed

The compilation constants must account for all resources involved in the model transformation. If they depend on system size, this dependence remains in Eq.~\eqref{eq:A.73}. If reverse compilation obeys bounds of the same kind, applying Proposition A.2 in both directions gives a two-sided comparison.

If the pure-state invocation above is realized by a constant-length program, its reference-domination parameter relative to the maximally mixed structured vacuum grows with size under the stated error conditions. This low-entropy structure must be accounted for by the reference background or resource cost. For controlled reference transformations, Lemma A.7 and Proposition A.2 give comparison bounds on state-side readings and optimal generation costs, respectively.

\subsection{Main Conclusions}
\label{app:A.11}

\indent
This appendix has rigorously established the lower bound connecting final-state information to universal optimal quantum circuit complexity. In any fixed RCC-admissible model with exact faithful transcription, every admissible circuit generating the target state $\rho$ within error $\epsilon$ satisfies
\begin{equation*}
\Lambda_R(L_{\mathcal C})
\ge
D_{\max}^{\epsilon}(\rho\Vert\sigma_R)-\chi_U.
\end{equation*}

\indent
Monotone inversion of the transcription envelope followed by the infimum over feasible costs therefore gives
\begin{equation*}
\boxed{
C_{\rm opt}^{(\epsilon)}(\rho;\mathfrak M_R)
\ge
\Lambda_R^{\leftarrow}\!\left[
D_{\max}^{\epsilon}(\rho\Vert\sigma_R)-\chi_U
\right].
}
\end{equation*}

\indent
This common lower bound constrains all admissible generation paths in the fixed model. Equation~\eqref{eq:A.47} gives a directly computable nonrecursive form for the canonical atomic-slot model. Under smoothing regularity, Appendix~\ref{app:A.8} displays $C_R$ as its linear leading term with unit coefficient. Equation~\eqref{eq:A.48} retains the compilation error and code-length overhead of approximate transcription.

Theorem A.1 reduces smoothed information relative to the structured vacuum to an exact spectral clipping problem. Corollary A.4 further proves that ideal optimization over all effects recovers the same complexity lower bound. Equations~\eqref{eq:A.62} and \eqref{eq:A.63} give hypothesis-testing certificates for exact and approximate transcription. Appendix~\ref{app:A.10} establishes information comparisons under reference-state domination and optimal-complexity comparisons under uniform resource compilation.

\indent
The native constructions in Appendix~\ref{app:F} realize both $\mathsf{TC}$ and $\mathsf{RA}$. The finite-control reference certificates of Appendix~\ref{app:H}, together with $\mathsf{TC}$ for the corresponding model, provide certifiable realizations.

With a common reference, output semantics, and atomic cost convention, final-state structure beyond that supplied by the background thus imposes a computable, auditable minimum cost on every admissible generation history.

\endgroup

\begingroup

\setlength{\parindent}{1.5em}
\setlength{\parskip}{1pt plus 0.3pt minus 0.2pt}
\frenchspacing
\linespread{1.005}\selectfont
\setlength{\abovedisplayskip}{8.5pt plus 1pt minus 2pt}
\setlength{\belowdisplayskip}{8.5pt plus 1pt minus 2pt}
\setlength{\abovedisplayshortskip}{5.5pt plus 1pt minus 1pt}
\setlength{\belowdisplayshortskip}{6.5pt plus 1pt minus 1pt}
\setlength{\jot}{2pt}
\setlength{\emergencystretch}{1.25em}
\clubpenalty=10000
\widowpenalty=10000
\displaywidowpenalty=10000
\brokenpenalty=5000
\raggedbottom

\newcommand{\Bstatement}[1]{%
  \par\addvspace{0.4\baselineskip}%
  \Needspace{4\baselineskip}%
  \indent\textbf{#1}\hspace{0.5em}\ignorespaces
}
\newcommand{\Bproof}{%
  \par\addvspace{0.25\baselineskip}%
  \Needspace{3\baselineskip}%
  \indent\textbf{Proof.}\hspace{0.5em}\ignorespaces
}
\newcommand{\Bqed}{%
  \unskip\nobreak\hspace{0.35em}\ensuremath{\square}%
  \par\addvspace{0.3\baselineskip}%
}
\makeatletter
\renewcommand{\subsection}{\@startsection{subsection}{2}{\z@}%
  {14pt plus 1pt minus 1pt}{8pt plus 0.5pt minus 1pt}%
  {\normalfont\small\bfseries\centering}}
\makeatother
\let\BOriginalSubsection\subsection
\renewcommand{\subsection}[1]{%
  \par\Needspace{3\baselineskip}%
  \BOriginalSubsection{#1}%
}

\section{Support Leakage, Support Regularization, and Conservative Corrections}
\label{app:B}

\indent
The main proof in Appendix~\ref{app:A} assumes ideal reference support, ${\rm supp}(\rho)\subseteq\mathcal H_R$, so that the target state lies within the support of the structured vacuum $\sigma_R=\Pi_R/d_R$. In actual preparation, numerical simulation, or experimental readout, part of the physical output state's weight may lie in $\mathcal H_R^\perp$. The original reference state's zero eigenvalues on this orthogonal complement then produce support singularities in both relative entropy and max-relative entropy. This appendix makes that boundary explicit and develops mathematical routes for extracting conditional-state values, support-regularized values, and leakage-discounted values from the physical output.

\indent
Throughout, we fix $R,\Pi_R,\mathcal H_R,d_R$, $\sigma_R$, and the atomic control bandwidth $\log\Gamma_R$, and assume a finite-dimensional total Hilbert space, $d=\dim\mathcal H<\infty$. Appendix~\ref{app:F} treats reference selection and upgrading, non-Abelian structure, energy or mode truncation, and Gibbs-type reference states. Appendix~\ref{app:D} treats finite-sample leakage estimation and joint statistical propagation, and Appendix~\ref{app:C.6} combines the final certificates. Unless stated otherwise, distances and logarithms follow the conventions of Appendix~\ref{app:A}.

\subsection{Ideal Support and Reference Leakage}
\label{app:B.1}

\indent
The weight retained in the reference subspace quantifies the physical output's departure from ideal support. Let $\rho_{\rm phys}$ denote the physical output state, and define the retention probability, leakage probability, and orthogonal-complement projection by
\begin{equation}
\label{eq:B.1}
\begin{aligned}
q_R
&=
{\rm Tr}(\Pi_R\rho_{\rm phys}),
\\
p_{\rm leak}
&=
1-q_R
=
{\rm Tr}\big[(I-\Pi_R)\rho_{\rm phys}\big],
\\
\Pi_R^\perp
&=
I-\Pi_R .
\end{aligned}
\end{equation}

\indent
When $p_{\rm leak}=0$, $\rho_{\rm phys}$ is fully supported on $\mathcal H_R$, and Appendix~\ref{app:A} applies directly. If $p_{\rm leak}>0$, then
\({\rm supp}(\rho_{\rm phys})
\not\subseteq
{\rm supp}(\sigma_R)
=
\mathcal H_R,
\)
and hence
\(D(\rho_{\rm phys}\Vert\sigma_R)
=
+\infty .
\)
The same support obstruction makes max-relative entropy diverge: no finite $\lambda$ satisfies
\(\rho_{\rm phys}
\le
2^\lambda\sigma_R,
\)
because the right-hand side vanishes on $\mathcal H_R^\perp$. Thus
\(D_{\max}(\rho_{\rm phys}\Vert\sigma_R)
=
+\infty .
\)
This divergence signals a mismatch with the reference domain: the physical output extends beyond the original reference state's support. We resolve this support singularity below, both for the conditional state within the reference domain and for the full physical state.

\subsection{Normalized Projected Conditional States and Conditional Values}
\label{app:B.2}

\indent
To distinguish structure within the reference domain from the weight with which it occurs, for $q_R>0$ define the normalized projected conditional state
\begin{equation}
\label{eq:B.2}
\rho_R
=
\frac{\Pi_R\rho_{\rm phys}\Pi_R}{q_R}.
\end{equation}
It satisfies
\(\rho_R\ge0,
{\rm Tr}(\rho_R)=1,
{\rm supp}(\rho_R)\subseteq\mathcal H_R .
\)
Its RCC linear leading term relative to the structured vacuum is
\begin{equation}
\label{eq:B.3}
C_R(\rho_R)
=
\frac{D(\rho_R\Vert\sigma_R)}{\log\Gamma_R}.
\end{equation}

\indent
Equation~\eqref{eq:B.3} measures the conditional state's structural gap relative to the structured vacuum within the reference subspace, given that the output lies in that subspace. The probability $q_R$ gives the weight of this conditional branch in the full physical output. Together, they specify the physical meaning of the conditional value. If $q_R=0$, $\rho_R$ is undefined. If $q_R$ is small, the conditional state remains well defined but describes only a branch of small weight.

\subsection{Support-Regularized Reference States and Full-State Values}
\label{app:B.3}

\indent
To retain the full physical state, extend the reference support to the nonempty orthogonal complement $\mathcal H_R^\perp$. Let $\tau_\perp$ be a normalized reference state on this complement. For $0<\delta<1$, define the support-regularized reference state
\begin{equation}
\label{eq:B.4}
\widetilde\sigma_{R,\delta}
=
(1-\delta)\sigma_R+\delta\tau_\perp .
\end{equation}

\indent
The corresponding unbiased choice is the maximally mixed state on the orthogonal complement,
\begin{equation}
\label{eq:B.5}
\tau_\perp
=
\frac{\Pi_R^\perp}{{\rm Tr}(\Pi_R^\perp)}
=
\frac{I-\Pi_R}{d-d_R}.
\end{equation}

\indent
More generally, $\tau_\perp$ may be specified by the audit protocol's background model on the reference complement. For the full-state relative entropy to be finite, it suffices that its support cover the actual leakage block:
\begin{equation*}
{\rm supp}(\Pi_R^\perp\rho_{\rm phys}\Pi_R^\perp)
\subseteq
{\rm supp}(\tau_\perp).
\end{equation*}

\indent
The support of a positive semidefinite block matrix is contained in the direct sum of the supports of its diagonal blocks. This condition therefore also covers any interblock coherence; the maximally mixed state on the complement satisfies it automatically.

Support regularization addresses zero eigenvalues of the reference state, whereas one-shot smoothing in $D_{\max}^{\epsilon}$ optimizes over a neighborhood of the target state. Here we keep $\sigma_R$ and the atomic control bandwidth fixed, and use $\widetilde\sigma_{R,\delta}$ as a comparison reference covering the physical output's support. Promoting it to a new structured vacuum also requires a corresponding reference-consistent quantum circuit--description model.
Define
\begin{equation}
\label{eq:B.6}
C_{R,\delta}^{\rm reg}(\rho_{\rm phys})
=
\frac{
D(\rho_{\rm phys}\Vert\widetilde\sigma_{R,\delta})
}{
\log\Gamma_R
}.
\end{equation}

\indent
This quantity measures the full physical state's structural gap relative to the support-regularized background. The parameter $\delta$ and complementary reference state $\tau_\perp$ specify that comparison background and are therefore part of the definition of this value.

\subsection{Leakage Decomposition and Conservative Discounting}
\label{app:B.4}

\indent
To separate leakage weight, structure within each domain, and coherence between the domains, introduce the pinching map that removes interblock coherence,
\begin{equation}
\label{eq:B.7}
\Delta_R(X)
=
\Pi_RX\Pi_R
+
\Pi_R^\perp X\Pi_R^\perp .
\end{equation}

\indent
When $q_R>0$ and $p_{\rm leak}>0$, write
\begin{equation}
\label{eq:B.8}
\rho_\perp
=
\frac{
\Pi_R^\perp\rho_{\rm phys}\Pi_R^\perp
}{
p_{\rm leak}
}.
\end{equation}
Then
\begin{equation}
\label{eq:B.9}
\Delta_R(\rho_{\rm phys})
=
q_R\rho_R
\oplus
p_{\rm leak}\rho_\perp
\end{equation}
is block diagonal with respect to
\(\mathcal H
=
\mathcal H_R
\oplus
\mathcal H_R^\perp
\)
and the support-regularized reference state is
\begin{equation}
\label{eq:B.10}
\widetilde\sigma_{R,\delta}
=
(1-\delta)\sigma_R
\oplus
\delta\tau_\perp .
\end{equation}

\Bstatement{Proposition B.1 (Relative-entropy decomposition and conservative discounting for leakage).}
If $q_R,p_{\rm leak}>0$ and $\tau_\perp$ covers the support of the actual leakage block, then
\(\Delta_R(\widetilde\sigma_{R,\delta})
=
\widetilde\sigma_{R,\delta},
\)
and relative entropy obeys the Pythagorean decomposition for the pinching conditional expectation,
\begin{equation}
\label{eq:B.11}
D(\rho_{\rm phys}\Vert\widetilde\sigma_{R,\delta})
=
D(\rho_{\rm phys}\Vert\Delta_R(\rho_{\rm phys}))
+
D(\Delta_R(\rho_{\rm phys})
\Vert\widetilde\sigma_{R,\delta}).
\end{equation}

\indent
The full-state relative entropy is therefore at least the conditional relative entropy weighted by the retention probability; see Eq.~\eqref{eq:B.18}.

\Bproof
On their respective supports, $\log\Delta_R(\rho_{\rm phys})$ and $\log\widetilde\sigma_{R,\delta}$ are block diagonal with respect to $\mathcal H_R\oplus\mathcal H_R^\perp$. Their trace pairings with $\rho_{\rm phys}$ therefore equal those with $\Delta_R(\rho_{\rm phys})$. Expanding the three relative entropies makes the cross terms cancel exactly, proving Eq.~\eqref{eq:B.11}. The first term is nonnegative, so
\begin{equation}
\label{eq:B.12}
D(\rho_{\rm phys}\Vert\widetilde\sigma_{R,\delta})
\ge
D(\Delta_R(\rho_{\rm phys})
\Vert\widetilde\sigma_{R,\delta}).
\end{equation}

\indent
Evaluating the two direct-sum blocks separately gives
\begin{equation}
\label{eq:B.13}
\begin{aligned}
D(\Delta_R(\rho_{\rm phys})
\Vert\widetilde\sigma_{R,\delta})
={}&\;
q_R\log\frac{q_R}{1-\delta}
+
p_{\rm leak}\log\frac{p_{\rm leak}}{\delta}
\\
&+
q_R D(\rho_R\Vert\sigma_R)
+
p_{\rm leak}D(\rho_\perp\Vert\tau_\perp).
\end{aligned}
\end{equation}

\Needspace{4\baselineskip}
\indent
The first two terms form the Bernoulli relative entropy
\begin{equation}
\label{eq:B.14}
D_{\rm Bern}(q_R\Vert1-\delta)
=
q_R\log\frac{q_R}{1-\delta}
+
(1-q_R)\log\frac{1-q_R}{\delta}.
\end{equation}
Hence
\begin{equation}
\label{eq:B.15}
\begin{aligned}
D(\Delta_R(\rho_{\rm phys})
\Vert\widetilde\sigma_{R,\delta})
={}&
D_{\rm Bern}(q_R\Vert1-\delta)
+
q_R D(\rho_R\Vert\sigma_R)
\\
&+
p_{\rm leak}D(\rho_\perp\Vert\tau_\perp).
\end{aligned}
\end{equation}

\indent
Equations~\eqref{eq:B.11} and \eqref{eq:B.15} decompose the full-state relative entropy into four physical contributions: coherence between the reference domain and its complement, mismatch between the total block weights, the conditional state's structural gap within the reference domain, and the leakage block's internal structural gap relative to $\tau_\perp$. Dropping the coherence term and the leakage block's internal term gives
\begin{equation}
\label{eq:B.16}
D(\rho_{\rm phys}\Vert\widetilde\sigma_{R,\delta})
\ge
D_{\rm Bern}(q_R\Vert1-\delta)
+
q_R D(\rho_R\Vert\sigma_R).
\end{equation}

\indent
Here $D_{\rm Bern}(q_R\Vert1-\delta)$ measures the classical distinguishability between the actual block weights and the regularized reference weights. The term $q_RD(\rho_R\Vert\sigma_R)$ discounts the conditional structural gap within the reference domain by the probability with which that branch physically occurs.

\indent
For $0<p_{\rm leak}<1$, the choice that matches the two block weights is
\begin{equation}
\label{eq:B.17}
\delta
=
p_{\rm leak}
=
1-q_R .
\end{equation}
Then $D_{\rm Bern}(q_R\Vert1-\delta)=0$. Equation~\eqref{eq:B.17} matches the block weights of the given output; a protocol using a fixed regularized background still sets $\delta$ by its predeclared rules. By nonnegativity of Bernoulli relative entropy, Eq.~\eqref{eq:B.16} gives, for every admissible $\delta$,
\begin{equation}
\label{eq:B.18}
D(\rho_{\rm phys}\Vert\widetilde\sigma_{R,\delta})
\ge
q_R D(\rho_R\Vert\sigma_R).
\end{equation}
This proves the conservative discounting bound for relative entropy.\Bqed
We can therefore define a leakage-discounted value independent of $\delta$,
\begin{equation}
\label{eq:B.19}
C_{\rm leak}(\rho_{\rm phys})
:=
q_R C_R(\rho_R)
\le
C_{R,\delta}^{\rm reg}(\rho_{\rm phys}).
\end{equation}

\indent
Equation~\eqref{eq:B.19} retains the structural gap within the reference domain, weighted by its actual occurrence probability. In a fixed reference model, small leakage gives $q_R\approx1$, so the weighted value is close to the conditional-state value. If $p_{\rm leak}=0$, then $C_{\rm leak}=C_R(\rho_{\rm phys})$ and support regularization is unnecessary. If $p_{\rm leak}=1$, $\rho_R$ is undefined and we set $C_{\rm leak}=0$.

\subsection{One-Shot Compatibility and Stability under Small Leakage}
\label{app:B.5}

The proof chain in Appendix~\ref{app:A} reaches the one-shot level through $D_{\max}^{\epsilon}$. We first establish unsmoothed divergence comparisons, then use trace distance to bound the effect of conditioning on target-state approximation when leakage is small.

\Bstatement{Proposition B.2 (Direct-sum comparison for unsmoothed max-relative entropy).}
Let $0<q_R<1$ and $0<\delta<1$. The block-diagonal states in Eqs.~\eqref{eq:B.9} and \eqref{eq:B.10} satisfy the exact direct-sum formula
\begin{equation}
\label{eq:B.20}
\begin{aligned}
&D_{\max}
\left(
\Delta_R(\rho_{\rm phys})
\middle\Vert
\widetilde\sigma_{R,\delta}
\right)
\\
={}&\;
\max
\left\{
D_{\max}(\rho_R\Vert\sigma_R)
+\log\frac{q_R}{1-\delta},
\;
D_{\max}(\rho_\perp\Vert\tau_\perp)
+\log\frac{p_{\rm leak}}{\delta}
\right\}.
\end{aligned}
\end{equation}

\Bproof
The block-diagonal operator inequality
\begin{equation*}
q_R\rho_R
\oplus
p_{\rm leak}\rho_\perp
\le
2^\lambda
\big[
(1-\delta)\sigma_R
\oplus
\delta\tau_\perp
\big]
\end{equation*}
is equivalent to the corresponding operator inequality on each block. The smallest $\lambda$ satisfying both block conditions is the maximum in Eq.~\eqref{eq:B.20}. If either block's support is not covered by its reference state, that block contributes $+\infty$, and the formula still holds in the extended sense.

\indent
The data-processing inequality for $D_{\max}$, together with
\(\Delta_R(\widetilde\sigma_{R,\delta})
=
\widetilde\sigma_{R,\delta},
\)
gives
\begin{equation}
\label{eq:B.21}
D_{\max}(\rho_{\rm phys}\Vert\widetilde\sigma_{R,\delta})
\ge
D_{\max}
\left(
\Delta_R(\rho_{\rm phys})
\middle\Vert
\widetilde\sigma_{R,\delta}
\right).
\end{equation}

\indent
In particular, when $0<p_{\rm leak}<1$ and $\delta=p_{\rm leak}$, the first block gives
\begin{equation}
\label{eq:B.22}
D_{\max}(\rho_{\rm phys}\Vert\widetilde\sigma_{R,\delta})
\ge
D_{\max}(\rho_R\Vert\sigma_R).
\end{equation}

\indent
This completes the proof of Proposition B.2.\Bqed

\indent
Equations~\eqref{eq:B.20}--\eqref{eq:B.22} compare the full and conditional states at the unsmoothed one-shot level. Appendix~\ref{app:B.6} aligns the smoothing error with the generation model. Small leakage also gives a direct trace-distance stability bound.

\Bstatement{Proposition B.3 (Trace-distance stability under small leakage).}
For any physical output state, removing coherence between the domains causes a trace-distance perturbation bounded by $T(\rho_{\rm phys},\Delta_R(\rho_{\rm phys}))\le\sqrt{q_Rp_{\rm leak}}$. If an ideal state $\rho$ supported on $\mathcal H_R$ also satisfies $T(\rho_{\rm phys},\rho)\le\epsilon$, then $p_{\rm leak}\le\epsilon$ and, for $q_R>0$, $T(\rho_R,\rho)\le\epsilon$.

\Bproof
Write $\rho_{\rm phys}$ with respect to $\mathcal H_R\oplus\mathcal H_R^\perp$ as
\begin{equation}
\label{eq:B.23}
\rho_{\rm phys}
=
\begin{pmatrix}
A & X\\
X^\dagger & B
\end{pmatrix},
\qquad
A=\Pi_R\rho_{\rm phys}\Pi_R,\quad
B=\Pi_R^\perp\rho_{\rm phys}\Pi_R^\perp ,
\end{equation}
where
\({\rm Tr}A=q_R,
{\rm Tr}B=p_{\rm leak}.
\)

\indent
Positive semidefiniteness of the block matrix gives $X=A^{1/2}KB^{1/2}$ with $\|K\|_\infty\le1$, and hence
\begin{equation}
\label{eq:B.24}
\|X\|_1
\le
\sqrt{{\rm Tr}A\,{\rm Tr}B}
=
\sqrt{q_Rp_{\rm leak}}.
\end{equation}
Thus
\begin{equation}
\label{eq:B.25}
T(\rho_{\rm phys},\Delta_R(\rho_{\rm phys}))
=
\frac12
\|\rho_{\rm phys}-\Delta_R(\rho_{\rm phys})\|_1
=
\|X\|_1
\le
\sqrt{q_Rp_{\rm leak}}
\le
\sqrt{p_{\rm leak}}.
\end{equation}

\indent
Small leakage therefore also controls the perturbation due to coherence between the reference subspace and its complement.

\indent
Further, if an ideal state $\rho$ satisfies
\begin{equation*}
{\rm supp}(\rho)\subseteq\mathcal H_R,
\qquad
T(\rho_{\rm phys},\rho)\le\epsilon,
\end{equation*}
then trace distance bounds the probability difference for the binary measurement $\{\Pi_R,\Pi_R^\perp\}$, giving
\begin{equation}
\label{eq:B.26}
p_{\rm leak}
=
{\rm Tr}(\Pi_R^\perp\rho_{\rm phys})
\le
\epsilon.
\end{equation}

\indent
Since $\Delta_R(\rho)=\rho$ and the complementary block has trace norm $p_{\rm leak}$, trace-distance contractivity under pinching gives
\begin{equation}
\label{eq:B.27}
T\!\left(\Delta_R(\rho_{\rm phys}),\rho\right) 
=
\frac12\|A-\rho\|_1+\frac{p_{\rm leak}}2
\le\epsilon .
\end{equation}

\indent
For $q_R>0$, the identity $A=q_R\rho_R$ gives $\rho_R-\rho=(A-\rho)+p_{\rm leak}\rho_R$. The triangle inequality and $\|\rho_R\|_1=1$ imply $T(\rho_R,\rho)\le\frac12\|A-\rho\|_1+p_{\rm leak}/2$. Combining this with Eq.~\eqref{eq:B.27} yields
\begin{equation}
\label{eq:B.28}
T(\rho_R,\rho)
\le
\epsilon.
\end{equation}

\indent
This completes the proof of Proposition B.3.\Bqed

\indent
Thus, for $q_R>0$, the normalized projected conditional state still approximates an ideal state supported on $\mathcal H_R$ within the original tolerance $\epsilon$. Equations~\eqref{eq:B.25}--\eqref{eq:B.28} establish geometric stability. The relative-entropy decomposition in Appendix~\ref{app:B.4} gives conservative discounting of the state-side linear leading term; Appendix~\ref{app:B.6} states the conditions for obtaining a process certificate.

\subsection{Levels of State-Side Values and the Certificate Interface}
\label{app:B.6}

\indent
The three state-side quantities in this appendix retain different forms of average information. The quantity $C_R(\rho_R)$ in Eq.~\eqref{eq:B.3} measures the conditional state's structural gap within the reference domain, with occurrence weight $q_R$. The quantity $C_{R,\delta}^{\rm reg}(\rho_{\rm phys})$ in Eq.~\eqref{eq:B.6} compares the full physical state with the regularized background specified by $\delta,\tau_\perp$. The quantity $C_{\rm leak}(\rho_{\rm phys})=q_RC_R(\rho_R)$ in Eq.~\eqref{eq:B.19} retains the structure within the reference domain at its actual weight, giving the conservative discounting bound in Eq.~\eqref{eq:B.18}.

\indent
A process certificate starts from a certified input $\underline{\mathcal I}_j(r)\le D_{\max}^{r}$ for the state--reference pair after treatment. This input may come from a direct one-shot witness, or from a conversion using relative entropy, spectral skew, and smoothing loss established for that pair.

Approximation errors for the full and conditional states must be aligned with the program radius $r$, smoothing radius, and preparation tolerance. The success probability and implementation cost of physical conditioning enter the corresponding generation model. Reference upgrading requires the model and program interface to be established together, as in Appendix~\ref{app:B.3}. Once the model, target state, and error semantics are aligned, Appendix~\ref{app:C} performs cost inversion and combines the bounds. Finite-sample propagation and audit records are treated in Appendices~\ref{app:D.5}--\ref{app:D.6}.

\indent
Certificate records retain $q_R$ or $p_{\rm leak}$ for the conditional-state route, and also $\delta$ and $\tau_\perp$ for the support-regularized route. Substantial leakage triggers a joint review of the reference subspace, preparation noise, and leakage model. If structure outside the reference domain is reproducible, Appendix~\ref{app:F} provides routes for testing or upgrading the reference model.

Appendix~\ref{app:A} establishes a rigorous complexity lower bound under ideal support. Appendix~\ref{app:B} decomposes the physical output into structure within the reference domain, leakage weight, structure in the complement, and coherence between the domains, and provides conservative interfaces through which they enter the auditing framework of the main text.

\endgroup

\begingroup

\setlength{\parindent}{1.5em}
\setlength{\parskip}{1pt plus 0.3pt minus 0.2pt}
\frenchspacing
\linespread{1.005}\selectfont
\setlength{\abovedisplayskip}{8.5pt plus 1pt minus 2pt}
\setlength{\belowdisplayskip}{8.5pt plus 1pt minus 2pt}
\setlength{\abovedisplayshortskip}{5.5pt plus 1pt minus 1pt}
\setlength{\belowdisplayshortskip}{6.5pt plus 1pt minus 1pt}
\setlength{\jot}{2pt}
\setlength{\emergencystretch}{1.25em}
\clubpenalty=10000
\widowpenalty=10000
\displaywidowpenalty=10000
\brokenpenalty=5000
\raggedbottom

\newcommand{\Cstatement}[1]{%
  \par\addvspace{0.4\baselineskip}%
  \Needspace{4\baselineskip}%
  \indent\textbf{#1}\hspace{0.5em}\ignorespaces
}
\newcommand{\Cproof}{%
  \par\addvspace{0.25\baselineskip}%
  \Needspace{3\baselineskip}%
  \indent\textbf{Proof.}\hspace{0.5em}\ignorespaces
}
\newcommand{\Cqed}{%
  \unskip\nobreak\hspace{0.35em}\ensuremath{\square}%
  \par\addvspace{0.3\baselineskip}%
}
\makeatletter
\renewcommand{\subsection}{\@startsection{subsection}{2}{\z@}%
  {14pt plus 1pt minus 1pt}{8pt plus 0.5pt minus 1pt}%
  {\normalfont\small\bfseries\centering}}
\makeatother
\let\COriginalSubsection\subsection
\renewcommand{\subsection}[1]{%
  \par\Needspace{3\baselineskip}%
  \COriginalSubsection{#1}%
}

\section{Explicit Evaluation of the Main Lower Bound and Combination of Audit Certificates}
\label{app:C}

\indent
Appendix~\ref{app:A} proves that faithful transcription constrains the cost of every admissible generation path by the target state's smooth one-shot information gap relative to the structured vacuum. Monotone inversion of the transcription envelope then gives a lower bound on universal optimal quantum circuit complexity. This appendix organizes that result into a certificate interface that supports explicit computation and combination of bounds.

The calculation starts from a certified information input: an analytical calculation, spectral data, or a finite-sample route supplies a conservative lower bound on $D_{\max}^{r}(\rho\Vert\sigma_R)$, and the transcription envelope converts that information into a lower bound on cost in the model's declared units. In the canonical atomic-slot model, cost is measured in atomic resource slots. General monotone envelopes and canonical atomic-slot models, under exact or approximate transcription, share this structure. Once their one-shot, model, and error conditions have been met, hypothesis testing, arbitrary effects, low-rank projections, reference-compatible measurements, and windowed RCC can therefore be compared on the same complexity scale and combined into a final certificate on a joint confidence event.

\indent
We use the ideal-support and normalized trace-distance smoothing conventions of Appendix~\ref{app:A}, fix the target state, structured vacuum, and model $\widehat{\mathfrak M}_R=(\mathfrak M_R,U_R)$, and write $g_R=\log\Gamma_R$. All logarithms are base 2.

\subsection{Certified Information Inputs and General Inversion}
\label{app:C.1}

\indent
To apply the same cost inversion to different forms of final-state evidence, fix a target generation tolerance $r\in[0,1]$. A quantity $\underline{\mathcal I}_j(r)$ supplied by route $j$ is a certified information input if, under its declared analytical conditions or on its declared statistical event,
\begin{equation}
\label{eq:C.1}
\underline{\mathcal I}_j(r)
\le
D_{\max}^{r}(\rho\Vert\sigma_R).
\end{equation}

\indent
It is measured in bits. It may equal the exact spectral value or be a conservative lower bound obtained from final-state effects, statistical endpoints, or the data-processing inequality.

\Cstatement{Corollary C.1 (Inversion certificates from certified information inputs).}
Suppose the fixed model satisfies $\mathsf{RCon}$, exact $\mathsf{TC}$, and semantic $\mathsf{RA}$, and the transcription envelope $\Lambda_R$ is nondecreasing on the admissible cost set $\mathcal L_R$. Every input satisfying Eq.~\eqref{eq:C.1} gives
\begin{equation}
\label{eq:C.2}
\boxed{
C_{\rm opt}^{(r)}(\rho;\mathfrak M_R)
\ge
\Lambda_R^{\leftarrow}
\!\left[
\underline{\mathcal I}_j(r)-\chi_U
\right].
}
\end{equation}

\Cproof
Theorem A.2 gives
\begin{equation*}
C_{\rm opt}^{(r)}
\ge
\Lambda_R^{\leftarrow}
\!\left[
D_{\max}^{r}(\rho\Vert\sigma_R)-\chi_U
\right].
\end{equation*}

\indent
The left generalized inverse is nondecreasing, so substituting Eq.~\eqref{eq:C.1} gives Eq.~\eqref{eq:C.2}.\Cqed

\indent
In the canonical atomic-slot model, $\Lambda_R(L)=\Phi_{g_R,a}(L)+\gamma$. Write
\begin{equation}
\label{eq:C.3}
\beta_U:=\gamma+\chi_U,
\qquad
Y_j(r)
:=
\underline{\mathcal I}_j(r)-\beta_U.
\end{equation}

\indent
The final route-level lower bound is then
\begin{equation}
\label{eq:C.4}
\boxed{
L_j^{\rm final}(r)
:=
F_{g_R,a}\!\left[Y_j(r)\right],
\qquad
C_{\rm opt}^{(r)}(\rho;\mathfrak M_R)
\ge
L_j^{\rm final}(r).
}
\end{equation}

\indent
Equations~\eqref{eq:C.3} and \eqref{eq:C.4} convert the certified information $\underline{\mathcal I}_j$ into the input $Y_j$ after subtracting model constants, then into the physical cost lower bound $L_j^{\rm final}$. Equation~\eqref{eq:C.2} gives the inversion for a general envelope.

\subsection{The Explicit Inverse in Canonical Models}
\label{app:C.2}

\indent
To evaluate the cost lower bound in Eq.~\eqref{eq:C.4} explicitly, consider the canonical code-length--cost function
\begin{equation}
\label{eq:C.5}
\Phi_{g,a}(L)
=
gL+a\log\max\{1,L\},
\qquad
g>0,
\quad
a\ge0.
\end{equation}

\indent
It is continuous and strictly increasing on $[0,\infty)$. Appendices~\ref{app:A.7}--\ref{app:A.8} give its exact inverse, conservative estimates, and asymptotic form. For use in certificate calculations, we collect the result as
\begin{equation}
\label{eq:C.6}
F_{g,a}(y)
=
\begin{cases}
0,
&y\le0,
\\[3pt]
y/g,
&0<y\le g,
\\[7pt]
\dfrac{a}{g\ln2}
W_0\!\left[
\dfrac{g\ln2}{a}
\exp\!\left(\dfrac{y\ln2}{a}\right)
\right],
&y>g,\ a>0,
\\[11pt]
y/g,
&y>g,\ a=0.
\end{cases}
\end{equation}

\indent
Here $W_0$ is the principal branch of Lambert-$W$. The breakpoint $y=g$ corresponds to $L=1$; nonpositive inputs return zero, and the extended-real input $+\infty$ returns $+\infty$. For numerical evaluation when $y>g$, interval root finding for the strictly monotone equation $gL+a\log L=y$ on $[1,y/g]$ avoids explicit exponentiation. Bounds on function values that include rounding error maintain an interval enclosing the root; its lower endpoint is a conservative inverse value.

\indent
Without special functions, one may use the conservative evaluator
\begin{equation}
\label{eq:C.7}
\underline F_{g,a}(y)
:=
\begin{cases}
0,
&y\le0,
\\[2pt]
y/g,
&0<y\le g,
\\[6pt]
\displaystyle
\max\left\{
1,
\frac{y-a\log(y/g)}{g}
\right\},
&y>g,
\end{cases}
\end{equation}
which satisfies
\begin{equation}
\label{eq:C.8}
F_{g,a}(y)
\ge
\underline F_{g,a}(y).
\end{equation}

\indent
Replacing $F_{g_R,a}$ in Eq.~\eqref{eq:C.4} by $\underline F_{g_R,a}$ therefore yields a valid, more conservative complexity certificate. For fixed $g>0$ and $a\ge0$, the exact inverse has the asymptotic form, as $y\to\infty$,
\begin{equation}
\label{eq:C.9}
F_{g,a}(y)
=
\frac{y}{g}
-
\frac{a}{g}\log\frac{y}{g}
+
O\!\left(\frac{\log y}{y}\right).
\end{equation}

\indent
This expansion shows that the atomic control bandwidth $g$ determines the leading conversion from information to physical cost, while self-delimiting code length contributes only a subleading logarithmic correction. If admissible costs are integer atomic slots, the continuous inverse may be rounded up. For a general discrete cost set, use the left generalized inverse in Eq.~\eqref{eq:C.2} directly.

\subsection{Spectral Values, Classical Measurements, and Window Inputs}
\label{app:C.3}

\indent
The target-state spectrum directly supplies an exact one-shot input. Accessible measurement or window values provide conservative inputs once spectral skew and smoothing loss are controlled. Let $t_r$ be the spectral clipping threshold of Theorem A.1. Then
\begin{equation}
\label{eq:C.10}
\underline{\mathcal I}_{\rm spec}(r)
=
D_{\max}^{r}(\rho\Vert\sigma_R)
=
\log(d_Rt_r).
\end{equation}

\indent
The same input can also be expressed in terms of average information, spectral skew, and smoothing loss as
\begin{equation}
\label{eq:C.11}
D_{\max}^{r}(\rho\Vert\sigma_R)
=
D(\rho\Vert\sigma_R)
+
\Delta_{\rm spec}(\rho)
-
\delta_{\rm sm}(\rho,r).
\end{equation}

\indent
Thus, if under the same analytical conditions or on the same joint statistical event,
\begin{equation*}
D(\rho\Vert\sigma_R)\ge D_L,
\qquad
\Delta_{\rm spec}(\rho)\ge\Delta_L,
\qquad
\delta_{\rm sm}(\rho,r)\le\delta_U(r),
\end{equation*}
we may take
\begin{equation}
\label{eq:C.12}
\underline{\mathcal I}_{\rm state}(r)
:=
D_L+\Delta_L-\delta_U(r).
\end{equation}

\indent
Appendix~\ref{app:A.8} states the smoothing-regularity conditions and their domain of applicability. Throughout the declared precision interval $0<r\le r_0<1$, a target-state family verified to satisfy Eq.~\eqref{eq:A.51} admits $\delta_U(r)=c_{\rm sm}\log(1/r)$. At $r=0$, use $\delta_{\rm sm}(\rho,0)=0$ directly, or the exact spectral input in Eq.~\eqref{eq:C.10}. When spectral information is available, the exact $\delta_{\rm sm}$ generally gives a stronger result. For pure states, one may also use the full smoothing formula in Eq.~\eqref{eq:A.53} directly.

\indent
A reference-compatible measurement gives, by data processing,
\begin{equation*}
D(\rho\Vert\sigma_R)
\ge
D_{{\rm cl},L}.
\end{equation*}

\indent
After including the corresponding smoothing loss and lower bound on spectral skew, its certified input is
\begin{equation}
\label{eq:C.13}
\underline{\mathcal I}_{\rm cl}(r)
:=
D_{{\rm cl},L}
+
\Delta_L
-
\delta_U(r).
\end{equation}

\indent
Without an independent lower bound on spectral skew, one may set $\Delta_L=0$. Until an upper bound on smoothing loss is available, $D_{{\rm cl},L}/g_R$ remains a conservative lower bound on the state-side leading term. A corresponding upper bound or a proof of smoothing regularity then gives a one-shot information input for Eq.~\eqref{eq:C.4}.

\indent
Windowed RCC resolves the same structure by object domain. If
\begin{equation*}
\underline C_{R,\Xi}
\le
C_{R,\Xi}(\rho),
\end{equation*}
then the data-processing relation in Appendix~\ref{app:E.9} gives
\begin{equation*}
D(\rho\Vert\sigma_R)
\ge
g_R\underline C_{R,\Xi}.
\end{equation*}

\indent
The corresponding certified input is
\begin{equation}
\label{eq:C.14}
\underline{\mathcal I}_{\Xi}(r)
:=
g_R\underline C_{R,\Xi}
+
\Delta_L
-
\delta_U(r).
\end{equation}

\indent
Equations~\eqref{eq:C.13} and \eqref{eq:C.14} convert the average information visible to classical measurements and windows into a one-shot lower bound for the original target state. The same inverse then yields a process-complexity certificate.

\subsection{Arbitrary Effects, Hypothesis Testing, and Projection Inputs}
\label{app:C.4}

\indent
One-shot routes can bypass separate estimates of average information and smoothing loss. Let $0\le Q\le I$ be a final-state effect declared in advance. Under the same analytical conditions or on the same joint confidence event, obtain
\begin{equation*}
{\rm Tr}(Q\rho)\ge a_L,
\qquad
{\rm Tr}(Q\sigma_R)\le b_U.
\end{equation*}

\indent
When $a_L>r$, Lemma A.6 gives, for every smoothing candidate satisfying $T(\tau,\rho)\le r$ and every $\lambda$ satisfying $\tau\le2^\lambda\sigma_R$,
\begin{equation*}
\lambda
\ge
\log\frac{a_L-r}{b_U}.
\end{equation*}

\indent
Taking the infimum over $\lambda$ and then over smoothing candidates $\tau$ gives the certified information input
\begin{equation}
\label{eq:C.15}
\boxed{
\underline{\mathcal I}_{Q}(r)
:=
\log\frac{a_L-r}{b_U}
\le
D_{\max}^{r}(\rho\Vert\sigma_R).
}
\end{equation}

\indent
Substituting Eq.~\eqref{eq:C.15} into Eq.~\eqref{eq:C.2} or Eq.~\eqref{eq:C.4} yields a complexity lower bound from any effect that reliably distinguishes the target state from the structured vacuum. Under the conditions of Corollary A.4, optimizing over all effects using exact responses and then applying the inversion in Eq.~\eqref{eq:C.4} recovers the main theorem's lower bound. A zero denominator with a positive numerator is treated in the extended-real sense. Finite-sample support mismatch and model diagnostics follow Appendix~\ref{app:D.5}.

\indent
The hypothesis-testing route is the standard optimized form of this effect certificate. If
\begin{equation*}
D_H^\eta(\rho\Vert\sigma_R)
\ge
D_{H,L}^{\eta},
\qquad
\eta+r<1,
\end{equation*}
then Eq.~\eqref{eq:A.59} gives
\begin{equation}
\label{eq:C.16}
\boxed{
\underline{\mathcal I}_{H}(r,\eta)
:=
D_{H,L}^{\eta}
-
\log\frac1{1-\eta-r}
\le
D_{\max}^{r}(\rho\Vert\sigma_R).
}
\end{equation}

\indent
For the same finite-sample effect, if $D_{H,L}^{\eta}=-\log b_U$ and $a_L\ge1-\eta$, then
\begin{equation}
\label{eq:C.17}
\underline{\mathcal I}_{Q}(r)
\ge
\underline{\mathcal I}_{H}(r,\eta).
\end{equation}

\indent
The direct-effect route retains the acceptance probability actually certified. The hypothesis-testing route provides a standard information-theoretic formulation that permits optimization over all tests. Corollary A.3 proves that the two formulations agree after full optimization using exact acceptance probabilities and reference responses.

Low-rank projections are a geometric special case of Eq.~\eqref{eq:C.15}. Let $P\le\Pi_R$ have rank $k>0$, and let $p_L$ be a lower endpoint for the target occupation probability. The structured-vacuum baseline occupation is $k/d_R$. Using $D_{\max}^{r}\ge0$ under normalized smoothing, the occupation input of Appendix~\ref{app:D.3} is written in terms of
\begin{equation*}
\Psi_{k,d}^{s}(x)
:=
\begin{cases}
\displaystyle
\left[
\log\dfrac{(x-s)d}{k}
\right]_+,
&x>s,
\\[3mm]
0,
&x\le s.
\end{cases}
\end{equation*}

\indent
For any $r\le r_{\rm sm}\le1$, Appendix~\ref{app:D.3} and the expansion of smoothing neighborhoods with increasing radius give
\begin{equation}
\label{eq:C.18}
\underline{\mathcal I}_{P}(r)
:=
\Psi_{k,d_R}^{r_{\rm sm}}(p_L)
=
D_{\max,L}^{r_{\rm sm}}(P)
\le
D_{\max}^{r_{\rm sm}}(\rho\Vert\sigma_R)
\le
D_{\max}^{r}(\rho\Vert\sigma_R).
\end{equation}

\indent
The standard choice is $r_{\rm sm}=r$. A positive one-shot input results when the occupation lower endpoint, after subtraction of the smoothing margin, exceeds $k/d_R$. If this input also exceeds the fixed model overhead $\beta_U$, Eq.~\eqref{eq:C.4} gives a positive route-level complexity lower bound.

\subsection{Approximate Transcription and Optimization of Compilation Precision}
\label{app:C.5}

\indent
Fix a circuit tolerance $\epsilon_{\rm circ}$ and choose compilation precision to balance information loss against description overhead. Approximate $\mathsf{TC}$ includes the compilation error $\delta_{\rm cmp}$ in the program radius
\(r_{\rm tot}
=
\epsilon_{\rm circ}+\delta_{\rm cmp},
\)
and incurs the code-length correction $b\log(1/\delta_{\rm cmp})$. For route $j$, let $\Delta_{j,{\rm cmp}}$ collect all compilation precisions satisfying the following conditions simultaneously:
\begin{equation*}
\delta_{\rm cmp}\in\Delta_{\rm TC},
\qquad
\epsilon_{\rm circ}+\delta_{\rm cmp}\le1,
\qquad
\underline{\mathcal I}_j
(\epsilon_{\rm circ}+\delta_{\rm cmp})
\text{ is certified on the same declared event}.
\end{equation*}

\Cstatement{Proposition C.2 (A route-level evaluator for approximate transcription).}
Suppose the fixed model satisfies $\mathsf{RCon}$, approximate $\mathsf{TC}$, and semantic $\mathsf{RA}$, and $\Delta_{j,{\rm cmp}}\neq\varnothing$. A general monotone envelope then gives
\begin{equation}
\label{eq:C.19}
\boxed{
C_{\rm opt}^{(\epsilon_{\rm circ})}
(\rho;\mathfrak M_R)
\ge
\sup_{\delta_{\rm cmp}\in\Delta_{j,{\rm cmp}}}
\Lambda_R^{\leftarrow}\!\Big[
\underline{\mathcal I}_j
(\epsilon_{\rm circ}+\delta_{\rm cmp})
-\chi_U
-b\log(1/\delta_{\rm cmp})
\Big].
}
\end{equation}

\indent
For the canonical atomic-slot model, the corresponding expression is
\begin{equation}
\label{eq:C.20}
\boxed{
L_{j,{\rm cmp}}^{\rm final}
:=
\sup_{\delta_{\rm cmp}\in\Delta_{j,{\rm cmp}}}
F_{g_R,a}\!\Big[
\underline{\mathcal I}_j
(\epsilon_{\rm circ}+\delta_{\rm cmp})
-\beta_U
-b\log(1/\delta_{\rm cmp})
\Big],
}
\end{equation}

\indent
with
\begin{equation*}
C_{\rm opt}^{(\epsilon_{\rm circ})}
(\rho;\mathfrak M_R)
\ge
L_{j,{\rm cmp}}^{\rm final}.
\end{equation*}

\Cproof
\looseness=-1
For every $\delta_{\rm cmp}\in\Delta_{j,{\rm cmp}}$, Eq.~\eqref{eq:A.27} gives the total radius and a code-length upper bound, while semantic $\mathsf{RA}$ constrains program length through the certified information input. Combining the two and applying monotone inversion gives a lower bound at fixed precision. Every admissible circuit obeys all bounds on the certified domain simultaneously. Taking the supremum over precision, then the infimum over admissible circuit costs, proves the claim.\Cqed

\indent
Equation~\eqref{eq:C.20} retains two competing effects: decreasing $\delta_{\rm cmp}$ reduces the total smoothing radius, so the exact one-shot gap cannot decrease, but it increases the cost of describing the precision. The precision dependence of the certified input $\underline{\mathcal I}_j$ depends on the route. In practice, the calculation uses the certified value at each precision and selects the strongest lower bound within the actual certified domain.

\indent
For the arbitrary-effect route, this balance can also be solved explicitly. Fix $a_L$ and $b_U>0$ valid on the same event, and write
\(A_Q
:=
a_L-\epsilon_{\rm circ}
>0.
\)

\indent
The input to the inverse in Eq.~\eqref{eq:C.20} is
\begin{equation}
\label{eq:C.21}
-\log b_U
-\beta_U
+\log(A_Q-\delta_{\rm cmp})
+b\log\delta_{\rm cmp}.
\end{equation}

\indent
If the certified domain covers the full interval $0<\delta_{\rm cmp}<A_Q$, then for $b>0$ the expression in Eq.~\eqref{eq:C.21} is strictly concave in $\delta_{\rm cmp}$. Setting its derivative to zero gives $b/\delta_{\rm cmp}=1/(A_Q-\delta_{\rm cmp})$, so the optimal precision is
\begin{equation}
\label{eq:C.22}
\delta_{\rm cmp}^{\ast}
=
\frac{b}{b+1}A_Q.
\end{equation}

\indent
Define the precision penalty
\begin{equation}
\label{eq:C.23}
\mathcal P_b(A)
:=
(b+1)\log\frac{b+1}{A}
-b\log b,
\end{equation}
with the continuous-limit convention $0\log0=0$, so that $\mathcal P_0(A)=\log(1/A)$. The optimized arbitrary-effect certificate is then
\begin{equation}
\label{eq:C.24}
\boxed{
C_{\rm opt}^{(\epsilon_{\rm circ})}
(\rho;\mathfrak M_R)
\ge
F_{g_R,a}\!\left[
-\log b_U
-\beta_U
-\mathcal P_b(A_Q)
\right].
}
\end{equation}

\indent
When $b=0$, the supremum is approached as $\delta_{\rm cmp}\downarrow0$. Setting $a_L=1-\eta$ and $-\log b_U=D_{H,L}^{\eta}$ in Eq.~\eqref{eq:C.24} immediately recovers the approximate-transcription hypothesis-testing certificate of Appendix~\ref{app:A.9}. For a restricted certified precision domain, Eq.~\eqref{eq:C.20} remains the complete result, with optimization confined to the actual domain.

\subsection{Joint Confidence and Certificates from Multiple Routes}
\label{app:C.6}

\indent
To combine multiple forms of evidence under a common coverage guarantee, use the joint statistical event constructed in Appendix~\ref{app:D},
\begin{equation}
\label{eq:C.25}
\mathcal E_{\rm stat}
=
\bigcap_{k=1}^{K}\mathcal E_k,
\qquad
{\rm Pr}(\mathcal E_{\rm stat})
\ge
1-\delta_{\rm stat}.
\end{equation}

\indent
On this event, Appendices~\ref{app:C.3}--\ref{app:C.4} convert the different forms of final-state evidence into certified information inputs. Leakage routes follow Appendix~\ref{app:B.6} in treating the state--reference pair, establishing the one-shot input, and aligning model and error conventions. Records retain the reference treatment actually used.

Let the finite set $\mathcal J_{\rm cert}$ collect all routes valid on the same event that lower-bound the same target quantity $C_{\rm opt}^{(\epsilon_\star)}(\rho;\mathfrak M_R)$. We use $L_j^{\rm final}$ for route-level lower bounds from either exact or approximate transcription, including general envelopes and their canonical special case. The final certificate is
\begin{equation}
\label{eq:C.26}
\boxed{
L_{\rm RCC}
=
\begin{cases}
\displaystyle
\max_{j\in\mathcal J_{\rm cert}}
L_j^{\rm final},
&\mathcal J_{\rm cert}\neq\varnothing,
\\[3mm]
0,
&\mathcal J_{\rm cert}=\varnothing,
\end{cases}
\qquad
C_{\rm opt}^{(\epsilon_\star)}
(\rho;\mathfrak M_R)
\ge
L_{\rm RCC}.
}
\end{equation}

\indent
On the event in Eq.~\eqref{eq:C.25}, every valid route-level value is bounded above by the same complexity. Their maximum is therefore still a lower bound, with coverage probability at least $1-\delta_{\rm stat}$. The routes may be statistically dependent. An empty route set returns zero, while retaining the raw information inputs, state-side values, and conditions still to be met.

\indent
Subtracting model constants and inverting the transcription envelope thus turns certified final-state information into a computable lower bound on universal optimal quantum circuit complexity. Approximate transcription also permits optimization of compilation precision within the certified domain. Equation~\eqref{eq:C.26} combines valid routes for the same generation task into a single RCC certificate; Appendix~\ref{app:D} supplies its statistical inputs and coverage guarantee.

\endgroup

\begingroup

\setlength{\parindent}{1.5em}
\setlength{\parskip}{1pt plus 0.3pt minus 0.2pt}
\frenchspacing
\linespread{1.005}\selectfont
\setlength{\abovedisplayskip}{8.5pt plus 1pt minus 2pt}
\setlength{\belowdisplayskip}{8.5pt plus 1pt minus 2pt}
\setlength{\abovedisplayshortskip}{5.5pt plus 1pt minus 1pt}
\setlength{\belowdisplayshortskip}{6.5pt plus 1pt minus 1pt}
\setlength{\jot}{2pt}
\setlength{\emergencystretch}{1.25em}
\clubpenalty=10000
\widowpenalty=10000
\displaywidowpenalty=10000
\brokenpenalty=5000
\raggedbottom

\newcommand{\Dstatement}[1]{%
  \par\addvspace{0.4\baselineskip}%
  \Needspace{4\baselineskip}%
  \indent\textbf{#1}\hspace{0.5em}\ignorespaces
}
\newcommand{\Dproof}{%
  \par\addvspace{0.25\baselineskip}%
  \Needspace{3\baselineskip}%
  \indent\textbf{Proof.}\hspace{0.5em}\ignorespaces
}
\newcommand{\Dqed}{%
  \unskip\nobreak\hspace{0.35em}\ensuremath{\square}%
  \par\addvspace{0.3\baselineskip}%
}
\makeatletter
\renewcommand{\subsection}{\@startsection{subsection}{2}{\z@}%
  {14pt plus 1pt minus 1pt}{8pt plus 0.5pt minus 1pt}%
  {\normalfont\small\bfseries\centering}}
\makeatother
\let\DOriginalSubsection\subsection
\renewcommand{\subsection}[1]{%
  \par\Needspace{3\baselineskip}%
  \DOriginalSubsection{#1}%
}

\section{Finite-Sample Statistics for Audit Protocols}
\label{app:D}

\indent
Section~\ref{sec:4} introduces three final-state audit routes, and Appendix~\ref{app:C} converts their theoretical inputs into nonrecursive complexity lower bounds. This appendix supplies the statistical layer between them: finite samples, readout calibration, and leakage observations define confidence sets covering the true physical parameters, from which the most conservative admissible inputs are extracted. Finite data thus determine the certificate's resolution while leaving the definition of RCC and the main theorem intact.

We take a normalized state $\rho$ supported on $\mathcal H_R$ and the structured vacuum $\sigma_R$ as the baseline; Appendix~\ref{app:D.5} handles readout and leakage. When using the reference treatments of Appendix~\ref{app:B}, general-effect and feasible-set methods apply to the corresponding state--reference pair. The projection baseline $k/d_R$ and entropy formulas for a uniform reference are used under their respective conditions. After a reference change, the relevant baselines are recalculated and the treatment is recorded.

\subsection{Statistical Inputs, One-Sided Confidence, and Sampling Assumptions}
\label{app:D.1}

\indent
To associate statistical endpoints with a definite physical generation task, fix the reference-consistent model and its audit parameters before sampling,
\begin{equation}
\label{eq:D.1}
\widehat{\mathfrak M}_R
=
(\mathfrak M_R,U_R),
\qquad
R,\qquad
\Pi_R,\qquad
\mathcal H_R,\qquad
d_R,\qquad
\sigma_R=\frac{\Pi_R}{d_R},\qquad
\Gamma_R .
\end{equation}

\indent
Model qualification uses the existing verification of $\mathsf{RCon}$, full $\mathsf{TC}$, and semantic $\mathsf{RA}$. When using the $\mathsf{RefCert}\Rightarrow\mathsf{RA}$ route, record the corresponding certificates and constants uniform across the model family. Qualification evidence is independent of the current final-state sampling; statistical failure probabilities apply to the subsequent parameter estimation.

\indent
The finite-sample formulas below assume independent, identically distributed repeated preparations and fixed measurement settings. Under drift, temporal dependence, or adaptive sampling, first declare the audited object, such as an average state or a family of states resolved in time. Then replace the relevant confidence sets with martingale bounds, confidence sequences, or robust intervals valid for that sampling model. The subsequent feasible-set propagation and theoretical conversion in Appendix~\ref{app:C} remain unchanged.

All sample sizes below are positive integers, in particular $N,N_R\ge1$, and all statistical failure parameters lie in $(0,1)$. Suppose a certificate uses $K$ statistical confidence events $\mathcal E_1,\ldots,\mathcal E_K$, with assigned failure probabilities $\delta_1,\ldots,\delta_K$. If
\begin{equation}
\label{eq:D.2}
\sum_{k=1}^{K}\delta_k
\le
\delta_{\rm stat},
\qquad
{\rm Pr}(\mathcal E_k)
\ge
1-\delta_k,
\end{equation}
then the union bound gives ${\rm Pr}(\bigcap_k\mathcal E_k)\ge1-\delta_{\rm stat}$, without requiring independence between the events.

\indent
Statistical failure probabilities control coverage under random sampling alone. Analytical truncation, floating-point rounding, deterministic calibration envelopes, and optimizer tolerances directly enlarge feasible sets or weaken numerical lower bounds; they do not consume $\delta_{\rm stat}$. The preparation tolerance $\epsilon$, smoothing radius $\epsilon_{\rm sm}$, and testing parameter $\eta$ enter the one-shot theorems or test definitions and are also recorded separately from statistical failure probabilities.

Acceptance, projection occupation, and retention probabilities can all be estimated from binary counts. For a Bernoulli event with success probability $p$, observe $k$ successes in $N$ trials and write
\begin{equation}
\label{eq:D.3}
\widehat p
=
\frac{k}{N}.
\end{equation}

\indent
One-sided endpoints $p_L(\delta)$ and $p_U(\delta)$ satisfy, respectively,
\begin{equation}
\label{eq:D.4}
{\rm Pr}\!\left[p\ge p_L(\delta)\right]
\ge
1-\delta,
\qquad
{\rm Pr}\!\left[p\le p_U(\delta)\right]
\ge
1-\delta .
\end{equation}

\indent
A direct conservative choice is given by the Hoeffding endpoints~\cite{Hoeffding1963_ProbabilityInequalities},
\begin{equation}
\label{eq:D.5}
p_L(\delta)
=
\max\!\left\{
0,\widehat p-\sqrt{\frac{\ln(1/\delta)}{2N}}
\right\},
\qquad
p_U(\delta)
=
\min\!\left\{
1,\widehat p+\sqrt{\frac{\ln(1/\delta)}{2N}}
\right\}.
\end{equation}

\indent
Here $\ln$ denotes the natural logarithm; $\log$ in information quantities throughout the paper remains base $2$. One-sided intervals with finite-sample coverage guarantees, such as Clopper--Pearson intervals~\cite{ClopperPearson1934_BinomialLimits}, may replace Eq.~\eqref{eq:D.5} directly.

All endpoints and feasible sets below follow the same principle of conservative propagation. If a random set $\mathcal F_\delta$ contains the true parameter $\theta$ with probability at least $1-\delta$, then every deterministic functional $f$ satisfies
\begin{equation*}
\inf_{\vartheta\in\mathcal F_\delta}f(\vartheta)
\le
f(\theta)
\le
\sup_{\vartheta\in\mathcal F_\delta}f(\vartheta)
\end{equation*}
on the same coverage event. The statistical task of Appendices~\ref{app:D.2}--\ref{app:D.5} is to construct an appropriate $\mathcal F_\delta$ for each physical route and take the most conservative extremum in the direction required by the lower or upper bound.

Test operators, candidate projection families, reference-compatible measurements, error parameters, and confidence budgets should be determined independently of the final audit samples. Witnesses may be learned from training data and audited on independent samples. Alternatively, simultaneous confidence bounds may cover a candidate family specified in advance. Selecting the strongest witness using audit data within such a simultaneously covered family preserves certificate validity.

\indent
These constructions provide a common coverage guarantee for $a_L,b_U$, $D_{H,L}^{\eta}$, $D_{\max,L}^{\epsilon_{\rm sm}}$, $D_{{\rm cl},L}$, and leakage endpoints. Appendix~\ref{app:C} then performs the corresponding one-shot conversion and cost inversion.

\subsection{The Hypothesis-Testing Route}
\label{app:D.2}

\indent
The hypothesis-testing route certifies the target acceptance probability while controlling false acceptance of the structured vacuum. Take $0\le\eta<1$ and a test effect $0\le Q\le I$ declared in advance, and write
\begin{equation}
\label{eq:D.6}
a={\rm Tr}(Q\rho),
\qquad
b={\rm Tr}(Q\sigma_R).
\end{equation}

\indent
Hypothesis-testing relative entropy is defined by
\begin{equation}
\label{eq:D.7}
D_H^\eta(\rho\Vert\sigma_R)
=
-\log
\inf_{0\le Q'\le I}
\left\{
{\rm Tr}(Q'\sigma_R):
{\rm Tr}(Q'\rho)\ge1-\eta
\right\}.
\end{equation}

\indent
Any test satisfying $a\ge1-\eta$ gives
\begin{equation}
\label{eq:D.8}
D_H^\eta(\rho\Vert\sigma_R)
\ge
-\log b.
\end{equation}

\indent
\looseness=-1
With finite samples, obtain $a_L(\delta_a)$ from target-state samples and construct a conservative upper bound $b_U$ on $b$. If
\begin{equation}
\label{eq:D.9}
a_L(\delta_a)\ge1-\eta,
\qquad
D_{H,L}^{\eta}:=-\log b_U,
\end{equation}

\indent
then, on the corresponding joint confidence event,
\begin{equation*}
D_H^\eta(\rho\Vert\sigma_R)
\ge
D_{H,L}^{\eta}.
\end{equation*}

\indent
The same endpoints also serve the direct-effect route of Appendix~\ref{app:C.4}: when $a_L>r$, $\log[(a_L-r)/b_U]$ directly lower-bounds $D_{\max}^{r}$. Equation~\eqref{eq:D.9} incorporates the acceptance requirement into the standard $D_H^\eta$ input.

When $b$ is known exactly from the reference model, it introduces no statistical failure probability. Certified numerical evaluation uses a deterministic upper envelope $b\le\overline b_{\rm det}$. If $b$ is estimated from finite reference samples or by Monte Carlo, use the statistical upper endpoint $b_U(\delta_b)$; the route's coverage probability is at least $1-\delta_a-\delta_b$. Zero false-acceptance counts still give a positive one-sided upper endpoint $b_U$, so $-\log b_U$ remains finite.

If $a_L<1-\eta$, the acceptance constraint of the current hypothesis-testing route has yet to be certified. A larger sample can improve resolution. Changing $\eta$ or switching routes must be covered by a new predeclared audit or a simultaneous-confidence protocol. Equation~\eqref{eq:D.9} measures how small the structured vacuum's false-acceptance probability can be certified to be while target acceptance remains high.

\subsection{Projection Witnesses and One-Shot Occupation Bounds}
\label{app:D.3}

\indent
To extract a one-shot lower bound from occupation of a low-rank subspace, let $P$ be a nonzero projection within the reference subspace and write
\begin{equation}
\label{eq:D.10}
P\le\Pi_R,
\qquad
k={\rm Tr}(P)>0,
\qquad
p={\rm Tr}(P\rho).
\end{equation}

\indent
The structured-vacuum baseline occupation of this projection is
\begin{equation}
\label{eq:D.11}
{\rm Tr}(P\sigma_R)
=
\frac{k}{d_R}.
\end{equation}

\indent
If $\rho\le2^\lambda\sigma_R$, then
\begin{equation}
\label{eq:D.12}
p
\le
2^\lambda{\rm Tr}(P\sigma_R)
=
2^\lambda\frac{k}{d_R}.
\end{equation}

\indent
Minimizing over $\lambda$ and using $D_{\max}\ge0$ for normalized states gives
\begin{equation}
\label{eq:D.13}
D_{\max}(\rho\Vert\sigma_R)
\ge
\left[
\log\frac{p\,d_R}{k}
\right]_+,
\end{equation}
with the extended-value convention $[\log0]_+=0$ when $p=0$.

\indent
To treat smoothing radii and endpoints together, define
\begin{equation}
\label{eq:D.14}
\Psi_{k,d}^{\epsilon}(x)
:=
\begin{cases}
\displaystyle
\left[
\log\frac{(x-\epsilon)d}{k}
\right]_+,
&x>\epsilon,
\\[2mm]
0,
&x\le\epsilon .
\end{cases}
\end{equation}

\Dstatement{Proposition D.1 (A one-shot witness from low-rank occupation).}
For any $0\le\epsilon_{\rm sm}\le1$ and every smoothing candidate satisfying $T(\tau,\rho)\le\epsilon_{\rm sm}$,
\begin{equation}
\label{eq:D.15}
{\rm Tr}(P\tau)
\ge
p-\epsilon_{\rm sm}.
\end{equation}
Hence
\begin{equation}
\label{eq:D.16}
D_{\max}^{\epsilon_{\rm sm}}(\rho\Vert\sigma_R)
\ge
\Psi_{k,d_R}^{\epsilon_{\rm sm}}(p).
\end{equation}

\indent
If $p_L(\delta_p)\le p$ on the corresponding confidence event, then
\begin{equation}
\label{eq:D.17}
D_{\max}^{\epsilon_{\rm sm}}(\rho\Vert\sigma_R)
\ge
\Psi_{k,d_R}^{\epsilon_{\rm sm}}
\!\bigl(p_L(\delta_p)\bigr)
=:
D_{\max,L}^{\epsilon_{\rm sm}}(P).
\end{equation}

\Dproof
Equation~\eqref{eq:D.15} follows from the trace-distance bound on measurement probability differences. Applying Eq.~\eqref{eq:D.13} to each smoothing candidate and taking the infimum over candidates gives Eq.~\eqref{eq:D.16}. The function $\Psi_{k,d_R}^{\epsilon_{\rm sm}}$ is nondecreasing in occupation probability. Substituting $p_L$ on the same event where $p_L\le p$ therefore gives Eq.~\eqref{eq:D.17}.\Dqed

\indent
The ratio $p/(k/d_R)$ directly compares the target state's occupation of the low-rank structural window with the structured-vacuum baseline. The one-shot value is positive when the occupation lower endpoint remains above that baseline after subtraction of the smoothing margin; otherwise the lower bound is zero.

\indent
For a projection family $P_1,\ldots,P_J$ specified in advance, write
\begin{equation}
\label{eq:D.18}
k_j={\rm Tr}(P_j)>0,
\qquad
p_j={\rm Tr}(P_j\rho).
\end{equation}

\indent
Construct simultaneously valid lower endpoints for the projections, with
\begin{equation}
\label{eq:D.19}
\sum_{j=1}^{J}\delta_j
\le
\delta_{\rm wit}.
\end{equation}

\indent
Then, with confidence at least $1-\delta_{\rm wit}$,
\begin{equation}
\label{eq:D.20}
D_{\max}^{\epsilon_{\rm sm}}(\rho\Vert\sigma_R)
\ge
\max_{1\le j\le J}
\Psi_{k_j,d_R}^{\epsilon_{\rm sm}}(p_{j,L}).
\end{equation}

\indent
Maximization takes place on the same simultaneous confidence event, so the witness statistics need not be independent. Projections generated from audit data can also enter Eq.~\eqref{eq:D.20} through an independent training--validation split or a simultaneous-confidence construction covering the entire selection procedure.

\subsection{Reference-Compatible Measurements, Classical Relative Entropy, and Entropy Upper Bounds}
\label{app:D.4}

\indent
To extract an average-information lower bound from accessible measurement distributions, let $\mathcal M$ be a fixed reference-compatible measurement or dephasing channel with $m$-dimensional classical outputs
\begin{equation}
\label{eq:D.21}
p=\mathcal M(\rho),
\qquad
u=\mathcal M(\sigma_R).
\end{equation}

\indent
The data-processing inequality for quantum relative entropy gives
\begin{equation}
\label{eq:D.22}
D(\rho\Vert\sigma_R)
\ge
D_{\rm cl}(p\Vert u),
\qquad
D_{\rm cl}(p\Vert u)
=
\sum_{i=1}^{m}p_i\log\frac{p_i}{u_i}.
\end{equation}

\indent
Classical relative entropy uses the extended-value conventions $0\log(0/v)=0$ for $v\ge0$ and $q\log(q/0)=+\infty$ for $q>0$. This route retains the quantum distinguishability visible to the measurement; any positive value is already a valid lower bound on the full quantum relative entropy.

\indent
Write $\mathcal M(\omega)_i={\rm Tr}(M_i\omega)$. Since $\sigma_R$ is full rank on $\mathcal H_R$ and $M_i\ge0$, the ideal-support model gives
\begin{equation}
\label{eq:D.23}
u_i=0
\quad\Longrightarrow\quad
\Pi_RM_i\Pi_R=0
\quad\Longrightarrow\quad
p_i=0.
\end{equation}

\indent
Positive observed mass on an outcome with zero reference probability therefore has a definite diagnostic meaning: it indicates support leakage, readout confusion, or incompatibility of the current measurement--reference model. The feasible sets of Appendix~\ref{app:D.5} address these cases.

\indent
Construct a joint confidence set $\mathcal F_\delta(\widehat p,\widehat u)$ from target-state and reference-state data such that
\begin{equation}
\label{eq:D.24}
{\rm Pr}\!\left[
(p,u)\in
\mathcal F_\delta(\widehat p,\widehat u)
\right]
\ge
1-\delta.
\end{equation}
Define
\begin{equation}
\label{eq:D.25}
D_{{\rm cl},L}(\delta)
:=
\inf_{(q,v)\in\mathcal F_\delta}
D_{\rm cl}(q\Vert v).
\end{equation}

\indent
Then, on that confidence event,
\begin{equation*}
D(\rho\Vert\sigma_R)
\ge
D_{\rm cl}(p\Vert u)
\ge
D_{{\rm cl},L}(\delta).
\end{equation*}

\indent
If $u$ is exactly determined by the known $\sigma_R$ and $\mathcal M$, take $\mathcal F_\delta=\mathcal C_p(\widehat p)\times\{u\}$. If $u$ is estimated from independent reference samples, one may use $\mathcal C_p\times\mathcal C_u$ with $\delta_p+\delta_u\le\delta$. With shared calibration data, retain the joint set directly. Relative entropy is jointly convex in $(q,v)$, so Eq.~\eqref{eq:D.25} is a convex optimization problem whenever $\mathcal F_\delta$ is convex. Numerical implementations use a certified lower bound on the exact infimum, such as a dual bound or an interval error bound; computational error is included in the deterministic error record.

\indent
When the reference output is uniform,
\begin{equation}
\label{eq:D.26}
u_i
=
\frac1m,
\qquad
1\le i\le m,
\end{equation}
classical relative entropy reduces to
\begin{equation}
\label{eq:D.27}
D_{\rm cl}(p\Vert u)
=
\log m-H(p),
\qquad
H(p)
=
-\sum_{i=1}^{m}p_i\log p_i.
\end{equation}

\indent
If $\mathcal M$ is complete reference-compatible dephasing on $\mathcal H_R$ and $m=d_R$, then
\begin{equation}
\label{eq:D.28}
D(\rho\Vert\sigma_R)
\ge
\log d_R-H(p).
\end{equation}

\indent
For finite samples, a total-variation confidence ball around the empirical distribution bounds the entropy estimation error. Write
\begin{equation}
\label{eq:D.29}
T_{\rm cl}(p,\widehat p)
=
\frac12\|p-\widehat p\|_1.
\end{equation}

\indent
For the finite outcome set $[m]$, we have the identity
\begin{equation}
\label{eq:D.30}
T_{\rm cl}(p,\widehat p)
=
\max_{A\subseteq[m]}
\bigl\{
p(A)-\widehat p(A)
\bigr\}.
\end{equation}

\indent
There are $2^m-2$ nonempty proper subsets. Applying the one-sided Hoeffding bound to the count for each subset $A$ and taking a union bound gives, for $m\ge2$ and $t>0$,
\begin{equation}
\label{eq:D.31}
{\rm Pr}\!\left[
T_{\rm cl}(p,\widehat p)\ge t
\right]
\le
(2^m-2)e^{-2Nt^2}.
\end{equation}

\indent
A directly computable radius is therefore
\begin{equation}
\label{eq:D.32}
t_N(\delta)
=
\min\!\left\{
1,
\sqrt{
\frac{\ln[(2^m-2)/\delta]}{2N}
}
\right\}.
\end{equation}

\indent
It guarantees $T_{\rm cl}(p,\widehat p)\le t_N(\delta)$ with probability at least $1-\delta$. When $m=1$, the distribution is unique and one may take $t_N=0$. Tighter multinomial confidence regions may replace this bound directly, provided they preserve the same finite-sample coverage.

\indent
To convert total-variation error into an entropy upper bound, define, for $m\ge2$, the modulus of continuity
\begin{equation}
\label{eq:D.33}
f_m(t)
=
t\log(m-1)+h_2(t),
\end{equation}
where $h_2$ is binary entropy. The Fannes--Audenaert bound~\cite{Audenaert2007_EntropyContinuity} gives $H(p)\le H(\widehat p)+f_m(T_{\rm cl})$ at the actual distance, while $f_m$ is increasing on $[0,1-1/m]$. This yields the piecewise entropy upper endpoint
\begin{equation}
\label{eq:D.34}
H_U(\delta)
=
\begin{cases}
\displaystyle
\min\!\left\{
\log m,\,
H(\widehat p)+f_m\!\bigl(t_N(\delta)\bigr)
\right\},
&
t_N(\delta)\le1-\dfrac1m,
\\[3mm]
\log m,
&
t_N(\delta)>1-\dfrac1m .
\end{cases}
\end{equation}

\indent
For a uniform reference output, one may take
\begin{equation}
\label{eq:D.35}
D_{{\rm cl},L}
=
\left[
\log m-H_U(\delta)
\right]_+.
\end{equation}

\indent
In particular, when $m=d_R$ and the measurement is complete reference dephasing,
\begin{equation*}
C_R(\rho)
\ge
\frac{
[\log d_R-H_U(\delta)]_+
}{
\log\Gamma_R
}.
\end{equation*}

\indent
Coarser windows or nonuniform reference outputs use Eq.~\eqref{eq:D.25}; data processing still guarantees that their observable values are conservative lower bounds.

\subsection{Readout Calibration, Leakage, and Feasible-Set Propagation}
\label{app:D.5}

\indent
Readout confusion and support leakage act at different physical levels. Readout confusion changes the channel from the ideal measurement distribution to the observed distribution; leakage changes the output state's actual weight in the reference subspace. They enter, respectively, the measurement feasible set and the reference treatment of Appendix~\ref{app:B}.

\indent
Let $p$ be the ideal distribution and $A$ a column-stochastic readout matrix. The observed distribution is
\begin{equation}
\label{eq:D.36}
y
=
Ap.
\end{equation}

\indent
Here $A_{ij}={\rm Pr}(i_{\rm obs}\mid j_{\rm ideal})$, so every column of $A$ sums to $1$.

\indent
From the data, obtain a confidence set $\mathcal C_{\rm obs}(\widehat y,\delta_y)$ for the observed distribution satisfying
\begin{equation}
\label{eq:D.37}
{\rm Pr}\!\left[
y\in\mathcal C_{\rm obs}(\widehat y,\delta_y)
\right]
\ge
1-\delta_y.
\end{equation}

\indent
When $A$ is known, the feasible set of ideal distributions is
\begin{equation}
\label{eq:D.38}
\mathcal C_{\rm read}
=
\left\{
q\in\Delta_m:
Aq\in\mathcal C_{\rm obs}
\right\}.
\end{equation}

\indent
The corresponding classical relative-entropy lower bound is
\begin{equation}
\label{eq:D.39}
D_{{\rm cl},L}^{\rm read}
=
\inf_{q\in\mathcal C_{\rm read}}
D_{\rm cl}(q\Vert u).
\end{equation}

\indent
If $A$ is estimated from finite calibration data, let $\mathcal A_{\rm adm}$ be the admissible matrix set incorporating both a statistical confidence set and a deterministic envelope for systematic error. Denote the statistical failure probability for coverage of the true $A$ by $\delta_A$; the deterministic envelope itself adds no failure probability. Then use
\begin{equation}
\label{eq:D.40}
\mathcal C_{\rm read}
=
\left\{
q\in\Delta_m:
\exists B\in\mathcal A_{\rm adm},
\;
Bq\in\mathcal C_{\rm obs}
\right\}.
\end{equation}

\indent
On the event covering both the true readout matrix and the observed distribution, the true $p$ belongs to the feasible set in Eq.~\eqref{eq:D.40}, with coverage probability at least $1-\delta_A-\delta_y$. If the reference distribution $u$ and target distribution $p$ share the same readout calibration, minimize $D_{\rm cl}(q\Vert v)$ over a joint feasible set in $(q,v,B)$. For an effect route, take $a_L=\inf a$ and $b_U=\sup b$ over the corresponding joint feasible set; for a projection route, take $p_L=\inf p$. These directions of optimization preserve the lower-bound direction of the information inputs required by Appendix~\ref{app:C}.

The data-compatible set also provides a direct check of model compatibility. If $\mathcal C_{\rm read}=\varnothing$, the observations are incompatible with the declared readout model and calibration envelope at the current confidence level. If the joint feasible set is nonempty but every feasible point violates the declared reference support and gives relative entropy $+\infty$, there is a support mismatch. Both cases are reported as model diagnostics, with recertification following review of the reference or readout model.

\indent
To discount conditional-state values by their actual occurrence weight, also estimate the retention and leakage probabilities of the physical state $\rho_{\rm phys}$:
\begin{equation}
\label{eq:D.41}
q_R
=
{\rm Tr}(\Pi_R\rho_{\rm phys}),
\qquad
p_{\rm leak}
=
1-q_R.
\end{equation}

\indent
If $N_R$ repeated measurements of $\Pi_R$ yield $k_R$ retention events, then
\begin{equation}
\label{eq:D.42}
\widehat q_R
=
\frac{k_R}{N_R}.
\end{equation}

\indent
Following Appendix~\ref{app:D.1}, construct one-sided endpoints in $[0,1]$,
\begin{equation}
\label{eq:D.43}
q_{R,L}(\delta_R)
\le
q_R,
\qquad
p_{{\rm leak},U}(\delta_R)
=
1-q_{R,L}(\delta_R).
\end{equation}

\indent
For $q_R>0$, let $\rho_R=\Pi_R\rho_{\rm phys}\Pi_R/q_R$, and let $C_{R,L}(\rho_R)$ be a conservative lower bound on the conditional state's structural gap. Since $C_R(\rho_R)\ge0$, define
\begin{equation}
\label{eq:D.44}
C_{R,L}^{+}(\rho_R)
:=
\left[
C_{R,L}(\rho_R)
\right]_+,
\qquad
0
\le
C_{R,L}^{+}(\rho_R)
\le
C_R(\rho_R).
\end{equation}

\indent
On the joint confidence event, the leakage-discounted value of Appendix~\ref{app:B} therefore has the finite-sample form
\begin{equation}
\label{eq:D.45}
C_{{\rm leak},L}
:=
q_{R,L}\,
C_{R,L}^{+}(\rho_R)
\le
C_{\rm leak}(\rho_{\rm phys})
=
q_R C_R(\rho_R).
\end{equation}

\indent
When $q_R=0$, $\rho_R$ is undefined and, following Appendix~\ref{app:B}, we set $C_{\rm leak}=C_{{\rm leak},L}=0$. When using samples from the retained branch, their effective measurement must correspond to the normalized projected state $\rho_R$ of Appendix~\ref{app:B}. Construct intervals with the same conditional coverage guarantee for every positive retained sample count; the law of total probability then gives an unconditional guarantee. Zero retained samples return an uncertified result. A joint construction that handles the random denominator may also be used directly. Both endpoints are used on the same joint event and may be statistically dependent. If $q_{R,L}$ falls below the protocol's declared reference-domain threshold, the certificate also includes a reference-leakage flag, reporting the conditional structural value together with its physical occurrence probability.

\subsection{Joint Confidence Events and Audit Records}
\label{app:D.6}

\indent
To combine and verify the final certificate, Table~\ref{tab:D.1} summarizes the statistical inputs and conversion conditions of Appendices~\ref{app:D.2}--\ref{app:D.5}. Here $r$ is the program radius entering the one-shot conversion: $r=\epsilon$ for exact transcription and $r=\epsilon_{\rm circ}+\delta_{\rm cmp}$ for approximate transcription, with the compilation code-length correction included as in Appendix~\ref{app:C.5}.

\begingroup
\renewcommand{\thetable}{D.1}
\begin{table}[H]
\caption{Finite-sample RCC audit routes}
\label{tab:D.1}
\centering
\footnotesize
\renewcommand{\arraystretch}{1.22}

\begin{tabular}{
@{}
>{\raggedright\arraybackslash}p{0.18\textwidth}
@{\hspace{1em}}
>{\raggedright\arraybackslash}p{0.14\textwidth}
@{\hspace{1em}}
>{\raggedright\arraybackslash}p{0.23\textwidth}
@{\hspace{1em}}
>{\raggedright\arraybackslash}p{0.37\textwidth}
@{}
}
\toprule
Route & Statistical input & Theoretical entry point & Completion conditions \\
\midrule

Arbitrary effects / hypothesis testing
&
$a_L,b_U$; $D_{H,L}^{\eta}$
&
Eq.~\eqref{eq:C.15} or Eq.~\eqref{eq:C.16} in Appendix~\ref{app:C.4}
&
Direct effects require $a_L>r$; hypothesis testing requires $a_L\ge1-\eta$ and $\eta+r<1$
\\[2pt]

Projection witnesses
&
$D_{\max,L}^{\epsilon_{\rm sm}}$
&
Eq.~\eqref{eq:C.18} in Appendix~\ref{app:C.4}
&
$\Psi_{k,d}^{\epsilon_{\rm sm}}$ already subtracts the smoothing radius; require $r\le\epsilon_{\rm sm}\le1$
\\[2pt]

Reference-compatible measurements
&
$D_{{\rm cl},L}$
&
Eq.~\eqref{eq:C.13} or Eq.~\eqref{eq:C.14} in Appendix~\ref{app:C.3}
&
Include spectral skew and smoothing control to form a one-shot input; until the conversion conditions are met, retain the corresponding state-side lower bound
\\[2pt]

Leakage discounting
&
$C_{{\rm leak},L}$
&
State-side discounting in Appendix~\ref{app:B}, followed by the appropriate one-shot route
&
Retain $q_{R,L}$ and the reference treatment; complete the one-shot and generation-model conversion of Appendix~\ref{app:B.6}
\\

\bottomrule
\end{tabular}
\end{table}
\endgroup

\indent
Write the intersection of all statistical confidence events as
\begin{equation}
\label{eq:D.46}
\mathcal E_{\rm stat}
=
\bigcap_{k=1}^{K}\mathcal E_k,
\qquad
{\rm Pr}(\mathcal E_{\rm stat})
\ge
1-\delta_{\rm stat}.
\end{equation}

\indent
On $\mathcal E_{\rm stat}$, all endpoint and feasible-set conclusions hold simultaneously. Outputs satisfying the conditions in Table~\ref{tab:D.1} yield $\underline{\mathcal I}_j(r)\le D_{\max}^{r}$ through Appendices~\ref{app:C.3}--\ref{app:C.4}, and are then inverted into cost lower bounds for the appropriate transcription branch. Routes that bound the cost for the same model, target state, and generation tolerance enter $\mathcal J_{\rm cert}$ and are combined into $L_{\rm RCC}$ by Appendix~\ref{app:C.6} and Proposition 4.1 of the main text.

\indent
Classical-measurement or leakage values that have yet to complete the one-shot conversion are reported as state-side results. To verify the final value and its conditions of applicability, the certificate record should include the lower bound $L_{\rm RCC}$ and confidence guarantee $1-\delta_{\rm stat}$, together with the model, qualification, path, parameters, calibration, reference treatment, and flags.

\indent
Here ${\rm model}$ uniquely specifies $\widehat{\mathfrak M}_R,R,\sigma_R,\Gamma_R$, the atomic physical interface, and the fixed unconditional readout. The field ${\rm qualification}$ records the certificates and family-level constants used for $\mathsf{RCon}$, $\mathsf{TC}$, and either direct semantic $\mathsf{RA}$ or the $\mathsf{RefCert}\Rightarrow\mathsf{RA}$ route. The field ${\rm parameters}$ records the generation tolerance, program and smoothing radii, $\eta$, and sample sizes; approximate transcription additionally records the chosen $\delta_{\rm cmp}$, code-length coefficient $b$, and certified precision domain.

The field ${\rm calibration}$ lists statistical confidence sets and deterministic error envelopes separately. The field ${\rm reference\ treatment}$ records the actual choice among ideal support, conditional projection, a support-regularized reference state, and leakage discounting. Leakage routes also retain $q_{R,L}$ and $p_{{\rm leak},U}$; support-regularized routes additionally record the weight $\delta$ and $\tau_\perp$. Route status identifies the output as a final complexity lower bound, one-shot input, state-side leading term, or model diagnostic.

Finite samples and calibration information thus define sets covering the true physical parameters, from which each route extracts conservative final-state information. Sample size and calibration precision determine the certificate's statistical resolution. Appendix~\ref{app:C} converts inputs satisfying the conversion conditions into lower bounds on universal optimal quantum circuit complexity within the same reference-consistent model.

\endgroup

\begingroup

\setlength{\parindent}{1.5em}
\setlength{\parskip}{1pt plus 0.3pt minus 0.2pt}
\frenchspacing
\linespread{1.005}\selectfont
\setlength{\abovedisplayskip}{8.5pt plus 1pt minus 2pt}
\setlength{\belowdisplayskip}{8.5pt plus 1pt minus 2pt}
\setlength{\abovedisplayshortskip}{5.5pt plus 1pt minus 1pt}
\setlength{\belowdisplayshortskip}{6.5pt plus 1pt minus 1pt}
\setlength{\jot}{2pt}
\setlength{\emergencystretch}{1.25em}
\clubpenalty=10000
\widowpenalty=10000
\displaywidowpenalty=10000
\brokenpenalty=5000
\raggedbottom

\newcommand{\Estatement}[1]{%
  \par\addvspace{0.4\baselineskip}%
  \Needspace{4\baselineskip}%
  \indent\textbf{#1}\hspace{0.5em}\ignorespaces
}
\newcommand{\Eproof}{%
  \par\addvspace{0.25\baselineskip}%
  \Needspace{3\baselineskip}%
  \indent\textbf{Proof.}\hspace{0.5em}\ignorespaces
}
\newcommand{\Eqed}{%
  \unskip\nobreak\hspace{0.35em}\ensuremath{\square}%
  \par\addvspace{0.3\baselineskip}%
}
\makeatletter
\renewcommand{\subsection}{\@startsection{subsection}{2}{\z@}%
  {14pt plus 1pt minus 1pt}{8pt plus 0.5pt minus 1pt}%
  {\normalfont\small\bfseries\centering}}
\makeatother
\let\EOriginalSubsection\subsection
\renewcommand{\subsection}[1]{%
  \par\Needspace{3\baselineskip}%
  \EOriginalSubsection{#1}%
}

\section{Properties at Fixed Reference and Windowed RCC}
\label{app:E}

\indent
This appendix provides the mathematical support for Sec.~\ref{sec:5}. In a fixed RCC-admissible reference-consistent quantum circuit--description model, the RCC state-side leading term $C_R(\rho)$ inherits the basic structure of relative entropy. The reference set, structured vacuum, and atomic control bandwidth give it a definite physical zero point and unit of measurement.

\indent
Section~\ref{sec:5} further restricts this leading term to accessible object domains. To formalize this restriction, we represent reference-compatible windows by finite-dimensional observable algebras and their trace-preserving conditional expectations. We prove that window effective states represent exactly the equivalence classes of global states that the algebra cannot distinguish, and establish monotone relations among the global RCC leading term, nested windowed RCC values, and classical audit lower bounds. The pure-state boundary is then resolved into a family of windowed complexity profiles that depend on the object domain.
Appendix~\ref{app:D} supplies the statistical bounds for window readouts; the one-shot conversion and cost inversion in Appendix~\ref{app:C} then yield process-complexity certificates. Unless stated otherwise, all logarithms are base $2$.

\subsection{Fixed Reference Model and Basic Notation}
\label{app:E.1}

\indent
To compare state structure using the same physical zero point and unit of measurement, fix a reference-consistent quantum circuit--description model $\widehat{\mathfrak M}_R=(\mathfrak M_R,U_R)$, and fix the reference set $R$ together with $\Pi_R$, $\mathcal H_R=\Pi_R\mathcal H$, and $d_R={\rm Tr}(\Pi_R)>0$.
The unbiased structured vacuum and atomic control bandwidth are, respectively, $\sigma_R=\frac{\Pi_R}{d_R}$ and $g_R:=\log\Gamma_R>0$.

\indent
Throughout, assume
\begin{equation}
\label{eq:E.4}
{\rm supp}(\rho)\subseteq\mathcal H_R.
\end{equation}

\indent
The RCC state-side linear leading term is defined by
\begin{equation}
\label{eq:E.5}
C_R(\rho)
=
\frac{D(\rho\Vert\sigma_R)}{g_R}.
\end{equation}

\indent
Since $\sigma_R$ is maximally mixed on $\mathcal H_R$,
\begin{equation}
\label{eq:E.6}
D(\rho\Vert\sigma_R)
=
\log d_R-S(\rho),
\qquad
C_R(\rho)
=
\frac{\log d_R-S(\rho)}{g_R}.
\end{equation}

\indent
The model $\widehat{\mathfrak M}_R$ remains fixed throughout this appendix, so we suppress the model argument in $C_R(\rho;\mathfrak M_R)$. Appendix~\ref{app:B} treats support leakage and its conservative corrections.

\subsection{Fixed Reference Structure}
\label{app:E.2}

\indent
We now determine the range of the state-side leading term and how mixing and reference-preserving operations affect it.

\Estatement{Proposition E.1 (Fixed reference structure).}
For states satisfying Eq.~\eqref{eq:E.4}, $C_R$ has the following properties:
\begin{equation}
\label{eq:E.7}
0
\le
C_R(\rho)
\le
\frac{\log d_R}{g_R},
\qquad
C_R(\rho)=0
\Longleftrightarrow
\rho=\sigma_R.
\end{equation}

\indent
The upper bound is attained exactly by pure states in $\mathcal H_R$. If $\rho=\sum_i p_i\rho_i$ is a convex combination of states in $\mathcal H_R$, then
\begin{equation}
\label{eq:E.8}
C_R(\rho)
\le
\sum_i p_iC_R(\rho_i).
\end{equation}

\indent
If a CPTP map $\mathcal E$ preserves the reference state, then
\begin{equation}
\label{eq:E.9}
\mathcal E(\sigma_R)=\sigma_R
\quad\Longrightarrow\quad
C_R(\mathcal E(\rho))
\le
C_R(\rho).
\end{equation}

\indent
If a unitary transformation $U$ preserves the reference state, then
\begin{equation}
\label{eq:E.10}
U\sigma_RU^\dagger=\sigma_R
\quad\Longrightarrow\quad
C_R(U\rho U^\dagger)
=
C_R(\rho).
\end{equation}

\Eproof
Equation~\eqref{eq:E.7} and its equality conditions follow from $0\le S(\rho)\le\log d_R$. Equation~\eqref{eq:E.8} follows from entropy concavity, Eq.~\eqref{eq:E.9} from the data-processing inequality for relative entropy, and Eq.~\eqref{eq:E.10} from invariance under simultaneous unitary conjugation. Moreover, since
\(0\le\rho\le d_R\sigma_R,
\)
a reference-preserving channel satisfies
\begin{equation*}
0
\le
\mathcal E(\rho)
\le
d_R\mathcal E(\sigma_R)
=
\Pi_R,
\end{equation*}
so the output in Eq.~\eqref{eq:E.9} remains within the reference support.\Eqed

\indent
For the unbiased structured vacuum, $U\sigma_RU^\dagger=\sigma_R$ is equivalent to $U\Pi_RU^\dagger=\Pi_R$. If $\mathcal E$ is actually invoked in an experiment or preparation protocol, its implementation cost is still counted under the resource rules of $\mathfrak M_R$; Eq.~\eqref{eq:E.9} states only the data-processing relation for the state-side leading term.

\subsection{The State-Side Leading Term, Windowed RCC, and Process Complexity}
\label{app:E.3}

\indent
To clarify how state values enter process-cost lower bounds, we distinguish three levels within the fixed reference model:
\begin{equation}
\label{eq:E.11}
C_R(\rho),
\qquad
C_{R,\Xi}(\rho),
\qquad
C_{\rm opt}^{(\epsilon)}(\rho;\mathfrak M_R).
\end{equation}

\indent
Here $C_R(\rho)$ is the global RCC state-side leading term, and $C_{R,\Xi}(\rho)$ is the windowed RCC defined below. Once the model, target state, and tolerance are fixed, $C_{\rm opt}^{(\epsilon)}(\rho;\mathfrak M_R)$ is the infimum of costs over all admissible successful paths in the model. It is jointly determined by the allowed operations, error conventions, and resource accounting. Through the description interface $U_R$, the main theorem converts valid inputs supplied by the first two quantities into a rigorous lower bound on this process optimum. The description interface provides transcription and certificate calibration while leaving the fixed definition of physical complexity unchanged.

\indent
The normalization factor $g_R$ sets the reporting scale within the fixed model. In binary information units,
\begin{equation}
\label{eq:E.12}
1\,{\rm st}_R
\equiv
\log_2\Gamma_R\ {\rm bits}
=
\ln\Gamma_R\ {\rm nats}.
\end{equation}

\indent
If the same information gap $I$ is held fixed and only the atomic control bandwidth used for reporting is changed, then
\begin{equation}
\label{eq:E.13}
C[{\rm st}_R]
=
\frac{I}{\log\Gamma_R},
\qquad
C[{\rm st}_{R'}]
=
\frac{I}{\log\Gamma_{R'}},
\qquad
C[{\rm st}_R]
=
\frac{\log\Gamma_{R'}}{\log\Gamma_R}
C[{\rm st}_{R'}].
\end{equation}

\indent
Equation~\eqref{eq:E.13} converts reporting units. Changing the reference set, structured vacuum, or resources that can be invoked changes the process model and may also change the information gap. Appendices~\ref{app:A.10} and~\ref{app:F.8} treat the corresponding relations across references; Appendices~\ref{app:A} and \ref{app:F} give the short-program gap bound and the resource conventions of structural fairness, respectively.

\subsection{Observable Algebras and Object-Domain Equivalence}
\label{app:E.4}

\indent
Assigning the same effective representative to operationally indistinguishable global states requires preserving every accessible expectation value in the object domain. Let $\mathcal A\subseteq B(\mathcal H_R)$ be a finite-dimensional von Neumann subalgebra containing the identity $\Pi_R$. A trace-preserving conditional expectation onto $\mathcal A$ is a linear map $E:B(\mathcal H_R)\to\mathcal A$ satisfying
\begin{equation}
\label{eq:E.14}
E^2=E,
\qquad
E\ {\rm CPTP},
\qquad
{\rm Ran}(E)=\mathcal A,
\qquad
E(AXB)=AE(X)B
\end{equation}
for all $A,B\in\mathcal A$ and $X\in B(\mathcal H_R)$. Trace preservation and the bimodule property give
\begin{equation}
\label{eq:E.15}
{\rm Tr}\!\left[A\,E(X)\right]
=
{\rm Tr}(AX),
\qquad
A\in\mathcal A.
\end{equation}

\indent
Thus ${\rm Tr}[A^\dagger(X-E(X))]=0$ for every $A\in\mathcal A$, so $E(X)$ is precisely the Hilbert--Schmidt orthogonal projection of $X$ onto $\mathcal A$. In particular, this conditional expectation is unique for the given trace and algebra.

\indent
The algebra $\mathcal A$ defines window equivalence between global states by
\begin{equation}
\label{eq:E.16}
\rho\sim_{\mathcal A}\tau
\quad\Longleftrightarrow\quad
{\rm Tr}(A\rho)
=
{\rm Tr}(A\tau)
\quad
\text{for all }A\in\mathcal A.
\end{equation}

\Estatement{Proposition E.2 (Canonical representatives of an object domain).}
For any states $\rho,\tau$ supported on $\mathcal H_R$,
\begin{equation}
\label{eq:E.17}
\rho\sim_{\mathcal A}\tau
\quad\Longleftrightarrow\quad
E(\rho)=E(\tau).
\end{equation}

\Eproof
Set $X=\rho-\tau$. Equation~\eqref{eq:E.16} states that $X$ is orthogonal to $\mathcal A$, while Eq.~\eqref{eq:E.15} identifies its orthogonal projection as $E(X)$. Hence Eq.~\eqref{eq:E.16} is equivalent to $E(X)=0$.\Eqed
Thus $E(\rho)$ is the canonical representative of the state $\rho$ in the object domain defined by $\mathcal A$: it preserves all expectation values in the algebra and identifies exactly those global states that the algebra cannot distinguish. What changes here is the object domain of windowed RCC; the target state in the main theorem remains the original $\rho$.

\subsection{Reference-Compatible Windows and Windowed RCC}
\label{app:E.5}

\indent
In the physical applications of the main text, the task, experiment, or readout protocol declares the window in advance. The corresponding resource rules fix the accessible observables in $\mathcal A_\Xi$ and the scope of its closure. The following definition gives the finite-dimensional mathematical structure of this choice.

\Estatement{Definition E.3 (Reference-compatible window).}
A reference-compatible window is denoted by $\Xi=(\mathcal A_\Xi,E_\Xi)$, where $\mathcal A_\Xi\subseteq B(\mathcal H_R)$ is a finite-dimensional von Neumann subalgebra and $E_\Xi$ is the trace-preserving conditional expectation onto $\mathcal A_\Xi$ defined in Appendix~\ref{app:E.4}. Its reference-compatibility condition is
\begin{equation}
\label{eq:E.18}
E_\Xi(\sigma_R)
=
\sigma_R.
\end{equation}

\indent
For the unbiased structured vacuum, this condition follows automatically from $E_\Xi(\Pi_R)=\Pi_R$. Define the window effective state, window entropy, and windowed RCC by
\begin{equation}
\label{eq:E.19}
\rho_\Xi
:=
E_\Xi(\rho),
\qquad
S_\Xi(\rho)
:=
S(\rho_\Xi),
\end{equation}
\begin{equation}
\label{eq:E.20}
C_{R,\Xi}(\rho)
=
\frac{D(\rho_\Xi\Vert\sigma_R)}{g_R}
=
\frac{\log d_R-S_\Xi(\rho)}{g_R}.
\end{equation}

\indent
Equation~\eqref{eq:E.20} replaces the global state by its canonical representative in the object domain, while retaining the same structured vacuum and atomic control bandwidth.

\indent
Local subsystems provide basic windows. Suppose the reference subspace has a task-dependent decomposition
\begin{equation}
\label{eq:E.21}
\mathcal H_R
=
\mathcal H_A\otimes\mathcal H_{\bar A},
\qquad
d_R=d_A d_{\bar A},
\qquad
\sigma_R
=
\frac{I_A}{d_A}
\otimes
\frac{I_{\bar A}}{d_{\bar A}}.
\end{equation}

\indent
Take
\begin{equation}
\label{eq:E.22}
\mathcal A_A
=
B(\mathcal H_A)\otimes I_{\bar A},
\qquad
E_A(X)
=
{\rm Tr}_{\bar A}(X)
\otimes
\frac{I_{\bar A}}{d_{\bar A}}.
\end{equation}

\indent
The map $E_A$ is a trace-preserving conditional expectation that preserves $\sigma_R$. Writing $\rho_A={\rm Tr}_{\bar A}\rho$, we have
\begin{equation}
\label{eq:E.23}
\rho_{\Xi_A}
=
\rho_A\otimes\frac{I_{\bar A}}{d_{\bar A}},
\qquad
C_{R,\Xi_A}(\rho)
=
\frac{\log d_A-S(\rho_A)}{g_R}.
\end{equation}

\indent
The complementary subsystem's maximally mixed entropy cancels exactly in $\log d_R-S(\rho_{\Xi_A})$, leaving the structural gap in the local object domain. Comparing the global leading term with the sum of the two complementary local windows further isolates the contribution from correlations.

\Estatement{Proposition E.4 (Local--correlation decomposition).}
Define the local window $\Xi_{\bar A}$ symmetrically for the complementary subsystem $\bar A$. Then
\begin{equation*}
C_R(\rho_{A\bar A})
=
C_{R,\Xi_A}(\rho)
+
C_{R,\Xi_{\bar A}}(\rho)
+
\frac{I(A:\bar A)_\rho}{g_R}.
\end{equation*}

\Eproof
Using $d_R=d_A d_{\bar A}$ and Eqs.~\eqref{eq:E.6} and \eqref{eq:E.23}, the logarithmic dimension terms cancel among the three terms, leaving
\begin{equation*}
S(\rho_A)
+
S(\rho_{\bar A})
-
S(\rho_{A\bar A})
=
I(A:\bar A)_\rho.
\end{equation*}

\indent
This is precisely the mutual information $I(A:\bar A)_\rho$.\Eqed
This identity is Eq.~\eqref{eq:5.10} of the main text. For a bipartite pure state, the correlation term reduces to $2S(\rho_A)/g_R$.

\subsection{Monotonicity and Hierarchy of Windowed RCC}
\label{app:E.6}

\indent
A smaller observable algebra identifies more global states as equivalent. Relative entropy under conditional expectation has a standard characterization as an entropy difference~\cite{GaoEtAl2020_SubalgebraRelativeEntropy}. We apply this structure to windowed RCC, proving that it decreases monotonically as the object domain contracts and expressing the loss as an increase in the effective state's entropy.

\Estatement{Proposition E.5 (Window hierarchy).}
For any reference-compatible window,
\begin{equation}
\label{eq:E.24}
0
\le
C_{R,\Xi}(\rho)
\le
C_R(\rho),
\end{equation}
and, for the unbiased structured vacuum, the coarse-graining loss is exactly
\begin{equation}
\label{eq:E.25}
C_R(\rho)-C_{R,\Xi}(\rho)
=
\frac{S(\rho_\Xi)-S(\rho)}{g_R}
\ge
0.
\end{equation}

\indent
If $\mathcal A_{\Xi_1}\subseteq\mathcal A_{\Xi_2}$, write $\Xi_1\preceq\Xi_2$. By the orthogonal-projection characterization in Appendix~\ref{app:E.4}, nested algebras automatically satisfy
\begin{equation}
\label{eq:E.26}
E_{\Xi_1}
=
E_{\Xi_1}\circ E_{\Xi_2}
=
E_{\Xi_2}\circ E_{\Xi_1}.
\end{equation}

\indent
Consequently,
\begin{equation}
\label{eq:E.27}
C_{R,\Xi_1}(\rho)
\le
C_{R,\Xi_2}(\rho)
\le
C_R(\rho).
\end{equation}

\indent
The identity window and the window of complete reference coarse-graining
\begin{equation}
\label{eq:E.28}
E_\infty
=
{\rm id},
\qquad
E_0(X)
=
{\rm Tr}(X)\sigma_R
\end{equation}
give the two endpoints
\begin{equation}
\label{eq:E.29}
C_{R,\infty}(\rho)
=
C_R(\rho),
\qquad
C_{R,0}(\rho)
=
0.
\end{equation}

\indent
Thus any nested family of windows containing both endpoints gives
\begin{equation}
\label{eq:E.30}
0
\le
C_{R,\Xi_1}(\rho)
\le
C_{R,\Xi_2}(\rho)
\le
C_R(\rho),
\qquad
\Xi_1\preceq\Xi_2.
\end{equation}

\Eproof
Nonnegativity follows from positivity of relative entropy. The remaining inequalities follow by applying the data-processing inequality to the reference-preserving maps $E_\Xi$ and $E_{\Xi_1}$ in turn. Subtracting Eqs.~\eqref{eq:E.6} and \eqref{eq:E.20} gives Eq.~\eqref{eq:E.25}, and the two endpoints follow directly from the definitions.\Eqed

\indent
For a general full-rank reference state $\sigma$, a trace-preserving conditional expectation that preserves it still gives monotonicity under nested windows by data processing. In this case, $\log\sigma\in\mathcal A_\Xi$, and Eq.~\eqref{eq:E.15} cancels the reference term in the relative-entropy difference, so the coarse-graining loss is again the entropy increase. Incomparable windows resolve the complexity gap along different observational directions and generally have no common ordering. If an experimental protocol actually implements $E_\Xi$, its physical cost still enters the process resource accounting of $\mathfrak M_R$. The full process model determines the ordering of $C_{\rm opt}^{(\epsilon)}$ for different target states.

\subsection{Window Pinching and Classically Auditable Lower Bounds}
\label{app:E.7}

\indent
To extract the part of a window accessible to projective readout, we decompose the effective state into blocks labeled by measurement outcomes, separating coherence between blocks, outcome probabilities, and structure within blocks. Let $\{P_i\}\subseteq\mathcal A_\Xi$ be a complete family of nonzero orthogonal projections on $\mathcal H_R$:
\begin{equation}
\label{eq:E.31}
P_iP_j
=
\delta_{ij}P_i,
\qquad
\sum_iP_i
=
\Pi_R.
\end{equation}

\indent
Write
\begin{equation}
\label{eq:E.32}
r_i
=
{\rm Tr}(P_i),
\qquad
p_i
=
{\rm Tr}(P_i\rho_\Xi)
=
{\rm Tr}(P_i\rho),
\qquad
u_i
=
{\rm Tr}(P_i\sigma_R)
=
\frac{r_i}{d_R}.
\end{equation}

\indent
The equality of the two trace expressions follows from $P_i\in\mathcal A_\Xi$ and Eq.~\eqref{eq:E.15}, so $p_i$ can be read directly from window observables on the original state. Define
\begin{equation}
\label{eq:E.33}
\Delta_P(X)
=
\sum_iP_iXP_i,
\end{equation}
and, for $p_i>0$, set
\begin{equation}
\label{eq:E.34}
\rho_{\Xi,i}
=
\frac{P_i\rho_\Xi P_i}{p_i}.
\end{equation}

\Estatement{Proposition E.6 (Exact decomposition under window pinching).}
For the projection family above,
\begin{equation}
\label{eq:E.35}
\begin{aligned}
D(\rho_\Xi\Vert\sigma_R)
&=
D(\rho_\Xi\Vert\Delta_P(\rho_\Xi))
+
D_{\rm cl}(p\Vert u)
\\
&\quad
+
\sum_{i:p_i>0}
p_i
D\!\left(
\rho_{\Xi,i}
\Big\Vert
\frac{P_i}{r_i}
\right).
\end{aligned}
\end{equation}

\Eproof
Pinching ensures ${\rm supp}(\rho_\Xi)\subseteq{\rm supp}(\Delta_P(\rho_\Xi))$, and $\Delta_P(\sigma_R)=\sigma_R$. On the respective supports, both $\log\Delta_P(\rho_\Xi)$ and $\log\sigma_R$ are block-diagonal operators. The self-adjointness of pinching with respect to the trace pairing gives
\begin{equation*}
D(\rho_\Xi\Vert\sigma_R)
-
D(\rho_\Xi\Vert\Delta_P(\rho_\Xi))
=
D(\Delta_P(\rho_\Xi)\Vert\sigma_R).
\end{equation*}

\indent
Substituting $\Delta_P(\rho_\Xi)=\bigoplus_i p_i\rho_{\Xi,i}$ and $\sigma_R=\bigoplus_i u_iP_i/r_i$ into the direct-sum decomposition of relative entropy then gives Eq.~\eqref{eq:E.35}.\Eqed

\indent
All three terms are nonnegative, so classical readout gives
\begin{equation}
\label{eq:E.36}
C_{R,\Xi}(\rho)
\ge
\frac{D_{\rm cl}(p\Vert u)}{g_R},
\qquad
u_i
=
\frac{r_i}{d_R}.
\end{equation}

\indent
The three terms in Eq.~\eqref{eq:E.35} resolve, in order, coherence between blocks, deviations of block weights from the structured vacuum, and nonuniformity within blocks. If all $P_i$ are rank-one projections, then
\begin{equation}
\label{eq:E.37}
u_i
=
\frac1{d_R},
\qquad
C_{R,\Xi}(\rho)
\ge
\frac{\log d_R-H(\{p_i\})}{g_R}.
\end{equation}

\indent
Equality in Eq.~\eqref{eq:E.36} holds if and only if $\rho_\Xi=\Delta_P(\rho_\Xi)$ and every occupied block satisfies $\rho_{\Xi,i}=P_i/r_i$.

\indent
With finite samples, on the corresponding joint coverage event for the confidence sets in Appendix~\ref{app:D},
\begin{equation}
\label{eq:E.38}
D_{\rm cl}(p\Vert u)
\ge
D_{{\rm cl},L},
\qquad
\underline C_{R,\Xi}
:=
\frac{D_{{\rm cl},L}}{g_R}
\le
C_{R,\Xi}(\rho).
\end{equation}

\indent
Selection among multiple windows or projection families based on the data uses simultaneous coverage, a training--validation split, or another predeclared selection protocol that preserves the stated coverage guarantee, as in Appendix~\ref{app:D}. Equation~\eqref{eq:E.38} then supplies a conservative lower-bound input for windowed RCC.

\subsection{The Pure-State Boundary and Window Profiles}
\label{app:E.8}

\indent
The global leading term takes the same value on all pure states; window profiles further resolve the structural gaps in individual object domains. For any pure state $\rho_\psi=|\psi\rangle\langle\psi|$ supported on $\mathcal H_R$,
\begin{equation}
\label{eq:E.39}
C_R(\rho_\psi)
=
\frac{\log d_R}{g_R}.
\end{equation}

\indent
The full process complexity of different pure states is determined by their admissible preparation paths in the fixed model, whereas their windowed leading terms are
\begin{equation}
\label{eq:E.40}
C_{R,\Xi}(\rho_\psi)
=
\frac{
\log d_R
-
S(\rho_{\psi,\Xi})
}{g_R},
\qquad
\rho_{\psi,\Xi}
:=
E_\Xi(\rho_\psi).
\end{equation}

\indent
For two pure states $\rho_\psi$ and $\rho_\phi$, the same window gives
\begin{equation}
\label{eq:E.41}
C_{R,\Xi}(\rho_\psi)
-
C_{R,\Xi}(\rho_\phi)
=
\frac{
S(\rho_{\phi,\Xi})
-
S(\rho_{\psi,\Xi})
}{g_R}.
\end{equation}

\indent
The larger the entropy gap between the window effective state and the structured vacuum, the higher the corresponding windowed RCC. The local window of Appendix~\ref{app:E.5} gives
\begin{equation}
\label{eq:E.42}
C_{R,\Xi_A}(\rho_\psi)
=
\frac{
\log d_A-S(\rho_{\psi,A})
}{g_R},
\end{equation}
thereby characterizing the complexity gap in the object domain through the reduced state of subsystem $A$. Nonlocal correlations are resolved by the correlation term of Proposition E.4 or by the corresponding joint windows.

\indent
Given a family $\mathcal W$ of reference-compatible windows, define the window profile by
\begin{equation}
\label{eq:E.43}
{\rm Prof}_{R,\mathcal W}(\rho)
:=
\bigl(
C_{R,\Xi}(\rho)
\bigr)_{\Xi\in\mathcal W},
\qquad
\Xi
\longmapsto
C_{R,\Xi}(\rho).
\end{equation}

\indent
For a nested $\mathcal W$, Eq.~\eqref{eq:E.30} makes the profile monotone along the window hierarchy; a general window family gives a multicomponent profile along different observational directions. The labeled family of effective states $\bigl(E_\Xi(\rho)\bigr)_{\Xi\in\mathcal W}$ retains all distinguishable information in the respective object domains. Windowed RCC extracts scalar components relative to the structured vacuum that can serve as inputs to process certificates.

\subsection{Windowed RCC and Final Complexity Certificates}
\label{app:E.9}

\indent
To use a window value for complexity certification, it must be converted into a one-shot lower-bound input for the original target state. Proposition E.5 gives
\begin{equation}
\label{eq:E.44}
D(\rho\Vert\sigma_R)
\ge
D(\rho_\Xi\Vert\sigma_R)
=
g_R C_{R,\Xi}(\rho).
\end{equation}

\indent
Each reference-compatible window therefore gives a conservative lower bound on the global state-side leading term, while the process target remains the original state $\rho$. Suppose a route provides, under the same analytical conditions or on the same joint statistical event,
\begin{equation}
\label{eq:E.45}
\underline C_{R,\Xi}
\le
C_{R,\Xi}(\rho),
\qquad
\Delta_{\rm spec}(\rho)\ge\Delta_L,
\qquad
\delta_{\rm sm}(\rho,\epsilon)
\le
\delta_U(\epsilon).
\end{equation}
Then the relative-entropy--one-shot bridge of Sec.~\ref{sec:3} gives
\begin{equation}
\label{eq:E.46}
\underline{\mathcal I}_\Xi(\epsilon)
:=
g_R\underline C_{R,\Xi}
+
\Delta_L
-
\delta_U(\epsilon)
\le
D_{\max}^{\epsilon}(\rho\Vert\sigma_R).
\end{equation}

\indent
Equation~\eqref{eq:E.46} is the one-shot information input that the window route supplies to Appendix~\ref{app:C}. Under exact transcription, apply the inversion in Eq.~\eqref{eq:C.2} or \eqref{eq:C.4}. Under approximate transcription, evaluate this input at the total radius $\epsilon=\epsilon_{\rm circ}+\delta_{\rm cmp}$ and include the compilation code-length correction over the certified precision domain as in Appendix~\ref{app:C.5}. When no independent lower bound on spectral skew is available, one may conservatively take $\Delta_L=0$. If no upper bound on the smoothing loss is available yet, $\underline C_{R,\Xi}$ remains a state-side audit lower bound. One-shot values obtained from hypothesis testing or projection witnesses enter the corresponding routes in Appendix~\ref{app:C.4}.

\indent
For a family of windows declared in advance, Appendix~\ref{app:D.6} supplies the joint statistical event. Appendix~\ref{app:C.6} takes the maximum over valid routes for which reference treatment, one-shot conversion, and cost inversion have all been completed. Window profiles thus resolve the structural gaps of the same target state in different object domains and enter a unified process-complexity certificate in nonrecursive form.

\endgroup
\begingroup

\setlength{\parindent}{1.5em}
\setlength{\parskip}{1pt plus 0.3pt minus 0.2pt}
\frenchspacing
\linespread{1.035}\selectfont
\setlength{\abovedisplayskip}{8.5pt plus 1pt minus 2pt}
\setlength{\belowdisplayskip}{8.5pt plus 1pt minus 2pt}
\setlength{\abovedisplayshortskip}{5.5pt plus 1pt minus 1pt}
\setlength{\belowdisplayshortskip}{6.5pt plus 1pt minus 1pt}
\setlength{\jot}{2.5pt}
\setlength{\emergencystretch}{1.25em}
\clubpenalty=10000
\widowpenalty=10000
\displaywidowpenalty=10000
\brokenpenalty=5000
\raggedbottom

\newcommand{\Fstatement}[1]{%
  \par\addvspace{0.4\baselineskip}%
  \Needspace{4\baselineskip}%
  \indent\textbf{#1}\hspace{0.5em}\ignorespaces
}
\newcommand{\Fproof}{%
  \par\addvspace{0.25\baselineskip}%
  \Needspace{3\baselineskip}%
  \indent\textbf{Proof.}\hspace{0.5em}\ignorespaces
}
\newcommand{\Fqed}{%
  \unskip\nobreak\hspace{0.35em}\ensuremath{\square}%
  \par\addvspace{0.3\baselineskip}%
}
\makeatletter
\renewcommand{\subsection}{\@startsection{subsection}{2}{\z@}%
  {14pt plus 1pt minus 1pt}{8pt plus 0.5pt minus 1pt}%
  {\normalfont\small\bfseries\centering}}
\makeatother
\let\FOriginalSubsection\subsection
\renewcommand{\subsection}[1]{%
  \par\Needspace{3\baselineskip}%
  \FOriginalSubsection{#1}%
}

\section{Structural Fairness, Reference-Admissible Constructions, and Generalized Reference Boundaries}
\label{app:F}

What qualifies a quantum circuit model to interpret instruction length as a generation cost relative to a common structural background? Starting from a fixed reference and atomic resources, this appendix constructs models that satisfy faithful transcription and reference domination at the program level, and connects these constructions to the general lower bound in Appendix~\ref{app:A} and finite-control certification in Appendix~\ref{app:H}.

Finite control, CPTP dynamics, and prefix-free descriptions still permit short commands to invoke global reset, or structural directions to be supplied in advance through duplicate labels, free pure ancillary states, and unrecorded control precision. Structural fairness requires the reference background, admissible resources, atomic actions, and program prior to be fixed together before the target state is chosen. We specify the semantic qualification of this interface, build concrete models through whole-word transcription, reference balance, and stationary envelopes, and then discuss the cost of reference gain, extensions of certified interpreters, and reference upgrading. All logarithms in this appendix are base 2, except in the Gibbs relations explicitly written with $\ln$.

\subsection{Finite-Resource Realization, Structural Fairness, and Fixed References}
\label{app:F.1}

The requirement that physical processes be realized with finite resources is a basic premise of quantum circuit models. It ensures that each concrete run uses only finite control, finite workspace, and finite precision, but does not by itself specify how much structure these resources carry. The structural fairness required by RCC further demands that any initial states, ancillary states, environments, control records, macroinstructions, addressing capabilities, postselection rules, and program priors that can supply a directional bias before target selection be incorporated into the same reference background or appear explicitly in the generation cost.

The reference set is therefore fixed before the target state is audited. Denote the independent physical priors by $\mathcal P_{\rm phys}$ and write
\begin{equation}
\label{eq:F.1}
R
=
R(\mathcal P_{\rm phys}).
\end{equation}

These priors may come from conservation laws, superselection sectors, stabilizer constraints, logical subspaces, symmetry-protected structures, Hamiltonians, control architectures, low-energy effective spaces, or task structures declared in advance. They determine
\begin{equation}
\label{eq:F.2}
\Pi_R,
\qquad
\mathcal H_R
=
\Pi_R\mathcal H,
\qquad
d_R
=
{\rm Tr}\,\Pi_R
>
0,
\end{equation}
and the finite-dimensional unbiased structured vacuum
\begin{equation}
\label{eq:F.3}
\sigma_R
=
\frac{\Pi_R}{d_R}.
\end{equation}

Reference selection follows minimal sufficiency: structure stably supplied by the background and available for repeated use enters $R$, while target-specific structure and features inferred from data after the fact remain part of the object to be generated. An insufficient reference misidentifies background structure as a generation burden; an excessively fine reference absorbs target information into the zero point in advance.

\Fstatement{Proposition F.1 (Maximum entropy of the unbiased structured vacuum).}
Among normalized states supported on finite-dimensional $\mathcal H_R$,
\begin{equation}
\label{eq:F.4}
\sigma_R
=
\operatorname*{arg\,max}_{\substack{
\omega\ge0,\ {\rm Tr}\omega=1\\
{\rm supp}(\omega)\subseteq\mathcal H_R
}}
S(\omega),
\qquad
S(\sigma_R)
=
\log d_R,
\end{equation}
and the maximizer is unique. Thus, for any state $\rho$ supported on $\mathcal H_R$,
\begin{equation}
\label{eq:F.5}
D(\rho\Vert\sigma_R)
=
\log d_R-S(\rho).
\end{equation}

\Fproof
Every normalized state supported on a $d_R$-dimensional space satisfies $S(\omega)\le\log d_R$, with equality if and only if all its nonzero eigenvalues equal $1/d_R$. Equation~\eqref{eq:F.5} follows directly from $\log\sigma_R=-(\log d_R)\Pi_R$ on the reference support.\Fqed

The maximum-entropy zero point fixes the state-side background; the generation side also requires admissible operations and their resource costs to be fixed. Structural fairness requires the structure supplied by local control, auxiliary resources, and description weights to be accounted for within the same specification. In the finite-dimensional models of this paper, this requirement is implemented as
\begin{equation}
\label{eq:F.6}
\mathsf{SF}
\leadsto
\mathsf{Acc}_{\rm RCC}
:=
\mathsf{RCon}
\wedge
\mathsf{TC}
\wedge
\mathsf{RA}.
\end{equation}

The arrow denotes the implementation of a physical principle through a model interface, rather than a logical equivalence. The condition $\mathsf{RCon}$ fixes a common reference for the structural zero point, initial states, admissible operations, and counting rules; the atomic interface and program interpreter also enter the verification of $\mathsf{TC}$ and $\mathsf{RA}$. Appendix~\ref{app:F.2} gives the full qualification conditions and corresponding sufficient certificates.

To give one step physical content, fix an atomic physical interface $\mathfrak I_{R,n}^{(1)}$ with equal costs before target selection, for each reference size $n$ in the model family. We write $\mathfrak I_R^{(1)}$ when suppressing the size index, and denote its set of raw control records by $\mathsf C_{R,n}^{(1)}$. An atomic resource slot specifies serial or parallel counting, spacetime support, control and addressing, parameter precision, ancillary states and environments, measurement records, and failure flags. For a raw command $c$, define its full composable semantics as
\begin{equation}
\label{eq:F.7}
\mathsf{FullSem}_{R,n}(c)
:=
\bigl(
\mathsf J_{R,n}(c),
\mathsf{sig}_{R,n}(c)
\bigr).
\end{equation}

Here $\mathsf J_{R,n}(c)$ is the total CPTP channel or quantum-instrument dynamical semantics acting on the declared system, finite workspace, retained classical control state, and outcome records. The static resource signature $\mathsf{sig}_{R,n}(c)$ records spacetime support, parallelism, reliable control resolution, preloaded structure, auxiliary-resource supply, unretained environments, and other cost data. Environmental or resource states that are retained and can affect subsequent execution enter the extended classical--quantum blocks of $\mathsf J$; discarded environments are recorded jointly by the reduced dynamical semantics and $\mathsf{sig}$.

Strict operational equivalence is defined by
\begin{equation}
\label{eq:F.8}
c\sim_{R,n}c'
\quad\Longleftrightarrow\quad
\mathsf{FullSem}_{R,n}(c)
=
\mathsf{FullSem}_{R,n}(c')
\end{equation}
for all allowed inputs and all retained outputs. The operationally reduced physical atomic alphabet, number of action classes, and single-step Hartley control bandwidth are, respectively,
\begin{equation}
\label{eq:F.9}
\mathcal A_{R,n}^{(1)}
:=
\mathsf C_{R,n}^{(1)}/\!\sim_{R,n},
\qquad
\Gamma_{R,n}
:=
\bigl|\mathcal A_{R,n}^{(1)}\bigr|,
\qquad
g_{R,n}
:=
\log_2\Gamma_{R,n}.
\end{equation}

Throughout this paper, a primitive is relative to the declared atomic resource slot, rather than to a globally minimal generating set of the channel algebra. Identity aliases, Kraus representations, and global phases with identical full semantics count as a single action class. Different addresses, accessible outcomes, control memory, or resource side effects define distinct actions when they can affect subsequent execution. Thus $\Gamma_R$ is a calibration quantity derived from the physical interface, not an opcode count that can be adjusted by copying labels or applying dyadic padding.

If the model retains failure flags, resource records, or control registers, denote the extended program output state by $\widehat\rho_p$. The model also fixes an unconditional CPTP readout $\mathcal R_{R,n}$ and an extended reference state $\widehat\sigma_{R,n}$, both target-independent and uniform across the model family. In addition to reference consistency, complete positivity gives, for any positive operator $\widehat M_n$ on the extended output space,
\begin{equation}
\label{eq:F.10}
\begin{gathered}
\rho_p
=
\mathcal R_{R,n}(\widehat\rho_p),
\qquad
\mathcal R_{R,n}(\widehat\sigma_{R,n})
=
\sigma_{R,n},
\\
\widehat M_n
\preceq
C\widehat\sigma_{R,n}
\Longrightarrow
\mathcal R_{R,n}(\widehat M_n)
\preceq
C\sigma_{R,n}.
\end{gathered}
\end{equation}

Complete positivity of $\mathcal R_{R,n}$ therefore carries operator domination on the extended space to the effective output space where acceptance is tested against the target state. Appendix~\ref{app:H} denotes the same readout by $\mathcal R_R$ when suppressing the size index. Program outputs always mean normalized unconditional outputs; conditional normalization of a success branch without accounting for its probability and resources is not part of the program semantics used here.

Finite-resource realization does not by itself turn continuous knobs into a finite alphabet. Continuous or noisy control requires a compact single-slot reachable domain, a nonzero reliable resolution, a finite discrete control codebook, and an indexing rule uniform across the model family to be fixed first. Alternatively, operational packing or covering numbers can define a precision-dependent capacity, with the precision overhead included explicitly in the model. The threshold relation $d(c,c')\le\delta$ is generally not transitive and cannot directly replace the exact quotient in Eq.~\eqref{eq:F.8}.

A simple counterexample shows why these specifications must be complete. Consider the one-state prefix grammar $\{1^k0:k\ge0\}$. On reading the terminator $0$, the controller performs a global reset of all $n$ qubits to $|0^n\rangle$. The grammar has a finite description and is Kraft-complete, every valid program has total output semantics, and the dynamics are CPTP; nevertheless,
\begin{equation}
\label{eq:F.11}
M_n
=
\sum_{k\ge0}
2^{-(k+1)}
|0^n\rangle\langle0^n|
=
|0^n\rangle\langle0^n|,
\qquad
C_n^\star
=
2^n.
\end{equation}

A finite controller has not prevented a macro of constant length from carrying extensive low-entropy structure. Model qualification therefore requires this structural generation capability to be constrained by the same reference specification over both the program family and the entire size-indexed family.

\subsection{Semantic Qualification, Verifiable Certificates, and Logical Levels}
\label{app:F.2}

This section gives the full mathematical characterization used in Proposition 2.1 of the main text. Within the specified prefix-free unconditional program semantics, the effect responses of the full prior relative to the structured vacuum are uniformly bounded if and only if the prior satisfies operator domination uniformly across the model family. We give the precise quantifiers, minimal domination constant, and verifiable certificates for this relation.

Fix a model family
\(\widehat{\mathfrak M}_{R,n}=(\mathfrak M_{R,n},U_{R,n}),\ n\in\mathbb N,\)
and let $\mathcal P_n$ be the full prefix-free valid program domain of $U_{R,n}$. All program outputs have been mapped, as in Eq.~\eqref{eq:F.10}, to the effective output space where acceptance is tested against the target state. Define the program semidensity
\begin{equation}
\label{eq:F.12}
M_n
:=
\sum_{p\in\mathcal P_n}
2^{-|p|}\rho_{p,n}.
\end{equation}

Taking the inverse on ${\rm supp}\,\sigma_{R,n}$, the minimal semantic domination constant is
\begin{equation}
\label{eq:F.13}
C_n^\star
:=
\left\|
\sigma_{R,n}^{-1/2}
M_n
\sigma_{R,n}^{-1/2}
\right\|_\infty.
\end{equation}

If ${\rm supp}\,M_n\nsubseteq{\rm supp}\,\sigma_{R,n}$, set $C_n^\star=\infty$. This constant also has a direct variational representation over effects:
\begin{equation*}
C_n^\star
=
\sup_{\substack{
0\preceq Q\preceq I\\
{\rm Tr}(Q\sigma_{R,n})>0
}}
\frac{
{\rm Tr}(QM_n)
}{
{\rm Tr}(Q\sigma_{R,n})
}.
\end{equation*}

If an effect satisfies ${\rm Tr}(Q\sigma_{R,n})=0$ but ${\rm Tr}(QM_n)>0$, the right-hand side is taken to be $+\infty$ under the support convention above. Indeed, $M_n\preceq C\sigma_{R,n}$ immediately bounds the response ratio by $C$ in every effect direction. Conversely, suppose these response ratios are uniformly bounded by $C<\infty$, but $M_n-C\sigma_{R,n}\npreceq0$. Choosing its positive spectral projection as $Q$ gives
\begin{equation*}
{\rm Tr}\!\left[
Q(M_n-C\sigma_{R,n})
\right]
>
0,
\end{equation*}
contradicting the upper bound on the response ratio. The supremum is therefore the smallest admissible domination constant and equals the operator norm in Eq.~\eqref{eq:F.13}. Semantic reference admissibility $\mathsf{RA}$ of the model family thus has the exact characterization
\begin{equation}
\label{eq:F.14}
\mathsf{RA}
\quad\Longleftrightarrow\quad
\sup_n C_n^\star<\infty
\quad\Longleftrightarrow\quad
\exists\,C<\infty:
M_n\preceq C\sigma_{R,n}
\ \text{for all }n.
\end{equation}

If the uniform bound across the family in Eq.~\eqref{eq:F.14} fails, then either, for every $k$, there is a size $n_k$ with $C_{n_k}^\star>2^k$, or support incompatibility already occurs at some size. In the former case, the variational characterization in Eq.~\eqref{eq:F.13} gives a nonzero vector $v_k$ satisfying
\begin{equation*}
\langle v_k,M_{n_k}v_k\rangle
>
2^k
\langle v_k,\sigma_{R,n_k}v_k\rangle;
\end{equation*}
setting $Q_k=|v_k\rangle\langle v_k|/\lVert v_k\rVert^2$ yields the effect witness used in Proposition 2.1 of the main text. In the support-incompatible case, one can choose $Q$ in $\ker\sigma_{R,n}$ such that
\begin{equation*}
{\rm Tr}(QM_n)
>
0
=
{\rm Tr}(Q\sigma_{R,n}).
\end{equation*}

Thus every failure of reference domination across the family has an explicit refuting witness in a finite measurement direction.

The constant $C_n^\star$ is the largest effect-response ratio of the full program prior relative to the structured vacuum. A single finite-dimensional model with a full-rank reference always has $C_n^\star<\infty$ when supports are compatible. Structural fairness requires these ratios to remain uniformly bounded over the entire size-indexed family, keeping the directional bias of the description prior on the same reference scale.
Equation~\eqref{eq:F.14} controls the directionality of the full program prior relative to the structured vacuum. Unused codewords and nonhalting mass can nevertheless leave $M_n$ subnormalized, with no weight in some directions allowed by the reference. To retain the same reference baseline explicitly in the certificate parameterization, fix $0<\alpha<1$ independently of the target state and system size, and set
\begin{equation*}
M_{\alpha,n}
:=
(1-\alpha)M_n
+
\alpha\sigma_{R,n}.
\end{equation*}

The operator $M_{\alpha,n}$ is an auxiliary anchored operator for certificate parameterization. The actual program semidensity remains $M_n$, and physical execution and resource counting follow the original model. Its relation to Eq.~\eqref{eq:F.14} is
\begin{equation}
\label{eq:F.15}
\begin{gathered}
M_n
\preceq
C\sigma_{R,n}
\ \Longrightarrow\
M_{\alpha,n}
\preceq
\bigl[
\alpha+(1-\alpha)C
\bigr]\sigma_{R,n},
\\[2pt]
M_{\alpha,n}
\preceq
2^{\kappa_U}\sigma_{R,n}
\ \Longrightarrow\
M_n
\preceq
\frac{
2^{\kappa_U}-\alpha
}{
1-\alpha
}
\sigma_{R,n}.
\end{gathered}
\end{equation}

Semantic $\mathsf{RA}$ and this $M_\alpha$ condition therefore admit a fully explicit conversion between their constants. One may take
\begin{equation*}
\kappa_U
=
\max\!\left\{
0,
\log_2\bigl[
\alpha+(1-\alpha)C
\bigr]
\right\},
\end{equation*}
while the program-output gap constant in Appendix~\ref{app:A} is
\begin{equation*}
\chi_U
=
\log_2
\frac{
2^{\kappa_U}-\alpha
}{
1-\alpha
}.
\end{equation*}

Reference domination controls only how the program prior is distributed over output structure; it does not yet establish that the same program interface faithfully represents every admissible circuit. Let $d_{R,n}^{\rm out}$ be the state distance on the effective output space specified by the model's error convention. Faithful transcription $\mathsf{TC}$ requires a target-independent compilation map, uniform across the model family, that covers all admissible circuits. Before target selection, the model declares which branch below it uses. The approximate branch also declares a nonempty certified precision domain $\Delta_{\rm TC}\subset(0,1)$, uniform across the model family, and must hold for every $\delta_{\rm cmp}$ in that domain. For a circuit with output state $\omega_{\mathcal C}$ and cost $L(\mathcal C)$, the two branches are
\begin{equation}
\label{eq:F.16}
\begin{aligned}
\rho_{p_{\mathcal C}}
&=
\omega_{\mathcal C},
\qquad
\text{exact transcription},
\\
|p_{\mathcal C}|
&\le
L(\mathcal C)\log_2\Gamma_{R,n}
+
a\log_2\max\{1,L(\mathcal C)\}
+
\gamma,
\\[2pt]
d_{R,n}^{\rm out}\!\left(
\rho_{p_{\mathcal C,\delta_{\rm cmp}}},
\omega_{\mathcal C}
\right)
&\le
\delta_{\rm cmp},
\qquad
\delta_{\rm cmp}\in\Delta_{\rm TC},
\\
|p_{\mathcal C,\delta_{\rm cmp}}|
&\le
L(\mathcal C)\log_2\Gamma_{R,n}
+
a\log_2\max\{1,L(\mathcal C)\}
+
b\log_2(1/\delta_{\rm cmp})
+
\gamma.
\end{aligned}
\end{equation}

Here $a,b,\gamma$ are uniform across the model family. Native discrete models use the exact branch with $b=0$. In the approximate branch, $\delta_{\rm cmp}$ is the compilation error, while $\epsilon$ denotes the target tolerance. If the circuit output approximates the target within error $\epsilon_{\rm circ}$, the final program error is composed according to the model's rule declared in advance; for a metric error criterion, it is at most $\epsilon_{\rm circ}+\delta_{\rm cmp}$. Addressing, precision, auxiliary resources, failure records, and control semantics that belong to circuit execution must be preserved by compilation or charged explicitly.

A model family satisfies the uniform reference-domination certificate condition $\mathsf{RefCert}$ if there is a sound proof object consisting of target-independent local data whose verification rigorously implies the domination across the family in Eq.~\eqref{eq:F.14}. Appendices~\ref{app:F.3}--\ref{app:F.4} give analytical certificates for native languages with equal costs. For finite controllers with measurement, classical control, and loops, Appendix~\ref{app:H} supplies another certificate route through least fixed points, Bellman--Choi supersolutions, or reference potentials uniform across the model family, once a synchronous prefix-CP realization has been established.
The condition $\mathsf{RefCert}_{\rm fc}$ in Appendix~\ref{app:H} is an instance of the general $\mathsf{RefCert}$ condition in the synchronous finite-control class. The subscript ${\rm fc}$ only identifies the certificate's model class; its logical role remains that of a sound sufficient condition for $\mathsf{RA}$.

\Fstatement{Proposition F.2 (Reference-qualification certificate interface).}
If a model family satisfies $\mathsf{RCon}$, full $\mathsf{TC}$, and $\mathsf{RefCert}$, then
\begin{equation}
\label{eq:F.17}
\boxed{
\mathsf{RCon}
\wedge
\mathsf{TC}
\wedge
\mathsf{RefCert}
\ \Longrightarrow\
\mathsf{RCon}
\wedge
\mathsf{TC}
\wedge
\mathsf{RA}.
}
\end{equation}

\Fproof
Soundness of $\mathsf{RefCert}$ gives a constant $C<\infty$, uniform across the model family, such that the full program semidensity satisfies $M_n\preceq C\sigma_{R,n}$. Equation~\eqref{eq:F.14} is precisely $\mathsf{RA}$; $\mathsf{RCon}$ and $\mathsf{TC}$ remain separate inputs.\Fqed

This proposition gives concrete models a sufficient certification route into Appendix~\ref{app:A}. The general lower bound applies on $\mathsf{RCon}\wedge\mathsf{TC}\wedge\mathsf{RA}$, including models that satisfy these three conditions but are not recognized by the certificate system used here.
Three levels enter in turn: a single path, the full program prior, and the size-indexed family. Passing from the reference gain of one path to Eq.~\eqref{eq:F.12} requires summing the Kraft weights of all descriptions; passing to Eq.~\eqref{eq:F.14} also requires control of the size dependence of the domination constant.

The finite-control route accordingly retains three proof obligations. Faithful transcription $\mathsf{TC}$ ensures circuit semantics, length bounds, and coverage. The synchronous realization and terminating-domain coverage in Appendix~\ref{app:H.1} reassemble terminating CP paths into the full program semidensity. The condition $\mathsf{RefCert}$ then certifies reference domination of this semidensity uniformly across the family. Code weights, Born weights, and runtime fuel record the description prior, physical branch probabilities, and resource cost, respectively.
We use a sound sufficient certificate system. Deciding reference domination across a family for a general program interface can encode the halting problem, as discussed in Appendix~\ref{app:H.6}. At a fixed size, for a fixed finite-control class and a given $C$, Appendix~\ref{app:H.3} instead gives a necessary and sufficient characterization through Bellman--Choi feasibility.

\subsection{Canonical \texorpdfstring{$\Gamma_R$}{Gamma R}, Elias Transcription, and Reference Balance}
\label{app:F.3}

This section encodes every finite atomic word as a program with controlled length, then certifies the full prior under reference balance. Suppress the size index $n$ temporarily, and order the operationally reduced atomic alphabet with equal costs by a fixed rule as
\begin{equation}
\label{eq:F.18}
\mathcal A_R
=
\{0,1,\ldots,\Gamma_R-1\},
\qquad
2\le\Gamma_R<\infty,
\qquad
\overline\Phi_R
:=
\frac1{\Gamma_R}
\sum_{a\in\mathcal A_R}\Phi_a,
\end{equation}
where $\Phi_a$ is the total CPTP semantics of the corresponding atomic action on the effective retained space. The alphabet is reference-balanced on $\sigma_R$ if
\begin{equation}
\label{eq:F.19}
\boxed{
\overline\Phi_R(\sigma_R)
=
\sigma_R.
}
\end{equation}

This condition constrains the uniform control prior before a label is selected. It does not require atomic gates to be executed randomly in an experiment, nor does it require each $\Phi_a$ to preserve $\sigma_R$ separately. Directional reset or cooling is allowed, with compensation supplied by physically distinct control actions, the reference background, or explicit cost.

Whole-word ranking asymptotically attains the Hartley code rate for any integer $\Gamma_R$. For a word $w=a_1\cdots a_L\in\mathcal A_R^L$ of length $L$, define its base-$\Gamma_R$ rank
\begin{equation}
\label{eq:F.20}
r_L(w)
:=
\sum_{j=1}^{L}
a_j\Gamma_R^{L-j},
\qquad
0
\le
r_L(w)
<
\Gamma_R^L.
\end{equation}

Define the binary code length and occupancy factor, respectively, by
\begin{equation}
\label{eq:F.21}
\ell_{R,L}
:=
\left\lceil
L\log_2\Gamma_R
\right\rceil,
\qquad
\theta_{R,L}
:=
\Gamma_R^L2^{-\ell_{R,L}}
\in
(1/2,1],
\end{equation}
with $\ell_{R,0}=0$ and $\theta_{R,0}=1$. At fixed $L$, binary prefix codes covering all $\Gamma_R^L$ words satisfy
\begin{equation}
\label{eq:F.22}
\min_{\substack{
\mathsf E_L:\mathcal A_R^L\hookrightarrow\{0,1\}^\ast\\
\mathsf E_L\ {\rm prefix\!-free}
}}
\max_{w\in\mathcal A_R^L}
|\mathsf E_L(w)|
=
\left\lceil
L\log_2\Gamma_R
\right\rceil.
\end{equation}

Kraft's inequality gives the lower bound, and fixed-length rank blocks attain it. Thus
\begin{equation}
\label{eq:F.23}
\boxed{
\log_2\Gamma_R
=
\lim_{L\to\infty}
\frac1L
\min_{\substack{
\mathsf E_L:\mathcal A_R^L\hookrightarrow\{0,1\}^\ast\\
\mathsf E_L\ {\rm prefix\!-free}
}}
\max_{w\in\mathcal A_R^L}
|\mathsf E_L(w)|
}
\end{equation}
gives a coding-independent minimax characterization of the single-step Hartley control bandwidth. The encoded objects are complete atomic control histories. Distinct histories are not identified again at this level, even if they happen to compose to the same terminal channel.
Use the Elias-gamma code~\cite{Elias1975_UniversalCodewordSets} $h(L)$ of the positive integer $L+1$ as the length header, and write
\begin{equation}
\label{eq:F.24}
s(L)
=
2\lfloor\log_2(L+1)\rfloor+1,
\qquad
q_L
:=
2^{-s(L)},
\qquad
\sum_{L\ge0}q_L
=
1.
\end{equation}

For length $L$, define the canonical program by
\begin{equation}
\label{eq:F.25}
p_w
=
h(L)\,
\operatorname{bin}_{\ell_{R,L}}
\bigl(r_L(w)\bigr).
\end{equation}

After decoding $L$, the decoder reads exactly $\ell_{R,L}$ bits. If the binary rank is at least $\Gamma_R^L$, the string is outside the valid program domain. Headers for different lengths are prefix-free, and payloads of the same length have equal size. The canonical valid domain is therefore prefix-free, with
\begin{equation}
\label{eq:F.26}
|p_w|
=
s(L)
+
\left\lceil
L\log_2\Gamma_R
\right\rceil
\le
L\log_2\Gamma_R
+
2\log_2(L+1)
+
2.
\end{equation}

For the native model whose admissible circuits are all finite atomic words and whose word semantics are sequential compositions of total CPTP atomic actions, this construction establishes full faithful transcription uniformly across the model family. In the non-dyadic case, a symbolwise prefix code has worst-case single-symbol length at least $\lceil\log_2\Gamma_R\rceil$. Repeating its longest codeword incurs linear redundancy, whereas whole-word encoding together with the length header adds only logarithmic overhead. Full $\mathsf{TC}$ for measurement, loops, or a general finite-control grammar still requires separate verification for the corresponding program interface.

Let the output state of word $w$ be
\begin{equation*}
\rho_w
=
\Phi_{a_L}
\circ
\cdots
\circ
\Phi_{a_1}(\sigma_R).
\end{equation*}

The program semidensity of the canonical rank interpreter has the exact depth decomposition
\begin{equation}
\label{eq:F.27}
\boxed{
M_{\rm rank}
=
\sum_{L\ge0}
q_L\theta_{R,L}
\overline\Phi_R^{\,L}(\sigma_R).
}
\end{equation}

\Fstatement{Theorem F.3 (Reference-balanced transcription for any finite alphabet).}
If Eq.~\eqref{eq:F.19} holds, then
\begin{equation}
\label{eq:F.28}
M_{\rm rank}
=
K_{\Gamma_R}\sigma_R,
\qquad
K_{\Gamma_R}
:=
\sum_{L\ge0}
q_L\theta_{R,L}
\le
1,
\qquad
M_{\rm rank}
\preceq
\sigma_R.
\end{equation}

Thus an interpreter whose full valid domain is the canonical rank domain may take $\kappa_{U_{\rm rank}}=0$. When $\Gamma_R$ is a power of two, $K_{\Gamma_R}=1$; in the non-dyadic case, $3/4<K_{\Gamma_R}<1$.

\Fproof
At fixed $L$, all valid rank codes have the same program weight $q_L2^{-\ell_{R,L}}$, so
\begin{equation*}
\sum_{w\in\mathcal A_R^L}
q_L2^{-\ell_{R,L}}\rho_w
=
q_L\theta_{R,L}
\overline\Phi_R^{\,L}(\sigma_R),
\end{equation*}
and summing over all depths gives Eq.~\eqref{eq:F.27}. Reference balance gives $\overline\Phi_R^{\,L}(\sigma_R)=\sigma_R$, yielding Eq.~\eqref{eq:F.28}. In the dyadic case, all $\theta_{R,L}=1$. In the non-dyadic case, $1/2<\theta_{R,L}<1$ for $L\ge1$, while $q_0=1/2$, so $3/4<K_{\Gamma_R}<1$.\Fqed

Strict subnormalization in the non-dyadic case comes from unused binary ranks, rather than missing physical actions. These ranks may remain invalid. If an exact upper envelope is needed, they may instead be assigned output $\sigma_R$ in code space alone. The resulting reference-null envelope machine satisfies $M_{\rm env}=\sigma_R$; deleting these programs recovers the transcription of actual circuits. Neither treatment adds an identity to the physical alphabet or changes $\Gamma_R$.
Theorem F.3 also gives a direct one-shot check within the native model. For every valid program,
\begin{equation}
\label{eq:F.29}
2^{-|p_w|}\rho_w
\preceq
M_{\rm rank}
\preceq
\sigma_R
\quad\Longrightarrow\quad
D_{\max}(\rho_w\Vert\sigma_R)
\le
|p_w|.
\end{equation}

If $T(\rho_w,\rho)\le\epsilon$, then
\begin{equation}
\label{eq:F.30}
D_{\max}^{\epsilon}(\rho\Vert\sigma_R)
\le
\Lambda_{\Gamma_R}^{\rm rank}(L),
\qquad
\Lambda_{\Gamma_R}^{\rm rank}(L)
:=
\left\lceil
L\log_2\Gamma_R
\right\rceil
+
s(L).
\end{equation}

This bound contains no unknown domination constant, but serves only as an exact cross-check of the general proof chain in Appendix~\ref{app:A} within the canonical native model.

Finally, if programs are deleted from an already certified domain $\mathcal P$ while retaining the original codewords, code lengths, and weights, then
\begin{equation}
\label{eq:F.31}
\mathcal P'
\subseteq
\mathcal P
\quad\Longrightarrow\quad
M_{\mathcal P'}
\preceq
M_{\mathcal P}.
\end{equation}

Deletion may destroy balance at each length, exact equalities, or circuit coverage, but preserves the existing operator domination. Shortening codewords, reweighting, normalization, or adding new program modes may change that domination and therefore requires recertification. If the canonical rank domain is only one branch of a larger interpreter, $\kappa_{U_{\rm rank}}=0$ likewise covers only that branch; reference domination must still be established uniformly across the model family on the enlarged interpreter's full program domain, either directly or through a sound $\mathsf{RefCert}$.

\subsection{Stationary Reference Envelopes, Depth Moments, and Natural \texorpdfstring{$C>1$}{C > 1} Models}
\label{app:F.4}

Reference balance gives a particularly tight $C=1$ certificate, but is not necessary for semantic $\mathsf{RA}$. Equation~\eqref{eq:F.27} allows us to study directly how the uniform control prior departs from the reference after multiple steps. Restore the size index $n$ and write the code-space occupancy factor of Eq.~\eqref{eq:F.21} as $\theta_{R,n,L}:=\Gamma_{R,n}^{L}2^{-\lceil L\log_2\Gamma_{R,n}\rceil}$. Define the corresponding depth weights and domination constant by
\begin{equation}
\label{eq:F.32}
\widetilde q_{n,L}
:=
q_L\theta_{R,n,L},
\qquad
C_{{\rm rank},n}^\star
:=
\left\|
\sigma_{R,n}^{-1/2}
M_{{\rm rank},n}
\sigma_{R,n}^{-1/2}
\right\|_\infty.
\end{equation}

The actual rank interpreter satisfies
\begin{equation*}
M_{{\rm rank},n}
=
\sum_L
\widetilde q_{n,L}
\overline\Phi_{R,n}^{\,L}(\sigma_{R,n}).
\end{equation*}

For an analytical upper envelope based on uniform control at each depth that is easier to construct, also define
\begin{equation}
\label{eq:F.33}
M_{{\rm unif},n}
:=
\sum_{L\ge0}
q_L
\overline\Phi_{R,n}^{\,L}(\sigma_{R,n}),
\qquad
C_{{\rm unif},n}^\star
:=
\left\|
\sigma_{R,n}^{-1/2}
M_{{\rm unif},n}
\sigma_{R,n}^{-1/2}
\right\|_\infty.
\end{equation}

Since $0<\theta_{R,n,L}\le1$ and each depth term is a positive operator,
\begin{equation}
\label{eq:F.34}
M_{{\rm rank},n}
\preceq
M_{{\rm unif},n},
\qquad
C_{{\rm rank},n}^\star
\le
C_{{\rm unif},n}^\star.
\end{equation}

The operator $M_{{\rm unif},n}$ is a convenient but generally looser analytical upper envelope. In the non-dyadic case, it is not the exact semidensity of the actual rank program domain.

\Fstatement{Proposition F.4 (Stationary reference envelope).}
Suppose that, for every size $n$, there are a positive operator $\Omega_{R,n}$ and a finite constant $C_{\Omega,n}$ satisfying
\begin{equation}
\label{eq:F.35}
\begin{aligned}
\overline\Phi_{R,n}(\sigma_{R,n})
&\preceq
\Omega_{R,n},
&
\overline\Phi_{R,n}(\Omega_{R,n})
&\preceq
\Omega_{R,n},
\\
\Omega_{R,n}
&\preceq
C_{\Omega,n}\sigma_{R,n},
&
\sup_n C_{\Omega,n}
&<
\infty,
\end{aligned}
\end{equation}
then, for all $n$,
\begin{equation}
\label{eq:F.36}
M_{{\rm rank},n}
\preceq
\bigl[
q_0
+
C_{\Omega,n}
(K_{\Gamma_{R,n}}-q_0)
\bigr]
\sigma_{R,n}.
\end{equation}

\Fproof
Fix $n$. The depth-zero term is $q_0\sigma_{R,n}$. Equation~\eqref{eq:F.35} and positivity of the channel give, for every $L\ge1$,
\begin{equation*}
\overline\Phi_{R,n}^{\,L}(\sigma_{R,n})
\preceq
\Omega_{R,n}
\preceq
C_{\Omega,n}\sigma_{R,n}.
\end{equation*}

Summing with the actual weights $\widetilde q_{n,L}$ gives the bound. Since $K_{\Gamma_{R,n}}\le1$, the domination constant in Eq.~\eqref{eq:F.36} is at most
\begin{equation*}
q_0
+
(1-q_0)
\sup_n C_{\Omega,n}
<
\infty,
\end{equation*}
so the certificate is uniform across the model family.\Fqed

The operator $\Omega_{R,n}$ gives a common upper envelope for the average outputs at every positive depth. The condition $\sup_n C_{\Omega,n}<\infty$ keeps this envelope on the same reference scale throughout the size-indexed family. Thus the one-step average output may depart from $\sigma_{R,n}$, provided subsequent evolution remains bounded by this envelope.
Depth gain gives another sufficient condition that is easy to audit. Define
\begin{equation*}
G_{R,n}(L)
:=
D_{\max}\!\left(
\overline\Phi_{R,n}^{\,L}(\sigma_{R,n})
\middle\Vert
\sigma_{R,n}
\right).
\end{equation*}

If there is a constant $C<\infty$, uniform across the model family, such that
\begin{equation}
\label{eq:F.37}
\sup_n
\sum_{L\ge0}
\widetilde q_{n,L}
2^{G_{R,n}(L)}
\le
C,
\end{equation}
then
$M_{{\rm rank},n}\preceq C\sigma_{R,n}$
for all $n$.
Replacing $\widetilde q_{n,L}$ by $q_L$ gives a looser but still sound corollary. Equation~\eqref{eq:F.37} takes the worst direction at each depth and is therefore only a sufficient condition: directions of high gain at different depths may be mutually orthogonal, making the norm of the total operator much smaller than the sum of the individual norms.

The following example shows that the native qualified class strictly contains the reference-balanced class. Take
\begin{equation}
\label{eq:F.38}
\sigma_n
=
\frac{I}{2^n},
\qquad
\mathcal A
=
\{\operatorname{id},\mathcal R_{1,0}\},
\end{equation}
where $\mathcal R_{1,0}$ resets only the first qubit to $|0\rangle$. Here $\Gamma=2$, and
\begin{equation}
\label{eq:F.39}
\overline\Phi^{\,L}(\sigma_n)
=
2^{-L}\sigma_n
+
(1-2^{-L})
\mathcal R_{1,0}(\sigma_n).
\end{equation}

Since the alphabet is dyadic,
$\widetilde q_{n,L}=q_L$.
Taking the largest eigenvalue in the $|0\rangle$ sector of the first qubit gives
\begin{equation}
\label{eq:F.40}
C_{{\rm rank},n}^\star
=
2
-
\sum_{L\ge0}
q_L2^{-L}
\approx
1.398804,
\end{equation}
independently of $n$. This example is certified by the depth-gain condition in Eq.~\eqref{eq:F.37}; its weighted sum equals the minimal domination constant in Eq.~\eqref{eq:F.40}. Thus, under the original equal-cost accounting, the class of reference-admissible models strictly contains the reference-balanced class. Stationary envelopes and depth-gain conditions provide two distinct sufficient certification routes.

\subsection{Nontrivial Qubit Models and Native Model Witnesses}
\label{app:F.5}

A qualified model containing only unital dynamics that preserve the maximally mixed state satisfies reference balance but cannot generate low-entropy states from the structured vacuum. We now give an explicit model family with both reference qualification and nontrivial generative capability.

Specifically, let the system space, dimension, and reference state be
\begin{equation}
\label{eq:F.41}
\mathcal H_n
=
(\mathbb C^2)^{\otimes n},
\qquad
d_n
=
2^n,
\qquad
\sigma_n
=
\frac{I}{2^n}.
\end{equation}

Fix a finite, location-addressed, operationally reduced atomic alphabet containing addressed universal discrete unitary gates, such as $H$, $T$, and controlled-NOT (CNOT)~\cite{BoykinEtAl2000_UniversalQuantumBasis}; the paired reset channels on each qubit $j$,
\begin{equation}
\label{eq:F.42}
\mathcal R_{j,b}(\rho)
=
|b\rangle\langle b|_j
\otimes
{\rm Tr}_j\rho,
\qquad
b\in\{0,1\};
\end{equation}
and the local complete dephasing channels
\begin{equation}
\label{eq:F.43}
\Delta_j(\rho)
=
\frac12
\bigl(
\rho
+
Z_j\rho Z_j
\bigr).
\end{equation}

The physical alphabet has no identity padding. If the platform has a genuine wait or identity control occupying the same resource slot as the other actions, it appears as one actual operational class. If additional action classes are included, the complete alphabet must still satisfy the reference-balance condition in Eq.~\eqref{eq:F.44}. $\Gamma_n$ is the actual number of classes of fully addressed actions after deduplication by Eq.~\eqref{eq:F.8}. Addressing enters $\Gamma_n$ when carried by the one-step control word; when loaded through an external register, its loading cost is charged separately.

Unitary channels and $\Delta_j$ each preserve $\sigma_n$, while the paired resets satisfy
\begin{equation*}
\frac12
\left[
\mathcal R_{j,0}(\sigma_n)
+
\mathcal R_{j,1}(\sigma_n)
\right]
=
\sigma_n.
\end{equation*}

Thus the entire actual alphabet satisfies
\begin{equation}
\label{eq:F.44}
\frac1{\Gamma_n}
\sum_{a\in\mathcal A_n}
\Phi_{n,a}(\sigma_n)
=
\sigma_n.
\end{equation}

Theorem F.3 then gives
\begin{equation}
\label{eq:F.45}
(M_n)_{\rm rank}
\preceq
\sigma_n,
\qquad
\kappa_{U_{\rm rank},n}
=
0
\quad
\text{for all }n.
\end{equation}

\Fstatement{Proposition F.5 (Nontrivial qubit models).}
The model family above has approximately universal state-generation capability together with native reference qualification that is uniform across the model family.

\Fproof
Starting from $\sigma_n$, execute
$\mathcal R_{1,0},\ldots,\mathcal R_{n,0}$
in sequence to obtain $|0^n\rangle$ in $n$ explicit atomic steps. Universal discrete unitary gates can then generate any pure state to a prescribed accuracy~\cite{DawsonNielsen2006_SolovayKitaev}.
For an arbitrary mixed state, take a spectral decomposition
\begin{equation}
\label{eq:F.46}
\rho
=
U\left(
\sum_x
\lambda_x
|x\rangle\langle x|
\right)
U^\dagger.
\end{equation}

Approximately prepare the amplitude state $\sum_x\sqrt{\lambda_x}|x\rangle$ from $|0^n\rangle$, apply $\Delta_j$ to all qubits to approximate $\sum_x\lambda_x|x\rangle\langle x|$, and then approximately implement $U$. Suppose the amplitude-state preparation has trace-distance error at most $\epsilon_1$ and the eigenbasis rotation has output error at most $\epsilon_2$ on the ideal spectral diagonal state. CPTP contractivity and the triangle inequality bound the final error by $\epsilon_1+\epsilon_2$. Both stages can be made arbitrarily accurate, so choosing $\epsilon_1+\epsilon_2\le\epsilon$ yields a circuit generating the mixed state within any positive tolerance. The actual control history carries the direction choices, addresses, and precision at each step.\Fqed

Alternatively, one may provide a workspace register of the same dimension, start from the corresponding maximally mixed reference, explicitly reset, prepare a purification of the target mixed state, and finally discard the workspace. Neither implementation supplies an undeclared pure ancilla for free.

This example shows how reference balance accommodates directional generation. The uniform control prior preserves the structured vacuum, while a specific program selects a structural direction through its reset-to-0 or reset-to-1 labels and execution steps. $\Gamma_n$ typically grows with system size, so the per-step bandwidth remains $g_n=\log_2\Gamma_n$; the reference-domination parameter uniform across the model family is $\kappa_{U_{\rm rank},n}=0$.

If one atomic slot permits parallel execution of disjoint local gates, the complete parallel pattern enters the action classes through Eq.~\eqref{eq:F.8}. Serial and parallel $L\log_2\Gamma$ scales must be compared using target-state-independent compilation bounds that preserve resource semantics. On platforms with asymmetric energy costs, the two resets may have different dynamical costs; Appendix~\ref{app:F.8} discusses the requirements for a corresponding weighted extension.

\subsection{Single-Path Reference Gain and Cost Calibration}
\label{app:F.6}

Reference balance constrains the full program prior, while an individual path may still generate low-entropy structure. This subsection quantifies single-path reference gain and uses it to construct the corresponding code costs. For a support-compatible CPTP atomic channel $\Phi_a$, define the one-step reference gain
\begin{equation}
\label{eq:F.47}
g_a
:=
D_{\max}\!\left(
\Phi_a(\sigma_R)
\middle\Vert
\sigma_R
\right),
\qquad
\Phi_a(\sigma_R)
\preceq
2^{g_a}\sigma_R.
\end{equation}

If support compatibility fails, set $g_a=\infty$. By recursively applying channel positivity, every word
$w=a_L\cdots a_1$
satisfies
\begin{equation}
\label{eq:F.48}
\Phi_w(\sigma_R)
\preceq
2^{\sum_{j=1}^{L}g_{a_j}}
\sigma_R.
\end{equation}

This is a compositional certificate for single-path reference gain; the full program prior still requires summing all descriptions with their code weights. In a reference-balanced alphabet, the positive-operator sum
\begin{equation*}
\sum_{a\in\mathcal A_R}
\Phi_a(\sigma_R)
=
\Gamma_R\sigma_R
\end{equation*}
also directly gives
\begin{equation}
\label{eq:F.49}
g_a
\le
\log_2\Gamma_R
\quad
\text{for every }a.
\end{equation}

This local bound shows that an individual action may generate reference structure, but its worst-case one-step gain cannot exceed the bandwidth of the control choices exposed in the same atomic slot.

\Needspace{4\baselineskip}
For an action set without reference balance, reference gain can be incorporated into action code lengths before the target state is selected. Let the finite action set be
$\{\Phi_a\}_{a=1}^{N}$,
and assign target-state-independent integer prefix-code lengths $\ell_a$ to its actions. Define the weighted one-step operator map
\begin{equation}
\label{eq:F.50}
\mathcal T_{\ell}
:=
\sum_{a=1}^{N}
2^{-\ell_a}\Phi_a.
\end{equation}

\Fstatement{Theorem F.6 (One-step reference gain--cost completion).}
If
\begin{equation}
\label{eq:F.51}
\mathcal T_{\ell}(\sigma_R)
\preceq
\sigma_R,
\end{equation}
then the interpreter for all finite words, using an explicit prefix-free length header followed by concatenated action codes, satisfies
$M\preceq\sigma_R$.

\Fproof
After the length header specifies $L$, the action prefix codes allow exactly $L$ actions to be parsed in sequence. The weighted program sum at fixed depth is
$\mathcal T_{\ell}^{\,L}(\sigma_R)$.
Equation~\eqref{eq:F.51} and positivity give
\begin{equation*}
\mathcal T_{\ell}^{\,L}(\sigma_R)
\preceq
\sigma_R.
\end{equation*}

Summing with weights $q_L$ and using
$\sum_Lq_L=1$
then shows that the full program semidensity is bounded above by $\sigma_R$.\Fqed

Equation~\eqref{eq:F.47} gives an actionwise sufficient condition:
\begin{equation}
\label{eq:F.52}
\sum_{a=1}^{N}
2^{-\ell_a+g_a}
\le
1
\quad\Longrightarrow\quad
\mathcal T_{\ell}(\sigma_R)
\preceq
\sigma_R.
\end{equation}

When all actions are support-compatible, that is,
$g_a<\infty$
for every $a$, an explicit choice is
\begin{equation}
\label{eq:F.53}
\ell_a
=
\lceil g_a\rceil
+
\lceil\log_2N\rceil
\end{equation}
which satisfies the Kraft condition and gives an explicit completion. If some $g_a=\infty$, every finite $\ell_a$ leaves positive weight in
$\mathcal T_\ell(\sigma_R)$
on $\ker\sigma_R$, so Eq.~\eqref{eq:F.51} cannot hold. The action must be removed from this effective alphabet or recertified after a coordinated change to the output space, extended reference state, and CPTP readout. Within the same effective output space and reference support, merely increasing code length or rewriting the action as a longer decomposition with unchanged terminal-channel semantics cannot eliminate the final support leakage.
If every action uses the same integer code length $b$, the minimal calibration is
\begin{equation}
\label{eq:F.54}
\begin{aligned}
b^\star
&=
\left\lceil
D_{\max}\!\left(
\sum_{a=1}^{N}
\Phi_a(\sigma_R)
\middle\Vert
\sigma_R
\right)
\right\rceil
\\
&=
\left\lceil
\log_2N
+
D_{\max}\!\left(
\frac1N
\sum_{a=1}^{N}
\Phi_a(\sigma_R)
\middle\Vert
\sigma_R
\right)
\right\rceil.
\end{aligned}
\end{equation}

Extensive one-step reference gain therefore requires extensive effective code cost. Theorem F.6 certifies a program model whose cost is reassigned according to the action code length $\sum_j\ell_{a_j}$; its correspondence with the original $L\log_2\Gamma_R$ scale must be established by an appropriate faithful transcription relation. If fuel or a lower-level action decomposition carries this burden, it should enter the effective description length through faithful compilation that is target-state-independent and uniform across the model family; alternatively, the new program interface must be recertified.

The same principle can calibrate high-level source code against the actual execution trace. Let $p$ belong to the halting program domain of a prefix-free high-level generator and deterministically produce a finite reference-balanced atomic circuit of length $L(p)$. Set
\begin{equation}
\label{eq:F.55}
f_R(p)
:=
\left\lceil
L(p)\log_2\Gamma_R
\right\rceil.
\end{equation}

By Eq.~\eqref{eq:F.49} and path composition, the output state $\rho_p$ satisfies
\begin{equation}
\label{eq:F.56}
\rho_p
\preceq
\Gamma_R^{L(p)}\sigma_R
\preceq
2^{f_R(p)}\sigma_R.
\end{equation}

The decoder first runs $p$ to generate the circuit without executing it, then checks the resource suffix once $f_R(p)$ is known. Define
\begin{equation}
\label{eq:F.57}
\widetilde p
=
p\,h\bigl(f_R(p)\bigr)0^{f_R(p)}.
\end{equation}

The effective length is then
\begin{equation}
\label{eq:F.58}
\ell_{\rm op}(p)
=
|p|
+
f_R(p)
+
s\bigl(f_R(p)\bigr),
\end{equation}
and the augmented program domain is prefix-free, with program semidensity satisfying
\begin{equation}
\label{eq:F.59}
\widetilde M_{\rm op}
=
\sum_p
2^{-|p|-s(f_R(p))-f_R(p)}
\rho_p
\preceq
\sigma_R.
\end{equation}

The last step applies Eq.~\eqref{eq:F.56} term by term, then uses the nonnegative resource-header length and the Kraft inequality for the source program domain to obtain
$\sum_p2^{-|p|-s(f_R(p))}
\le
\sum_p2^{-|p|}
\le1$.
This construction retains the source program's algorithmic compression of rules and loops while restoring the information volume of actual atomic execution to the effective physical length. \mbox{It handles deterministic finite execution;} adaptive protocols may encode a worst-case length valid for all branches. Random stopping times, repeat-until-success protocols, and expected costs belong to different resource models.
Appendix~\ref{app:H.6} extends the same cost completion to restricted finite controllers with syntax-state-dependent code lengths, measurement, and feed-forward. This route establishes reference domination after costs are reassigned; model qualification then also requires the independent full faithful transcription specified in Eq.~\eqref{eq:F.16}.

\subsection{Certified Mixtures, Low-Entropy No-Go Results, and Failure Boundaries}
\label{app:F.7}

The canonical rank interpreter proves that the class of reference-admissible models is nonempty. This subsection constructs an interpreter that is universal relative to a soundly certified class sharing the same reference and resource specification, then derives necessary constraints on program length from low-entropy generation.
Fix a sound verifier. Let $\{V_j\}_{j\ge0}$ be an effectively enumerable class of prefix-free interfaces certified by this verifier and uniformly simulatable with respect to the reference scale $R$. Each member carries a verifiable domination constant $0<C_j<\infty$ that is uniform across the model family. Set
$k_j:=\max\{0,\lceil\log_2C_j\rceil\}$,
so that
\begin{equation}
\label{eq:F.60}
M_{j,R}
\preceq
C_j\sigma_R
\preceq
2^{k_j}\sigma_R,
\qquad
k_j\in\mathbb N_0.
\end{equation}

Here $C_j$ and $k_j$ depend only on interface $j$, not on the target state, $d_R$, or system size. Let $g(j)$ be the Elias-gamma code of the positive integer $j+1$, and define the interface prefix
\begin{equation}
\label{eq:F.61}
e_j
:=
g(j)0^{k_j}.
\end{equation}

The set $\{e_j\}$ remains prefix-free, and
\begin{equation}
\label{eq:F.62}
\sum_{j\ge0}
2^{-|e_j|+k_j}
=
\sum_{j\ge0}
2^{-|g(j)|}
=
1.
\end{equation}

Define the mixture interpreter
\begin{equation}
\label{eq:F.63}
U_R(e_jp)
=
V_{j,R}(p).
\end{equation}

\Fstatement{Theorem F.7 (Universal mixture of a certified subclass).}
The interpreter $U_R$ above remains reference-admissible and simulates each $V_j$ with a fixed additive overhead depending only on the interface being simulated.

\Fproof
The concatenated program domain is prefix-free, and
\begin{equation}
\label{eq:F.64}
M_{U,R}
=
\sum_{j\ge0}
2^{-|e_j|}M_{j,R}
\preceq
\sum_{j\ge0}
2^{-|e_j|+k_j}\sigma_R
\preceq
\sigma_R.
\end{equation}

Thus one may take $\kappa_U=0$. For each fixed $j$,
\begin{equation}
\label{eq:F.65}
K_{\epsilon}^{U_R}(\rho)
\le
K_\epsilon^{V_{j,R}}(\rho)
+
|e_j|.
\end{equation}

This proves the universal-mixture property for the certified subclass.\Fqed

Taking $V_0$ to be the canonical rank transcription decoder of Appendix~\ref{app:F.3} also makes the mixture interpreter faithfully cover all native valid circuits. A finite-control interface certified by a strict proof object from Appendix~\ref{app:H} may likewise be included as a verified member, provided it also has independent full $\mathsf{TC}$ and satisfies the same reference, output-readout, and resource specifications. The theorem establishes relative universality within a soundly certified class. Algorithmic invariance for universal Turing machines (UTMs) and quantum Turing machines (QTMs) continues to serve its standard role for source-code descriptions.

Certified mixtures show how qualified interfaces can be extended under a shared reference and resource specification. The low-entropy no-go result identifies resource boundaries that such extensions cannot cross. Let
\begin{equation}
\label{eq:F.66}
P_n
=
|\psi_n\rangle\langle\psi_n|,
\qquad
\sigma_n
=
\frac{I}{d_n},
\qquad
T(\rho_{p_n},P_n)
\le
\epsilon_n.
\end{equation}

For $\epsilon_n<1$, if reference domination at each size is written as
\(M_n
\preceq
C_{U,n}\sigma_n
=
2^{\chi_{U,n}}\sigma_n,
\)
then taking the expectation in the $P_n$ direction gives
\begin{equation}
\label{eq:F.67}
\chi_{U,n}
\ge
\log_2 d_n
-
|p_n|
+
\log_2(1-\epsilon_n).
\end{equation}

\Fstatement{Proposition F.8 (Incompatibility of uncompensated low-entropy generation).}
If
$-\log_2(1-\epsilon_n)=O(1)$
and
$\log_2d_n-|p_n|\to\infty$,
then the semantic domination constant $\chi_{U,n}$ cannot remain uniformly bounded across the model family. In particular, when $d_n=2^n$,
\begin{equation}
\label{eq:F.68}
\chi_{U,n}
\ge
n
-
|p_n|
+
\log_2(1-\epsilon_n).
\end{equation}

\Fproof
Measurement monotonicity of trace distance gives
${\rm Tr}(P_n\rho_{p_n})\ge1-\epsilon_n$.
From
$M_n\succeq2^{-|p_n|}\rho_{p_n}$
and semantic reference domination, taking the expectation in the $P_n$ direction yields Eq.~\eqref{eq:F.67}. The stated asymptotic conditions then give the conclusion.\Fqed

If the anchored certificate parametrization of Eq.~\eqref{eq:F.15} is used instead, the same calculation gives
\begin{equation*}
\kappa_{U,n}
\ge
\log_2\!\left[
\alpha
+
(1-\alpha)(1-\epsilon_n)
d_n2^{-|p_n|}
\right].
\end{equation*}

Fixing $0<\alpha<1$ changes only the conversion between constants, leaving the family-level incompatibility conclusion of Proposition F.8 unchanged.
Equation~\eqref{eq:F.67} shows that, relative to the maximally mixed baseline, the information required for low-entropy generation must be jointly carried by program length and the reference-domination constant. To incorporate a pure-initial-state or global-reset model into this specification, one may explicitly charge entropy-removal steps, jointly upgrade the reference and resource rules, or establish incremental complexity conditional on the low-entropy initial state.

The other main failure modes can be understood by where structure enters the model.
At the atomic interface, deviations from exact reference balance generally accumulate with depth. If a weighted one-step map satisfies only
\begin{equation*}
\mathcal T(\sigma_R)
\preceq
2^\delta\sigma_R,
\end{equation*}
then one can generally infer only
\begin{equation}
\label{eq:F.69}
\mathcal T^{L}(\sigma_R)
\preceq
2^{L\delta}\sigma_R.
\end{equation}

Such depth-accumulating deviations can be controlled by establishing exact balance or a stable envelope, or by incorporating reference gain into the per-step cost as in Appendix~\ref{app:F.6}.
In execution semantics, Born weights of measurement outcomes are carried by the traces of the completely positive, trace-nonincreasing (TNI) branch maps of a quantum instrument. Different outcomes of the same program cannot each receive the program's code weight again. Conditional normalization in postselection, random stopping times, repeat-until-success (RUS) protocols, hidden clocks, undeclared auxiliary states, and unreturned catalysts must enter unconditional output semantics or resource costs whenever they change the attainable structure. Continuous control, addressing, and precision return to the complete atomic interface of Appendix~\ref{app:F.1}.

At the size limit, finite-control semidefinite programming (SDP) can discover candidate certificates at a fixed size. A family-level conclusion additionally requires controlling $\sup_n C_n^\star$ through the uniform reference potentials of Appendix~\ref{app:H.4} or the analytical envelopes of Appendix~\ref{app:F.4}.

In the reference-balanced class, Eq.~\eqref{eq:F.49} directly gives a local calibration for bare atomic actions. More generally, exact
$\mathsf{TC}+\mathsf{RA}$
gives an unsmoothed bound of the same kind for unit-cost actions. Approximate transcription gives the corresponding smoothed bound at any certified precision
$\delta_{\rm cmp}\in\Delta_{\rm TC}$
fixed in advance. In both cases, the additional terms are uniformly $O(1)$ across the model family. Thus, if a family of unit-cost actions satisfies, in the exact branch,
\begin{equation}
\label{eq:F.70}
D_{\max}\!\left(
\Phi_{a_n}(\sigma_{R,n})
\middle\Vert
\sigma_{R,n}
\right)
-
\log_2\Gamma_{R,n}
\longrightarrow
\infty,
\end{equation}
then this action family is incompatible with exact $\mathsf{TC}$ and $\mathsf{RA}$ that are uniform across the model family. The approximate branch uses the corresponding $D_{\max}^{\delta_{\rm cmp}}$ criterion, whose conclusion applies at that certified smoothing radius. The required reference gain may be charged as cost following Appendix~\ref{app:F.6}, or incorporated into a new resource specification through a reference upgrade. A strong operation directly implemented by hardware within one atomic slot may remain a primitive, with its complete spacetime support, control, auxiliary-state, and environment semantics all entering Eq.~\eqref{eq:F.7}.
Reference balance, stationary envelopes, and finite-control reference potentials provide sufficient certification routes in Appendix~\ref{app:F.3}, Appendix~\ref{app:F.4}, and Appendix~\ref{app:H}, respectively, turning the corresponding resource requirements into verifiable reference domination.

\subsection{Reference Upgrades, Robustness, and Generalized Boundaries}
\label{app:F.8}

New conserved quantities, stable logical constraints, or reproducible structural residuals can support an upgrade of the reference background. This subsection compares state-side readings under the old and new references and gives the corresponding relations for symmetry, truncation, and weighted references. Denote the old and new models by
\begin{equation}
\label{eq:F.71}
\widehat{\mathfrak M}_R
=
(\mathfrak M_R,U_R)
\longrightarrow
\widehat{\mathfrak M}_{R'}
=
(\mathfrak M_{R'},U_{R'}).
\end{equation}

A reference upgrade requires $R$, $\sigma_R$, callable resources, allowed operations, the atomic interface, $\Gamma_R$, the program interpreter, and the audit protocol to be reassessed together. The new model must be rechecked for $\mathsf{RCon}$, $\mathsf{TC}$, and $\mathsf{RA}$ or $\mathsf{RefCert}$.
Structural residuals incorporated into $R'$ must be confirmed by independent data, reproducible stable mechanisms, or new physical priors and entered into the new reference specification before the next target state is selected. This upgrade defines a new reference-consistent model. It does not retroactively rewrite the background of the original model's certificate or permit a fit to an existing target to be fed back as that target's own zero-cost prior. The original certificate remains valid only within $\widehat{\mathfrak M}_R$.

Reference refinement has a transparent state-side relation. Suppose
$\mathcal H_{R'}\subseteq\mathcal H_R$
and
${\rm supp}(\rho)\subseteq\mathcal H_{R'}$,
and write
$g_R=\log_2\Gamma_R$
and
$g_{R'}=\log_2\Gamma_{R'}$.
The difference between the two state-side readings is
\begin{equation}
\label{eq:F.72}
C_R(\rho)-C_{R'}(\rho)
=
\frac{\log_2 d_R}{g_R}
-
\frac{\log_2 d_{R'}}{g_{R'}}
-
S(\rho)
\left(
\frac1{g_R}
-
\frac1{g_{R'}}
\right).
\end{equation}

If $\Gamma_R=\Gamma_{R'}$, then
\begin{equation}
\label{eq:F.73}
C_R(\rho)-C_{R'}(\rho)
=
\frac{
\log_2 d_R-\log_2 d_{R'}
}{
\log_2\Gamma_R
}
\ge
0.
\end{equation}

These equalities compare only the state-side readings of two already fixed models. When resource and operation rules also change, process-side comparison uses the target-state-independent bidirectional compilation conditions of Appendix~\ref{app:A.10}.

Reference mismatch within a fixed support can also be controlled. Let strictly positive reference states $\sigma$ and $\tau$ lie on the same finite-dimensional support and satisfy
\begin{equation}
\label{eq:F.74}
2^{-a_-}\sigma
\preceq
\tau
\preceq
2^{a_+}\sigma,
\qquad
a_-,a_+\ge0.
\end{equation}

Operator monotonicity of the logarithm and operator-order comparison over a common smoothing set give
\begin{equation}
\label{eq:F.75}
\begin{aligned}
-a_+
&\le
D(\rho\Vert\tau)
-
D(\rho\Vert\sigma)
\le
a_-,
\\
-a_+
&\le
D_{\max}^{\epsilon}(\rho\Vert\tau)
-
D_{\max}^{\epsilon}(\rho\Vert\sigma)
\le
a_-.
\end{aligned}
\end{equation}

Equation~\eqref{eq:F.75} controls state-side changes within the same support and object domain. If a reference change also alters the generation rules, model qualification and process compilation must still be addressed anew.

Non-Abelian symmetry does not obstruct an unbiased structured vacuum. Suppose a finite-dimensional representation of a compact group $G$ decomposes as
\begin{equation}
\label{eq:F.76}
\mathcal H
\simeq
\bigoplus_{\lambda\in\widehat G}
\bigl(
V_\lambda\otimes M_\lambda
\bigr).
\end{equation}

Choose $\Lambda\subseteq\widehat G$ and multiplicity-space projections $Q_\lambda$, and define
\begin{equation}
\label{eq:F.77}
\Pi_R
=
\bigoplus_{\lambda\in\Lambda}
\bigl(
I_{V_\lambda}\otimes Q_\lambda
\bigr),
\qquad
d_R
=
\sum_{\lambda\in\Lambda}
(\dim V_\lambda)
{\rm Tr}\,Q_\lambda.
\end{equation}

\Fstatement{Proposition F.9 (Non-Abelian unbiased block reference).}
The reference state
$\sigma_R=\Pi_R/d_R$
satisfies
$U_g\sigma_RU_g^\dagger=\sigma_R$,
and for $\rho$ supported on $\Pi_R$,
\begin{equation}
\label{eq:F.78}
\begin{aligned}
D(\rho\Vert\sigma_R)
&=
\log_2 d_R-S(\rho),
\\
D_{\max}(\rho\Vert\sigma_R)
&=
\log_2 d_R-H_{\min}(\rho),
\\
D_{\max}(\rho\Vert\sigma_R)
-
D(\rho\Vert\sigma_R)
&=
S(\rho)-H_{\min}(\rho).
\end{aligned}
\end{equation}

\Fproof
$\Pi_R$ commutes with the group action, and $\sigma_R$ is maximally mixed on its support. The three identities then follow from the definitions of relative entropy and max-relative entropy.\Fqed
The non-Abelian representation decomposition itself does not change the unbiased leading term. The corresponding atomic interface, program transcription, and qualification certificates must still be reestablished on the new $\Pi_R$; centrally weighted block references belong to the weighted extension.

For infinite-dimensional or continuous-variable systems, independent physical priors may first select a finite truncation
\begin{equation}
\label{eq:F.79}
\Pi_{R,\Lambda},
\qquad
d_{R,\Lambda}
:=
{\rm Tr}\,\Pi_{R,\Lambda}
<
\infty,
\qquad
\sigma_{R,\Lambda}
:=
\frac{
\Pi_{R,\Lambda}
}{
d_{R,\Lambda}
}.
\end{equation}

The retention probability of the physical state $\rho_{\rm phys}$ in this subspace and its conditional state are
\begin{equation}
\label{eq:F.80}
q_\Lambda
:=
{\rm Tr}
\bigl(
\Pi_{R,\Lambda}\rho_{\rm phys}
\bigr),
\qquad
\rho_{R,\Lambda}
:=
\frac{
\Pi_{R,\Lambda}\rho_{\rm phys}\Pi_{R,\Lambda}
}{
q_\Lambda
},
\quad
q_\Lambda>0.
\end{equation}

The retention probability and conditional-state treatment follow Appendix~\ref{app:B}. Its full-state regularization formulas apply within their stated finite-dimensional domain; an infinite-dimensional complement requires a separately specified normalized reference and additional finiteness analysis. When $q_\Lambda=0$, this truncation produces no certificate within the subspace. A cutoff-independent infinite-dimensional limit requires new uniform control and does not follow automatically from the finite-dimensional results.

When the Hamiltonian and thermal-bath background are part of the physical prior, one may also define, for $\beta>0$, a Gibbs reference on a finite-dimensional support:
\begin{equation}
\label{eq:F.81}
\sigma_{R,\beta}
:=
\frac{
e^{-\beta H_R}
}{
Z_{R,\beta}
},
\qquad
Z_{R,\beta}
:=
{\rm Tr}_{\mathcal H_R}
e^{-\beta H_R}.
\end{equation}

This paragraph uses natural logarithms. Write
$S_{\rm nat}(\rho):=-{\rm Tr}(\rho\ln\rho)$
and
$D_{\rm nat}(\rho\Vert\sigma):={\rm Tr}[\rho(\ln\rho-\ln\sigma)]$,
and define
\begin{equation}
\label{eq:F.82}
F_\beta(\rho)
:=
{\rm Tr}(\rho H_R)
-
\beta^{-1}
S_{\rm nat}(\rho).
\end{equation}

The definitions directly give
\begin{equation}
\label{eq:F.83}
D_{\rm nat}
\bigl(
\rho\Vert\sigma_{R,\beta}
\bigr)
=
\beta
\bigl[
F_\beta(\rho)
-
F_\beta(\sigma_{R,\beta})
\bigr].
\end{equation}

Equation~\eqref{eq:F.83} provides the state-side entry point for weighted RCC. General weighted references, asymmetric energy costs, open systems with bath memory or non-Markovian structure, expected costs under random stopping times, infinite-dimensional limits, and full covariant resource theories each require corresponding program qualification and dynamical resource specifications, together with renewed checks of spectral structure, one-shot corrections, and audit formulas.

Structural fairness is thus implemented through a jointly fixed reference, atomic interface, control bandwidth, and program prior. Native transcription, stationary envelopes, and certified mixtures give concrete qualified models. Among them, the model family fixed in advance in Appendix~\ref{app:F.5} can approximately generate arbitrary pure and mixed states, providing nontrivial instances to which the universal lower bound on optimal complexity in Appendix~\ref{app:A} applies. Reference upgrades and weighted extensions begin by reestablishing the corresponding model qualification and resource specification.
\par
\endgroup

\begingroup

\setlength{\parindent}{1.5em}
\setlength{\parskip}{1pt plus 0.3pt minus 0.2pt}
\frenchspacing
\linespread{1.025}\selectfont
\setlength{\abovedisplayskip}{8.5pt plus 1pt minus 2pt}
\setlength{\belowdisplayskip}{8.5pt plus 1pt minus 2pt}
\setlength{\abovedisplayshortskip}{5.5pt plus 1pt minus 1pt}
\setlength{\belowdisplayshortskip}{6.5pt plus 1pt minus 1pt}
\setlength{\jot}{2.25pt}
\setlength{\emergencystretch}{1.25em}
\clubpenalty=10000
\widowpenalty=10000
\displaywidowpenalty=10000
\brokenpenalty=5000
\raggedbottom

\newcommand{\Gstatement}[1]{%
  \par\addvspace{0.4\baselineskip}%
  \Needspace{4\baselineskip}%
  \indent\textbf{#1}\hspace{0.5em}\ignorespaces}
\newcommand{\Gproof}{%
  \par\addvspace{0.25\baselineskip}%
  \Needspace{3\baselineskip}%
  \indent\textbf{Proof.}\hspace{0.5em}\ignorespaces}
\newcommand{\Gqed}{%
  \unskip\nobreak\hspace{0.35em}\ensuremath{\square}%
  \par\addvspace{0.3\baselineskip}}
\makeatletter
\renewcommand{\subsection}{\@startsection{subsection}{2}{\z@}%
  {14pt plus 1pt minus 1pt}{8pt plus 0.5pt minus 1pt}%
  {\normalfont\small\bfseries\centering}}
\makeatother
\let\GOriginalSubsection\subsection
\renewcommand{\subsection}[1]{%
  \par\Needspace{3\baselineskip}%
  \GOriginalSubsection{#1}}

\Needspace{5\baselineskip}
\section{Near-Tightness Criteria and Constructive Evidence for Reference-Aligned Structured-State Families}
\label{app:G}

\indent
Appendix~\ref{app:A} converts the target state's smoothed one-shot information gap into a common cost floor for all admissible generation paths; Appendix~\ref{app:C} gives its general inversion and canonical computational forms. Near-tightness requires constructing feasible short circuits from the other side, so that the rigorous lower bound and an explicit upper bound meet at the same asymptotic scale.

Starting from two-sided bounds under a general transcription envelope, this appendix gives remainder conditions for canonical exact and approximate transcription. It then connects construction costs to the same one-shot scale through three mechanisms: classical probability concentration, contraction along nested uniform projections, and uniform logical encoding. Under the uniform construction assumptions and asymptotic conditions of the respective propositions, the RCC lower bound captures the leading growth of universal optimal quantum circuit complexity in three reference-aligned structured-state subfamilies.

\indent
Near-tightness is an asymptotic statement about a model family. Let $n$ index the scale, and fix a family of reference-consistent models
\begin{equation*}
\widehat{\mathfrak M}_{R_n}
=
(\mathfrak M_{R_n},U_{R_n}),
\end{equation*}
and retain the family-level qualification conditions of Appendix~\ref{app:A}. Every construction produces a normalized unconditional output in the current model and approximates the target within the declared error. Initial states, auxiliary resources, atomic operations, addressing, and precision compilation enter the fixed model or the circuit cost. The constants below may depend on the model family and state-family parameters declared in advance, but do not grow with $n$ or the particular target state. The nonnegative constants $A,B,E$ reused in different subsections may take different values.

\subsection{The One-Shot Scale, Transcription Inversion, and Near-Tightness Criteria}
\label{app:G.1}

\indent
We first compare the one-shot cost floor with explicit construction lengths under exact transcription, then account for the radius shift and code-length cost introduced by approximate transcription. Write
\begin{equation}
\label{eq:G.1}
\sigma_n
:=
\sigma_{R_n}
=
\frac{\Pi_{R_n}}{d_n},
\qquad
d_n
:=
{\rm Tr}\,\Pi_{R_n},
\qquad
g_n
:=
\log\Gamma_{R_n},
\qquad
\chi_n
:=
\chi_{U_{R_n}},
\end{equation}
and take $0<\epsilon_n\le1/2$. Define
\begin{equation}
\label{eq:G.2}
C_n
:=
C_{\rm opt}^{(\epsilon_n)}
(\rho_n;\mathfrak M_{R_n}),
\qquad
Q_n^{(\epsilon_n)}
:=
\frac{
D_{\max}^{\epsilon_n}
(\rho_n\Vert\sigma_n)
}{
g_n
}.
\end{equation}

The quantity $Q_n^{(\epsilon_n)}$ is the smoothed one-shot RCC scale directly constrained by the main proof in Appendix~\ref{app:A}. Write
$\Lambda_n:=\Lambda_{R_n}$.
For a general nondecreasing transcription envelope, define the cost-floor function
\begin{equation}
\label{eq:G.3}
\mathscr L_n^{\Lambda}(q)
:=
\Lambda_n^{\leftarrow}
\!\left[
g_nq-\chi_n
\right],
\qquad
q\ge0.
\end{equation}

\Gstatement{Proposition G.1 (Constructive bounds for general envelopes and canonical models).}
Suppose an explicit admissible circuit outputs $\rho_n$ within error $\epsilon_n$, at an atomic resource-slot cost $L_n$. Then
\begin{equation}
\label{eq:G.4}
\mathscr L_n^{\Lambda}
\!\left(
Q_n^{(\epsilon_n)}
\right)
\le
C_n
\le
L_n.
\end{equation}

Thus, if
$Q_n^{(\epsilon_n)}\to\infty$,
the general-envelope floor satisfies
\begin{equation*}
\mathscr L_n^{\Lambda}
\!\left(
Q_n^{(\epsilon_n)}
\right)
=
\Omega\!\left(
Q_n^{(\epsilon_n)}
\right),
\end{equation*}
and the construction length satisfies
$L_n=O(Q_n^{(\epsilon_n)})$,
then
\begin{equation*}
C_n
=
\Theta\!\left(
Q_n^{(\epsilon_n)}
\right).
\end{equation*}

If, in addition,
\begin{equation*}
\mathscr L_n^{\Lambda}
\!\left(
Q_n^{(\epsilon_n)}
\right)
=
Q_n^{(\epsilon_n)}
-
o\!\left(
Q_n^{(\epsilon_n)}
\right)
\qquad\text{and}\qquad
L_n
=
\bigl(1+o(1)\bigr)
Q_n^{(\epsilon_n)},
\end{equation*}
then
\begin{equation*}
C_n
=
\bigl(1+o(1)\bigr)
Q_n^{(\epsilon_n)}.
\end{equation*}

These conditions provide sufficient criteria for near-tightness in scaling and asymptotic saturation under a general transcription envelope.
In a canonical atomic-slot model, write
\begin{equation}
\label{eq:G.5}
\Lambda_n(L)
=
\Phi_{g_n,a}(L)
+
\gamma,
\qquad
\beta_n
:=
\gamma+\chi_n,
\qquad
\mathfrak r_n
:=
\frac{
\beta_n
+
a\log\max\{1,Q_n^{(\epsilon_n)}\}
}{
g_n
},
\end{equation}
where $a,\gamma\ge0$ are transcription constants uniform across the model family and $\beta_n$ is uniformly bounded. Equation~\eqref{eq:G.4} then strengthens to
\begin{equation}
\label{eq:G.6}
Q_n^{(\epsilon_n)}
-
\mathfrak r_n
\le
F_{g_n,a}\!\left[
g_nQ_n^{(\epsilon_n)}-\beta_n
\right]
\le
C_n
\le
L_n.
\end{equation}

\Gproof
The left inequality in Eq.~\eqref{eq:G.4} follows from the general-envelope lower bound of Theorem A.2; the right inequality follows from the definition of optimal complexity. For the canonical model, write
$q=Q_n^{(\epsilon_n)}$
and
$r=\mathfrak r_n$.
If $q-r\le0$, the first inequality in Eq.~\eqref{eq:G.6} follows from
$F_{g_n,a}\ge0$.
If $q-r>0$, then
$0<q-r\le q$,
and
\begin{equation*}
\Phi_{g_n,a}(q-r)
\le
g_n(q-r)
+
a\log\max\{1,q\}
=
g_nq-\beta_n.
\end{equation*}

Monotonicity of $\Phi_{g_n,a}$ gives the first inequality, and the canonical inversion in Theorem A.2 gives the second.\Gqed

Since
$g_n=\log\Gamma_{R_n}\ge1$,
whenever
$Q_n^{(\epsilon_n)}\to\infty$,
a canonical model automatically satisfies
\begin{equation*}
\mathfrak r_n
=
O\!\left(
1+\log Q_n^{(\epsilon_n)}
\right)
=
o\!\left(
Q_n^{(\epsilon_n)}
\right).
\end{equation*}

The inversion remainder under exact transcription is therefore automatically subleading; near-tightness then requires only the uniform upper construction bound to be checked. If a state family satisfies
\begin{equation}
\label{eq:G.7}
Q_n^{(\epsilon_n)}
\longrightarrow
\infty,
\qquad
L_n
\le
A\,Q_n^{(\epsilon_n)}
+
o\!\left(
Q_n^{(\epsilon_n)}
\right),
\end{equation}
with $A$ independent of $n$, then
\begin{equation}
\label{eq:G.8}
C_n
=
\Theta\!\left(
Q_n^{(\epsilon_n)}
\right).
\end{equation}

We call Eq.~\eqref{eq:G.7} the constructive criterion for near-tightness in scaling for this state family. Taking $A=1$ and
$L_n=(1+o(1))Q_n^{(\epsilon_n)}$,
Eq.~\eqref{eq:G.6} further proves that the construction asymptotically saturates the main RCC lower bound.

\indent
The approximate-transcription branch retains the distinction between circuit tolerance and program radius. Let $b\ge0$ be the compilation-precision code-length coefficient uniform across the model family, and set
\begin{equation*}
\mathcal D_n
:=
\left\{
\delta\in\Delta_{\rm TC}:
\epsilon_n+\delta\le1
\right\}
\neq
\varnothing.
\end{equation*}

The general-envelope cost floor under approximate transcription is
\begin{equation*}
\mathscr L_{n,{\rm cmp}}^{\Lambda}
:=
\sup_{\delta\in\mathcal D_n}
\Lambda_n^{\leftarrow}\!\left[
D_{\max}^{\epsilon_n+\delta}
(\rho_n\Vert\sigma_n)
-
\chi_n
-
b\log(1/\delta)
\right],
\qquad
C_n
\ge
\mathscr L_{n,{\rm cmp}}^{\Lambda}.
\end{equation*}

In the canonical model, define the effective one-shot scale
\begin{equation*}
\widetilde Q_n^{(\epsilon_n)}
:=
\sup_{\delta\in\mathcal D_n}
\frac{
D_{\max}^{\epsilon_n+\delta}
(\rho_n\Vert\sigma_n)
-
b\log(1/\delta)
}{
g_n
}.
\end{equation*}

Equation~\eqref{eq:A.48} and continuity and monotonicity of $F_{g_n,a}$ give
\begin{equation*}
C_n
\ge
F_{g_n,a}\!\left[
g_n\widetilde Q_n^{(\epsilon_n)}
-
\beta_n
\right].
\end{equation*}

Monotonicity in the smoothing radius ensures
$\widetilde Q_n^{(\epsilon_n)}
\le
Q_n^{(\epsilon_n)}$.
Thus
\begin{equation*}
\Delta_{n,{\rm cmp}}
:=
Q_n^{(\epsilon_n)}
-
\widetilde Q_n^{(\epsilon_n)}
\ge
0
\end{equation*}
records both the radius-shift loss and the compilation code-length penalty. If
$Q_n^{(\epsilon_n)}\to\infty$
and
$\Delta_{n,{\rm cmp}}
=
o(Q_n^{(\epsilon_n)})$,
canonical inversion still gives
$C_n\ge Q_n^{(\epsilon_n)}-o(Q_n^{(\epsilon_n)})$.
Together with Eq.~\eqref{eq:G.7}, this yields Eq.~\eqref{eq:G.8} in the approximate-transcription branch as well. This condition collects all additional requirements of approximate transcription into a directly verifiable scale deficit.

\indent
The atomic control bandwidth $g_n$ remains explicit throughout the comparison. If a natural construction gives only
$L_n=O(g_nQ_n^{(\epsilon_n)})$,
Eq.~\eqref{eq:G.7} still holds for model families with $g_n=O(1)$. When $g_n$ grows with $n$, the upper bound retains the corresponding bandwidth factor. Near-tightness on the atomic-slot scale then requires bandwidth-saturating compilation that achieves
$L_n=O(Q_n^{(\epsilon_n)})$.

\subsection{A Smoothed One-Shot Bridge from Low-Rank Occupation}
\label{app:G.2}

\indent
The three constructions share a geometric starting point: a low-rank projection declared in advance converts stable occupation into a smoothed one-shot gap. The following lemma is the family-level analytical form of Eq.~\eqref{eq:C.18} and Eq.~\eqref{eq:D.16}. We retain its short proof to fix the rank, occupation, and smoothing margin as they vary with $n$.
Dephasing, random mixing, reset, measurement--feedback, and discarding auxiliary systems in this subsection and Appendices~\ref{app:G.3}--\ref{app:G.5} all mean admissible unconditional CPTP channels in the current model. The costs of auxiliary systems, random sources, classical records, feedback, addressing, and finite-precision compilation are included in the corresponding $L_n$.

\Gstatement{Lemma G.2 (Smoothed occupation bridge).}
Let $P_n\le\Pi_{R_n}$ be a projection of rank $s_n\ge1$, and write
\begin{equation}
\label{eq:G.9}
p_n
:=
{\rm Tr}(P_n\rho_n).
\end{equation}

If $0\le\epsilon_n<p_n$, then
\begin{equation}
\label{eq:G.10}
D_{\max}^{\epsilon_n}(\rho_n\Vert\sigma_n)
\ge
\max\left\{
0,\,
\log\frac{(p_n-\epsilon_n)d_n}{s_n}
\right\}.
\end{equation}

\Gproof
Take any normalized smoothing candidate $\tau$ satisfying
$T(\tau,\rho_n)\le\epsilon_n$.
The variational characterization of trace distance gives
\begin{equation*}
{\rm Tr}(P_n\tau)
\ge
p_n-\epsilon_n.
\end{equation*}

If
$\tau\le2^\lambda\sigma_n$,
then
\begin{equation*}
{\rm Tr}(P_n\tau)
\le
2^\lambda{\rm Tr}(P_n\sigma_n)
=
2^\lambda\frac{s_n}{d_n}.
\end{equation*}

Hence
\begin{equation*}
\lambda
\ge
\log\frac{(p_n-\epsilon_n)d_n}{s_n}.
\end{equation*}

Taking the infimum over all smoothing candidates and using
$D_{\max}\ge0$
for normalized states yields Eq.~\eqref{eq:G.10}.\Gqed

Set
\begin{equation}
\label{eq:G.11}
q_n
:=
\log\frac{d_n}{s_n},
\qquad
\zeta_n
:=
p_n-\epsilon_n
>
0.
\end{equation}

Lemma G.2 gives
\begin{equation}
\label{eq:G.12}
q_n
\le
g_nQ_n^{(\epsilon_n)}
+
\log\frac1{\zeta_n}.
\end{equation}

If an admissible construction satisfies
\begin{equation}
\label{eq:G.13}
L_n
\le
Aq_n
+
B\log(1/\epsilon_n)
+
E,
\end{equation}
then
\begin{equation}
\label{eq:G.14}
L_n
\le
A g_nQ_n^{(\epsilon_n)}
+
B\log(1/\epsilon_n)
+
A\log\frac1{\zeta_n}
+
E.
\end{equation}

Equation~\eqref{eq:G.14} is a common upper-bound template for the three structured subfamilies below. It explicitly retains the occupation margin after smoothing, $\zeta_n$, precision costs, and bandwidth growth. When these terms satisfy the scaling conditions of Appendix~\ref{app:G.1}, the main-theorem lower bound and the constructive upper bound give asymptotic two-sided bounds.
Here
$B\log(1/\epsilon_n)$
is the implementation-precision cost of the explicit upper construction. Models with exact implementation, or with this cost already absorbed into the atomic resource-slot count, may take $B=0$. This construction cost is distinct from the inversion remainder in Appendix~\ref{app:G.1}, which is already automatically subleading under exact transcription.
In each subsection below, $Q_n^{(\epsilon_n)}$ is evaluated for the corresponding target state using Eq.~\eqref{eq:G.2}.

\subsection{Classically Concentrated States with Efficient Amplitude Preparation}
\label{app:G.3}

\indent
Stable concentration of a classical distribution on a small label set gives an information lower bound, while uniform amplitude preparation and dephasing give a constructive upper bound. We compare their scales to determine when near-tightness follows.

Suppose the current reference model has a reference-compatible classical label basis $\{|x\rangle\}_{x\in\Omega_n}, \, |\Omega_n| = d_n$, and consider
\begin{equation}
\label{eq:G.15}
\rho_n^{\rm cl}
=
\sum_{x\in\Omega_n}
p_{n,x}
|x\rangle\langle x|,
\qquad
|\psi_n\rangle
=
\sum_{x\in\Omega_n}
\sqrt{p_{n,x}}\,
|x\rangle.
\end{equation}

Assume that the distribution family admits uniform amplitude-preparation circuits $V_n^{\rm amp}$ satisfying
\begin{equation}
\label{eq:G.16}
T\!\left(
V_n^{\rm amp}
|0_n\rangle\langle0_n|
(V_n^{\rm amp})^\dagger,
|\psi_n\rangle\langle\psi_n|
\right)
\le
\epsilon_n/2.
\end{equation}

The cost of generating $|0_n\rangle$ is included in the construction length. Applying reference-compatible complete dephasing to $|\psi_n\rangle$ in the label basis gives
\begin{equation}
\label{eq:G.17}
\Delta_n
\!\left(
|\psi_n\rangle\langle\psi_n|
\right)
=
\rho_n^{\rm cl}.
\end{equation}

By contractivity of trace distance under CPTP maps, amplitude preparation followed by dephasing gives an $\epsilon_n$-approximate realization of $\rho_n^{\rm cl}$. Suppose its total length satisfies
\begin{equation}
\label{eq:G.18}
L_n^{\rm cl}
\le
A\log d_n
+
B\log(1/\epsilon_n)
+
E.
\end{equation}

Assume further that a uniform structural rule supplies the distribution family with sets
$S_n\subseteq\Omega_n$
such that
\begin{equation}
\label{eq:G.19}
w_n
:=
\sum_{x\in S_n}
p_{n,x},
\qquad
|S_n|
\ge
1,
\qquad
|S_n|
\le
d_n^{\,1-\nu},
\qquad
\zeta_n
:=
w_n-\epsilon_n
>
0,
\end{equation}
where $0<\nu\le1$ is independent of $n$.

\Gstatement{Proposition G.3 (Matching scales for classically concentrated states).}
Under the conditions of Eqs.~\eqref{eq:G.16}--\eqref{eq:G.19},
\begin{equation}
\label{eq:G.20}
Q_n^{(\epsilon_n)}
\ge
\frac{
\nu\log d_n+\log\zeta_n
}{
g_n
},
\end{equation}
and
\begin{equation}
\label{eq:G.21}
L_n^{\rm cl}
\le
\frac{A}{\nu}
g_nQ_n^{(\epsilon_n)}
+
B\log(1/\epsilon_n)
+
\frac{A}{\nu}
\log\frac1{\zeta_n}
+
E.
\end{equation}

In the canonical exact-transcription branch, if additionally
$Q_n^{(\epsilon_n)}\to\infty$,
$g_n=O(1)$,
$\inf_n\zeta_n>0$,
and
$B\log(1/\epsilon_n)
=
o(Q_n^{(\epsilon_n)})$,
then
\begin{equation*}
C_n
=
\Theta\!\left(
Q_n^{(\epsilon_n)}
\right).
\end{equation*}

\Gproof
Take
\begin{equation*}
P_{S_n}
=
\sum_{x\in S_n}
|x\rangle\langle x|.
\end{equation*}

Its rank is at most $d_n^{1-\nu}$ and its occupation in $\rho_n^{\rm cl}$ is $w_n$. Lemma G.2 gives
\begin{equation*}
D_{\max}^{\epsilon_n}
(\rho_n^{\rm cl}\Vert\sigma_n)
\ge
\log\frac{\zeta_nd_n}{|S_n|}
\ge
\nu\log d_n+\log\zeta_n,
\end{equation*}
which is Eq.~\eqref{eq:G.20}. Rearranging gives
\begin{equation*}
\log d_n
\le
\frac{
g_nQ_n^{(\epsilon_n)}
+
\log(1/\zeta_n)
}{
\nu
},
\end{equation*}
and substitution into Eq.~\eqref{eq:G.18} yields Eq.~\eqref{eq:G.21}. Under the stated asymptotic conditions, Eq.~\eqref{eq:G.21} reduces to
$L_n^{\rm cl}=O(Q_n^{(\epsilon_n)})$,
so Proposition G.1 gives matching scales.\Gqed

Among classical structures such as low-treewidth probabilistic models and autoregressive distributions, Proposition G.3 applies to subfamilies satisfying the amplitude-preparation, total-length, and concentration-set conditions in Eqs.~\eqref{eq:G.16}--\eqref{eq:G.19}.
Equation~\eqref{eq:G.19} also identifies the structural core of this criterion: concentration of probability mass on a lower-dimensional support makes $Q_n^{(\epsilon_n)}$ grow on the same scale as amplitude compilation. As a diagonal distribution approaches uniformity, $Q_n^{(\epsilon_n)}$ can become much smaller than
$\log d_n/g_n$.
Equation~\eqref{eq:G.18} then certifies constructibility; the corresponding near-tight scaling requires additional structural parameters.

\subsection{Nested Uniform Projection States and Mixtures with Low-Cost Backgrounds}
\label{app:G.4}

\indent
A nested projection chain decomposes support contraction into a sequence of admissible channels. When the cost per level and the chain length are uniformly controlled, the occupation bridge connects uniform projection states and their mixtures with low-cost backgrounds to the near-tightness criterion.

Let
\begin{equation}
\label{eq:G.22}
\Pi_{R_n}
=
P_{n,0}
\ge
P_{n,1}
\ge
\cdots
\ge
P_{n,J_n}
=
P_n
\end{equation}
be a nested projection chain declared in advance, and write
$s_{n,j}:={\rm Tr}\,P_{n,j}$,
$s_n:=s_{n,J_n}\ge1$.
Assume that at every level there is an admissible channel $\mathcal E_{n,j}$ in the current reference-consistent model satisfying
\begin{equation}
\label{eq:G.23}
\mathcal E_{n,j}
\!\left(
\frac{P_{n,j-1}}{s_{n,j-1}}
\right)
=
\frac{P_{n,j}}{s_{n,j}},
\end{equation}
and that constants $\ell,\mu,\mu_0$, independent of $n$, satisfy
\begin{equation}
\label{eq:G.24}
{\rm len}(\mathcal E_{n,j})
\le
\ell,
\qquad
J_n
\le
\mu\log\frac{d_n}{s_n}
+
\mu_0.
\end{equation}

Equations~\eqref{eq:G.23}--\eqref{eq:G.24} already give an explicit construction of the uniform projection state, with
\begin{equation*}
L_n^{\Pi}
\le
\ell J_n
\le
\ell\mu
\log\frac{d_n}{s_n}
+
\ell\mu_0.
\end{equation*}

Take a background state $\tau_n$ that can be realized uniformly at low cost in this model, and consider the subfamily of background mixtures
\begin{equation}
\label{eq:G.25}
\rho_n^{\rm occ}
=
w_n\frac{P_n}{s_n}
+
(1-w_n)\tau_n,
\qquad
0<w_n\le1,
\qquad
\zeta_n
:=
w_n-\epsilon_n
>
0,
\end{equation}
and assume that, after accounting for the background branch, specification of the mixing probability, and finite-precision compilation, the total construction length satisfies
\begin{equation}
\label{eq:G.26}
L_n^{\rm occ}
\le
A\log\frac{d_n}{s_n}
+
B\log(1/\epsilon_n)
+
E.
\end{equation}

\Gstatement{Proposition G.4 (Matching scales for nested projection mixtures).}
Under the conditions of Eqs.~\eqref{eq:G.22}--\eqref{eq:G.26},
\begin{equation}
\label{eq:G.27}
L_n^{\rm occ}
\le
A g_nQ_n^{(\epsilon_n)}
+
B\log(1/\epsilon_n)
+
A\log\frac1{\zeta_n}
+
E.
\end{equation}

In the canonical exact-transcription branch, if additionally
$Q_n^{(\epsilon_n)}\to\infty$,
$g_n=O(1)$,
$\inf_n\zeta_n>0$,
and
$B\log(1/\epsilon_n)
=
o(Q_n^{(\epsilon_n)})$,
then
\begin{equation*}
C_n
=
\Theta\!\left(
Q_n^{(\epsilon_n)}
\right).
\end{equation*}

\Gproof
Equation~\eqref{eq:G.23} gives
\begin{equation*}
\sigma_n
\longmapsto
\frac{P_{n,1}}{s_{n,1}}
\longmapsto
\cdots
\longmapsto
\frac{P_n}{s_n},
\end{equation*}

while Eq.~\eqref{eq:G.24} controls the length of the uniform projection branch. For the target state in Eq.~\eqref{eq:G.25},
\begin{equation*}
{\rm Tr}
\bigl(
P_n\rho_n^{\rm occ}
\bigr)
\ge
w_n.
\end{equation*}

Apply Lemma G.2 with $P_n$, then substitute
$\zeta_n=w_n-\epsilon_n$
and Eq.~\eqref{eq:G.26} into the general template, Eq.~\eqref{eq:G.14}, to obtain Eq.~\eqref{eq:G.27}. Under the stated asymptotic conditions, Eq.~\eqref{eq:G.27} reduces to
$L_n^{\rm occ}=O(Q_n^{(\epsilon_n)})$,
and Proposition G.1 gives matching scales.\Gqed

When $w_n=1$, the proposition directly covers uniform projection states, and one may take
$L_n^{\rm occ}=L_n^{\Pi}$.
One concrete realization fixes logical bits selected in advance, one at a time: each local reset halves the rank of the current uniform block. The paired-reset model of Appendix~\ref{app:F.5} supplies a reference-balanced realization. Measurement--feedback repreparation, cooling, and rank-reduction CPTP channels satisfying the same uniform length conditions also fall within this criterion.
The information lower bound in Proposition G.4 comes from stable occupation of the final low-rank subspace; the upper cost bound comes from successive contractions along the same projection chain. When the occupation margin vanishes with $n$, the term
$\log(1/\zeta_n)$
in Eq.~\eqref{eq:G.27} records the degradation in scaling caused by smoothing.

\subsection{Uniformly Encodable Logical Pure States}
\label{app:G.5}

\indent
Under a common unbiased reference and tolerance, logical pure states share the same one-shot scale. We use uniform encoding circuits to identify subfamilies with matching generation costs.

Suppose the reference subspace is the logical space of $k_n\ge1$ logical qubits, $d_n = 2^{k_n}$, and consider logical pure states
\begin{equation}
\label{eq:G.28}
\rho_n^{\rm log}
=
|\psi_{L,n}\rangle
\langle\psi_{L,n}|.
\end{equation}

By Eq.~\eqref{eq:A.53} of Appendix~\ref{app:A},
\begin{equation}
\label{eq:G.29}
D_{\max}^{\epsilon_n}
(\rho_n^{\rm log}\Vert\sigma_n)
=
\log\!\left[
d_n(1-\epsilon_n)
\right]
=
k_n+\log(1-\epsilon_n),
\end{equation}
hence
\begin{equation}
\label{eq:G.30}
Q_n^{(\epsilon_n)}
=
\frac{
k_n+\log(1-\epsilon_n)
}{
g_n
}.
\end{equation}

Assume this logical-state family admits uniform reference-consistent encoding and logical-implementation circuits satisfying
\begin{equation}
\label{eq:G.31}
L_n^{\rm log}
\le
A k_n
+
B\log(1/\epsilon_n)
+
E.
\end{equation}

\Gstatement{Proposition G.5 (Matching scales under uniform logical encoding).}
Under the condition in Eq.~\eqref{eq:G.31},
\begin{equation}
\label{eq:G.32}
L_n^{\rm log}
\le
A g_nQ_n^{(\epsilon_n)}
+
B\log(1/\epsilon_n)
+
A\log\frac1{1-\epsilon_n}
+
E.
\end{equation}

In the canonical exact-transcription branch, if additionally
$Q_n^{(\epsilon_n)}\to\infty$,
$g_n=O(1)$,
and
$B\log(1/\epsilon_n)
=
o(Q_n^{(\epsilon_n)})$,
then
\begin{equation*}
C_n
=
\Theta\!\left(
Q_n^{(\epsilon_n)}
\right).
\end{equation*}

\Gproof
Equation~\eqref{eq:G.30} gives
\begin{equation*}
k_n
=
g_nQ_n^{(\epsilon_n)}
+
\log\frac1{1-\epsilon_n}.
\end{equation*}

Substituting into Eq.~\eqref{eq:G.31} yields Eq.~\eqref{eq:G.32}. Since
$0<\epsilon_n\le1/2$,
the last smoothing term is at most $A$. Under the stated asymptotic conditions,
$L_n^{\rm log}
=
O(Q_n^{(\epsilon_n)})$,
and Proposition G.1 gives matching scales.\Gqed

If $g_n$ grows with $n$, Eq.~\eqref{eq:G.32} retains the corresponding bandwidth factor. Obtaining
$C_n=\Theta(Q_n^{(\epsilon_n)})$
then requires bandwidth-saturating compilation with
$L_n^{\rm log}=O(k_n/g_n)$,
or an atomic resource-slot specification whose control bandwidth is uniformly bounded.
Sparse graph states, linearly encodable stabilizer states, logical states of low-density parity-check (LDPC) codes at fixed rate that satisfy the uniform length condition, and geometrically sparse logical states can provide such evidence. This evidence class is characterized by a uniform serial-length upper bound for the entire encoding--implementation family: pure-state entropy supplies the one-shot floor, and uniform encoding structure closes the circuit upper bound.

\indent
For a fixed unbiased logical reference and common tolerance, Eq.~\eqref{eq:G.29} shows that all logical pure states share the same $Q_n^{(\epsilon_n)}$ scale. Equation~\eqref{eq:G.31}, however, provides a matching upper bound only for subfamilies with uniform $O(k_n)$ encoding. Proposition G.5 therefore applies to uniformly encodable subfamilies of compressible pure states. Haar-typical pure states, or states incompressible over a fixed gate set, discussed in Sec.~\ref{sec:3}, may retain process complexity far above this one-shot floor and lie outside the criterion's scope.

\subsection{Near-Tightness Results, Construction Requirements, and Open Boundaries}
\label{app:G.6}

\indent
Propositions G.3--G.5 establish matching scales for three conditionally near-tight subfamilies based on classical probability concentration, nested uniform projection contraction, and coherent logical encoding, respectively. Under the construction assumptions and asymptotic conditions of each proposition, Eq.~\eqref{eq:G.7} and Proposition G.1 give the common canonical exact-transcription form
\begin{equation}
\label{eq:G.33}
Q_n^{(\epsilon_n)}
-
o\!\left(
Q_n^{(\epsilon_n)}
\right)
\le
C_n
\le
A Q_n^{(\epsilon_n)}
+
o\!\left(
Q_n^{(\epsilon_n)}
\right).
\end{equation}

The canonical inversion remainder is subleading when $Q_n^{(\epsilon_n)}\to\infty$. The substantive requirement in Eq.~\eqref{eq:G.33} is therefore the upper construction bound: can a circuit generate the target structure at atomic-slot cost
$O(Q_n^{(\epsilon_n)})$?
For sufficiently large $n$, the construction length and true optimal complexity satisfy
\begin{equation*}
1
\le
\frac{L_n}{C_n}
\le
A+o(1);
\end{equation*}
for $A=1$, the construction is certified to be asymptotically optimal.

Under exact transcription, Eq.~\eqref{eq:G.4} gives the same criterion for a general envelope:
$\mathscr L_n^{\Lambda}(Q_n^{(\epsilon_n)})
=
\Omega(Q_n^{(\epsilon_n)})$
establishes near-tightness up to a constant factor, while
$\mathscr L_n^{\Lambda}(Q_n^{(\epsilon_n)})
=
Q_n^{(\epsilon_n)}
-
o(Q_n^{(\epsilon_n)})$
recovers the left side of Eq.~\eqref{eq:G.33}. Approximate transcription is controlled by the effective scale
$\widetilde Q_n^{(\epsilon_n)}$.
In the same limit
$Q_n^{(\epsilon_n)}\to\infty$,
if
$\Delta_{n,{\rm cmp}}
=
o(Q_n^{(\epsilon_n)})$,
the combined deficit from the radius shift and compilation code length remains subleading, and the three constructive criteria continue to give the same scaling conclusions.

The three constructions all require a uniform generation mechanism, a smoothing margin that preserves the leading scale, and actual compilation costs matched to the model bandwidth. The reference-consistent model fixes resources, errors, and counting conventions, so the final-state information lower bound and the constructive upper bound constrain the same $C_{\rm opt}^{(\epsilon_n)}$.

\indent
These results provide three rigorous starting points for the classification sought in Open Problem 3.1.
That classification concerns the stability and scope of these mechanisms: whether random and chaotic pure states admit corresponding uniform generation structures; how finer scaling emerges when the occupation margin is small; when variable-bandwidth models permit saturating compilation; and how concentration, uniform projection contraction, and encodability persist under perturbations, mixing, and closure operations on state families. The answers will determine whether near-tight state families form broader physical classes that can be classified systematically.

\endgroup

\begingroup

\setlength{\parindent}{1.5em}
\setlength{\parskip}{1pt plus 0.3pt minus 0.2pt}
\frenchspacing
\linespread{1.005}\selectfont
\setlength{\abovedisplayskip}{8.5pt plus 1pt minus 2pt}
\setlength{\belowdisplayskip}{8.5pt plus 1pt minus 2pt}
\setlength{\abovedisplayshortskip}{5.5pt plus 1pt minus 1pt}
\setlength{\belowdisplayshortskip}{6.5pt plus 1pt minus 1pt}
\setlength{\jot}{2pt}
\setlength{\emergencystretch}{1.25em}
\clubpenalty=10000
\widowpenalty=10000
\displaywidowpenalty=10000
\brokenpenalty=5000
\raggedbottom

\newcommand{\Hstatement}[1]{%
  \par\addvspace{0.4\baselineskip}%
  \Needspace{4\baselineskip}%
  \indent\textbf{#1}\hspace{0.5em}\ignorespaces}
\newcommand{\Hproof}{%
  \par\addvspace{0.25\baselineskip}%
  \Needspace{3\baselineskip}%
  \indent\textbf{Proof.}\hspace{0.5em}\ignorespaces}
\newcommand{\Hqed}{%
  \unskip\nobreak\hspace{0.35em}\ensuremath{\square}%
  \par\addvspace{0.3\baselineskip}}
\makeatletter
\renewcommand{\subsection}{\@startsection{subsection}{2}{\z@}%
  {14pt plus 1pt minus 1pt}{8pt plus 0.5pt minus 1pt}%
  {\normalfont\small\bfseries\centering}}
\makeatother
\let\HOriginalSubsection\subsection
\renewcommand{\subsection}[1]{%
  \par\Needspace{3\baselineskip}%
  \HOriginalSubsection{#1}}

\section{Reference Qualification and Certification for Finite-Control Generation--Description Interfaces}
\label{app:H}

\indent
Reference qualification of a finite-control model depends on the full program semidensity jointly generated by measurement, classical control, and loops. The code tree assigns the program prior, quantum instruments carry Born weights, and nonhalting paths contribute to no finite program output. This appendix combines these components within a common unconditional semantics and constructs verifiable certificates for the family-level reference domination characterized by Proposition 2.1.

\indent
We inherit the operationally reduced atomic interface $\mathfrak I_R^{(1)}$ from Appendix~\ref{app:F}. Raw commands have already been deduplicated exactly according to their full dynamical semantics and static resource signatures, and $\Gamma_R$ and its Hartley bandwidth have been derived from that interface. The CP maps below carry only dynamical records that can still affect execution or output; static cost information, including spacetime support, parallelism, control precision, and auxiliary resources, remains fixed by the resource signatures. Appendix~\ref{app:F} continues to supply the definition of physical $\Gamma_R$ and transcription over arbitrary finite alphabets. We build finite-control qualification certificates on this established interface.
The model qualification for the general RCC main theorem remains
$\mathsf{RCon}\wedge\mathsf{TC}\wedge\mathsf{RA}$.
This appendix constructs a class of sound finite-control certificates
\begin{equation*}
\mathsf{RefCert}_{\rm fc}
\Longrightarrow
\mathsf{RA}.
\end{equation*}

\indent
Three kinds of proof objects remain distinct. Faithful transcription $\mathsf{TC}$ supplies circuit compilation, semantic preservation, length bounds, and circuit coverage. The synchronous realization and terminating-domain coverage of Appendix~\ref{app:H.1} unambiguously reassemble all terminating CP paths into the program semidensity. The condition $\mathsf{RefCert}_{\rm fc}$ then proves reference domination of that semidensity uniformly across the family. This appendix supplies the latter two components; full $\mathsf{TC}$ remains an independent qualification condition.

The companion library rcc-refcert provides program semidensity calculations, numerical checks of candidate certificates, and reproducible examples for the finite models in this appendix \cite{LiuTianWu2026_RCCRefcert}.

\indent
We first fix the model-family index $n$ and suppress it where no ambiguity arises. For linear maps $\Phi,\Psi$, write
\begin{equation*}
\Phi
\succeq_{\rm CP}
\Psi
\quad\Longleftrightarrow\quad
\Phi-\Psi
\ \text{is completely positive}.
\end{equation*}

\indent
We use the Choi-matrix convention~\cite{Choi1975_CPLinearMaps}
\begin{equation*}
J(\Phi)
=
\sum_{ij}
|i\rangle\langle j|
\otimes
\Phi(|i\rangle\langle j|),
\end{equation*}
with direct-sum algebras interpreted blockwise on the input. The structured vacuum $\sigma_R$ has full rank on the declared output reference support; every inverse below is restricted to that support.

\subsection{Finite-Control Semantics and Synchronous Prefix--CP Realization}
\label{app:H.1}

\indent
We record code parsing and physical branches separately, then sum all finite halting paths with their code weights. Syntax states fix program boundaries, while physical control states carry measurement and feed-forward, combining the two kinds of weights within one unconditional semantics.

Fix a finite set $S$ of nonterminal syntax states and a finite set $Q$ of physical control states, and write $X = S\times Q, \, x = (s,q)$. Each $q$ carries a finite-dimensional Hilbert space $\mathcal K_q$. The classical--quantum (CQ) operator algebra of the transient extension is
\begin{equation*}
\mathfrak X
=
\bigoplus_{(s,q)\in X}
\mathcal B(\mathcal K_q),
\qquad
Z
=
\bigoplus_{s,q}
Z_{s,q},
\qquad
\operatorname{Tr}Z
=
\sum_{s,q}
\operatorname{Tr}Z_{s,q}.
\end{equation*}

\indent
Any classical records, flags, workspace, environment states, or resource states that can still affect subsequent execution enter these CQ blocks. Data used only to determine atomic costs remain in the static resource signature. 

\indent
In Appendices~\ref{app:H.1}--\ref{app:H.6}, we take
$\mathcal H_{\rm out}\equiv\mathcal H$
to be the effective output space used by the main theorem of Appendix~\ref{app:A}. If the terminal semantics first takes values in a finite-dimensional extended space
$\widehat{\mathcal H}_{\rm out}$
containing failure flags or resource records, the model must fix in advance an unconditional CPTP readout $\mathcal R_R$, generated by the same target-state-independent rule across the model family. The terminal maps below are then understood to have already been composed with this readout. If the model's fixed reference state $\widehat{\sigma}_R$ is first certified on the extended space, the full interface is
\begin{equation*}
\begin{gathered}
\mathcal R_R:
\mathcal B(\widehat{\mathcal H}_{\rm out})
\longrightarrow
\mathcal B(\mathcal H_{\rm out}),
\qquad
\mathcal R_R(\widehat{\sigma}_R)
=
\sigma_R,
\\
\widehat{M}
\preceq
C\widehat{\sigma}_R
\quad\Longrightarrow\quad
M
:=
\mathcal R_R(\widehat{M})
\preceq
C\sigma_R.
\end{gathered}
\end{equation*}

\indent
The last step follows from CP positivity. Records retained after readout must explicitly enter $\mathcal H_{\rm out}$ and its reference structure. The same readout also defines the program output $\rho_p$ used for target-state comparison and the error interface of full $\mathsf{TC}$. Thus the domination results in H and the subsequent complexity bounds always act on the same effective output space.

\indent
For each syntax state $s$, specify a finite action set $A_s$ and actual binary codewords
\begin{equation*}
c_{s,a}
\in
\{0,1\}^{\ell_{s,a}},
\qquad
\ell_{s,a}\ge1,
\qquad
a\in A_s.
\end{equation*}

\indent
The local codebook
$\{c_{s,a}:a\in A_s\}$
is prefix-free, hence
\begin{equation}
\label{eq:H.1}
\sum_{a\in A_s}
2^{-\ell_{s,a}}
\le
1.
\end{equation}

\indent
The syntax successor is given by the deterministic map
\begin{equation}
\label{eq:H.2}
\nu(s,a)
\in
S\cup\{\dagger\}
\end{equation}
where $\dagger$ denotes termination. Codewords, code lengths, the next syntax state, the termination decision, and the remaining code tree depend only on the current syntax state and the action already read, independently of Born measurement outcomes that have not been encoded. Measurement outcomes update the physical control block $q$ or explicit flags to implement feed-forward control of subsequent encoded actions. Program boundaries, Kraft weights, and parsing rules remain determined by the code tree.

\indent
If $\nu(s,a)=s'\in S$, represent the action's unconditional dynamical semantics by a sum of CP-TNI branches
\begin{equation*}
\Phi_{r\leftarrow q}^{s,a}:
\mathcal B(\mathcal K_q)
\longrightarrow
\mathcal B(\mathcal K_r)
\end{equation*}
and require, for every $Y\succeq0$,
\begin{equation}
\label{eq:H.3}
\sum_r
\operatorname{Tr}
\Phi_{r\leftarrow q}^{s,a}(Y)
=
\operatorname{Tr}Y.
\end{equation}

\indent
Equation~\eqref{eq:H.3} already sums over all quantum-instrument outcomes of the action, with each outcome's Born probability carried by the trace of its branch~\cite{DaviesLewis1970_QuantumProbability}. If $\nu(s,a)=\dagger$, represent the unconditional output by a CP map
\begin{equation*}
\Psi_q^{s,a}:
\mathcal B(\mathcal K_q)
\longrightarrow
\mathcal B(\mathcal H_{\rm out})
\end{equation*}
and require
\begin{equation}
\label{eq:H.4}
\operatorname{Tr}
\Psi_q^{s,a}(Y)
=
\operatorname{Tr}Y.
\end{equation}

\indent
Every accepted program therefore produces a total, normalized unconditional output. If a protocol retains a conditional success state, its failure flag, success probability, and execution cost must be included separately in the semantics.

\indent
Fix the initial syntax state $s_0$ and take a normalized initial block
\begin{equation}
\label{eq:H.5}
\Omega_0
=
\bigoplus_q
\Omega_{s_0,q},
\qquad
\Omega_{s,q}
=
0
\quad
(s\ne s_0),
\qquad
\operatorname{Tr}\Omega_0
=
1.
\end{equation}

\indent
Starting from $s_0$, take an action sequence
$a_1,\ldots,a_m$
and recursively set
$s_j=\nu(s_{j-1},a_j)$.
If all intermediate states lie in $S$ and $s_m=\dagger$, then
\begin{equation}
\label{eq:H.6}
p
=
c_{s_0,a_1}
c_{s_1,a_2}
\cdots
c_{s_{m-1},a_m}
\end{equation}
is an accepted program of length
\begin{equation}
\label{eq:H.7}
|p|
=
\sum_{j=1}^{m}
\ell_{s_{j-1},a_j}.
\end{equation}

\indent
Local prefix-freeness gives unique action-by-action parsing. Trailing bits are rejected after $\dagger$ is reached, so the accepted program set $\mathcal P$ is globally prefix-free. Summing unconditionally over all physical branches along a fixed program gives total CPTP semantics $\mathcal E_p$. Define
\begin{equation}
\label{eq:H.8}
\rho_p
:=
\mathcal E_p(\Omega_0),
\qquad
\operatorname{Tr}\rho_p
=
1.
\end{equation}

\indent
The code tree and physical dynamics can now be combined into two block CP maps. Define
$\mathbb T:\mathfrak X\to\mathfrak X$
and
$\mathbb H:\mathfrak X\to\mathcal B(\mathcal H_{\rm out})$
by
\begin{gather}
\label{eq:H.9}
[\mathbb T(Z)]_{s',r}
=
\sum_{\substack{
s,q,a\\
\nu(s,a)=s'
}}
2^{-\ell_{s,a}}
\Phi_{r\leftarrow q}^{s,a}
(Z_{s,q}),
\\
\label{eq:H.10}
\mathbb H(Z)
=
\sum_{\substack{
s,q,a\\
\nu(s,a)=\dagger
}}
2^{-\ell_{s,a}}
\Psi_q^{s,a}
(Z_{s,q}).
\end{gather}

\indent
Each action's code factor is inserted only once, in either Eq.~\eqref{eq:H.9} or Eq.~\eqref{eq:H.10}; Born weights are already included in the CP-TNI branches. Equations~\eqref{eq:H.1}, \eqref{eq:H.3}, and \eqref{eq:H.4} imply that every $Z\succeq0$ satisfies
\begin{equation}
\label{eq:H.11}
\operatorname{Tr}\mathbb T(Z)
+
\operatorname{Tr}\mathbb H(Z)
\le
\operatorname{Tr}Z,
\end{equation}

\indent
or, equivalently,
\begin{equation}
\label{eq:H.12}
\mathbb T^*(I_{\mathfrak X})
+
\mathbb H^*(I_{\rm out})
\preceq
I_{\mathfrak X}.
\end{equation}

\Hstatement{Lemma H.1 (Synchronous prefix--CP realization and terminating-domain coverage).}
Under the conditions above, for every $m\ge1$,
\begin{equation}
\label{eq:H.13}
\mathbb H\mathbb T^{m-1}(\Omega_0)
=
\sum_{\substack{
p\in\mathcal P\\
p\text{ has }m\text{ syntax actions}
}}
2^{-|p|}\rho_p.
\end{equation}

\indent
Consequently, the semidensity of all terminating trajectories is exactly the program semidensity:
\begin{equation}
\label{eq:H.14}
\sum_{k\ge0}
\mathbb H\mathbb T^k(\Omega_0)
=
\sum_{p\in\mathcal P}
2^{-|p|}\rho_p.
\end{equation}

\Hproof
Proceed by induction on nonterminal action prefixes. The identity $\mathbb T^0(\Omega_0)=\Omega_0$ corresponds to the empty prefix. Each application of $\mathbb T$ adds one continuing action along the deterministic syntax successor and inserts its code weight. Thus $\mathbb T^m(\Omega_0)$ sums all prefixes with $m\ge0$ continuing actions, with each action's physical branches already summed unconditionally.

\indent
One further application of $\mathbb H$ adds a terminal action, giving accepted programs with $m+1$ actions. Uniqueness of local parsing ensures that each accepted program appears exactly once, and the product of code factors is
\begin{equation*}
\prod_{j=1}^{m+1}
2^{-\ell_{s_{j-1},a_j}}
=
2^{-|p|}.
\end{equation*}

\indent
The sum of CP branches along that program is precisely $\rho_p$, so Eq.~\eqref{eq:H.13} holds for every positive integer action count. Summing over all finite terminating depths gives Eq.~\eqref{eq:H.14}.\Hqed

\indent
Lemma H.1 establishes the correspondence between terminating paths and the program semidensity for finite synchronous syntax. Unbounded storage or length counters, measurement-dependent code trees, free conditional normalization, and random stopping times or RUS protocols without a finite worst-case branch length require separate program and cost semantics. Their semantic $\mathsf{RA}$ may be proved for the corresponding interfaces; continuous control is discretized according to the resolution rules of Appendix~\ref{app:F}.

\subsection{The Least CP Fixed Point of the Program Semidensity}
\label{app:H.2}

\indent
Lemma H.1 identifies each finite terminating depth with the corresponding contribution to the program semidensity. We now sum all finite halting contributions through an infinite-depth limit and establish their least-fixed-point characterization. Define the truncated value map
\begin{equation}
\label{eq:H.15}
\mathbb V_N
:=
\sum_{k=0}^{N}
\mathbb H\circ\mathbb T^k,
\qquad
N\ge0.
\end{equation}

\Hstatement{Theorem H.2 (The least CP fixed point of the program semidensity).}
Under the mass-dissipation condition in Eq.~\eqref{eq:H.11}:

\begin{enumerate}
\setlength{\topsep}{0pt}
\setlength{\partopsep}{0pt}
\setlength{\itemsep}{0.4\baselineskip}
\setlength{\parsep}{0pt}

\item
$J(\mathbb V_N)$ increases monotonically in the positive semidefinite (PSD) order and converges in trace norm. Its limit defines the CP map
\begin{equation}
\label{eq:H.16}
\mathbb V
:=
\sum_{k\ge0}
\mathbb H\circ\mathbb T^k;
\end{equation}

\item
$\mathbb V$ satisfies the Bellman fixed-point equation
\begin{equation}
\label{eq:H.17}
\mathbb V
=
\mathbb H
+
\mathbb V\circ\mathbb T;
\end{equation}

\item
If a CP map $\mathbb W$ satisfies
\begin{equation}
\label{eq:H.18}
\mathbb W
\succeq_{\rm CP}
\mathbb H
+
\mathbb W\circ\mathbb T,
\end{equation}
then
$\mathbb W\succeq_{\rm CP}\mathbb V$.
Thus $\mathbb V$ is the least fixed point in the CP order and is bounded above by every CP supersolution;

\item
Under the realization and terminating-domain coverage conditions of Lemma H.1, the full program semidensity is
\begin{equation}
\label{eq:H.19}
M
=
\mathbb V(\Omega_0)
=
\sum_{p\in\mathcal P}
2^{-|p|}
\rho_p.
\end{equation}

\end{enumerate}

\Hproof
Equation~\eqref{eq:H.11} gives, for every $Z\succeq0$,
\begin{equation*}
\operatorname{Tr}\mathbb H(Z)
\le
\operatorname{Tr}Z
-
\operatorname{Tr}\mathbb T(Z).
\end{equation*}

\indent
Applying this successively to
$Z,\mathbb T(Z),\ldots,\mathbb T^N(Z)$
and summing gives the telescoping bound
\begin{equation}
\label{eq:H.20}
\sum_{k=0}^{N}
\operatorname{Tr}\!\left[
\mathbb H\mathbb T^k(Z)
\right]
\le
\operatorname{Tr}Z
-
\operatorname{Tr}\mathbb T^{N+1}(Z)
\le
\operatorname{Tr}Z.
\end{equation}

\indent
Since
$\mathbb V_{N+1}-\mathbb V_N
=
\mathbb H\circ\mathbb T^{N+1}$
is CP, the Choi matrices of the truncated value maps increase monotonically. Applying Eq.~\eqref{eq:H.20} to each diagonal basis input in the Choi construction and summing gives
\begin{equation}
\label{eq:H.21}
\operatorname{Tr}J(\mathbb V_N)
\le
d_{\rm in},
\end{equation}
where $d_{\rm in} = \sum_{(s,q)} \dim\mathcal K_q$. In finite dimensions, a monotone PSD sequence with bounded trace converges in trace norm to a PSD limit, proving Eq.~\eqref{eq:H.16}. Taking the limit in
\begin{equation*}
\mathbb V_N
=
\mathbb H
+
\mathbb V_{N-1}\circ\mathbb T
\end{equation*}
gives Eq.~\eqref{eq:H.17}.

\indent
If $\mathbb W$ satisfies Eq.~\eqref{eq:H.18}, then
$\mathbb W\succeq_{\rm CP}\mathbb V_0$.
Assuming
$\mathbb W\succeq_{\rm CP}\mathbb V_N$,
CP positivity gives
\begin{equation*}
\mathbb W
\succeq_{\rm CP}
\mathbb H+\mathbb W\circ\mathbb T
\succeq_{\rm CP}
\mathbb H+\mathbb V_N\circ\mathbb T
=
\mathbb V_{N+1}.
\end{equation*}

\indent
Induction and passage to the limit establish minimality. Equation~\eqref{eq:H.19} then follows directly from Lemma H.1.\Hqed

\indent
If the spectral radius of $\mathbb T$ on the full transient operator space satisfies
$r(\mathbb T)<1$,
the Neumann series rewrites Eq.~\eqref{eq:H.16} as
\begin{equation}
\label{eq:H.22}
\mathbb V
=
\mathbb H\circ(I-\mathbb T)^{-1}.
\end{equation}

\indent
Theorem H.2 remains valid when the full space contains a recurrent sector of spectral radius one. Equation~\eqref{eq:H.22} applies on a reachable, $\mathbb T$-invariant productive subspace with a spectral gap. Sectors that transport no mass to $\mathbb H$ leave only a nonhalting residual.

\indent
In particular, Eq.~\eqref{eq:H.20} gives
$\operatorname{Tr}M\le1$.
Infinite code-reading paths and dark nonhalting loops correspond to no finite accepted program and therefore do not enter $M$. Renormalizing this residual mass and assigning it to successful outputs changes both program semantics and Kraft weights.

\subsection{Bellman--Choi Qualification at Fixed Size}
\label{app:H.3}

\indent
Fix a finite-dimensional finite-control model, and assume Lemma H.1 has identified
$M=\mathbb V(\Omega_0)$
as the full program semidensity. The minimality in Theorem H.2 turns domination over all paths into a CP-supersolution problem.

\Hstatement{Theorem H.3 (CP-supersolution characterization of qualification for a fixed model).}
For any candidate constant $C\ge0$, the following are equivalent:

\begin{enumerate}
\setlength{\topsep}{0pt}
\setlength{\partopsep}{0pt}
\setlength{\itemsep}{0.4\baselineskip}
\setlength{\parsep}{0pt}

\item
The program semidensity satisfies
\begin{equation}
\label{eq:H.23}
M
\preceq
C\sigma_R;
\end{equation}

\item
There exists a CP map
$\mathbb W:\mathfrak X\to\mathcal B(\mathcal H_{\rm out})$
such that
\begin{equation}
\label{eq:H.24}
\mathbb W
-
\mathbb H
-
\mathbb W\circ\mathbb T
\succeq_{\rm CP}
0,
\end{equation}
and
\begin{equation}
\label{eq:H.25}
\mathbb W(\Omega_0)
\preceq
C\sigma_R.
\end{equation}

\end{enumerate}

\Hproof
If such a $\mathbb W$ exists, Theorem H.2 gives
$\mathbb V\preceq_{\rm CP}\mathbb W$.
Applying this to $\Omega_0\succeq0$ yields
\begin{equation*}
M
=
\mathbb V(\Omega_0)
\preceq
\mathbb W(\Omega_0)
\preceq
C\sigma_R.
\end{equation*}

\indent
Conversely, if Eq.~\eqref{eq:H.23} holds, take
$\mathbb W=\mathbb V$.
The fixed-point equation, Eq.~\eqref{eq:H.17}, makes Eq.~\eqref{eq:H.24} an equality, and Eq.~\eqref{eq:H.25} is then precisely Eq.~\eqref{eq:H.23}.\Hqed

\indent
For a fixed model, Theorem H.3 can be written as a finite set of explicit linear matrix inequalities (LMIs). Decompose $\mathbb W$ over transient blocks as
\begin{equation}
\label{eq:H.26}
\mathbb W(Z)
=
\sum_{x\in X}
\mathbb W_x(Z_x),
\qquad
X_x
:=
J(\mathbb W_x).
\end{equation}

\indent
Similarly, denote the halting blocks by $\mathbb H_x$, and choose a Kraus representation for each transition block,
\begin{equation}
\label{eq:H.27}
\mathbb T_{y\leftarrow x}(Y)
=
\sum_\mu
K_\mu^{y\leftarrow x}
Y
(K_\mu^{y\leftarrow x})^\dagger.
\end{equation}

\indent
The code factors have been absorbed into the corresponding Kraus operators as
$2^{-\ell/2}$.
For a fixed transition, define the linear map on Choi variables
\begin{equation}
\label{eq:H.28}
\Lambda_{y\leftarrow x}(X_y)
:=
\sum_\mu
\bigl[
(K_\mu^{y\leftarrow x})^T
\otimes
I_{\rm out}
\bigr]
X_y
\bigl[
\overline{K_\mu^{y\leftarrow x}}
\otimes
I_{\rm out}
\bigr].
\end{equation}

\indent
Direct expansion of the Choi definition gives
\begin{equation}
\label{eq:H.29}
J(\mathbb W_y\circ\mathbb T_{y\leftarrow x})
=
\Lambda_{y\leftarrow x}(X_y).
\end{equation}

\indent
The same relation can also be written as the standard link product~\cite{ChiribellaEtAl2009_QuantumNetworks}:
\begin{equation}
\label{eq:H.30}
J(\mathbb W_y\circ\mathbb T_{y\leftarrow x})
=
\operatorname{Tr}_{\mathcal K_y}
\left[
\bigl(
J(\mathbb T_{y\leftarrow x})^{T_{\mathcal K_y}}
\otimes
I_{\rm out}
\bigr)
\bigl(
I_{\mathcal K_x}
\otimes
X_y
\bigr)
\right].
\end{equation}

\indent
Thus Eqs.~\eqref{eq:H.24}--\eqref{eq:H.25} are equivalent to
\begin{gather}
\label{eq:H.31}
X_x
\succeq
0
\qquad
(\forall x\in X),
\\
\label{eq:H.32}
X_x
-
J(\mathbb H_x)
-
\sum_{y\in X}
\Lambda_{y\leftarrow x}(X_y)
\succeq
0
\qquad
(\forall x\in X),
\\
\label{eq:H.33}
C\sigma_R
-
\sum_{x\in X}
\operatorname{Tr}_{\mathcal K_x}
\left[
\bigl(
\Omega_{0,x}^{T}
\otimes
I_{\rm out}
\bigr)
X_x
\right]
\succeq
0.
\end{gather}

\indent
The control blocks, Hilbert spaces, syntax graph, and action sets are all finite at a fixed size. Equations~\eqref{eq:H.31}--\eqref{eq:H.33} therefore form a finite SDP feasibility problem for a given $C$. This compresses the joint contributions of arbitrary terminating depths and all loop paths into a finite set of Choi variables. Each $X_x$ is an envelope for future terminating outputs starting from control block $x$; the Bellman--Choi inequalities control all its successors at once, without enumerating programs or expanding loop depths.

\Hstatement{Corollary H.3.1 (Exact domination constant and refuting witnesses).}
If
$\operatorname{supp}M\subseteq\operatorname{supp}\sigma_R$,
the smallest feasible constant is
\begin{equation}
\label{eq:H.34}
C^\star
=
\left\|
\sigma_R^{-1/2}
M
\sigma_R^{-1/2}
\right\|_\infty.
\end{equation}

\indent
If the support condition fails, let $\Pi_0$ project onto $\ker\sigma_R$. Then
$\Pi_0 M\Pi_0\neq0$.
Taking $P$ to be the support projection of this positive operator gives
\begin{equation*}
0
\preceq
P
\preceq
I,
\qquad
\operatorname{Tr}(PM)
>
0,
\qquad
\operatorname{Tr}(P\sigma_R)
=
0.
\end{equation*}

\indent
This measurement direction therefore refutes every finite domination constant, and
$C^\star=+\infty$.
If the support condition holds but $C<C^\star$, take an eigendirection $|u\rangle$ of
$\sigma_R^{-1/2}M\sigma_R^{-1/2}$
with eigenvalue greater than $C$, and set
\begin{equation*}
P
=
\sigma_R^{-1/2}
|u\rangle\langle u|
\sigma_R^{-1/2}.
\end{equation*}

\indent
If necessary, rescale $P$ by a positive factor so that
$0\preceq P\preceq I$.
It still satisfies
\begin{equation}
\label{eq:H.35}
\operatorname{Tr}(PM)
>
C\operatorname{Tr}(P\sigma_R).
\end{equation}

\indent
Thus every candidate below $C^\star$ when the support condition holds, and every finite candidate when it fails, has a corresponding refuting witness in a finite measurement direction.

\indent
Theorem H.3 gives a necessary and sufficient characterization for a fixed finite model and a given $C$. A floating-point solution must undergo rational, algebraic, or interval PSD verification to yield a rigorous certificate. Qualification of a model family additionally requires proving $\sup_n C_n<\infty$; the next subsection gives a reference-potential construction that provides this uniformity.

\subsection{Uniform Reference Potentials across a Model Family}
\label{app:H.4}

\indent
RCC reference qualification requires a domination constant uniformly bounded across the model family. We use local operator envelopes to control one-step amplification of reference load, then sum all future halting contributions through scalar Bellman inequalities. Fix a model family
$\{\widehat{\mathfrak M}_{R,n}\}_{n\in\mathbb N}$,
and assume Appendix~\ref{app:H.1} has supplied, for each $n$, finite control blocks $X_n$, code-weighted transitions $\mathbb T_n$, halting emissions $\mathbb H_n$, and the initial block
\begin{equation}
\label{eq:H.36}
\Omega_{0,n}
=
\bigoplus_{x\in X_n}
\Omega_{0,x,n}.
\end{equation}

\indent
Write the block maps as
\begin{equation}
\label{eq:H.37}
\mathbb T_{y\leftarrow x,n}:
\mathcal B(\mathcal K_{x,n})
\longrightarrow
\mathcal B(\mathcal K_{y,n}),
\qquad
\mathbb H_{x,n}:
\mathcal B(\mathcal K_{x,n})
\longrightarrow
\mathcal B(\mathcal H_{{\rm out},n}).
\end{equation}

\indent
For each $x,n$, choose a normalized reference envelope, fixed in advance independently of the target state and full-rank on its declared support,
\begin{equation}
\label{eq:H.38}
\Theta_{x,n}
\succ
0,
\qquad
\operatorname{Tr}\Theta_{x,n}
=
1.
\end{equation}

\indent
Normalization fixes the envelope scale, giving control blocks and system sizes a common load convention.

\Hstatement{Theorem H.4 (Qualification by uniform reference potentials across a model family).}
Suppose there are nonnegative numbers $a_{x,n}, \, b_{x,n}, \, A_{x\to y,n}, \, v_{x,n}$, satisfying
\begin{gather}
\label{eq:H.39}
\Omega_{0,x,n}
\preceq
a_{x,n}\Theta_{x,n},
\\
\label{eq:H.40}
\mathbb T_{y\leftarrow x,n}
(\Theta_{x,n})
\preceq
A_{x\to y,n}
\Theta_{y,n},
\\
\label{eq:H.41}
\mathbb H_{x,n}
(\Theta_{x,n})
\preceq
b_{x,n}
\sigma_{R,n},
\end{gather}
and the Bellman supersolution inequality
\begin{equation}
\label{eq:H.42}
v_{x,n}
\ge
b_{x,n}
+
\sum_{y\in X_n}
A_{x\to y,n}
v_{y,n}.
\end{equation}

\indent
Then the full program semidensity at that size satisfies
\begin{equation}
\label{eq:H.43}
M_n
\preceq
C_n\sigma_{R,n},
\qquad
C_n
:=
\sum_{x\in X_n}
a_{x,n}v_{x,n}.
\end{equation}

\indent
If
\begin{equation}
\label{eq:H.44}
C
:=
\sup_n C_n
<
\infty,
\end{equation}
then the entire model family satisfies semantic $\mathsf{RA}$. For the anchored certificate parameterization with fixed $0<\alpha<1$,
\begin{equation}
\label{eq:H.45}
M_{\alpha,n}
\preceq
\bigl[
\alpha+(1-\alpha)C
\bigr]
\sigma_{R,n},
\end{equation}
so one may take
\begin{equation}
\label{eq:H.46}
\kappa_U
=
\max\!\left\{
0,\,
\log_2\!\bigl[
\alpha+(1-\alpha)C
\bigr]
\right\}.
\end{equation}

\indent
Here $\Theta_{x,n}$ is the normalized reference envelope fixed in advance on control block $x$, and $a_{x,n}$ measures the initial-state load relative to that envelope. The coefficients $A_{x\to y,n}$ and $b_{x,n}$ bound the maximal amplification of reference load through a one-step continuing transition and a terminal output, respectively. The Bellman reference potential $v_{x,n}$ bounds all future terminating outputs starting from that block. The uniform bound $\sup_nC_n<\infty$ keeps the controller's reference-structure load on terminating outputs within a size-independent range.

\Hproof
For each starting block $x$, define the depth-truncated value maps
\begin{gather}
\label{eq:H.47}
\mathbb V_{x,n}^{(0)}
:=
\mathbb H_{x,n},
\\
\label{eq:H.48}
\mathbb V_{x,n}^{(N+1)}
:=
\mathbb H_{x,n}
+
\sum_{y\in X_n}
\mathbb V_{y,n}^{(N)}
\circ
\mathbb T_{y\leftarrow x,n}.
\end{gather}

\indent
We prove that, for all $N\ge0$,
\begin{equation}
\label{eq:H.49}
\mathbb V_{x,n}^{(N)}
(\Theta_{x,n})
\preceq
v_{x,n}
\sigma_{R,n}.
\end{equation}

\indent
For $N=0$, the claim follows from Eqs.~\eqref{eq:H.41} and \eqref{eq:H.42}. If Eq.~\eqref{eq:H.49} holds at depth $N$, CP positivity and Eq.~\eqref{eq:H.40} give
\begin{equation*}
\begin{aligned}
\mathbb V_{x,n}^{(N+1)}
(\Theta_{x,n})
&\preceq
b_{x,n}\sigma_{R,n}
+
\sum_y
A_{x\to y,n}
\mathbb V_{y,n}^{(N)}
(\Theta_{y,n})
\\
&\preceq
\left(
b_{x,n}
+
\sum_y
A_{x\to y,n}
v_{y,n}
\right)
\sigma_{R,n}
\\
&\preceq
v_{x,n}
\sigma_{R,n}.
\end{aligned}
\end{equation*}

\indent
Thus Eq.~\eqref{eq:H.49} follows by induction. Equation~\eqref{eq:H.39} then gives
\begin{equation*}
\sum_x
\mathbb V_{x,n}^{(N)}
(\Omega_{0,x,n})
\preceq
\sum_x
a_{x,n}
v_{x,n}
\sigma_{R,n}
=
C_n
\sigma_{R,n}.
\end{equation*}

\indent
Letting $N\to\infty$, monotone convergence from Theorem H.2 and the realization in Lemma H.1 yield Eq.~\eqref{eq:H.43}. Equation~\eqref{eq:H.44} establishes domination uniformly across the family; Eqs.~\eqref{eq:H.45}--\eqref{eq:H.46} follow directly from the definition of the anchored operator.\Hqed
If the initial state lies only in block $x_0$, then
$C_n=a_{0,n}v_{x_0,n}$.
In particular, if
$a_{0,n}=v_{x_0,n}=1$
for every $n$, then
\begin{equation*}
M_n
\preceq
\sigma_{R,n},
\qquad
\kappa_U
=
0.
\end{equation*}

\indent
For fixed $n$, view $A_{x\to y,n}$ as a nonnegative matrix $A_n$ and $b_{x,n}$ as a vector $b_n$. If $\rho(A_n)<1$, then
\begin{equation}
\label{eq:H.50}
v_n
=
(I-A_n)^{-1}b_n
=
\sum_{k\ge0}
A_n^k b_n
\end{equation}
is the least nonnegative Bellman solution. Model-family qualification still requires the separate check
$\sup_n a_n^T v_n<\infty$.
The spectral-radius condition is not necessary for Theorem H.4: a critical recurrent class that transports no reference load to the halting output may still admit a finite supersolution.

\indent
Suppose all $X_n$ embed in the same finite master control set $X_\star$ independently of the target state, with missing blocks padded by zero, and there are data independent of $n$ satisfying
\begin{equation}
\label{eq:H.51}
\overline v
\ge
0,
\qquad
A_n
\le
\overline A,
\qquad
b_n
\le
\overline b,
\qquad
a_n
\le
\overline a,
\qquad
\overline v
\ge
\overline b
+
\overline A\,\overline v.
\end{equation}

\indent
Then
\begin{equation}
\label{eq:H.52}
C
\le
\overline a^{\,T}
\overline v
<
\infty.
\end{equation}

\indent
A fixed master control set is only a sufficient template. The control blocks may also grow with $n$, provided the same parametrized rule directly proves Eq.~\eqref{eq:H.44}.

\indent
We can now specify the proof object for $\mathsf{RefCert}_{\rm fc}$ rigorously. A synchronous finite-control model family satisfies $\mathsf{RefCert}_{\rm fc}$ if Appendix~\ref{app:H.1} establishes realization and terminating-domain coverage of the program semidensity at every size, and a target-state-independent rule, uniform across the model family, supplies a rigorous proof object. This object either constructs the CP supersolutions of Theorem H.3 with a single finite constant controlling all sizes, or constructs the uniform reference potentials of Theorem H.4 and proves Eq.~\eqref{eq:H.44}. The corresponding PSD relations must be established analytically or by numerical verification with rigorous error bounds.
Theorem H.3 gives a necessary and sufficient characterization for a fixed finite model; Theorem H.4 directly establishes family-level uniformity through a compositional sufficient certificate.

\subsection{Certificate Slack, Implementation Stability, and Rigorous Proof Objects}
\label{app:H.5}

\indent
Whether a finite-precision implementation can use the same reference potential depends on Bellman slack and amplification of errors along reference directions. Define
\begin{equation}
\label{eq:H.53}
\eta_{x,n}
:=
v_{x,n}
-
b_{x,n}
-
\sum_y
A_{x\to y,n}
v_{y,n}
\ge
0.
\end{equation}

\indent
The perturbed model must still satisfy the synchronous syntax, total-output semantics, and separation of code and Born weights in Appendix~\ref{app:H.1}. Suppose its normalized envelopes, initial blocks, and nonnegative coefficients satisfy
\begin{gather}
\label{eq:H.54}
\widetilde\Omega_{0,x,n}
\preceq
\widetilde a_{x,n}
\widetilde\Theta_{x,n},
\\
\label{eq:H.55}
\widetilde{\mathbb T}_{y\leftarrow x,n}
(\widetilde\Theta_{x,n})
\preceq
\widetilde A_{x\to y,n}
\widetilde\Theta_{y,n},
\\
\label{eq:H.56}
\widetilde{\mathbb H}_{x,n}
(\widetilde\Theta_{x,n})
\preceq
\widetilde b_{x,n}
\sigma_{R,n}.
\end{gather}

\Hstatement{Theorem H.5 (Stability from certificate slack).}
If there are nonnegative error budgets
$\delta A,\delta b,\delta a$
such that
\begin{equation}
\label{eq:H.57}
\widetilde A_{x\to y,n}
\le
A_{x\to y,n}
+
\delta A_{x\to y,n},
\qquad
\widetilde b_{x,n}
\le
b_{x,n}
+
\delta b_{x,n},
\end{equation}
and
\begin{equation}
\label{eq:H.58}
\delta b_{x,n}
+
\sum_y
\delta A_{x\to y,n}
v_{y,n}
\le
\eta_{x,n}
\qquad
(\forall x,n),
\end{equation}
then the same potential $v_{x,n}$ remains a Bellman supersolution for the perturbed model. If, in addition,
\begin{equation}
\label{eq:H.59}
\widetilde a_{x,n}
\le
a_{x,n}
+
\delta a_{x,n},
\end{equation}
and
\begin{equation}
\label{eq:H.60}
\widetilde C
:=
\sup_n
\sum_x
\bigl(
a_{x,n}
+
\delta a_{x,n}
\bigr)
v_{x,n}
<
\infty,
\end{equation}
then the perturbed full program semidensity satisfies
\begin{equation}
\label{eq:H.61}
\widetilde M_n
\preceq
\widetilde C\,
\sigma_{R,n}
\qquad
(\forall n).
\end{equation}

\Hproof
Equations~\eqref{eq:H.53}, \eqref{eq:H.57}, and \eqref{eq:H.58} give
\begin{equation*}
\begin{aligned}
\widetilde b_{x,n}
+
\sum_y
\widetilde A_{x\to y,n}
v_{y,n}
&\le
b_{x,n}
+
\sum_y
A_{x\to y,n}
v_{y,n}
\\
&\quad+
\delta b_{x,n}
+
\sum_y
\delta A_{x\to y,n}
v_{y,n}
\\
&\le
v_{x,n}.
\end{aligned}
\end{equation*}

\indent
Thus $v$ still satisfies the perturbed Bellman inequality. Equations~\eqref{eq:H.54}--\eqref{eq:H.56} and Eq.~\eqref{eq:H.59} allow Theorem H.4 to apply directly; Eq.~\eqref{eq:H.60} then gives Eq.~\eqref{eq:H.61}.\Hqed

\indent
If the envelopes remain unchanged, define the reference-relative errors
\begin{gather}
\label{eq:H.62}
\varepsilon^{T}_{x\to y,n}
:=
\left\|
\Theta_{y,n}^{-1/2}
\bigl[
(\widetilde{\mathbb T}_{y\leftarrow x,n}
-
\mathbb T_{y\leftarrow x,n})
(\Theta_{x,n})
\bigr]
\Theta_{y,n}^{-1/2}
\right\|_\infty,
\\
\label{eq:H.63}
\varepsilon^{H}_{x,n}
:=
\left\|
\sigma_{R,n}^{-1/2}
\bigl[
(\widetilde{\mathbb H}_{x,n}
-
\mathbb H_{x,n})
(\Theta_{x,n})
\bigr]
\sigma_{R,n}^{-1/2}
\right\|_\infty.
\end{gather}

\indent
The operators in brackets are Hermitian, hence
\begin{gather}
\label{eq:H.64}
(\widetilde{\mathbb T}_{y\leftarrow x,n}
-
\mathbb T_{y\leftarrow x,n})
(\Theta_{x,n})
\preceq
\varepsilon^T_{x\to y,n}
\Theta_{y,n},
\\
\label{eq:H.65}
(\widetilde{\mathbb H}_{x,n}
-
\mathbb H_{x,n})
(\Theta_{x,n})
\preceq
\varepsilon^H_{x,n}
\sigma_{R,n}.
\end{gather}

\indent
Thus $\varepsilon^T,\varepsilon^H$ may serve as $\delta A,\delta b$, respectively. Changes to the envelopes or output reference require two-sided operator bounds uniform across the family, or renewed verification of Eqs.~\eqref{eq:H.54}--\eqref{eq:H.56}.

\indent
When estimating these relative errors from ordinary channel norms, the smallest reference eigenvalue determines the amplification factor. For example,
\begin{equation*}
\left\|
\widetilde{\mathbb T}_{y\leftarrow x,n}
-
\mathbb T_{y\leftarrow x,n}
\right\|_\diamond
\le
\delta_{x\to y,n}
\end{equation*}
implies
\begin{equation}
\label{eq:H.66}
\varepsilon^T_{x\to y,n}
\le
\|\Theta_{y,n}^{-1}\|_\infty
\delta_{x\to y,n}.
\end{equation}

\indent
The halting map similarly introduces
$\|\sigma_{R,n}^{-1}\|_\infty$.
For
$\sigma_{R,n}=I/d_n$,
an absolute diamond-norm error may be amplified to
$O(d_n\delta)$.
Uniform stability therefore requires joint control of the slack and the norms of the inverse reference operators. When retaining the same reference potential $v$ and using the worst-case one-sided nonnegative budgets of Eqs.~\eqref{eq:H.57}--\eqref{eq:H.58}, zero slack requires every error contribution that increases the Bellman left-hand side to vanish. Perturbations in favorable directions and changes leading to zero-potential sectors may still be allowed. Choosing new envelopes and reference potentials requires the certificate to be rechecked.

A rigorous finite-control certificate suitable for third-party verification should record the operationally reduced action classes, full resource semantics, and the syntax graph, actual codewords, successors, termination rules, CP edges, quantum-instrument completeness, initial blocks, output reference, and fixed readout of Appendix~\ref{app:H.1}. Reference domination is supplied by a Bellman supersolution from Appendix~\ref{app:H.3} or the $\Theta,A,b,a,v$ data of Appendix~\ref{app:H.4}, together with an analytical proof of $\sup_nC_n<\infty$ and PSD verification with rigorous error bounds.

The full complexity interface additionally carries an independent $\mathsf{TC}$ compilation and length certificate, with explicit scopes for nonhalting behavior, postselection, random stopping times, auxiliary resources, precision, addressing, and variable costs.

\indent
Violation of a candidate local inequality shows only that this proof object fails to certify the model. Within the fixed finite-model class characterized by the equivalence in Theorem H.3, exact infeasibility of the LMIs for a given $C$ means that the fixed model does not satisfy the corresponding domination statement.

\subsection{Reference Gain--Cost Completion for Finite Control}
\label{app:H.6}

\indent
Appendices~\ref{app:H.3} and \ref{app:H.4} certify models whose code weights are already fixed. We now incorporate an action's reference-generating capability into its effective code length, so that code weights absorb the corresponding gain and yield an interface with unit reference potential.
The construction applies to synchronous, total-output, support-compatible finite-control syntax. Conditions for incorporating separate fuel accounting and lower-level action decompositions into effective code length are stated after the construction.

\indent
Retain the syntax states $s$, physical control states $q$, and actions of Appendix~\ref{app:H.1}, writing $A_s$ and $\nu$ for
$A_{s,n}$ and $\nu_n$.
Temporarily separate code factors from the dynamical maps. If
$\nu(s,a)=s'$,
denote the CP block transition before code weighting by $\mathcal T^{s,a}_{(s',r)\leftarrow(s,q),n}$; and denote the unweighted halting map for a terminal action by $\mathcal H^{s,a}_{(s,q),n}$. Choose target-state-independent, normalized, support-compatible envelopes
$\Theta_{s,q,n}$,
and define the minimal reference-gain coefficients
\begin{gather}
\label{eq:H.67}
g^{s,a}_{(s,q)\to(s',r),n}
:=
\inf\left\{
g\ge0:
\mathcal T^{s,a}_{(s',r)\leftarrow(s,q),n}
(\Theta_{s,q,n})
\preceq
g\Theta_{s',r,n}
\right\},
\\
\label{eq:H.68}
h^{s,a}_{s,q,n}
:=
\inf\left\{
h\ge0:
\mathcal H^{s,a}_{(s,q),n}
(\Theta_{s,q,n})
\preceq
h\sigma_{R,n}
\right\}.
\end{gather}

\indent
Set inapplicable terms to zero. Full-rank envelopes and support compatibility ensure that the coefficients are finite. The common codeword of a syntax action must cover every physical control block $q$ that measurement outcomes may reach, so take
\begin{equation}
\label{eq:H.69}
G_{s,a,n}
:=
\max_q
\left[
h^{s,a}_{s,q,n}
+
\sum_r
g^{s,a}_{(s,q)\to(\nu(s,a),r),n}
\right].
\end{equation}

\indent
When $\nu(s,a)=\dagger$, the continuation sum is empty; when the action continues, the halting term is zero.

\Hstatement{Theorem H.6 (Reference gain--cost completion for finite control).}
For each size $n$ and syntax state $s$, if the same target-state-independent rule generates positive integer code lengths $\ell_{s,a,n}$ and an actual local prefix-free code satisfying
\begin{equation}
\label{eq:H.70}
\sum_{a\in A_s}
2^{-\ell_{s,a,n}}
\max\{1,G_{s,a,n}\}
\le
1,
\end{equation}
then the code-weighted transitions satisfy the conditions of Theorem H.4, with
\begin{equation}
\label{eq:H.71}
v_{s,q,n}
=
1.
\end{equation}

\indent
If the initial blocks also satisfy
\begin{equation}
\label{eq:H.72}
\Omega_{0,s,q,n}
\preceq
a_{s,q,n}
\Theta_{s,q,n},
\qquad
C_0
:=
\sup_n
\sum_{s,q}
a_{s,q,n}
<
\infty,
\end{equation}
then the full program semidensity satisfies
\begin{equation}
\label{eq:H.73}
M_n
\preceq
C_0
\sigma_{R,n}.
\end{equation}

\indent
In particular, if the initial state is a single canonical envelope block, one may take
$C_0=1$
and
$\kappa_U=0$.

\Hproof
After code weighting, define the local coefficients of Theorem H.4 by
\begin{gather}
\label{eq:H.74}
A_{(s,q)\to(s',r),n}
:=
\sum_{\substack{
a\in A_s\\
\nu(s,a)=s'
}}
2^{-\ell_{s,a,n}}
g^{s,a}_{(s,q)\to(s',r),n},
\\
\label{eq:H.75}
b_{s,q,n}
:=
\sum_{\substack{
a\in A_s\\
\nu(s,a)=\dagger
}}
2^{-\ell_{s,a,n}}
h^{s,a}_{s,q,n}.
\end{gather}

\indent
By Eqs.~\eqref{eq:H.67}--\eqref{eq:H.68}, these coefficients satisfy the local operator inequalities. Equations~\eqref{eq:H.69}--\eqref{eq:H.70} then give
\begin{equation*}
\begin{aligned}
b_{s,q,n}
+
\sum_{s',r}
A_{(s,q)\to(s',r),n}
&=
\sum_{a\in A_s}
2^{-\ell_{s,a,n}}
\left[
h^{s,a}_{s,q,n}
+
\sum_r
g^{s,a}_{(s,q)\to(\nu(s,a),r),n}
\right]
\\
&\le
\sum_{a\in A_s}
2^{-\ell_{s,a,n}}
G_{s,a,n}
\le
1.
\end{aligned}
\end{equation*}

\indent
Thus $v_{s,q,n}=1$ is a Bellman supersolution. Equation~\eqref{eq:H.72} and Theorem H.4 yield Eq.~\eqref{eq:H.73}. The factor
$\max\{1,G\}$
in Eq.~\eqref{eq:H.70} also ensures the ordinary Kraft inequality, so the integer code lengths can be realized by an actual local prefix-free code.\Hqed

\Hstatement{Corollary H.6.1 (Explicit calibration of finite syntax).}
Let
$N_{s,n}:=|A_{s,n}|<\infty$,
and assume all
$G_{s,a,n}<\infty$.
Choosing
\begin{equation}
\label{eq:H.76}
\ell_{s,a,n}
=
\left\lceil
\log_2
\max\{2,N_{s,n}\}
\right\rceil
+
\left\lceil
\log_2
\max\{1,G_{s,a,n}\}
\right\rceil
\end{equation}
satisfies Eq.~\eqref{eq:H.70}. Thus the local transition part of any finite, synchronous, total-output, support-compatible physical-action syntax can be calibrated once its one-step reference-generating capability is incorporated into actual code lengths in advance. Together with the family-level initial-state condition in Eq.~\eqref{eq:H.72}, this gives a uniformly reference-admissible interface.

\indent
Corollary H.6.1 establishes reference qualification of the recoded program interface; the new actual code lengths determine the corresponding Kraft weights and program semidensity. A platform may carry this burden through separate fuel accounting, an additive resource coordinate distinct from bare instruction length, and incorporate it into description weights through an explicit resource suffix, equivalent effective code lengths, or another faithful compilation. When using lower-level action decompositions, the certificate of this subsection is reapplied to the decomposed transition family.

Connecting this result back to the atomic-step scale $L\log_2\Gamma_R$ requires a fixed, target-state-independent full $\mathsf{TC}$ compilation calibration that is uniform across the family. Changing process costs alone leaves the original program semidensity $M_n$ in place; the original interface's $\mathsf{RA}$ and $\mathsf{TC}$ are certified separately according to its actual semantics.
If an action family $a_n$ satisfies, on a block whose input envelope and output reference both equal $\sigma_{R,n}$,
\begin{equation*}
D_{\max}\!\left(
\Phi_{a_n}(\sigma_{R,n})
\middle\Vert
\sigma_{R,n}
\right)
=
\Theta(f(n)),
\end{equation*}
then Eq.~\eqref{eq:H.70} requires an effective description length of $\Omega(f(n))$. Through the compilation relations above, this burden may enter fuel accounting or a lower-level decomposition, or be reassessed after adjusting the reference background. Global reset on a maximally mixed reference therefore carries a reference cost that grows with system size.

\indent
To connect with the single-control calibration in Appendix~\ref{app:F.6}, consider one control state and $N\ge2$ total CPTP actions $\Phi_a$, with a uniform integer code length $b$ for each action under an independent self-delimiting length header. The one-step reference-domination condition is
\begin{equation}
\label{eq:H.77}
2^{-b}
\sum_{a=1}^{N}
\Phi_a(\sigma_R)
\preceq
\sigma_R.
\end{equation}

\indent
The minimal uniform code length is therefore
\begin{align}
\label{eq:H.78}
b_\star
&=
\left\lceil
D_{\max}\!\left(
\sum_{a=1}^{N}
\Phi_a(\sigma_R)
\middle\Vert
\sigma_R
\right)
\right\rceil
\\
\label{eq:H.79}
&=
\left\lceil
\log_2 N
+
D_{\max}\!\left(
\frac{1}{N}
\sum_{a=1}^{N}
\Phi_a(\sigma_R)
\middle\Vert
\sigma_R
\right)
\right\rceil.
\end{align}

\indent
Equation~\eqref{eq:H.79} separates action-selection bandwidth from average reference bias. When $N=\Gamma_R$ is a power of two, setting $b=\log_2\Gamma_R$ makes Eq.~\eqref{eq:H.77} equivalent to the reference-balance condition below:
\begin{equation*}
\frac{1}{\Gamma_R}
\sum_{a=1}^{\Gamma_R}
\Phi_a(\sigma_R)
=
\sigma_R.
\end{equation*}

\indent
Equality of traces turns the operator inequality into an equality. Thus this bandwidth-only code length satisfies Eq.~\eqref{eq:H.77} if and only if the action alphabet is reference-balanced on $\sigma_R$. For a general finite $\Gamma_R$ that is not a power of two, the whole-word rank-block encoding of Appendix~\ref{app:F.3} achieves $L\log_2\Gamma_R+O(\log L)$.
With variable-length codes, the effective description length of a history is
\begin{equation*}
\sum_{t=1}^{L}
\ell_{s_{t-1},a_t},
\end{equation*}
which accumulates term by term according to the actual variable-length code. Its comparison with $L\log_2\Gamma_R$ or physical execution cost uses the compilation calibration and qualification conditions above.

\indent
The following four boundaries delimit the scope of finite-control certificates.

\begin{enumerate}
\setlength{\topsep}{0pt}
\setlength{\partopsep}{0pt}
\setlength{\itemsep}{0.4\baselineskip}
\setlength{\parsep}{0pt}

\item
\textbf{Finite control still requires reference-gain calibration.}
A one-state prefix grammar
$\{1^k0:k\ge0\}$
performs a global reset on reading $0$. It is Kraft-complete and its programs are total, yet
\begin{equation*}
M_n
=
|0^n\rangle\langle0^n|,
\qquad
C_n^\star
=
2^n
\end{equation*}
relative to
$\sigma_n=I/2^n$.
The extensive reference gain must still enter the cost.

\item
\textbf{Deviations from exact balance can accumulate step by step.}
If the known relation is only
\begin{equation*}
\mathcal T(\sigma_R)
\preceq
2^\delta
\sigma_R,
\end{equation*}
then, in general, one obtains only
\begin{equation*}
\mathcal T^L(\sigma_R)
\preceq
2^{L\delta}
\sigma_R.
\end{equation*}

\indent
Keeping $\kappa_U=O(1)$ requires exact closure, a contractive envelope, or charging $\delta$ at each step.

\item
\textbf{No complete uniform decision procedure exists for general interfaces.}
Let a constant-length program first simulate a classical machine $e$. If it halts, the program outputs
$|0^n\rangle$
with a fixed code weight; otherwise, it never produces output. Relative to $I/2^n$, whether the uniform domination bound receives an extensive contribution depends on whether $e$ halts. A model family containing universal interpreters and unbounded storage therefore admits no complete uniform decision procedure for $\mathsf{RA}$. The certificate system here is required only to be sound.

\item
\textbf{Random stopping times and RUS require new cost objects.}
Almost-sure termination, expected length, worst-case branch length, and conditional success states belong to different resource semantics.

\end{enumerate}

\Needspace{3\baselineskip}
\HOriginalSubsection*{Closing Results and the Main-Theorem Interface}

\indent
The core dependencies of Appendices~\ref{app:H.1}--\ref{app:H.4} are
\begin{equation}
\label{eq:H.80}
\begin{gathered}
\text{Synchronous prefix syntax and total CP instruments}
\\
\Downarrow\quad
\text{H.1 realization / terminating-domain coverage}
\\
M_n
=
\mathbb V_n(\Omega_{0,n})
\\
\Downarrow\quad
\text{H.2 least CP fixed point}
\\
\begin{cases}
\text{H.3: fixed-size Bellman--Choi equivalence},\\
\text{H.4: sufficient certificate via uniform reference potentials}
\end{cases}
\\
\Downarrow
\\
\mathsf{RefCert}_{\rm fc}
\Longrightarrow
\mathsf{RA}.
\end{gathered}
\end{equation}

\indent
Appendix~\ref{app:H.5} gives conditions for certificate stability under implementation errors, and Appendix~\ref{app:H.6} absorbs one-step reference gain into effective code lengths. These certificates establish the $\mathsf{RA}$ required by the general main theorem. If the same model family also satisfies reference consistency and full $\mathsf{TC}$, including semantic preservation, circuit coverage, length control, and the error interface, then
\begin{equation}
\label{eq:H.81}
\mathsf{RCon}
\wedge
\mathsf{TC}
\wedge
\mathsf{RefCert}_{\rm fc}
\Longrightarrow
\mathsf{RCon}
\wedge
\mathsf{TC}
\wedge
\mathsf{RA}.
\end{equation}

\indent
The model family therefore inherits the RCC lower bound for the appropriate transcription branch in Sec.~\ref{sec:3} and Appendix~\ref{app:A}.
This appendix combines prefix-code weights and Born weights into unconditional program semidensity semantics. The structured vacuum connects the least CP fixed point, uniform reference potentials, certificate slack, and cost calibration into a verifiable finite-control qualification route, $\mathsf{RefCert}_{\rm fc}\Rightarrow\mathsf{RA}$. The general main theorem also applies to models whose full $\mathsf{TC}$ and $\mathsf{RA}$ are established by other methods.

\par
\endgroup

\begingroup
\setlength{\parindent}{1.5em}
\setlength{\parskip}{1pt plus 0.3pt minus 0.2pt}
\frenchspacing
\linespread{1.008}\selectfont
\setlength{\abovedisplayskip}{8.5pt plus 1pt minus 2pt}
\setlength{\belowdisplayskip}{8.5pt plus 1pt minus 2pt}
\setlength{\abovedisplayshortskip}{5.5pt plus 1pt minus 1pt}
\setlength{\belowdisplayshortskip}{6.5pt plus 1pt minus 1pt}
\setlength{\jot}{2pt}
\emergencystretch=1.25em
\clubpenalty=10000
\widowpenalty=10000
\displaywidowpenalty=10000
\brokenpenalty=5000
\raggedbottom

\newcommand{\Istatement}[1]{%
  \par\addvspace{0.4\baselineskip}%
  \Needspace{4\baselineskip}%
  \indent\textbf{#1}\hspace{0.5em}\ignorespaces}
\newcommand{\Iproof}{%
  \par\addvspace{0.25\baselineskip}%
  \Needspace{3\baselineskip}%
  \indent\textbf{Proof.}\hspace{0.5em}\ignorespaces}
\newcommand{\Iqed}{%
  \unskip\nobreak\hspace{0.35em}\ensuremath{\square}%
  \par\addvspace{0.3\baselineskip}}
\newcommand{\Ilead}[1]{%
  \par\addvspace{0.3\baselineskip}\Needspace{3\baselineskip}%
  \indent\textit{#1}\hspace{0.5em}\ignorespaces}
\makeatletter
\renewcommand{\subsection}{\@startsection{subsection}{2}{\z@}%
  {14pt plus 1pt minus 1pt}{8pt plus 0.5pt minus 1pt}%
  {\normalfont\small\bfseries\centering}}
\makeatother
\let\IOriginalSubsection\subsection
\renewcommand{\subsection}[1]{%
  \Needspace{3\baselineskip}%
  \IOriginalSubsection{#1}}

\section{Interfaces with Dynamics and Complexity Window Thermodynamics}
\label{app:I}

\indent
The main RCC theorem provides a complexity lower bound from the final state, while a concrete implementation can supply an upper bound from the process side. This appendix first proves that a dynamical path compiled into an admissible circuit for the same target, tolerance, and cost function is a feasible upper-bound witness for the optimal complexity within the model; together with the final-state certificate, it yields a two-sided interval. For continuous pure-state unitary records, the Hall decomposition and exact quantum speed limits (QSLs) further relate probability motion in a fixed basis, relative-phase motion, and energy fluctuations exactly, providing an independent geometric check of the same path.

\indent
Finite generative capacity also restricts the effective object domain of a thermodynamic description. Complexity Window Thermodynamics (CWT) characterizes this restriction through optimal quantum circuit complexity and a finite budget; windowed RCC fixes the reference zero point, atomic scale, and state-side readout within the same object domain. Appendices~\ref{app:I.2}--\ref{app:I.4} construct shared finite-dimensional windows based on conditional expectations and establish the finite staircase structure of exact windows and the fixed-reference capacity relation. Appendix~\ref{app:I.5} derives thermodynamic responses on an independently established effective smooth branch, and Appendix~\ref{app:I.6} incorporates reference and bandwidth changes and nonsmooth jump terms. These two connections support the dynamical and thermodynamic interfaces in Sec.~\ref{sec:6}.

\subsection{Path-Witness Complexity Intervals and Geometric Checks via Exact Quantum Speed Limits}
\label{app:I.1}

\Ilead{Model-consistent path compilation and upper-bound witnesses.}
 A dynamical path that has been executed or can be explicitly compiled can supplement the RCC lower bound to form a two-sided interval. Both endpoints use the same target state, error semantics, reference-consistent model, and cost function. The lower certificate $L_{\rm RCC}$ has been calibrated by $\log_2\Gamma_R$ and includes one-shot, smoothing, and coding corrections; the upper path is likewise measured in atomic resource slots under the same serial and parallel counting rules.
On the respective certificate events and under the tolerance ordering below, the interval to be established is
\begin{equation*}
\underline L_{\rm RCC}^{(\epsilon_L)}
\le
C_{\rm opt}^{(\epsilon_*)}
(\rho;\mathfrak M_R)
\le
C_{\gamma,R}^{(\epsilon_U)}(\rho),
\qquad
0
\le
\epsilon_U
\le
\epsilon_*
\le
\epsilon_L
\le
\frac12.
\end{equation*}

\indent
The left endpoint comes from final-state auditing and the right endpoint from a feasible implementation. We first define the path compilation cost and then incorporate device calibration and statistical errors into a joint interval.

\indent
Fix a reference-consistent quantum circuit model $\mathfrak M_R = (\mathcal H,\mathcal G,\mathcal F_R,R,\mathcal D_R,\Gamma_R),$ and let $\widehat{\mathfrak M}_R=(\mathfrak M_R,U_R)$ be the RCC-admissible quantum circuit--description model used in Appendices~\ref{app:A} and \ref{app:D}. The lower certificate is constructed relative to the full model $\widehat{\mathfrak M}_R$, whereas the dynamical compiler outputs admissible circuits in its physical circuit component $\mathfrak M_R$.

\indent
Write ${\rm Out}_R(\mathcal C)$ and $L_R(\mathcal C)$ for the normalized unconditional output state and the cost, measured in atomic resource slots, of an admissible circuit $\mathcal C$, respectively. Then
\begin{equation}
\label{eq:I.1}
C_{\rm opt}^{(\epsilon)}(\rho;\mathfrak M_R)
=
\inf\!\left\{
L_R(\mathcal C):
\mathcal C\ \text{is admissible in}\ \mathfrak M_R,\
\mathcal D_R\!\left(
{\rm Out}_R(\mathcal C),\rho
\right)
\le
\epsilon
\right\}.
\end{equation}

\indent
Whenever the triangle inequality is needed below, we take $\mathcal D_R=T$ to be trace distance. The conclusions extend in the same way to a general error specification with a corresponding error composition rule.

\Istatement{Definition I.1 (Model-consistent path compilation cost).}
A dynamical implementation record $\mathfrak r_\gamma$ specifies a state-space curve $\gamma$ and its complete generation protocol. Process data include initial resources, the Hamiltonian or control sequence, locality and addressing, auxiliary systems, measurements, and feedforward branches; certification data include calibration, precision, success conditions, and conventions for handling the final workspace. The model uses these data to determine the normalized unconditional output and to charge the predeclared finite worst-case cost of the complete protocol in atomic resource slots.

\indent
Fix a model-consistent compiler ${\rm Comp}_R$, and collect compilation parameters for discretization, Hamiltonian simulation, gate synthesis, and truncation of finite failure tails in $\boldsymbol\eta$. The compiler either converts $\mathfrak r_\gamma$ into a finite admissible circuit in $\mathfrak M_R$,
\begin{equation*}
\mathcal C_{\gamma,R}(\boldsymbol\eta)
=
{\rm Comp}_R[
\mathfrak r_\gamma;
\boldsymbol\eta
],
\end{equation*}
or reports compilation failure.

\indent
To keep the different error sources traceable, suppose there is a finite certification chain $\omega_0,\omega_1,\ldots,\omega_m$. Each $\omega_j$ is a normalized unconditional state on the same target Hilbert space after the output map prescribed by the corresponding model, including any permitted discarding of workspace, and
\begin{equation*}
\omega_0
=
{\rm Out}_R(
\mathcal C_{\gamma,R}(\boldsymbol\eta)
),
\qquad
\omega_m
=
\rho,
\end{equation*}

\indent
with
\begin{equation}
\label{eq:I.2}
T(\omega_{j-1},\omega_j)
\le
\eta_j
\qquad
(j=1,\ldots,m).
\end{equation}

\indent
These radii can accommodate compilation or synthesis error, dynamical-record and calibration error, finite failure tails, and mismatch between the implemented final state and the target state. A common three-layer notation is $\epsilon_{\rm cmp}+\epsilon_{\rm path}+\epsilon_{\rm state}$; intermediate layers that are unnecessary in a specific implementation can be merged. Given a total tolerance $\epsilon$, denote the path compilation cost of this record relative to the fixed compiler and compilation parameters by $C_{\mathfrak r_\gamma,R}^{(\epsilon)}(\rho)$. When the path label is unambiguous, abbreviate it as $C_{\gamma,R}^{(\epsilon)}(\rho)$:
\begin{equation}
\label{eq:I.3}
C_{\mathfrak r_\gamma,R}^{(\epsilon)}(\rho)
\equiv
C_{\gamma,R}^{(\epsilon)}(\rho)
=
\begin{cases}
L_R(
\mathcal C_{\gamma,R}(\boldsymbol\eta)
),
&
\mathcal C_{\gamma,R}(\boldsymbol\eta)\ \text{is admissible and}\
\displaystyle
\sum_{j=1}^{m}
\eta_j
\le
\epsilon,
\\[1mm]
+\infty,
&
\text{otherwise}.
\end{cases}
\end{equation}

\indent
Equation~\eqref{eq:I.3} abbreviates as a path cost the cost jointly determined by the complete implementation record, target model, compilation rules, and precision allocation.

\Istatement{Proposition I.1 (Path upper-bound witness).}
For any dynamical implementation record satisfying Definition~I.1,
\begin{equation}
\label{eq:I.4}
C_{\rm opt}^{(\epsilon)}(\rho;\mathfrak M_R)
\le
C_{\gamma,R}^{(\epsilon)}(\rho).
\end{equation}

\Iproof
The claim is trivial when $C_{\gamma,R}^{(\epsilon)}(\rho)=+\infty$. Otherwise, the triangle inequality for trace distance and Eq.~\eqref{eq:I.2} give
\begin{equation*}
T\!\left(
{\rm Out}_R(
\mathcal C_{\gamma,R}(\boldsymbol\eta)
),
\rho
\right)
\le
\sum_{j=1}^{m}
\eta_j
\le
\epsilon.
\end{equation*}

\indent
Thus $\mathcal C_{\gamma,R}(\boldsymbol\eta)$ belongs to the feasible set in Eq.~\eqref{eq:I.1}. Taking the infimum of the lengths over this set yields Eq.~\eqref{eq:I.4}.\Iqed

\indent
The certified compiled circuit is therefore a feasible upper-bound witness for the same optimization problem. If the path is implemented in another dynamical model $\mathfrak M_{\rm dyn}$, first compile the continuous control record into a complete unconditional protocol in that model. When this model is a reference-consistent discrete quantum circuit model as specified in Appendix~\ref{app:A.10}, then apply resource compilation across models. Suppose the normalized unconditional outputs of both models have been identified with the same target domain, and $\mathcal C_\gamma$ is a finite admissible protocol in $\mathfrak M_{\rm dyn}$ with discrete length $L_{\rm dyn}(\mathcal C_\gamma)$ satisfying
\begin{equation*}
T\!\left(
{\rm Out}_{\rm dyn}(\mathcal C_\gamma),
\rho
\right)
\le
\epsilon_{\rm dyn}.
\end{equation*}

\indent
If resource compilation satisfies the output-error and length bounds of Appendix~\ref{app:A.10},
\begin{equation*}
\begin{aligned}
T\!\left(
{\rm Out}_R\!\left(
{\rm Comp}_{{\rm dyn}\to R}(\mathcal C_\gamma)
\right),
{\rm Out}_{\rm dyn}(\mathcal C_\gamma)
\right)
&\le
\eta,
\\
L_R\!\left(
{\rm Comp}_{{\rm dyn}\to R}(\mathcal C_\gamma)
\right)
&\le
s_{{\rm dyn}\to R}
L_{\rm dyn}(\mathcal C_\gamma)
+
b_{{\rm dyn}\to R},
\end{aligned}
\end{equation*}
and $\epsilon_{\rm dyn}+\eta\le1$, the same feasible-witness argument gives
\begin{equation}
\label{eq:I.5}
C_{\rm opt}^{(\epsilon_{\rm dyn}+\eta)}
(\rho;\mathfrak M_R)
\le
s_{{\rm dyn}\to R}
L_{\rm dyn}(\mathcal C_\gamma)
+
b_{{\rm dyn}\to R}.
\end{equation}

\indent
The compilation constants in Eq.~\eqref{eq:I.5} must retain their size dependence. When they are uniformly controlled across the model family under consideration, the path implementation supplies a corresponding family-level complexity upper bound. Combining Eq.~\eqref{eq:I.5} with an RCC lower bound at tolerance $\epsilon_L$ further requires $\epsilon_{\rm dyn}+\eta\le\epsilon_L\le1/2$.

\Ilead{Device calibration, path cost, and joint intervals.}
 The dynamical upper readout comes from model-consistent compilation. A concrete construction can directly output $\mathcal C_{\gamma,R}(\boldsymbol\eta)$ and count $L_R(\mathcal C_{\gamma,R}(\boldsymbol\eta))$, or establish both precision and length guarantees for the same output circuit:
\begin{equation}
\label{eq:I.6}
\begin{aligned}
T\!\left(
{\rm Out}_R(
\mathcal C_{\gamma,R}(\boldsymbol\eta)
),
\omega_{\rm impl}
\right)
&\le
\eta_{\rm cmp},
\\
L_R(
\mathcal C_{\gamma,R}(\boldsymbol\eta)
)
&\le
\mathcal K_R(
\mathfrak r_\gamma,
\boldsymbol\eta
).
\end{aligned}
\end{equation}

\indent
Here $\omega_{\rm impl}$ is the implemented final state in the certification chain of Eq.~\eqref{eq:I.2}. Device calibration and control records then provide a conservative upper endpoint for $\mathcal K_R$. The function $\mathcal K_R$ must inherit all conditions of the Hamiltonian simulation or gate synthesis theorem used, including interaction locality, the control dictionary and addressing, control amplitudes and time dependence, the requisite norm and commutator bounds, synthesis precision, auxiliary states and reset, complete adaptive branches, and output cleanup.

\indent
Equation~\eqref{eq:I.1} measures the physical execution cost $L_R$ of the compiled atomic sequence; the source-code length and classical runtime of the offline compiler lie outside this cost. Compressed programs, macros, or target-specific data invoked at runtime are implemented through admissible interfaces of $\mathfrak M_R$ and charged to the same cost function. The quantity $\Gamma_R$ calibrates the action-choice capacity per atomic resource slot, and $L_R\log_2\Gamma_R$ converts the length into action-choice capacity. This follows the full-trajectory resource calibration in Appendix~\ref{app:F.6}.

Let dynamical data $\mathsf D_{\rm dyn}$ generate a joint confidence set $\mathfrak R_{\rm adm}(\mathsf D_{\rm dyn})$ for implementation records and device parameters, and define the coverage event
\begin{equation*}
\mathcal E_{\rm dyn}
:=
\left\{
\mathfrak r_\gamma^{\rm true}
\in
\mathfrak R_{\rm adm}(\mathsf D_{\rm dyn})
\right\},
\qquad
{\rm Pr}(\mathcal E_{\rm dyn})
\ge
1-\delta_{\rm dyn}.
\end{equation*}

\indent
For every record in the set, check model compatibility, the complete unconditional output, error composition, and size-dependent cost according to Definition I.1, and compute $C_{\mathfrak r,R}^{(\epsilon_U)}(\rho)$. If any certification step fails, this cost is $+\infty$ by Eq.~\eqref{eq:I.3}. Define
\begin{equation*}
\overline U_{\rm dyn}^{(\epsilon_U)}
=
\sup_{\mathfrak r\in
\mathfrak R_{\rm adm}(\mathsf D_{\rm dyn})}
C_{\mathfrak r,R}^{(\epsilon_U)}(\rho).
\end{equation*}

\indent
For an empty joint feasible set, set $\overline U_{\rm dyn}^{(\epsilon_U)}=+\infty$. The supremum is also infinite if any admissible record has infinite certified cost. On the coverage event $\mathcal E_{\rm dyn}$, Proposition I.1 gives
\begin{equation}
\label{eq:I.7}
C_{\rm opt}^{(\epsilon_U)}
(\rho;\mathfrak M_R)
\le
C_{\gamma,R}^{(\epsilon_U)}(\rho)
\le
\overline U_{\rm dyn}^{(\epsilon_U)}.
\end{equation}

\indent
Thus $\overline U_{\rm dyn}^{(\epsilon_U)}$ is a dynamical upper endpoint that can be combined with the RCC lower bound. If the compiled circuit, complete output, and length are all fixed, the joint feasible set can reduce to a singleton, and the upper endpoint is the circuit's $L_R$ length. The case $\delta_{\rm dyn}=0$ corresponds to deterministic certification of the entire path, its error, and its cost; when measurement or calibration uncertainty is present, use the joint coverage probability above.

\indent
For a postselection protocol repeated at most $N$ times, the full cost of the $N$ attempts enters $L_R$, and the failure-tail probability enters the total error in Eq.~\eqref{eq:I.2}. Specifically, if the normalized unconditional output is
\begin{equation*}
\omega_{\rm uncond}
=
(1-p_{\rm tail})
\omega_{\rm succ}
+
p_{\rm tail}
\omega_{\rm fail},
\end{equation*}
then
\begin{equation*}
T(\omega_{\rm uncond},\omega_{\rm succ})
=
p_{\rm tail}
T(\omega_{\rm fail},\omega_{\rm succ})
\le
p_{\rm tail}.
\end{equation*}

\indent
Hence $p_{\rm tail}$ can serve as a conservative trace-distance radius in the certification chain. A bounded random stopping time is charged by the predeclared finite worst-case length; an unbounded repeat-until-success protocol has cost $+\infty$ under this definition. Expected cost defines a different complexity object.

\Istatement{Corollary I.1 (Two-sided complexity interval with joint confidence).}
Let $\underline L_{\rm RCC}^{(\epsilon_L)}:=L_{\rm RCC}^{(\epsilon_L)}$ be the final combined certificate of Proposition 4.1 and Appendix~\ref{app:C.6}, with its joint statistical event fixed by Appendix~\ref{app:D.6}. Suppose $\mathcal E_{\rm stat}$ and $\mathcal E_{\rm dyn}$ are defined on the same joint probability space, and the RCC audit gives, on $\mathcal E_{\rm stat}$,
\begin{equation*}
\underline L_{\rm RCC}^{(\epsilon_L)}
\le
C_{\rm opt}^{(\epsilon_L)}
(\rho;\mathfrak M_R),
\qquad
{\rm Pr}(\mathcal E_{\rm stat})
\ge
1-\delta_{\rm stat},
\end{equation*}

\indent
while Eq.~\eqref{eq:I.7} holds on $\mathcal E_{\rm dyn}$. If
\begin{equation*}
0
\le
\epsilon_U
\le
\epsilon_*
\le
\epsilon_L
\le
\frac12,
\qquad
\delta_{\rm stat}
+
\delta_{\rm dyn}
<
1,
\end{equation*}
and both sides correspond to the same true target state $\rho$, target domain, reference and support treatment, RCC-admissible reference-consistent model, normalized unconditional output semantics, error metric, and cost function in atomic resource slots, then
\begin{equation}
\label{eq:I.8}
{\rm Pr}\!\left[
\underline L_{\rm RCC}^{(\epsilon_L)}
\le
C_{\rm opt}^{(\epsilon_*)}
(\rho;\mathfrak M_R)
\le
\overline U_{\rm dyn}^{(\epsilon_U)}
\right]
\ge
1-\delta_{\rm stat}-\delta_{\rm dyn}.
\end{equation}

\indent
In particular, for $0\le\epsilon\le1/2$, setting $\epsilon_U=\epsilon_*=\epsilon_L=\epsilon$ gives
\begin{equation}
\label{eq:I.9}
{\rm Pr}\!\left[
\underline L_{\rm RCC}^{(\epsilon)}
\le
C_{\rm opt}^{(\epsilon)}
(\rho;\mathfrak M_R)
\le
\overline U_{\rm dyn}^{(\epsilon)}
\right]
\ge
1-\delta_{\rm stat}-\delta_{\rm dyn}.
\end{equation}

\Iproof
Complexity is nonincreasing in the allowed error. Thus $\epsilon_U\le\epsilon_*\le\epsilon_L$ implies
\begin{equation*}
C_{\rm opt}^{(\epsilon_L)}
\le
C_{\rm opt}^{(\epsilon_*)}
\le
C_{\rm opt}^{(\epsilon_U)}.
\end{equation*}

\indent
On the intersection $\mathcal E_{\rm stat}\cap\mathcal E_{\rm dyn}$, combining this monotone chain with the lower and upper endpoints yields Eq.~\eqref{eq:I.8}. Finally, the union bound gives
\begin{equation*}
{\rm Pr}
(\mathcal E_{\rm stat}\cap\mathcal E_{\rm dyn})
\ge
1-\delta_{\rm stat}-\delta_{\rm dyn}.
\end{equation*}

\indent
The argument does not require statistical independence of the two events.\Iqed

\Ilead{Extension to different target states.}
 In experiments, the lower and upper endpoints are sometimes constructed around $\rho_{\rm stat}$ and $\rho_{\rm dyn}$, respectively. If there is a common operational target $\rho$ with certified bounds
\begin{equation*}
T(\rho_{\rm dyn},\rho)
\le
\zeta_U,
\qquad
T(\rho_{\rm stat},\rho)
\le
\zeta_L,
\end{equation*}
then the tolerance conditions
\begin{equation*}
\epsilon_U+\zeta_U
\le
\epsilon_*,
\qquad
\epsilon_*+\zeta_L\le\epsilon_L,
\end{equation*}

\indent
give the following interval for different target states:
\begin{equation*}
{\rm Pr}\!\left[
\underline L_{\rm RCC}^{(\epsilon_L)}
(\rho_{\rm stat})
\le
C_{\rm opt}^{(\epsilon_*)}
(\rho;\mathfrak M_R)
\le
\overline U_{\rm dyn}^{(\epsilon_U)}
(\rho_{\rm dyn})
\right]
\ge
1-\delta_{\rm stat}-\delta_{\rm dyn}.
\end{equation*}

\indent
The first condition makes the dynamical circuit a feasible upper-bound witness for the common target; the second makes every $\epsilon_*$-implementation of the common target an $\epsilon_L$-implementation of $\rho_{\rm stat}$. If the target-state discrepancies are themselves estimated from data, their coverage events and failure probabilities must also enter the joint probability space and union bound above.

Equation~\eqref{eq:I.8} gives an operational two-sided complexity interval. Controlling the endpoint difference gives additive bounds; when the lower endpoint is positive, controlling the endpoint ratio gives multiplicative bounds, and a ratio tending to one establishes asymptotic saturation. For state families with near-tight constructions in Appendix~\ref{app:G}, the dynamical upper bound can check the construction scale and compilation constants; elsewhere, it supplies a new finite upper endpoint. Once the object, model, and tolerances are aligned, Eq.~\eqref{eq:I.8} also implies
\begin{equation*}
{\rm Pr}\!\left[
\underline L_{\rm RCC}^{(\epsilon_L)}
>
\overline U_{\rm dyn}^{(\epsilon_U)}
\right]
\le
\delta_{\rm stat}+\delta_{\rm dyn}.
\end{equation*}

\indent
An empty reported interval should therefore be retained as a diagnostic signal: at the stated confidence level, it indicates failure of a confidence event or incomplete alignment of the target state, error, reference and support treatment, resource compilation, or counting units.

\Ilead{Geometric compatibility checks for continuous dynamics.}
 Implementations such as analog quantum computation and adiabatic evolution typically supply a time-dependent Hamiltonian, control waveforms, and calibration errors, while state estimates along the path provide the corresponding energy fluctuations. Model-consistent compilation converts these records into path costs. We now use the Hall decomposition and exact quantum speed limits to relate probability motion, relative phases, and energy fluctuations and check the geometric compatibility of the same record. For the decomposition and exact QSLs, see Hall~\cite{Hall2001_ExactUncertaintyRelations} and Pati, Mohan, Sahil, and Braunstein~\cite{PatiEtAl2025_ExactQuantumSpeedLimits}.

\indent
We restrict the following discussion to finite-dimensional pure-state unitary paths. Let the total evolution time be $t_{\rm f}>0$, and let $t\mapsto|\psi_t\rangle$ be absolutely continuous on $[0,t_{\rm f}]$ and satisfy almost everywhere
\begin{equation*}
i\hbar
|\dot\psi_t\rangle
=
H_t
|\psi_t\rangle,
\end{equation*}
where $H_t$ is a Hermitian Hamiltonian. Fix a time-independent orthonormal basis $\mathcal B=\{|a_i\rangle\}_{i=1}^{d}$, and write
\begin{equation}
\label{eq:I.10}
c_i(t)
=
\langle a_i|\psi_t\rangle,
\qquad
p_i(t)
=
|c_i(t)|^2,
\qquad
z_i(t)
=
\langle a_i|H_t|\psi_t\rangle.
\end{equation}

\indent
On the support where $p_i(t)>0$, Hall's basis-dependent classical part is defined by
\begin{equation}
\label{eq:I.11}
h_{i,\mathcal B}^{\rm cl}(t)
=
\frac{
\langle a_i|
\{H_t,|\psi_t\rangle\langle\psi_t|\}
|a_i\rangle
}{
2p_i(t)
}
=
{\rm Re}\!\left(
\frac{z_i(t)}{c_i(t)}
\right),
\end{equation}
and
\begin{equation}
\label{eq:I.12}
H_{{\rm cl},\mathcal B}(t)
=
\sum_i
h_{i,\mathcal B}^{\rm cl}(t)
|a_i\rangle\langle a_i|,
\qquad
H_{{\rm nc},\mathcal B}(t)
=
H_t
-
H_{{\rm cl},\mathcal B}(t).
\end{equation}

\indent
On zero-probability components, $h_{i,\mathcal B}^{\rm cl}$ may be any finite real number, leaving $H_{{\rm cl},\mathcal B}|\psi_t\rangle$ unchanged. Expressions containing $\dot p_i^2/p_i$ are interpreted through the absolutely continuous extension of $q_i=\sqrt{p_i}=|c_i|$.

\Istatement{Proposition I.2 (Exact QSL path identity).}
For the pure-state unitary path above, $H_{{\rm nc},\mathcal B}$ has zero expectation in $|\psi_t\rangle$, and almost everywhere,
\begin{equation}
\label{eq:I.13}
\bigl(
\Delta H_{{\rm nc},\mathcal B}(t)
\bigr)^2
=
\frac{\hbar^2}{4}
\sum_{i:p_i(t)>0}
\frac{\dot p_i(t)^2}{p_i(t)}.
\end{equation}

\indent
The Wootters path length of the probability curve $p(t)=(p_1(t),\ldots,p_d(t))$ therefore satisfies the exact identity
\begin{equation}
\label{eq:I.14}
\begin{aligned}
\ell_{W,\mathcal B}[\gamma]
&=
\int_0^{t_{\rm f}}
\sqrt{
\sum_i
\dot q_i(t)^2
}\,dt
=
\frac12
\int_0^{t_{\rm f}}
\sqrt{
\sum_{i:p_i(t)>0}
\frac{\dot p_i(t)^2}{p_i(t)}
}\,dt
\\
&=
\frac1\hbar
\int_0^{t_{\rm f}}
\Delta H_{{\rm nc},\mathcal B}(t)\,dt.
\end{aligned}
\end{equation}

\Iproof
The Schrödinger equation gives
\begin{equation*}
\dot p_i(t)
=
\frac{2}{\hbar}
{\rm Im}\!\left[
z_i(t)c_i(t)^*
\right].
\end{equation*}

\indent
On the other hand, for $p_i(t)>0$,
\begin{equation*}
\langle a_i|
H_{{\rm nc},\mathcal B}(t)
|\psi_t\rangle
=
z_i(t)
-
h_{i,\mathcal B}^{\rm cl}(t)
c_i(t),
\end{equation*}
and hence
\begin{equation*}
\left|
\langle a_i|
H_{{\rm nc},\mathcal B}(t)
|\psi_t\rangle
\right|^2
=
\frac{
\bigl[
{\rm Im}(z_i(t)c_i(t)^*)
\bigr]^2
}{
p_i(t)
}
=
\frac{\hbar^2}{4}
\frac{\dot p_i(t)^2}{p_i(t)}.
\end{equation*}
Moreover,
\begin{equation*}
\langle
H_{{\rm cl},\mathcal B}(t)
\rangle_{\psi_t}
=
\sum_i
p_i(t)
h_{i,\mathcal B}^{\rm cl}(t)
=
\langle H_t\rangle_{\psi_t},
\end{equation*}
so
\begin{equation*}
\langle
H_{{\rm nc},\mathcal B}(t)
\rangle_{\psi_t}
=
0.
\end{equation*}

\indent
On the zero set $Z_i=\{t:c_i(t)=0\}$, absolute continuity ensures that $\dot c_i=0$ almost everywhere. Since $z_i=i\hbar\dot c_i$, zero-probability components therefore contribute no omitted norm terms. Summing the squared moduli over the positive support gives Eq.~\eqref{eq:I.13}. Furthermore, $q_i=|c_i|$ is absolutely continuous, and $\dot q_i^2=\dot p_i^2/(4p_i)$ holds where $p_i>0$, with extension to the boundary of the probability simplex through the metric derivative. Possible nondifferentiability at isolated zeros does not affect the integral identity in Eq.~\eqref{eq:I.14}.\Iqed

\indent
With the Fubini--Study convention
\begin{equation*}
ds_{\rm FS}^2
=
\langle d\psi|d\psi\rangle
-
|\langle\psi|d\psi\rangle|^2
\end{equation*}

\indent
define the instantaneous Fubini--Study speed by
\begin{equation*}
v_{\rm FS}(t)
:=
\sqrt{
\langle\dot\psi_t|\dot\psi_t\rangle
-
|\langle\psi_t|\dot\psi_t\rangle|^2
}
=
\frac{\Delta H_t}{\hbar}.
\end{equation*}

\indent
The corresponding standard geometric length is
\begin{equation}
\label{eq:I.15}
\ell_{\rm FS}[\gamma]
=
\int_0^{t_{\rm f}}
v_{\rm FS}(t)\,dt
=
\frac1\hbar
\int_0^{t_{\rm f}}
\Delta H_t\,dt.
\end{equation}

\indent
Writing $c_i=\sqrt{p_i}e^{i\phi_i}$ on the instantaneous occupied support $p_i(t)>0$, we have almost everywhere
\begin{equation}
\label{eq:I.16}
\begin{aligned}
v_{\rm FS}(t)^2
&=
\sum_i
\dot q_i(t)^2
+
\sum_{i:p_i(t)>0}
p_i(t)
\left[
\dot\phi_i(t)
-
\sum_{j:p_j(t)>0}
p_j(t)\dot\phi_j(t)
\right]^2
\\
&=
\frac14
\sum_{i:p_i(t)>0}
\frac{\dot p_i(t)^2}{p_i(t)}
+
\sum_{i:p_i(t)>0}
p_i(t)
\left[
\dot\phi_i(t)
-
\sum_{j:p_j(t)>0}
p_j(t)\dot\phi_j(t)
\right]^2.
\end{aligned}
\end{equation}

\indent
Equation~\eqref{eq:I.16} separates two types of motion in projective state space: the first term records redistribution of occupation probabilities in the fixed basis, and the second records relative-phase motion among occupied components. Nonclassical energy fluctuations read only the former; total energy fluctuations read the Fubini--Study speed formed by both contributions.

\indent
From $i\hbar\dot c_i=z_i$ and $c_i=\sqrt{p_i}e^{i\phi_i}$, we have almost everywhere on the occupied support
\begin{equation*}
h_{i,\mathcal B}^{\rm cl}
=
{\rm Re}\!\left(
\frac{z_i}{c_i}
\right)
=
-\hbar\dot\phi_i.
\end{equation*}

\indent
Thus the second term in Eq.~\eqref{eq:I.16} is precisely $(\Delta H_{{\rm cl},\mathcal B}/\hbar)^2$, and
\begin{equation}
\label{eq:I.17}
(\Delta H_t)^2
=
\bigl(
\Delta H_{{\rm nc},\mathcal B}(t)
\bigr)^2
+
\bigl(
\Delta H_{{\rm cl},\mathcal B}(t)
\bigr)^2,
\qquad
\ell_{W,\mathcal B}[\gamma]
\le
\ell_{\rm FS}[\gamma].
\end{equation}

\indent
The length inequality in Eq.~\eqref{eq:I.17} is an equality if and only if $\Delta H_{{\rm cl},\mathcal B}(t)=0$ almost everywhere. Equivalently, all components in the occupied support have a common phase velocity, and relative phases are constant on each connected occupied interval. This condition constrains fluctuations of the classical part in the current state; the Fubini--Study length of a general path also includes relative-phase motion.
The Wootters geodesic distance between the probability endpoints is
\begin{equation*}
D_{W,\mathcal B}
\bigl(
p(0),p(t_{\rm f})
\bigr)
=
\arccos\!\left(
\sum_i
\sqrt{
p_i(0)p_i(t_{\rm f})
}
\right)
\le
\ell_{W,\mathcal B}[\gamma].
\end{equation*}

\indent
Equality requires the square-root probability vector $q(t)=(\sqrt{p_1(t)},\ldots,\sqrt{p_d(t)})$ to follow a shortest spherical geodesic between the endpoints, up to monotone reparametrization. On the other hand, the triangle inequality for complex amplitudes gives
\begin{equation*}
D_{W,\mathcal B}
\bigl(
p(0),p(t_{\rm f})
\bigr)
\le
\theta_{0{\rm f}}
:=
\arccos
|\langle\psi_0|\psi_{t_{\rm f}}\rangle|.
\end{equation*}

\indent
If the chosen fixed basis contains the initial state, $|a_1\rangle=|\psi_0\rangle$, then $D_{W,\mathcal B}(p(0),p(t_{\rm f}))=\theta_{0{\rm f}}$, and hence
\begin{equation}
\label{eq:I.18}
\theta_{0{\rm f}}
\le
\ell_{W,\mathcal B}[\gamma]
\le
\ell_{\rm FS}[\gamma].
\end{equation}

\indent
Define the time-averaged fluctuations
\begin{equation*}
\overline{\Delta H_{{\rm nc},\mathcal B}}
=
\frac1{t_{\rm f}}
\int_0^{t_{\rm f}}
\Delta H_{{\rm nc},\mathcal B}(t)\,dt,
\qquad
\overline{\Delta H}
=
\frac1{t_{\rm f}}
\int_0^{t_{\rm f}}
\Delta H_t\,dt.
\end{equation*}

\indent
If $\overline{\Delta H_{{\rm nc},\mathcal B}}>0$ and the fixed basis contains the initial state, Eq.~\eqref{eq:I.14} can be rearranged as
\begin{equation}
\label{eq:I.19}
t_{\rm f}
=
\frac{
\hbar\,\ell_{W,\mathcal B}[\gamma]
}{
\overline{\Delta H_{{\rm nc},\mathcal B}}
}
\ge
\frac{
\hbar\,\theta_{0{\rm f}}
}{
\overline{\Delta H_{{\rm nc},\mathcal B}}
}
\ge
\frac{
\hbar\,\theta_{0{\rm f}}
}{
\overline{\Delta H}
}.
\end{equation}

\indent
The equality is the exact path identity; the two inequalities follow from the endpoint geodesic distance and $\Delta H_{{\rm nc},\mathcal B}\le\Delta H$, respectively. When $\theta_{0{\rm f}}=0$, the endpoint lower bound vanishes, although a nontrivial closed loop may still have positive path length. When $\overline{\Delta H_{{\rm nc},\mathcal B}}=0$, Eq.~\eqref{eq:I.14} gives $\ell_{W,\mathcal B}=0$; retain the original identity in this case without dividing by zero.

Equations~\eqref{eq:I.14}--\eqref{eq:I.19} convert the same control record into dimensionless geometric path readouts. Calibrated $\Delta H_t$ gives $\ell_{\rm FS}$, while the fixed-basis probability trajectory $p_i(t)$ or $\Delta H_{{\rm nc},\mathcal B}(t)$ gives $\ell_{W,\mathcal B}$. Their exact relation can check compatibility among the controls, final-state overlap, and energy fluctuations. Quantitative conversion between geometric length and gate cost requires a separate compilation theorem within the model: the same endpoint angle can correspond to different local circuit complexities, and the same projective-state path can be implemented by Hamiltonians with different control overheads.

\indent
Proposition I.1 and Corollary I.1 apply to finite process records that can be compiled into admissible circuits, including channels, measurements, and adaptive control allowed by the model; adaptive control is charged by the complete control policy or finite branching tree. Proposition I.2 establishes exact geometric identities for finite-dimensional pure-state unitary paths. Final-state RCC, path compilation, and exact QSLs thus provide, respectively, a lower certificate, an upper witness, and a geometric check of the same path.

\subsection{Scope, Notation, and Units}
\label{app:I.2}

\indent
To determine the thermodynamic consequences of a complexity budget, window entropy and the complexity readout must be defined on the same effective state. CWT starts from the bounded dynamical complexity postulate~\cite{LiuTianWu2025_CWT}: with the reference state, elementary operations, precision specification, and finite physical resources fixed, realizable target states or transformations satisfy
\begin{equation*}
C^*(\rho_{\rm tar})
\le
\xi,
\qquad
C^*(U)
\le
\xi.
\end{equation*}

\indent
The budget $\xi$ collects resource restrictions such as evolution time, action, control precision, energy supply, and system size; the task model uses these restrictions to determine the controllable and distinguishable object domain. Within this domain, windowed RCC reads the leading complexity term of the effective state relative to the structured vacuum.
The original scalar budget of CWT is denoted by $\Xi$, while Appendix~\ref{app:E} already uses $\Xi=(\mathcal A_\Xi,E_\Xi)$ for a complete window. Here we denote the scalar budget by $\xi$ and the induced window by $\Xi_\xi$, retaining CWT's budget direction and first-law sign convention.

CWT leaves freedom in the implementation of a particular budget window. Its original construction organizes windows through basic observables, transformations realizable within the budget, and their von Neumann algebra. Here we choose shared windows $(\mathcal A_\xi,E_\xi)$ based on conditional expectations as a sharp, idempotent, and directly auditable finite-dimensional realization. Physical protocols first determine what can be generated, controlled, or distinguished under the complexity budget; the algebra and conditional expectation then represent this effective structure.

\indent
CWT takes a well-defined optimal quantum circuit complexity $C^*$ as input. RCC fixes the corresponding model for reference-consistent tasks and supplies per-circuit and optimal-complexity lower bounds from the final state. We allow a general family $\rho(E,V,N,\ldots)$ parametrized by macroscopic variables. Taking the original CWT microcanonical state $\rho_E=\Pi_E/\omega(E)$ and choosing $(\mathcal A_\xi,E_\xi)$ as the restriction or isometric identification of $(\mathcal A^{\rm CWT}(\xi),\mathcal E_\xi^{\rm CWT})$ within the current reference support recovers its microcanonical state-side construction. This appendix derives capacity and response relations for the interface; CWT's dynamical bounds and action relations retain their original physical assumptions.

\indent
Throughout Appendices~\ref{app:I.2}--\ref{app:I.5}, we fix one finite-dimensional RCC-admissible reference-consistent quantum circuit--description model and its unbiased reference data,
\begin{equation}
\label{eq:I.20}
\widehat{\mathfrak M}_R
=
(\mathfrak M_R,U_R),
\qquad
\mathcal H_R
=
\Pi_R\mathcal H,
\qquad
d_R
=
{\rm Tr}\,\Pi_R,
\qquad
\sigma_R
=
\frac{\Pi_R}{d_R},
\qquad
\Gamma_R
>
1.
\end{equation}

\indent
All normalized states under consideration are supported in $\mathcal H_R$; support leakage is handled according to Appendix~\ref{app:B}. Weighted references, window-dependent reference spaces, and varying atomic control bandwidths are treated separately in Appendix~\ref{app:I.6}.

Let $\rho(E,V,N,\ldots)$ denote the normalized input state selected by CWT under the given macroscopic constraints. For each budget value $\xi\in J\subseteq\mathbb R$, specify a reference-compatible window
\begin{equation}
\label{eq:I.21}
\xi
\longmapsto
\Xi_\xi
=
(\mathcal A_\xi,E_\xi),
\qquad
\tau_\xi(E,V,N,\ldots)
:=
E_\xi\!\left[
\rho(E,V,N,\ldots)
\right].
\end{equation}

\indent
Here $E$ is the thermodynamic energy variable, while $E_\xi$ is the trace-preserving conditional expectation onto $\mathcal A_\xi$, with its action written as $E_\xi[\,\cdot\,]$. Each window satisfies the CPTP, idempotence, and reference-preservation conditions of Appendix~\ref{app:E}; nesting and composition compatibility across budgets are specified in Appendix~\ref{app:I.3}.
The macroscopic variables $E,V,N,\ldots$ label the input-state family. When the corresponding Hamiltonian, particle-number operator, and other macroscopic observables belong to every $\mathcal A_\xi$, or when the window maps separately guarantee exact preservation of their expectation values, these labels also give the same macroscopic expectation values for the effective state $\tau_\xi$.

Below, we suppress the macroscopic arguments of $\tau_\xi$, $s_\xi$, and $C_\xi$ when no ambiguity arises.
Equation~\eqref{eq:I.21} distinguishes resource supply, the object domain, and the effective state. A concrete protocol specifies $\xi\mapsto\Xi_\xi$, the reference $R$ fixes the zero point for state comparison, and $\widehat{\mathfrak M}_R$ fixes the error semantics and atomic scale. The budget $\xi$ is an admission ceiling; windowed RCC is the readout of the target state's leading complexity term in that object domain.

\indent
To put the budget and RCC readout on a common scale, the atomic resource slot scale can be used directly if both employ the same cost function with the same serial and parallel counting rules. If the budget uses circuit depth, total gate count, or another unit, a monotone calibration $\widetilde\xi=\chi_R(\xi)$ or corresponding two-sided resource compilation bounds must be supplied. The derivative $\partial_\xi C_\xi$ depends on the declared budget coordinate. If $\chi_R$ is invertible and differentiable on a regular branch, the corresponding conjugate potential transforms by the chain rule as
\begin{equation*}
\Pi_{\widetilde\xi}
=
\Pi_\xi
\frac{d\xi}{d\widetilde\xi},
\qquad
\Pi_{\widetilde\xi}\,d\widetilde\xi
=
\Pi_\xi\,d\xi.
\end{equation*}

\indent
Thus the physical work one-form is invariant under smooth reparametrization of the budget coordinate. This derivative conversion requires invertible calibration; general resource compilation bounds instead compare budget intervals.
From this subsection onward, information quantities in the thermodynamic interface use natural logarithms by default. Define
\begin{equation}
\label{eq:I.22}
s_\xi
:=
-\operatorname{Tr}\!\left(
\tau_\xi\ln\tau_\xi
\right),
\qquad
\mathscr S_\xi
:=
k_B s_\xi,
\qquad
r_R
:=
\ln d_R,
\qquad
g_R^{({\rm nat})}
:=
\ln\Gamma_R
=
(\ln2)g_R.
\end{equation}

\indent
Here $s_\xi$ is the dimensionless von Neumann entropy reported in nats, and $\mathscr S_\xi=k_Bs_\xi$ is the CWT window entropy in thermodynamic entropy units. Its further use as a differentiable thermodynamic state function depends on the regularity and reversible-interface conditions of Appendix~\ref{app:I.5}. The quantity $r_R$ is the total information capacity of the fixed reference support, and $g_R^{({\rm nat})}$ expresses the global atomic control bandwidth $g_R=\log_2\Gamma_R$ in natural information units. Relative entropy is finite because ${\rm supp}(\tau_\xi)\subseteq\mathcal H_R={\rm supp}(\sigma_R)$. Using the standard convention $0\ln0=0$ for zero eigenvalues of $\tau_\xi$ and noting that $\ln\sigma_R=-r_R I_{\mathcal H_R}$ on the reference support, we obtain
\begin{equation}
\label{eq:I.23}
D_{\rm nat}(\tau_\xi\Vert\sigma_R)
:=
\operatorname{Tr}\!\left[
\tau_\xi
\bigl(
\ln\tau_\xi-\ln\sigma_R
\bigr)
\right]
=
-s_\xi+r_R.
\end{equation}

\indent
Accordingly, following the definition of windowed RCC in Appendix~\ref{app:E} and suppressing $R$ and the target-state argument at fixed reference, write
\begin{equation}
\label{eq:I.24}
C_\xi
:=
C_{R,\Xi_\xi}(\rho)
=
\frac{
D_{\rm nat}(\tau_\xi\Vert\sigma_R)
}{
g_R^{({\rm nat})}
}
=
\frac{
r_R-s_\xi
}{
g_R^{({\rm nat})}
}.
\end{equation}

\indent
Equation~\eqref{eq:I.24} uses the atomic control bandwidth to convert the information gap into a complexity readout in atomic resource slots for $\mathfrak M_R$, with the unit relation
\begin{equation}
\label{eq:I.25}
1\,{\rm st}_R
\equiv
g_R^{({\rm nat})}\ {\rm nats}
=
g_R\ {\rm bits}
=
\log_2\Gamma_R\ {\rm bits}.
\end{equation}

\indent
The quantity $C_\xi$ is the windowed linear leading term of complexity, measured in equivalent atomic resource slots, and need not be an integer; the actual circuit cost $L_R$ remains a discrete slot count. The per-path theorems in Sec.~\ref{sec:3} and Appendix~\ref{app:A} connect this leading term to optimal-complexity lower bounds, while the full certificate retains one-shot, coding, error, and statistical corrections. The quantity $\mathscr S_\xi$ is reported in thermodynamic entropy units.

\indent
This complexity readout is independent of the logarithm base. In binary information units, let
\begin{equation}
\label{eq:I.26}
s_\xi^{(2)}
=
\frac{s_\xi}{\ln2},
\qquad
r_R^{(2)}
=
\frac{r_R}{\ln2},
\qquad
g_R
=
\frac{
g_R^{({\rm nat})}
}{
\ln2
}
=
\log_2\Gamma_R.
\end{equation}
Then
\begin{equation}
\label{eq:I.27}
C_\xi
=
\frac{
r_R^{(2)}-s_\xi^{(2)}
}{
g_R
},
\qquad
\mathscr S_\xi
=
k_Bs_\xi
=
k_B\ln2\,s_\xi^{(2)}.
\end{equation}

\indent
Conversion between bits and nats therefore acts on both the information gap and atomic bandwidth, leaving the numerical value of $C_\xi$ unchanged; converting physical entropy from bits retains the factor $k_B\ln2$. The thermodynamic derivatives below use $\mathscr S_\xi$ as the entropy variable, with $k_B$ included explicitly when $s_\xi$ is used instead. Appendices~\ref{app:I.2}--\ref{app:I.5} hold $r_R,g_R^{({\rm nat})}$ fixed, concentrating the window dependence in $\tau_\xi,s_\xi,C_\xi$; reference or bandwidth changes are treated in Appendix~\ref{app:I.6}.

\subsection{Compatibility Conditions for Shared Windows}
\label{app:I.3}

\indent
We now organize pointwise windows into an ordered budget family. Increasing the budget refines the observable algebra, and the tower property of trace-preserving conditional expectations passes this ordering to effective states and their information readouts.

\indent
The following sufficient conditions provide a finite-dimensional realization of RCC--CWT shared windows. Controls, measurements, and composite protocols callable within the budget determine the window, with $\mathcal A_\xi$ representing its effective observable structure. Resource costs for products, long compositions, and high-precision functional calculus are assessed under their respective protocols.

\Istatement{Definition I.2 (RCC--CWT shared window family).}
Let $J\subseteq\mathbb R$ be a totally ordered budget set. We call
\begin{equation}
\label{eq:I.28}
\{
\Xi_\xi
=
(\mathcal A_\xi,E_\xi)
\}_{\xi\in J}
\end{equation}
an RCC--CWT shared window family at fixed reference $R$ if, for every $\xi\in J$, $\mathcal A_\xi\subseteq B(\mathcal H_R)$ is a finite-dimensional von Neumann subalgebra with identity $\Pi_R$, and
\begin{equation}
\label{eq:I.29}
\begin{gathered}
E_\xi:
B(\mathcal H_R)
\longrightarrow
\mathcal A_\xi,
\qquad
E_\xi^2
=
E_\xi,
\qquad
E_\xi\ \text{is CPTP},
\\
{\rm Ran}(E_\xi)
=
\mathcal A_\xi,
\qquad
E_\xi(AXB)
=
A E_\xi(X)B,
\qquad
E_\xi(\sigma_R)
=
\sigma_R
\end{gathered}
\end{equation}
for all $A,B\in\mathcal A_\xi$ and $X\in B(\mathcal H_R)$. Each $E_\xi$ is a trace-preserving conditional expectation with respect to the same trace. Since the unbiased structured vacuum satisfies $\sigma_R=\Pi_R/d_R$, reference preservation follows from $E_\xi(\Pi_R)=\Pi_R$. Equation~\eqref{eq:I.29} retains this condition explicitly to indicate that the subsequent data-processing comparisons always use the same reference state.

\indent
Increasing the budget is defined to refine the object domain: for any $\xi_1\le\xi_2$, require
\begin{equation}
\label{eq:I.30}
\mathcal A_{\xi_1}
\subseteq
\mathcal A_{\xi_2}.
\end{equation}

\indent
This direction agrees with CWT's existing definition of the complexity budget. A larger $\xi$ admits a richer set of observables into the effective object domain; a smaller $\xi$ corresponds to stronger coarse-graining. All algebras, maps, and effective states remain within the common reference support of Eq.~\eqref{eq:I.20}.

\Istatement{Proposition I.3 (Tower compatibility and window monotonicity).}
For any shared window family in Definition I.2 and $\xi_1\le\xi_2$,
\begin{equation}
\label{eq:I.31}
E_{\xi_1}\circ E_{\xi_2}
=
E_{\xi_1}
=
E_{\xi_2}\circ E_{\xi_1},
\qquad
\tau_{\xi_1}
=
E_{\xi_1}(\tau_{\xi_2}).
\end{equation}
Consequently,
\begin{equation}
\label{eq:I.32}
D_{\rm nat}
(\tau_{\xi_1}\Vert\sigma_R)
\le
D_{\rm nat}
(\tau_{\xi_2}\Vert\sigma_R),
\qquad
C_{\xi_1}
\le
C_{\xi_2},
\qquad
s_{\xi_1}
\ge
s_{\xi_2}.
\end{equation}

\Iproof
Since ${\rm Ran}(E_{\xi_1})=\mathcal A_{\xi_1}\subseteq\mathcal A_{\xi_2}$ and a conditional expectation is the identity on its range, we immediately have
\begin{equation*}
E_{\xi_2}\circ E_{\xi_1}
=
E_{\xi_1}.
\end{equation*}

\indent
On the other hand, for any $A\in\mathcal A_{\xi_1}$, trace preservation and the bimodule property of conditional expectations give
\begin{equation*}
\begin{aligned}
{\rm Tr}\!\left[
A E_{\xi_1}(E_{\xi_2}(X))
\right]
&=
{\rm Tr}\!\left[
A E_{\xi_2}(X)
\right]
\\
&=
{\rm Tr}(AX)
=
{\rm Tr}\!\left[
A E_{\xi_1}(X)
\right].
\end{aligned}
\end{equation*}
Let
\begin{equation*}
Y
=
E_{\xi_1}(E_{\xi_2}(X))
-
E_{\xi_1}(X)
\in
\mathcal A_{\xi_1}.
\end{equation*}

\indent
The preceding equality holds for all $A\in\mathcal A_{\xi_1}$. Taking $A=Y^\dagger$ gives ${\rm Tr}(Y^\dagger Y)=0$, hence $Y=0$. This proves tower compatibility, $E_{\xi_1}\circ E_{\xi_2}=E_{\xi_1}$. Applying this relation to $\rho$ yields the effective-state relation in Eq.~\eqref{eq:I.31}.
Finally, the data-processing inequality for quantum relative entropy and $E_{\xi_1}(\sigma_R)=\sigma_R$ give
\begin{equation*}
\begin{aligned}
D_{\rm nat}
(\tau_{\xi_2}\Vert\sigma_R)
&\ge
D_{\rm nat}\!\left(
E_{\xi_1}(\tau_{\xi_2})
\middle\Vert
E_{\xi_1}(\sigma_R)
\right)
\\
&=
D_{\rm nat}
(\tau_{\xi_1}\Vert\sigma_R).
\end{aligned}
\end{equation*}

\indent
Dividing by the fixed $g_R^{({\rm nat})}>0$ gives complexity monotonicity; Eq.~\eqref{eq:I.24}, with $r_R,g_R^{({\rm nat})}$ fixed, then gives entropy monotonicity.\Iqed
Equation~\eqref{eq:I.32} shows that the same target state reveals more structure in a finer object domain: windowed RCC is nondecreasing, and window entropy is nonincreasing. Algebra nesting and the tower property guarantee this ordering of the state-side readouts across windows. If $E_\xi$ is physically executed in an experiment, its controls, auxiliary resources, and channel implementation are still charged according to the cost function of $\mathfrak M_R$.

\Istatement{Lemma I.1 (Staircase structure of finite-dimensional exact windows).}
In the fixed finite-dimensional reference of Definition I.2, the totally ordered nested family $\{\mathcal A_\xi\}_{\xi\in J}$ contains at most $d_R^2$ distinct von Neumann subalgebras. For a fixed input state $\rho$, fixed reference data, and the same trace, $E_\xi,\tau_\xi,\mathscr S_\xi$ and $C_\xi$ therefore also take only finitely many distinct values. If $J$ is a real interval, they form a staircase with at most $d_R^2-1$ strict refinement thresholds.

\Iproof
Along a totally ordered inclusion chain, every strict inclusion increases the complex vector-space dimension of the subalgebra by at least one, while
\begin{equation*}
1
\le
\dim_{\mathbb C}\mathcal A_\xi
\le
\dim_{\mathbb C}B(\mathcal H_R)
=
d_R^2.
\end{equation*}

\indent
Hence there are at most $d_R^2$ distinct algebras. For a fixed trace, the trace-preserving conditional expectation onto a given subalgebra is the unique orthogonal projection in the Hilbert--Schmidt inner product. The remaining claims follow directly from $\tau_\xi=E_\xi(\rho)$ and the definitions.\Iqed
Each strict refinement threshold marks a new distinguishable structural sector, observation channel, or control protocol crossing the complexity admission threshold. Equation~\eqref{eq:I.32} and the finite capacity differences are therefore the basic responses of exact finite-dimensional windows. Quasistatic operation moves the system slowly across thresholds; at fixed input, the window family retains its staircase structure.

\indent
Finite experimental resolution can induce an effective protocol or ensemble interpolation between sharp windows. For example, if
\begin{equation*}
\widetilde E_\lambda
=
\sum_a
w_a(\lambda)
E_{\xi_a},
\qquad
w_a(\lambda)\ge0,
\qquad
\sum_a
w_a(\lambda)
=
1,
\end{equation*}
then $\widetilde E_\lambda$ remains a CPTP map preserving $\sigma_R$, but generally loses idempotence and the conditional-expectation structure. Such outputs still satisfy the pointwise capacity identity, but belong to an effective interpolation branch; tower compatibility, monotonicity, and differentiability must be established separately for the protocol. A controlled thermodynamic limit can also produce a smooth branch. If it changes $d_R$ or $\Gamma_R$ at the same time, use the variable-reference or variable-bandwidth formulas of Appendix~\ref{app:I.6}.

\subsection{Window Capacity Identity}
\label{app:I.4}

\indent
Shared windows define CWT entropy and windowed RCC on the same effective state $\tau_\xi$. We now convert them to common units to determine how the fixed reference capacity $r_R=\ln d_R$ is distributed between the two readouts.

\Istatement{Proposition I.4 (Window capacity identity at fixed reference).}
Under the fixed-reference conventions of Appendix~\ref{app:I.2} and the shared-window conditions of Definition I.2, for every $\xi\in J$ and every normalized input state supported in $\mathcal H_R$,
\begin{equation}
\label{eq:I.33}
\mathscr S_\xi
+
k_B g_R^{({\rm nat})} C_\xi
=
k_B r_R
=
k_B\ln d_R.
\end{equation}

\indent
Equivalently, expressing all quantities in reference atomic resource slots gives
\begin{equation}
\label{eq:I.34}
\frac{\mathscr S_\xi}
{k_B g_R^{({\rm nat})}}
+
C_\xi
=
\frac{r_R}
{g_R^{({\rm nat})}}.
\end{equation}

\Iproof
Substituting Eqs.~\eqref{eq:I.22} and \eqref{eq:I.24} into the left-hand side gives Eq.~\eqref{eq:I.33}. Dividing by $k_B g_R^{({\rm nat})}>0$ yields Eq.~\eqref{eq:I.34}.\Iqed

\indent
Since $0\le s_\xi\le r_R$, the ranges over which the fixed reference capacity is distributed between the two readouts are
\begin{equation*}
0
\le
\mathscr S_\xi
\le
k_B r_R,
\qquad
0
\le
C_\xi
\le
\frac{r_R}
{g_R^{({\rm nat})}}.
\end{equation*}

\indent
Equation~\eqref{eq:I.33} gives two complementary readouts of the same effective state within the fixed reference capacity. The quantity $\mathscr S_\xi$ reads window entropy, while $C_\xi$ reads the leading complexity term calibrated in atomic resource slots; the product $k_Bg_R^{({\rm nat})}C_\xi$ converts the latter back into thermodynamic entropy units. Its complexity interpretation follows the calibration and per-path lower-bound relation of Appendix~\ref{app:I.2}.

\Istatement{Corollary I.2 (Capacity difference identity for nested windows).}
For $\xi_1\le\xi_2$ in a shared window family,
\begin{equation}
\label{eq:I.35}
\mathscr S_{\xi_1}
-
\mathscr S_{\xi_2}
=
k_B g_R^{({\rm nat})}
\bigl(
C_{\xi_2}
-
C_{\xi_1}
\bigr)
\ge
0.
\end{equation}

\Iproof
Subtract Eq.~\eqref{eq:I.33} for the two windows and use Eq.~\eqref{eq:I.32}.\Iqed

Equation~\eqref{eq:I.35} describes the complementary changes under window refinement: more structure enters the observable object domain, window entropy decreases, and the leading complexity term increases by the same amount after conversion with the same bandwidth. If the window family contains the two limiting endpoints of Appendix~\ref{app:E}, complete reference coarse-graining $E_0(X)={\rm Tr}(X)\sigma_R$ gives $\tau_0=\sigma_R$, $\mathscr S_0=k_Br_R$, and $C_0=0$. The identity window $E_\infty={\rm id}$ recovers the global leading term, with $\mathscr S_\infty=0$ and $C_\infty=r_R/g_R^{({\rm nat})}$ for a pure input state. The next subsection connects this capacity allocation to thermodynamic response on an effective smooth branch.

\subsection{Fixed-Reference Responses and Path Integrals on Effective Smooth Branches}
\label{app:I.5}

\indent
The basic response of exact finite-dimensional windows is the finite difference in Eq.~\eqref{eq:I.35}. On an effective smooth branch established through a calibrated protocol, ensemble interpolation, or controlled continuous limit, we now use the capacity identity to determine the complexity-generation potential and its path integral. Branches with changing references or bandwidths are treated in Appendix~\ref{app:I.6}.

\indent
Fix a local regular region $\mathcal U$ in which $\mathscr S_\xi(E,\xi,V,N,\ldots)$ and $C_\xi(E,\xi,V,N,\ldots)$, defined by the continuum outputs, are at least $C^1$, $(\partial\mathscr S_\xi/\partial E)_{\xi,V,N,\ldots}\ne0$, and the reference data $R,\sigma_R,d_R,\Gamma_R$ remain fixed. The continuum construction or model limit used must establish these conditions and CWT's reversible thermodynamic interface. Following CWT's sign convention, define
\begin{equation}
\label{eq:I.36}
\frac{1}{T_\xi}
:=
\left(
\frac{\partial\mathscr S_\xi}{\partial E}
\right)_{\xi,V,N,\ldots},
\qquad
\Pi_\xi
:=
\left(
\frac{\partial E}{\partial\xi}
\right)_{\mathscr S_\xi,V,N,\ldots}
=
-
T_\xi
\left(
\frac{\partial\mathscr S_\xi}{\partial\xi}
\right)_{E,V,N,\ldots}.
\end{equation}

\indent
The corresponding reversible extended first law is
\begin{equation}
\label{eq:I.37}
dE
=
T_\xi\,d\mathscr S_\xi
+
\Pi_\xi\,d\xi
-
p\,dV
+
\sum_a
\mu_a\,dN_a.
\end{equation}

\indent
If the declared effective smooth branch preserves the monotone window ordering of Appendix~\ref{app:I.3}, Eq.~\eqref{eq:I.32} further gives
\begin{equation}
\label{eq:I.38}
\left(
\frac{\partial\mathscr S_\xi}{\partial\xi}
\right)_{E,V,N,\ldots}
\le
0.
\end{equation}

\indent
On a positive-temperature branch $T_\xi>0$, Eq.~\eqref{eq:I.36} then gives $\Pi_\xi\ge0$; on a negative-temperature branch, the sign is still determined by $-T_\xi\partial_\xi\mathscr S_\xi$. Equation~\eqref{eq:I.38} describes a smooth branch, whereas threshold responses of exact windows are handled through Eq.~\eqref{eq:I.32} and the nonsmooth relations in Appendix~\ref{app:I.6}.

\Istatement{Proposition I.5 (Fixed-reference response identity).}
In the regular region above,
\begin{equation}
\label{eq:I.39}
\left(
\frac{\partial\mathscr S_\xi}{\partial\xi}
\right)_{E,V,N,\ldots}
=
-
k_B g_R^{({\rm nat})}
\left(
\frac{\partial C_\xi}{\partial\xi}
\right)_{E,V,N,\ldots},
\end{equation}
and
\begin{equation}
\label{eq:I.40}
\Pi_\xi
=
k_B T_\xi g_R^{({\rm nat})}
\left(
\frac{\partial C_\xi}{\partial\xi}
\right)_{E,V,N,\ldots}.
\end{equation}

\Iproof
Differentiate Eq.~\eqref{eq:I.33} at fixed $E,V,N,\ldots$. Since $r_R$ and $g_R^{({\rm nat})}$ are constant, this gives Eq.~\eqref{eq:I.39}; substitution into Eq.~\eqref{eq:I.36} yields Eq.~\eqref{eq:I.40}.\Iqed

Equation~\eqref{eq:I.40} expresses the complexity-generation potential as the budget response of windowed RCC. The partial derivative measures the leading complexity term revealed per unit increase in budget; $g_R^{({\rm nat})}$ converts it into an information increment, and $k_BT_\xi$ supplies the corresponding energy scale. For a dimensionless budget, $\Pi_\xi$ has energy units; other budget units give energy per unit budget.
We next consider the path form of this pointwise relation. Let
\begin{equation*}
\gamma:
\lambda\in[0,1]
\longmapsto
\bigl(
E(\lambda),
\xi(\lambda),
V(\lambda),
N(\lambda),
\ldots
\bigr)
\subset
\mathcal U
\end{equation*}
be a piecewise $C^1$ reversible CWT path, and define the oriented reversible generalized work contributed by the complexity coordinate according to Eq.~\eqref{eq:I.37} as
\begin{equation*}
W_{\rm info}[\gamma]
:=
\int_\gamma
\Pi_\xi\,d\xi.
\end{equation*}

\indent
Under the $+\Pi_\xi d\xi$ convention, a positive value denotes complexity-generation work supplied to the system in the chosen direction. Total device work and dissipation require a separate hardware energy model.

\indent
All thermodynamic quantities below are evaluated at the state point corresponding to $\gamma(\lambda)$, with undisplayed macroscopic arguments suppressed. Integrating Eq.~\eqref{eq:I.40} along the chosen path gives
\begin{equation}
\label{eq:I.41}
W_{\rm info}[\gamma]
=
k_B g_R^{({\rm nat})}
\int_0^1
T_{\xi(\lambda)}
\left(
\frac{\partial C_\xi}{\partial\xi}
\right)_{E,V,N,\ldots}
\dot\xi(\lambda)\,d\lambda.
\end{equation}

\indent
Equation~\eqref{eq:I.41} extracts only the partial change of the complexity readout in the budget direction. Along a general path,
\begin{equation}
\label{eq:I.42}
\begin{aligned}
dC_\xi
&=
\left(
\frac{\partial C_\xi}{\partial E}
\right)_{\xi,V,N,\ldots}
dE
+
\left(
\frac{\partial C_\xi}{\partial\xi}
\right)_{E,V,N,\ldots}
d\xi
\\
&\quad+
\left(
\frac{\partial C_\xi}{\partial V}
\right)_{E,\xi,N,\ldots}
dV
+
\sum_a
\left(
\frac{\partial C_\xi}{\partial N_a}
\right)_{E,\xi,V,N_{b\ne a},\ldots}
dN_a
+\cdots .
\end{aligned}
\end{equation}

\indent
With all macroscopic variables fixed, this expression reduces to the total differential in the budget direction, yielding the following integral relation.

\Istatement{Corollary I.3 (Integral relation along paths with fixed macroscopic variables).}
If $\gamma$ further satisfies
\begin{equation}
\label{eq:I.43}
E(\lambda)
=
E_0,
\qquad
V(\lambda)
=
V_0,
\qquad
N_a(\lambda)
=
N_{a,0}
\quad
\text{for all }a,
\end{equation}
and all undisplayed macroscopic constraints also remain fixed along the path, then
\begin{equation*}
\frac{d}{d\lambda}
C_{\xi(\lambda)}
=
\left(
\frac{\partial C_\xi}{\partial\xi}
\right)_{E,V,N,\ldots}
\dot\xi(\lambda),
\end{equation*}
and therefore
\begin{equation}
\label{eq:I.44}
W_{\rm info}[\gamma]
=
k_B g_R^{({\rm nat})}
\int_\gamma
T_\xi\,dC_\xi.
\end{equation}

\Iproof
Equation~\eqref{eq:I.43} makes every term in Eq.~\eqref{eq:I.42} outside the budget direction vanish. Substituting the resulting total differential into Eq.~\eqref{eq:I.41} gives Eq.~\eqref{eq:I.44}.\Iqed
Equation~\eqref{eq:I.44} weights complexity increments by the temperature along the path. If the path is isothermal, with $T_\xi=T_0$ throughout $\gamma$, then
\begin{equation}
\label{eq:I.45}
W_{\rm info}[\gamma]
=
k_B T_0 g_R^{({\rm nat})}
\bigl(
C_{\xi_f}
-
C_{\xi_i}
\bigr)
=
T_0
\bigl(
\mathscr S_{\xi_i}
-
\mathscr S_{\xi_f}
\bigr).
\end{equation}

\indent
Here $\xi_i=\xi(0)$ and $\xi_f=\xi(1)$. A more general sufficient condition is that, on the branch under consideration, $T_\xi$ is a single-valued function $\Theta(C_\xi)$ of $C_\xi$ with an antiderivative satisfying $F'(C)=\Theta(C)$. In that case,
\begin{equation*}
W_{\rm info}[\gamma]
=
k_B g_R^{({\rm nat})}
\bigl[
F(C_{\xi_f})
-
F(C_{\xi_i})
\bigr].
\end{equation*}

\indent
General paths are still evaluated through the parametrized oriented integral in Eq.~\eqref{eq:I.41} or Eq.~\eqref{eq:I.44}. If the same $C_\xi$ corresponds to different temperatures on different branches, retain each branch's temperature profile. Contributions along a nonmonotone path accumulate with the actual direction of travel.
Equations~\eqref{eq:I.40}--\eqref{eq:I.45} turn fixed-reference capacity allocation into responses and reversible budget work on the declared smooth branch. We now incorporate reference and bandwidth changes and nonsmooth thresholds.

\subsection{Variable References, Bandwidths, and Nonsmooth Window Corrections}
\label{app:I.6}

\indent
This subsection treats reference support, atomic bandwidth, reference weights, and window jumps in turn, determining how they modify the response relations of Appendix~\ref{app:I.5}. After the state-side decomposition, we give the conversion conditions needed for comparisons across references. Unless stated otherwise, $\partial_\xi$ holds $E,V,N,\ldots$ and all other undisplayed macroscopic constraints fixed.

\Ilead{Smooth branches of variable unbiased references.}
 To avoid confusion with CWT's complexity-generation potential $\Pi_\xi$, denote the reference-support projection by $P_\xi$. First suppose that the different references have been embedded in a common finite-dimensional ambient space $\mathcal H_*$ through an explicit set of isometric identifications, and define
\begin{equation}
\label{eq:I.46}
\begin{gathered}
\mathcal H_{R_\xi}
=
P_\xi\mathcal H_*,
\qquad
d_\xi
=
{\rm Tr}\,P_\xi,
\qquad
\sigma_\xi
=
\frac{P_\xi}{d_\xi},\\
r_\xi
=
\ln d_\xi,
\qquad
g_\xi
=
\log_2\Gamma_{R_\xi},
\qquad
g_\xi^{({\rm nat})}
=
\ln\Gamma_{R_\xi}
=
(\ln2)g_\xi
>
0.
\end{gathered}
\end{equation}

\indent
A variable reference must also specify how the fixed input becomes a normalized effective state within the current support. For each $\xi$, specify an allowed domain $\mathcal D_\xi\subseteq\mathsf S(\mathcal H_*)$ and an explicit effective-state map
\begin{equation*}
\mathcal J_\xi:
\mathcal D_\xi
\longrightarrow
\mathsf S(\mathcal H_{R_\xi}),
\qquad
\tau_\xi
:=
\mathcal J_\xi(\rho),
\end{equation*}
where $\mathsf S(\mathcal K)$ denotes the normalized states on $\mathcal K$. If $\mathcal J_\xi$ is a reference-compatible CPTP window channel after the common embedding, one may take $\mathcal D_\xi=\mathsf S(\mathcal H_*)$. An externally specified family of normalized states supports only state-side comparisons. Projection conditioning is generally a nonlinear conditional-state map, for which one must instead take
\begin{equation*}
\mathcal D_\xi^{\rm proj}
:=
\left\{
\rho\in\mathsf S(\mathcal H_*):
{\rm Tr}(P_\xi\rho)>0
\right\}
\end{equation*}
and write
\begin{equation*}
q_\xi
:=
{\rm Tr}(P_\xi\rho)
>
0,
\qquad
\tau_\xi
=
\frac{
P_\xi\rho P_\xi
}{
q_\xi
}.
\end{equation*}

\indent
Here $1-q_\xi$ is the support-leakage or postselection-failure probability, included in the leakage reduction, conditional-state confidence accounting, and process cost according to Appendix~\ref{app:B}.
At each budget value, first define the state-side readout calibrated in the current atomic resource slots:
\begin{equation}
\label{eq:I.47}
C_\xi^{\rm state}
:=
\frac{
D_{\rm nat}
(\tau_\xi\Vert\sigma_\xi)
}{
g_\xi^{({\rm nat})}
}
=
\frac{
r_\xi-s_\xi
}{
g_\xi^{({\rm nat})}
}.
\end{equation}

\indent
If $\mathcal J_\xi$ comes from a certified reference-compatible window $\Xi_\xi$, and $R_\xi$ and the corresponding program model satisfy the RCC reference-admissibility conditions, then $C_\xi^{\rm state}=C_{R_\xi,\Xi_\xi}(\rho)$ is the windowed leading complexity term at the current reference. The superscript ${\rm state}$ also accommodates state-side extensions awaiting model certification; the algebraic decomposition in Eq.~\eqref{eq:I.47} already holds at this level.
The fixed-reference identity in Eq.~\eqref{eq:I.33} thus extends pointwise to
\begin{equation}
\label{eq:I.48}
\mathscr S_\xi
+
k_B g_\xi^{({\rm nat})}
C_\xi^{\rm state}
=
k_B r_\xi.
\end{equation}

\indent
Equation~\eqref{eq:I.48} holds pointwise. On a regular branch where $\mathscr S_\xi,C_\xi^{\rm state},g_\xi^{({\rm nat})},r_\xi$ are all $C^1$ and the thermodynamic partial derivatives in Eq.~\eqref{eq:I.36} exist, the product rule gives
\begin{equation}
\label{eq:I.49}
\partial_\xi
\mathscr S_\xi
=
k_B
\left[
\partial_\xi r_\xi
-
g_\xi^{({\rm nat})}
\partial_\xi C_\xi^{\rm state}
-
C_\xi^{\rm state}
\partial_\xi g_\xi^{({\rm nat})}
\right],
\end{equation}
and hence
\begin{equation}
\label{eq:I.50}
\Pi_\xi
=
k_B T_\xi
\left[
g_\xi^{({\rm nat})}
\partial_\xi C_\xi^{\rm state}
+
C_\xi^{\rm state}
\partial_\xi g_\xi^{({\rm nat})}
-
\partial_\xi r_\xi
\right].
\end{equation}

\indent
The three terms give the responses of the state readout, atomic bandwidth, and reference capacity, respectively. The complexity term in the entropy balance is the bandwidth-calibrated information gap $g_\xi^{({\rm nat})}C_\xi^{\rm state}=D_{\rm nat}(\tau_\xi\Vert\sigma_\xi)$. If changing $g_\xi^{({\rm nat})}$ merely reparametrizes the reporting unit while leaving the information gap unchanged, then
\begin{equation*}
g_\xi^{({\rm nat})}
\partial_\xi C_\xi^{\rm state}
+
C_\xi^{\rm state}
\partial_\xi g_\xi^{({\rm nat})}
=
\partial_\xi
\bigl(
g_\xi^{({\rm nat})}
C_\xi^{\rm state}
\bigr)
=
0.
\end{equation*}

\indent
The two terms cancel, as required by invariance of the information gap under a change of units. When the actual action dictionary or callable resources change, compare process costs through the cross-model interface below.

\Istatement{Proposition I.6 (Smooth supports of finite-dimensional unbiased references).}
If $P_\xi\ne0$ and $\xi\mapsto P_\xi$ is an operator-norm $C^1$ family of orthogonal projections in a common finite-dimensional ambient space, then $d_\xi$ and $r_\xi$ are constant on each connected regular branch. Thus an exact finite-dimensional unbiased reference satisfies, on a smooth branch,
\begin{equation}
\label{eq:I.51}
\Pi_\xi
=
k_B T_\xi
\left[
g_\xi^{({\rm nat})}
\partial_\xi C_\xi^{\rm state}
+
C_\xi^{\rm state}
\partial_\xi g_\xi^{({\rm nat})}
\right].
\end{equation}

\Iproof
The continuous integer-valued function $d_\xi={\rm Tr}\,P_\xi={\rm rank}\,P_\xi$ is constant on a connected branch, so $\partial_\xi r_\xi=0$. Equation~\eqref{eq:I.51} then follows from Eq.~\eqref{eq:I.50}.\Iqed

Proposition I.6 shows that smooth motion of an exact finite-dimensional support preserves capacity. A constant-rank support can rotate; if the supports are also nested by inclusion, the projections coincide on the same connected branch. The corresponding isometric identifications serve for state comparison, while the implementation cost of physical rotations is charged separately. A nonzero $\partial_\xi r_\xi$ corresponds to a finite effective capacity defined by truncation, thermodynamic scaling, or explicit renormalization; changes in the exact rank are treated as jump terms.

\indent
Atomic control bandwidth has the same discreteness: the exact cardinality $\Gamma_{R_\xi}$ of a finite set of alias-free action classes is locally constant on a continuous regular branch, and so is $g_\xi^{({\rm nat})}=\ln\Gamma_{R_\xi}$. Nonzero bandwidth derivatives arise from declared effective or asymptotic interpolations, continuous resource calibration, or generalized capacity definitions. Changes in the exact action-class count enter the full product jump term.

\Ilead{State-side extension to weighted references.}
 Reference weighting changes the structural zero point through $\sigma_\xi$, while protocol weighting changes observable admission and coarse-graining through $(\mathcal A_\xi,E_\xi)$. We now consider reference weighting: take $\sigma_\xi$ strictly positive on the common fixed support $\mathcal H_*$, and denote the corresponding readout by $C_{\xi,{\rm wt}}^{\rm state}$. Connecting this readout to a full RCC certificate requires reference consistency, faithful transcription, and semantic reference admissibility, together with spectral and one-shot bounds established for the weighted reference. Write
\begin{equation}
\label{eq:I.52}
K_\xi
:=
-\ln\sigma_\xi,
\qquad
h_\xi^{\rm ref}
:=
{\rm Tr}(\tau_\xi K_\xi)
=
-{\rm Tr}(\tau_\xi\ln\sigma_\xi),
\qquad
C_{\xi,{\rm wt}}^{\rm state}
=
\frac{
D_{\rm nat}(\tau_\xi\Vert\sigma_\xi)
}{
g_\xi^{({\rm nat})}
}.
\end{equation}

\indent
Here $K_\xi$ is the reference surprisal operator, also called the modular Hamiltonian. Its expectation gives the cross-entropy of the effective state relative to the reference weights. By the definition of relative entropy,
\begin{equation}
\label{eq:I.53}
\mathscr S_\xi
+
k_B g_\xi^{({\rm nat})}
C_{\xi,{\rm wt}}^{\rm state}
=
k_B h_\xi^{\rm ref}.
\end{equation}

\indent
The reference cross-entropy on the right-hand side, expressed in nats before conversion to thermodynamic entropy units, reduces for all effective states to the state-independent $\ln d$ precisely when the reference is unbiased on its support, $\sigma_\xi=P/d$. On the common fixed support, suppose $\sigma_\xi$ is locally uniformly strictly positive, $\tau_\xi,\sigma_\xi,g_\xi^{({\rm nat})}$ are $C^1$, $g_\xi^{({\rm nat})}$ is locally bounded away from zero, and $s_\xi$ and $C_{\xi,{\rm wt}}^{\rm state}$ are differentiable on the branch under consideration, for example when $\tau_\xi$ has fixed rank. Then Eq.~\eqref{eq:I.53} gives the general response formula
\begin{equation}
\label{eq:I.54}
\Pi_\xi
=
k_B T_\xi
\left[
g_\xi^{({\rm nat})}
\partial_\xi C_{\xi,{\rm wt}}^{\rm state}
+
C_{\xi,{\rm wt}}^{\rm state}
\partial_\xi g_\xi^{({\rm nat})}
-
\partial_\xi h_\xi^{\rm ref}
\right].
\end{equation}

\indent
The first two terms in Eq.~\eqref{eq:I.54} combine into $\partial_\xi(g_\xi^{({\rm nat})}C_{\xi,{\rm wt}}^{\rm state})$, while the last subtracts the response of the reference cross-entropy. The latter includes both redistribution of the effective state among reference weights and changes in the reference weights themselves. We separate these contributions explicitly below.

\indent
For noncommuting references, compute the cross-entropy variation using the Fréchet derivative of the operator logarithm~\cite{Higham2008_FunctionsMatrices}:
\begin{equation}
\label{eq:I.55}
D(\ln)_\sigma[X]
=
\mathcal T_\sigma(X)
:=
\int_0^\infty
(\sigma+tI)^{-1}
X
(\sigma+tI)^{-1}
\,dt.
\end{equation}
Then
\begin{equation}
\label{eq:I.56}
\partial_\xi h_\xi^{\rm ref}
=
-{\rm Tr}\!\left[
\partial_\xi\tau_\xi
\ln\sigma_\xi
\right]
-
{\rm Tr}\!\left[
\tau_\xi\,
\mathcal T_{\sigma_\xi}
\!\left(
\partial_\xi\sigma_\xi
\right)
\right].
\end{equation}

\indent
The two terms in Eq.~\eqref{eq:I.56} give the effective-state and reference-weight contributions, respectively. These derivatives are valid when the reference state is strictly positive on the common support. When supports change, reference eigenvalues approach zero, or relative entropy diverges, use one-sided limits, piecewise formulas, or support-leakage treatment.

\indent
A fixed weighted reference admits a useful cancellation. If $\sigma$ and $g^{({\rm nat})}$ are fixed, $\tau_\xi=E_\xi(\rho)$, and every $E_\xi$ is a trace-preserving conditional expectation with $E_\xi(\sigma)=\sigma$, then $\sigma\in\mathcal A_\xi$, and functional calculus gives $K_\sigma=-\ln\sigma\in\mathcal A_\xi$. Thus
\begin{equation}
\label{eq:I.57}
h_\xi^{\rm ref}
=
{\rm Tr}\!\left[
E_\xi(\rho)K_\sigma
\right]
=
{\rm Tr}(\rho K_\sigma).
\end{equation}

\indent
The reference cross-entropy is constant in the budget direction at fixed macroscopic input. The pointwise balance identity along the fixed input state then becomes
\begin{equation}
\label{eq:I.58}
\mathscr S_\xi
+
k_B g^{({\rm nat})}
C_{\xi,{\rm wt}}^{\rm state}
=
k_B{\rm Tr}(\rho K_\sigma).
\end{equation}

\indent
A fixed weighted reference therefore also recovers smooth responses of the same form as Eqs.~\eqref{eq:I.39}--\eqref{eq:I.40} and the difference equality in Eq.~\eqref{eq:I.35}, with a reference cross-entropy constant depending on $\rho,\sigma$. If the windows also satisfy the nesting and tower compatibility of Definition I.2, nonnegativity of the difference and window monotonicity are recovered as well. Reference preservation further requires every algebra to contain ${\rm alg}(\sigma)$, which determines the coarsest algebra compatible with that weighted reference.

\indent
For a Gibbs reference, the reference term in Eq.~\eqref{eq:I.54} takes a more explicit form. Let
\begin{equation}
\label{eq:I.59}
\sigma_\xi
=
\frac{e^{-A_\xi}}{Z_\xi},
\qquad
A_\xi
=
\beta_\xi H_\xi,
\qquad
Z_\xi
=
{\rm Tr}\,e^{-A_\xi},
\end{equation}
where the exponential requires $A_\xi$ to be dimensionless. If $H_\xi$ has energy units, $\beta_\xi$ has inverse-energy units; for a physical Gibbs reference, it is usually written as $\beta_\xi=1/(k_B T_\xi^{\rm ref})$. Then
\begin{equation}
\label{eq:I.60}
h_\xi^{\rm ref}
=
{\rm Tr}(\tau_\xi A_\xi)
+
\ln Z_\xi,
\qquad
\partial_\xi h_\xi^{\rm ref}
=
{\rm Tr}\!\left[
(\partial_\xi\tau_\xi)A_\xi
\right]
+
{\rm Tr}\!\left[
(\tau_\xi-\sigma_\xi)
\partial_\xi A_\xi
\right].
\end{equation}

\indent
The second equality uses
\begin{equation*}
\partial_\xi\ln Z_\xi
=
-
{\rm Tr}\!\left(
\sigma_\xi
\partial_\xi A_\xi
\right),
\end{equation*}
which follows from the Duhamel formula and cyclicity of the trace, without requiring $A_\xi$ and $\partial_\xi A_\xi$ to commute. All partial derivatives still hold $E,V,N,\ldots$ fixed. Expanding $A_\xi=\beta_\xi H_\xi$ gives
\begin{equation}
\label{eq:I.61}
\begin{aligned}
\partial_\xi h_\xi^{\rm ref}
&=
\beta_\xi
{\rm Tr}\!\left[
H_\xi
\partial_\xi\tau_\xi
\right]
+
\beta_\xi
{\rm Tr}\!\left[
(\tau_\xi-\sigma_\xi)
\partial_\xi H_\xi
\right]
\\
&\quad+
(\partial_\xi\beta_\xi)
{\rm Tr}\!\left[
(\tau_\xi-\sigma_\xi)
H_\xi
\right].
\end{aligned}
\end{equation}

\indent
Changes in the effective state, reference Hamiltonian, and Gibbs weights thus all enter the response through Eq.~\eqref{eq:I.54}. The temperature $T_\xi$ is determined by the CWT state function in Eq.~\eqref{eq:I.36}, while $\beta_\xi$ is determined by the reference weights; any connection between them in a physical implementation must be specified.

\indent
Furthermore, if $\tau_\xi=E_\xi(\rho)$, $E_\xi$ is a trace-preserving conditional expectation preserving the current $\sigma_\xi$, and $\rho$ is independent of $\xi$ at fixed macroscopic input, then $A_\xi\in\mathcal A_\xi$ and ${\rm Tr}[E_\xi(\rho)A_\xi]={\rm Tr}(\rho A_\xi)$, so Eq.~\eqref{eq:I.60} reduces to
\begin{equation}
\label{eq:I.62}
\partial_\xi h_\xi^{\rm ref}
=
{\rm Tr}\!\left[
(\rho-\sigma_\xi)
\partial_\xi A_\xi
\right].
\end{equation}

\indent
Equation~\eqref{eq:I.62} further reduces the reference term to a difference of operator expectations between the fixed input and the Gibbs reference. We now incorporate nonsmooth responses arising from window thresholds and switches of reference or bandwidth.

\Ilead{Bounded variation, jumps, and discrete budgets.}
 In what follows, hold $E,V,N,\ldots$ and the other macroscopic constraints fixed, and express the pointwise balance as difference and measure relations along the budget variable $\xi$. For a general path on which macroscopic variables also change, include the other thermodynamic directions according to Eq.~\eqref{eq:I.42}.

\indent
The language of bounded variation (BV) retains the jump responses produced by finite-dimensional window refinement. Abbreviate $\xi\mapsto g_\xi^{({\rm nat})}C_\xi^{\rm state}$ as $gC^{\rm state}$, and $\xi\mapsto g_\xi^{({\rm nat})}C_{\xi,{\rm wt}}^{\rm state}$ as $gC_{\rm wt}^{\rm state}$. On a closed bounded budget interval in the common embedding space, if the relevant quantities are finite and $\mathscr S$, $gC^{\rm state}$, and $r\in BV$, interpret $df$ as the finite signed Radon derivative measure of the standard precise representative. Then
\begin{equation}
\label{eq:I.63}
d\mathscr S
+
k_B\,d(gC^{\rm state})
=
k_B\,dr.
\end{equation}

\indent
If $h^{\rm ref}$, $\mathscr S$, and $gC_{\rm wt}^{\rm state}$ all belong to $BV$ and the weighted relative entropy is finite everywhere, Eq.~\eqref{eq:I.53} likewise gives the full measure identity
\begin{equation*}
d\mathscr S
+
k_B\,d(gC_{\rm wt}^{\rm state})
=
k_B\,dh^{\rm ref}.
\end{equation*}

\indent
Both measure identities contain absolutely continuous, singular continuous, and atomic jump parts. The jump set of a $BV$ function is at most countable. Writing $d^c f$ for the singular continuous part of the derivative measure and $\delta_{\xi_j}$ for the Dirac measure at $\xi_j$, we have
\begin{equation}
\label{eq:I.64}
df
=
f'_{\rm ac}\,d\xi
+
d^c f
+
\sum_j
[f]_j\,\delta_{\xi_j},
\qquad
[f]_j
:=
f(\xi_j^+)
-
f(\xi_j^-).
\end{equation}

\indent
Whenever componentwise left and right traces are used below, assume that $g^{({\rm nat})}$ and $C^{\rm state}$ have finite one-sided limits at the relevant jump, as holds, for example, if both are also in BV. In general, retain the jump $[gC^{\rm state}]_j$ of the product as a whole. The same convention applies in the weighted case. At jumps where the component traces exist,
\begin{equation}
\label{eq:I.65}
[\mathscr S]_j
+
k_B
\left(
(g_j^{({\rm nat})})^+
(C^{\rm state})_j^+
-
(g_j^{({\rm nat})})^-
(C^{\rm state})_j^-
\right)
=
k_B[r]_j.
\end{equation}

\indent
When the reference support undergoes a finite-dimensional rank jump,
\begin{equation*}
[r]_j
=
\ln\!\left(
\frac{d_j^+}{d_j^-}
\right).
\end{equation*}

\indent
The product jump must be retained as a whole, $[gC^{\rm state}]_j$. In terms of left and right traces, it has the unambiguous expansions
\begin{equation*}
\begin{aligned}
\relax[gC^{\rm state}]_j
&=
(g_j^{({\rm nat})})^-
[C^{\rm state}]_j
+
(C^{\rm state})_j^+
[g^{({\rm nat})}]_j
\\
&=
(g_j^{({\rm nat})})^+
[C^{\rm state}]_j
+
(C^{\rm state})_j^-
[g^{({\rm nat})}]_j.
\end{aligned}
\end{equation*}

\indent
Pairing left and right traces retains the cross-contribution from simultaneous bandwidth and complexity jumps. For a weighted reference, the corresponding atomic part is
\begin{equation}
\label{eq:I.66}
[\mathscr S]_j
+
k_B
[gC_{\rm wt}^{\rm state}]_j
=
k_B
[h^{\rm ref}]_j.
\end{equation}

\indent
Capacity differences describe the state relation across a window threshold; transition work is determined by the reversible protocol connecting the two sides. Suppose a protocol $\gamma_j$ preserves the macroscopic constraints above and defines $\Pi_\xi$ along the path. The transition work through the budget channel is
\begin{equation}
\label{eq:I.67}
W_{{\rm info},j}[\gamma_j]
:=
\int_{\gamma_j}
\Pi_\xi\,d\xi
=
-
\int_{\gamma_j}
T\,d\mathscr S.
\end{equation}

\indent
If the transition is isothermal, or if the temperature has a unique continuous value $T_j$ at the jump and the reversible differential law has explicitly been extended to atomic measures, this becomes
\begin{equation}
\label{eq:I.68}
W_{{\rm info},j}[\gamma_j]
=
-
T_j[\mathscr S]_j
=
k_B T_j
\left(
[gC^{\rm state}]_j
-
[r]_j
\right).
\end{equation}

\indent
The simplification in Eq.~\eqref{eq:I.68} relies on a constant temperature along the transition or an explicitly declared convention for extension to atomic measures. A general transition still uses its actual temperature profile in Eq.~\eqref{eq:I.67}; identical endpoint temperatures can accompany different intermediate profiles and work values.

More generally, if the same macroscopic variables remain fixed along a $BV$ path, $T$ is continuous, and the reversible relation is explicitly assumed to extend to the entire path, one may conditionally define measure-valued information-processing work by
\begin{equation}
\label{eq:I.69}
d\mathcal W_{\rm info}
:=
-
T\,d\mathscr S
=
k_B T
\left[
d(gC^{\rm state})
-
dr
\right],
\end{equation}
and the corresponding Lebesgue--Stieltjes integral. Equation~\eqref{eq:I.69} includes absolutely continuous, singular continuous, and jump contributions. An ordinary function $\Pi_\xi$ supplies the absolutely continuous density with respect to $d\xi$; the full measure retains the other parts. If temperature is also discontinuous at an atom, specify its representative or a physical interpolation.
If the budget is given only at ordered discrete values $\xi_0<\xi_1<\cdots<\xi_n$, the rigorous quantity to report is the finite difference
\begin{equation}
\label{eq:I.70}
\mathscr S_{j+1}
-
\mathscr S_j
+
k_B
\left(
g_{j+1}^{({\rm nat})}
C_{j+1}^{\rm state}
-
g_j^{({\rm nat})}
C_j^{\rm state}
\right)
=
k_B
(r_{j+1}-r_j).
\end{equation}

\indent
For fixed reference and bandwidth, Eq.~\eqref{eq:I.70} reduces to
\begin{equation}
\label{eq:I.71}
-
(\mathscr S_{j+1}-\mathscr S_j)
=
k_B
g_R^{({\rm nat})}
\left(
C_{j+1}^{\rm state}
-
C_j^{\rm state}
\right).
\end{equation}

\indent
If these discrete windows are also nested according to Definition I.2 and every point belongs to the same certified fixed-reference branch, then $C_{j+1}^{\rm state}-C_j^{\rm state}\ge0$ and $\mathscr S_{j+1}-\mathscr S_j\le0$.

The finite differences above relate the discrete endpoints. Local response requires a calibrated smooth interpolation. Budget work for a particular step is computed from its reversible transition protocol using Eq.~\eqref{eq:I.67}, while the secant readout changes with the calibration of the budget coordinate.

Rank changes along continuous effective-state paths must be distinguished from jumps of the window or state. Finite-dimensional entropy is continuous along trace-norm-continuous state paths, but may be nondifferentiable when eigenvalues reach or leave zero. Use the smooth formulas on piecewise $C^1$ regular segments of fixed rank and one-sided limits at their boundaries; a measure formulation additionally requires checking $\mathscr S\in BV$. An actual jump of a window or state can produce an atomic entropy term as in Eq.~\eqref{eq:I.65}. If a reference change gives ${\rm supp}(\tau_\xi)\nsubseteq{\rm supp}(\sigma_\xi)$, relative entropy is infinite; report support leakage, the conditional state, and confidence costs according to Appendix~\ref{app:B}.
The term $[r]_j$ records a change in support dimension. A rotation between equal-rank supports has $[r]_j=0$; the corresponding model conversion and implementation costs are described separately by compilation.

\Ilead{Comparability across references and certificate conditions.}
 To compare readouts under different references, place the same normalized effective state $\tau$ in a common declared domain, require ${\rm supp}(\tau)\subseteq{\rm supp}(\sigma_A)$, and use the operator domination $\sigma_A\le2^{a_{A\to B}}\sigma_B$ from Appendix~\ref{app:A.10}. In natural logarithmic units, define
\begin{equation*}
a_{A\to B}^{({\rm nat})}
:=
(\ln2)
a_{A\to B}.
\end{equation*}
Then
\begin{equation*}
D_{\rm nat}(\tau\Vert\sigma_B)
\le
D_{\rm nat}(\tau\Vert\sigma_A)
+
a_{A\to B}^{({\rm nat})}.
\end{equation*}

\indent
Define the state-side readout
\begin{equation*}
C_X^{\rm state}(\tau)
:=
\frac{
D_{\rm nat}(\tau\Vert\sigma_X)
}{
g_X^{({\rm nat})}
}.
\end{equation*}

\indent
Converting with the bandwidths at both ends gives
\begin{equation}
\label{eq:I.72}
C_B^{\rm state}(\tau)
\le
\frac{
g_A^{({\rm nat})}
}{
g_B^{({\rm nat})}
}
C_A^{\rm state}(\tau)
+
\frac{
a_{A\to B}^{({\rm nat})}
}{
g_B^{({\rm nat})}
}.
\end{equation}

\indent
Equation~\eqref{eq:I.72} uses the bandwidth ratio to express both sides in model $B$'s atomic resource slots. When comparing $D_{\max}^{\epsilon}$ or the full $L_{\rm RCC}$, the smoothing sets must also satisfy the common, reference-independent convention of Appendix~\ref{app:A.10}. A strict support change generally prevents finite domination in both directions. Comparisons of rotated equal-rank supports require an isometric identification first; the controls and resources for its physical implementation enter the compilation cost.

For variable references, suppose there is a CPTP coarse-graining channel $\Phi_{\xi_2\to\xi_1}$ connecting the common embedding domains at the two ends and satisfying both
\begin{equation}
\label{eq:I.73}
\Phi_{\xi_2\to\xi_1}
(\tau_{\xi_2})
=
\tau_{\xi_1},
\qquad
\Phi_{\xi_2\to\xi_1}
(\sigma_{\xi_2})
=
\sigma_{\xi_1}.
\end{equation}

\indent
The data-processing inequality then gives
\begin{equation*}
D_{\rm nat}
(\tau_{\xi_1}\Vert\sigma_{\xi_1})
\le
D_{\rm nat}
(\tau_{\xi_2}\Vert\sigma_{\xi_2}).
\end{equation*}

\indent
Equation~\eqref{eq:I.73} orders the information gap $g_\xi^{({\rm nat})}C_\xi^{\rm state}$. When bandwidth changes, the ordering of complexity readouts also depends on the bandwidth ratio. If the effective-state windows additionally satisfy tower compatibility in Eq.~\eqref{eq:I.31} and the corresponding entropy ordering, then $\Pi_\xi\ge0$ on a positive-temperature branch follows from $-T_\xi\partial_\xi\mathscr S_\xi$.
Process-side $C_{\rm opt}^{(\epsilon)}$ values are compared through the resource compilation maps, output-error bounds, and length bounds of Appendix~\ref{app:A.10}, retaining the size dependence of compilation constants. The full model, resource, and audit data for a reference upgrade follow the reporting conventions of Appendix~\ref{app:F.8}.

Support leakage, explicit cost inversion, finite samples, window structure, and reference upgrades retain the respective conditions of Appendices~\ref{app:B}--\ref{app:F}. This appendix thus establishes two model-constrained connections for finite-dimensional conditional-expectation windows and their variable-reference extensions: Appendix~\ref{app:I.1} provides model-consistent complexity intervals and path geometry checks, while Appendices~\ref{app:I.2}--\ref{app:I.6} provide window capacity, conditional responses, and their nonsmooth corrections.

\par
\endgroup

\endgroup

\bibliographystyle{apsrev4-2}
\bibliography{references}

@article{Nielsen2006_GeometricCircuitLowerBounds,
  author  = {Nielsen, Michael A.},
  title   = {A Geometric Approach to Quantum Circuit Lower Bounds},
  journal = {Quantum Information and Computation},
  volume  = {6},
  number  = {3},
  pages   = {213--262},
  year    = {2006},
  doi     = {10.26421/QIC6.3-2},
  url     = {https://doi.org/10.26421/QIC6.3-2}
}

@article{NielsenEtAl2006_QuantumComputationGeometry,
  author  = {Nielsen, Michael A. and Dowling, Mark R. and Gu, Mile and Doherty, Andrew C.},
  title   = {Quantum Computation as Geometry},
  journal = {Science},
  volume  = {311},
  number  = {5764},
  pages   = {1133--1135},
  year    = {2006},
  doi     = {10.1126/science.1121541},
  url     = {https://doi.org/10.1126/science.1121541}
}

@article{Gacs2001_QuantumAlgorithmicEntropy,
  author  = {G{\'a}cs, Peter},
  title   = {Quantum Algorithmic Entropy},
  journal = {Journal of Physics A: Mathematical and General},
  volume  = {34},
  number  = {35},
  pages   = {6859--6880},
  year    = {2001},
  doi     = {10.1088/0305-4470/34/35/312},
  url     = {https://doi.org/10.1088/0305-4470/34/35/312}
}

@article{Vitanyi2001_QuantumKolmogorovComplexity,
  author  = {Vit{\'a}nyi, Paul M. B.},
  title   = {Quantum Kolmogorov Complexity Based on Classical Descriptions},
  journal = {IEEE Transactions on Information Theory},
  volume  = {47},
  number  = {6},
  pages   = {2464--2479},
  year    = {2001},
  doi     = {10.1109/18.945258},
  url     = {https://doi.org/10.1109/18.945258}
}

@article{Datta2009_MinMaxRelativeEntropies,
  author  = {Datta, Nilanjana},
  title   = {Min- and Max-Relative Entropies and a New Entanglement Monotone},
  journal = {IEEE Transactions on Information Theory},
  volume  = {55},
  number  = {6},
  pages   = {2816--2826},
  year    = {2009},
  doi     = {10.1109/TIT.2009.2018325},
  url     = {https://doi.org/10.1109/TIT.2009.2018325}
}

@article{ChitambarGour2019_QRT,
  author  = {Chitambar, Eric and Gour, Gilad},
  title   = {Quantum Resource Theories},
  journal = {Reviews of Modern Physics},
  volume  = {91},
  number  = {2},
  pages   = {025001},
  year    = {2019},
  doi     = {10.1103/RevModPhys.91.025001},
  url     = {https://doi.org/10.1103/RevModPhys.91.025001}
}

@misc{Knill1995_ApproximationQuantumCircuits,
  author        = {Knill, Emanuel},
  title         = {Approximation by Quantum Circuits},
  year          = {1995},
  eprint        = {quant-ph/9508006},
  archivePrefix = {arXiv},
  primaryClass  = {quant-ph},
  note          = {{LANL} report {LAUR}-95-2225},
  doi           = {10.48550/arXiv.quant-ph/9508006},
  url           = {https://doi.org/10.48550/arXiv.quant-ph/9508006}
}

@article{ZhuEtAl2026_BipartiteEntanglementSurfaceCriticality,
  author  = {Zhu, Yanzhang and Liu, Zenan and Wang, Zhe and Wang, Yan-Cheng and Yan, Zheng},
  title   = {Bipartite Entanglement and Surface Criticality: The Extra Contribution of the Nonordinary Edge in Entanglement},
  journal = {Physical Review Letters},
  volume  = {136},
  number  = {4},
  pages   = {046501},
  year    = {2026},
  doi     = {10.1103/tkx5-kzhh},
  url     = {https://doi.org/10.1103/tkx5-kzhh}
}

@article{Hall2001_ExactUncertaintyRelations,
  author  = {Hall, Michael J. W.},
  title   = {Exact Uncertainty Relations},
  journal = {Physical Review A},
  volume  = {64},
  number  = {5},
  pages   = {052103},
  year    = {2001},
  doi     = {10.1103/PhysRevA.64.052103},
  url     = {https://doi.org/10.1103/PhysRevA.64.052103}
}

@article{PatiEtAl2025_ExactQuantumSpeedLimits,
  author  = {Pati, Arun K. and Mohan, Brij and Sahil and Braunstein, Samuel L.},
  title   = {Exact Quantum Speed Limits},
  journal = {Physical Review A},
  volume  = {112},
  number  = {6},
  pages   = {062209},
  year    = {2025},
  doi     = {10.1103/mmzc-fyr9},
  url     = {https://doi.org/10.1103/mmzc-fyr9}
}

@misc{LiuTianWu2025_CostNonlocality,
  author        = {Liu, HongZheng and Tian, YiNuo and Wu, Zhiyue},
  title         = {The Cost of Nonlocality: A Dynamical Performance Equation of Energy-Entanglement-Complexity},
  year          = {2025},
  eprint        = {2508.03781v1},
  url           = {https://arxiv.org/abs/2508.03781v1},
  archivePrefix = {arXiv},
  primaryClass  = {quant-ph}
}

@misc{LiuTianWu2025_CWT,
  author        = {Liu, HongZheng and Tian, YiNuo and Wu, Zhiyue},
  title         = {A Heuristic Study of Temperature: Quantum Circuitry in Thermal Systems},
  year          = {2025},
  eprint        = {2506.06994v3},
  url           = {https://arxiv.org/abs/2506.06994v3},
  archivePrefix = {arXiv},
  primaryClass  = {cond-mat.stat-mech}
}

@article{Brown2024_PolynomialEquivalenceComplexityGeometries,
  author  = {Brown, Adam R.},
  title   = {Polynomial Equivalence of Complexity Geometries},
  journal = {Quantum},
  volume  = {8},
  pages   = {1391},
  year    = {2024},
  doi     = {10.22331/q-2024-07-02-1391},
  url     = {https://doi.org/10.22331/q-2024-07-02-1391}
}

@article{ParkerEtAl2019_UniversalOperatorGrowth,
  author  = {Parker, Daniel E. and Cao, Xiangyu and Avdoshkin, Alexander and Scaffidi, Thomas and Altman, Ehud},
  title   = {A Universal Operator Growth Hypothesis},
  journal = {Physical Review X},
  volume  = {9},
  number  = {4},
  pages   = {041017},
  year    = {2019},
  doi     = {10.1103/PhysRevX.9.041017},
  url     = {https://doi.org/10.1103/PhysRevX.9.041017}
}

@article{RabinoviciEtAl2021_OperatorComplexity,
  author  = {Rabinovici, E. and S{\'a}nchez-Garrido, A. and Shir, R. and Sonner, J.},
  title   = {Operator Complexity: A Journey to the Edge of Krylov Space},
  journal = {Journal of High Energy Physics},
  volume  = {2021},
  number  = {6},
  pages   = {062},
  year    = {2021},
  doi     = {10.1007/JHEP06(2021)062},
  url     = {https://doi.org/10.1007/JHEP06(2021)062}
}

@article{BalasubramanianEtAl2022_SpreadComplexity,
  author  = {Balasubramanian, Vijay and Caputa, Pawel and Magan, Javier M. and Wu, Qingyue},
  title   = {Quantum Chaos and the Complexity of Spread of States},
  journal = {Physical Review D},
  volume  = {106},
  number  = {4},
  pages   = {046007},
  year    = {2022},
  doi     = {10.1103/PhysRevD.106.046007},
  url     = {https://doi.org/10.1103/PhysRevD.106.046007}
}

@article{StanfordSusskind2014_ComplexityShockWaves,
  author  = {Stanford, Douglas and Susskind, Leonard},
  title   = {Complexity and Shock Wave Geometries},
  journal = {Physical Review D},
  volume  = {90},
  number  = {12},
  pages   = {126007},
  year    = {2014},
  doi     = {10.1103/PhysRevD.90.126007},
  url     = {https://doi.org/10.1103/PhysRevD.90.126007}
}

@article{BrownEtAl2016_ComplexityActionBlackHoles,
  author  = {Brown, Adam R. and Roberts, Daniel A. and Susskind, Leonard and Swingle, Brian and Zhao, Ying},
  title   = {Complexity, Action, and Black Holes},
  journal = {Physical Review D},
  volume  = {93},
  number  = {8},
  pages   = {086006},
  year    = {2016},
  doi     = {10.1103/PhysRevD.93.086006},
  url     = {https://doi.org/10.1103/PhysRevD.93.086006}
}

@article{JeffersonMyers2017_CircuitComplexityQFT,
  author  = {Jefferson, Ro and Myers, Robert C.},
  title   = {Circuit Complexity in Quantum Field Theory},
  journal = {Journal of High Energy Physics},
  volume  = {2017},
  number  = {10},
  pages   = {107},
  year    = {2017},
  doi     = {10.1007/JHEP10(2017)107},
  url     = {https://doi.org/10.1007/JHEP10(2017)107}
}

@book{Tomamichel2016_FiniteResources,
  author    = {Tomamichel, Marco},
  title     = {Quantum Information Processing with Finite Resources: Mathematical Foundations},
  series    = {SpringerBriefs in Mathematical Physics},
  volume    = {5},
  publisher = {Springer},
  address   = {Cham},
  year      = {2016},
  doi       = {10.1007/978-3-319-21891-5},
  eprint    = {1504.00233},
  archivePrefix = {arXiv},
  primaryClass  = {quant-ph},
  url       = {https://doi.org/10.1007/978-3-319-21891-5}
}

@book{LiVitanyi2019_KolmogorovComplexity,
  author    = {Li, Ming and Vit{\'a}nyi, Paul M. B.},
  title     = {An Introduction to Kolmogorov Complexity and Its Applications},
  edition   = {4},
  publisher = {Springer},
  address   = {Cham},
  year      = {2019},
  doi       = {10.1007/978-3-030-11298-1},
  url       = {https://doi.org/10.1007/978-3-030-11298-1}
}

@article{WangRenner2012_HypothesisTesting,
  author  = {Wang, Ligong and Renner, Renato},
  title   = {One-Shot Classical-Quantum Capacity and Hypothesis Testing},
  journal = {Physical Review Letters},
  volume  = {108},
  number  = {20},
  pages   = {200501},
  year    = {2012},
  doi     = {10.1103/PhysRevLett.108.200501},
  eprint  = {1007.5456},
  archivePrefix = {arXiv},
  primaryClass  = {quant-ph},
  url     = {https://doi.org/10.1103/PhysRevLett.108.200501}
}

@article{Lindblad1975_CPMapsEntropy,
  author  = {Lindblad, G{\"o}ran},
  title   = {Completely Positive Maps and Entropy Inequalities},
  journal = {Communications in Mathematical Physics},
  volume  = {40},
  number  = {2},
  pages   = {147--151},
  year    = {1975},
  doi     = {10.1007/BF01609396},
  url     = {https://doi.org/10.1007/BF01609396}
}

@article{Umegaki1954_ConditionalExpectation,
  author  = {Umegaki, Hisaharu},
  title   = {Conditional Expectation in an Operator Algebra},
  journal = {Tohoku Mathematical Journal, Second Series},
  volume  = {6},
  number  = {2--3},
  pages   = {177--181},
  year    = {1954},
  doi     = {10.2748/tmj/1178245177},
  url     = {https://doi.org/10.2748/tmj/1178245177}
}

@article{Wootters1981_StatisticalDistance,
  author  = {Wootters, William K.},
  title   = {Statistical Distance and Hilbert Space},
  journal = {Physical Review D},
  volume  = {23},
  number  = {2},
  pages   = {357--362},
  year    = {1981},
  doi     = {10.1103/PhysRevD.23.357},
  url     = {https://doi.org/10.1103/PhysRevD.23.357}
}

@article{AnandanAharonov1990_GeometryQuantumEvolution,
  author  = {Anandan, Jeeva and Aharonov, Yakir},
  title   = {Geometry of Quantum Evolution},
  journal = {Physical Review Letters},
  volume  = {65},
  number  = {14},
  pages   = {1697--1700},
  year    = {1990},
  doi     = {10.1103/PhysRevLett.65.1697},
  url     = {https://doi.org/10.1103/PhysRevLett.65.1697}
}

@incollection{MandelstamTamm1945_EnergyTime,
  author    = {Mandelstam, Leonid and Tamm, Igor},
  title     = {The Uncertainty Relation Between Energy and Time in Non-relativistic Quantum Mechanics},
  booktitle = {Selected Papers},
  editor    = {Bolotovskii, Boris M. and Frenkel, Victor Ya. and Peierls, Rudolf},
  publisher = {Springer},
  address   = {Berlin, Heidelberg},
  pages     = {115--123},
  year      = {1991},
  doi       = {10.1007/978-3-642-74626-0_8},
  url       = {https://doi.org/10.1007/978-3-642-74626-0_8},
  note      = {Originally published in J. Phys. (USSR) 9, 249--254 (1945)}
}

@article{LiebRobinson1972_FiniteGroupVelocity,
  author  = {Lieb, Elliott H. and Robinson, Derek W.},
  title   = {The Finite Group Velocity of Quantum Spin Systems},
  journal = {Communications in Mathematical Physics},
  volume  = {28},
  number  = {3},
  pages   = {251--257},
  year    = {1972},
  doi     = {10.1007/BF01645779},
  url     = {https://doi.org/10.1007/BF01645779}
}

@article{BravyiHastingsVerstraete2006_LRGeneration,
  author  = {Bravyi, Sergey and Hastings, Matthew B. and Verstraete, Frank},
  title   = {Lieb-Robinson Bounds and the Generation of Correlations and Topological Quantum Order},
  journal = {Physical Review Letters},
  volume  = {97},
  number  = {5},
  pages   = {050401},
  year    = {2006},
  doi     = {10.1103/PhysRevLett.97.050401},
  eprint  = {quant-ph/0603121},
  archivePrefix = {arXiv},
  url     = {https://doi.org/10.1103/PhysRevLett.97.050401}
}

@article{BrownEtAl2016_HolographicComplexityBulkAction,
  author  = {Brown, Adam R. and Roberts, Daniel A. and Susskind, Leonard and Swingle, Brian and Zhao, Ying},
  title   = {Holographic Complexity Equals Bulk Action?},
  journal = {Physical Review Letters},
  volume  = {116},
  number  = {19},
  pages   = {191301},
  year    = {2016},
  doi     = {10.1103/PhysRevLett.116.191301},
  eprint  = {1509.07876},
  archivePrefix = {arXiv},
  primaryClass  = {hep-th},
  url     = {https://doi.org/10.1103/PhysRevLett.116.191301}
}

@article{Hoeffding1963_ProbabilityInequalities,
  author  = {Hoeffding, Wassily},
  title   = {Probability Inequalities for Sums of Bounded Random Variables},
  journal = {Journal of the American Statistical Association},
  volume  = {58},
  number  = {301},
  pages   = {13--30},
  year    = {1963},
  doi     = {10.1080/01621459.1963.10500830},
  url     = {https://doi.org/10.1080/01621459.1963.10500830}
}

@article{ClopperPearson1934_BinomialLimits,
  author  = {Clopper, C. J. and Pearson, E. S.},
  title   = {The Use of Confidence or Fiducial Limits Illustrated in the Case of the Binomial},
  journal = {Biometrika},
  volume  = {26},
  number  = {4},
  pages   = {404--413},
  year    = {1934},
  doi     = {10.1093/biomet/26.4.404},
  url     = {https://doi.org/10.1093/biomet/26.4.404}
}

@article{Audenaert2007_EntropyContinuity,
  author  = {Audenaert, Koenraad M. R.},
  title   = {A Sharp Continuity Estimate for the von Neumann Entropy},
  journal = {Journal of Physics A: Mathematical and Theoretical},
  volume  = {40},
  number  = {28},
  pages   = {8127--8136},
  year    = {2007},
  doi     = {10.1088/1751-8113/40/28/S18},
  eprint  = {quant-ph/0610146},
  archivePrefix = {arXiv},
  url     = {https://doi.org/10.1088/1751-8113/40/28/S18}
}

@article{Elias1975_UniversalCodewordSets,
  author  = {Elias, Peter},
  title   = {Universal Codeword Sets and Representations of the Integers},
  journal = {IEEE Transactions on Information Theory},
  volume  = {21},
  number  = {2},
  pages   = {194--203},
  year    = {1975},
  doi     = {10.1109/TIT.1975.1055349},
  url     = {https://doi.org/10.1109/TIT.1975.1055349}
}

@article{BoykinEtAl2000_UniversalQuantumBasis,
  author  = {Boykin, P. Oscar and Mor, Tal and Pulver, Matthew and Roychowdhury, Vwani and Vatan, Farrokh},
  title   = {A New Universal and Fault-Tolerant Quantum Basis},
  journal = {Information Processing Letters},
  volume  = {75},
  number  = {3},
  pages   = {101--107},
  year    = {2000},
  doi     = {10.1016/S0020-0190(00)00084-3},
  eprint  = {quant-ph/9906054},
  archivePrefix = {arXiv},
  url     = {https://doi.org/10.1016/S0020-0190(00)00084-3}
}

@article{Choi1975_CPLinearMaps,
  author  = {Choi, Man-Duen},
  title   = {Completely Positive Linear Maps on Complex Matrices},
  journal = {Linear Algebra and Its Applications},
  volume  = {10},
  number  = {3},
  pages   = {285--290},
  year    = {1975},
  doi     = {10.1016/0024-3795(75)90075-0},
  url     = {https://doi.org/10.1016/0024-3795(75)90075-0}
}

@article{CorlessEtAl1996_LambertW,
  author  = {Corless, Robert M. and Gonnet, Gaston H. and Hare, David E. G. and Jeffrey, David J. and Knuth, Donald E.},
  title   = {On the Lambert W Function},
  journal = {Advances in Computational Mathematics},
  volume  = {5},
  number  = {1},
  pages   = {329--359},
  year    = {1996},
  doi     = {10.1007/BF02124750},
  url     = {https://doi.org/10.1007/BF02124750}
}

@book{Higham2008_FunctionsMatrices,
  author    = {Higham, Nicholas J.},
  title     = {Functions of Matrices: Theory and Computation},
  publisher = {Society for Industrial and Applied Mathematics},
  address   = {Philadelphia},
  year      = {2008},
  doi       = {10.1137/1.9780898717778},
  url       = {https://doi.org/10.1137/1.9780898717778}
}

@article{Stinespring1955_PositiveFunctions,
  author  = {Stinespring, W. Forrest},
  title   = {Positive Functions on {C*}-Algebras},
  journal = {Proceedings of the American Mathematical Society},
  volume  = {6},
  number  = {2},
  pages   = {211--216},
  year    = {1955},
  doi     = {10.1090/S0002-9939-1955-0069403-4},
  url     = {https://doi.org/10.1090/S0002-9939-1955-0069403-4}
}

@article{Hartley1928_TransmissionInformation,
  author  = {Hartley, Ralph V. L.},
  title   = {Transmission of Information},
  journal = {Bell System Technical Journal},
  volume  = {7},
  number  = {3},
  pages   = {535--563},
  year    = {1928},
  doi     = {10.1002/j.1538-7305.1928.tb01236.x},
  url     = {https://doi.org/10.1002/j.1538-7305.1928.tb01236.x}
}

@article{BerthiaumeEtAl2001_QuantumKolmogorov,
  author  = {Berthiaume, Andr{\'e} and van Dam, Wim and Laplante, Sophie},
  title   = {Quantum Kolmogorov Complexity},
  journal = {Journal of Computer and System Sciences},
  volume  = {63},
  number  = {2},
  pages   = {201--221},
  year    = {2001},
  doi     = {10.1006/jcss.2001.1765},
  eprint  = {quant-ph/0005018},
  archivePrefix = {arXiv},
  url     = {https://doi.org/10.1006/jcss.2001.1765}
}

@article{ChiribellaEtAl2009_QuantumNetworks,
  author  = {Chiribella, Giulio and D'Ariano, Giacomo M. and Perinotti, Paolo},
  title   = {Theoretical Framework for Quantum Networks},
  journal = {Physical Review A},
  volume  = {80},
  number  = {2},
  pages   = {022339},
  year    = {2009},
  doi     = {10.1103/PhysRevA.80.022339},
  eprint  = {0904.4483},
  archivePrefix = {arXiv},
  primaryClass  = {quant-ph},
  url     = {https://doi.org/10.1103/PhysRevA.80.022339}
}

@misc{Aaronson2016_ComplexityStatesTransformations,
  author = {Aaronson, Scott},
  title = {The Complexity of Quantum States and Transformations: From Quantum Money to Black Holes},
  year = {2016},
  eprint = {1607.05256},
  archivePrefix = {arXiv},
  primaryClass = {quant-ph},
  url = {https://arxiv.org/abs/1607.05256}
}

@article{BrandaoGour2015_ReversibleResourceTheories,
  author = {Brand{\~a}o, Fernando G. S. L. and Gour, Gilad},
  title = {Reversible Framework for Quantum Resource Theories},
  journal = {Physical Review Letters},
  volume = {115},
  number = {7},
  pages = {070503},
  year = {2015},
  doi = {10.1103/PhysRevLett.115.070503},
  url = {https://doi.org/10.1103/PhysRevLett.115.070503},
  note = {Erratum: \href{https://doi.org/10.1103/PhysRevLett.115.199901}{ibid. \textbf{115}, 199901 (2015)}}
}

@misc{LiuTianWu2026_RCCRefcert,
  author = {Liu, HongZheng and Tian, YiNuo and Wu, Zhiyue},
  title = {{rcc-refcert}: a {Python} research toolkit for structure-fair quantum circuit complexity},
  year = {2026},
  howpublished = {Zenodo},
  note = {Version 0.1.3},
  doi = {10.5281/zenodo.22768920},
  url = {https://github.com/yeruciwenrou-dotcom/rcc-refcert}
}

@article{BrandaoEtAl2021_ModelsComplexityGrowth,
  author        = {Brand{\~a}o, Fernando G. S. L. and Chemissany, Wissam and Hunter-Jones, Nicholas and Kueng, Richard and Preskill, John},
  title         = {Models of Quantum Complexity Growth},
  journal       = {PRX Quantum},
  volume        = {2},
  number        = {3},
  pages         = {030316},
  year          = {2021},
  doi           = {10.1103/prxquantum.2.030316},
  url           = {https://doi.org/10.1103/prxquantum.2.030316},
  eprint        = {1912.04297},
  archivePrefix = {arXiv}
}

@article{HaferkampEtAl2022_LinearComplexityGrowth,
  author        = {Haferkamp, Jonas and Faist, Philippe and Kothakonda, Naga B. T. and Eisert, Jens and Yunger Halpern, Nicole},
  title         = {Linear growth of quantum circuit complexity},
  journal       = {Nature Physics},
  volume        = {18},
  number        = {5},
  pages         = {528--532},
  year          = {2022},
  doi           = {10.1038/s41567-022-01539-6},
  url           = {https://doi.org/10.1038/s41567-022-01539-6},
  eprint        = {2106.05305},
  archivePrefix = {arXiv}
}

@article{Eisert2021_EntanglingPowerComplexity,
  author        = {Eisert, Jens},
  title         = {Entangling Power and Quantum Circuit Complexity},
  journal       = {Physical Review Letters},
  volume        = {127},
  number        = {2},
  pages         = {020501},
  year          = {2021},
  doi           = {10.1103/physrevlett.127.020501},
  url           = {https://doi.org/10.1103/physrevlett.127.020501},
  eprint        = {2104.03332},
  archivePrefix = {arXiv}
}

@article{Landauer1961_Irreversibility,
  author        = {Landauer, Rolf},
  title         = {Irreversibility and Heat Generation in the Computing Process},
  journal       = {IBM Journal of Research and Development},
  volume        = {5},
  number        = {3},
  pages         = {183--191},
  year          = {1961},
  doi           = {10.1147/rd.53.0183},
  url           = {https://doi.org/10.1147/rd.53.0183}
}

@article{GourEtAl2015_InformationalNonequilibrium,
  author        = {Gour, Gilad and M{\"u}ller, Markus P. and Narasimhachar, Varun and Spekkens, Robert W. and Yunger Halpern, Nicole},
  title         = {The resource theory of informational nonequilibrium in thermodynamics},
  journal       = {Physics Reports},
  volume        = {583},
  pages         = {1--58},
  year          = {2015},
  doi           = {10.1016/j.physrep.2015.04.003},
  url           = {https://doi.org/10.1016/j.physrep.2015.04.003},
  eprint        = {1309.6586},
  archivePrefix = {arXiv}
}

@article{BartlettEtAl2007_ReferenceFrames,
  author        = {Bartlett, Stephen D. and Rudolph, Terry and Spekkens, Robert W.},
  title         = {Reference frames, superselection rules, and quantum information},
  journal       = {Reviews of Modern Physics},
  volume        = {79},
  number        = {2},
  pages         = {555--609},
  year          = {2007},
  doi           = {10.1103/revmodphys.79.555},
  url           = {https://doi.org/10.1103/revmodphys.79.555},
  eprint        = {quant-ph/0610030},
  archivePrefix = {arXiv}
}

@article{FaistEtAl2015_MinimalWorkCost,
  author        = {Faist, Philippe and Dupuis, Fr{\'e}d{\'e}ric and Oppenheim, Jonathan and Renner, Renato},
  title         = {The minimal work cost of information processing},
  journal       = {Nature Communications},
  volume        = {6},
  number        = {1},
  pages         = {7669},
  year          = {2015},
  doi           = {10.1038/ncomms8669},
  url           = {https://doi.org/10.1038/ncomms8669},
  eprint        = {1211.1037},
  archivePrefix = {arXiv}
}

@article{GaoEtAl2020_SubalgebraRelativeEntropy,
  author        = {Gao, Li and Junge, Marius and LaRacuente, Nicholas},
  title         = {Relative entropy for von Neumann subalgebras},
  journal       = {International Journal of Mathematics},
  volume        = {31},
  number        = {06},
  pages         = {2050046},
  year          = {2020},
  doi           = {10.1142/s0129167x20500469},
  url           = {https://doi.org/10.1142/s0129167x20500469},
  eprint        = {1909.01906},
  archivePrefix = {arXiv}
}

@article{DawsonNielsen2006_SolovayKitaev,
  author        = {Dawson, Christopher M. and Nielsen, Michael A.},
  title         = {The Solovay-Kitaev algorithm},
  journal       = {Quantum Information and Computation},
  volume        = {6},
  number        = {1},
  pages         = {81--95},
  year          = {2006},
  doi           = {10.26421/qic6.1-6},
  url           = {https://doi.org/10.26421/qic6.1-6},
  eprint        = {quant-ph/0505030},
  archivePrefix = {arXiv}
}

@article{DaviesLewis1970_QuantumProbability,
  author        = {Davies, E. B. and Lewis, J. T.},
  title         = {An operational approach to quantum probability},
  journal       = {Communications in Mathematical Physics},
  volume        = {17},
  number        = {3},
  pages         = {239--260},
  year          = {1970},
  doi           = {10.1007/bf01647093},
  url           = {https://doi.org/10.1007/bf01647093}
}

@article{PleschBrukner2011_StatePreparation,
  author = {Plesch, Martin and Brukner, {\v C}aslav},
  title = {Quantum-state preparation with universal gate decompositions},
  journal = {Physical Review A},
  volume = {83},
  number = {3},
  pages = {032302},
  year = {2011},
  doi = {10.1103/PhysRevA.83.032302},
  eprint = {1003.5760},
  archivePrefix = {arXiv},
  primaryClass = {quant-ph},
  url = {https://arxiv.org/abs/1003.5760}
}

@article{BarencoEtAl1995_ElementaryGates,
  author = {Barenco, Adriano and Bennett, Charles H. and Cleve, Richard and DiVincenzo, David P. and Margolus, Norman and Shor, Peter and Sleator, Tycho and Smolin, John A. and Weinfurter, Harald},
  title = {Elementary gates for quantum computation},
  journal = {Physical Review A},
  volume = {52},
  number = {5},
  pages = {3457--3467},
  year = {1995},
  doi = {10.1103/PhysRevA.52.3457},
  eprint = {quant-ph/9503016},
  archivePrefix = {arXiv},
  primaryClass = {quant-ph},
  url = {https://arxiv.org/abs/quant-ph/9503016}
}

@article{ShendeEtAl2006_Synthesis,
  author = {Shende, Vivek V. and Bullock, Stephen S. and Markov, Igor L.},
  title = {Synthesis of quantum-logic circuits},
  journal = {IEEE Transactions on Computer-Aided Design of Integrated Circuits and Systems},
  volume = {25},
  number = {6},
  pages = {1000--1010},
  year = {2006},
  doi = {10.1109/TCAD.2005.855930},
  eprint = {quant-ph/0406176},
  archivePrefix = {arXiv},
  primaryClass = {quant-ph},
  url = {https://arxiv.org/abs/quant-ph/0406176}
}

@article{DelRioEtAl2011_NegativeEntropy,
  author = {del Rio, L{\'i}dia and {\AA}berg, Johan and Renner, Renato and Dahlsten, Oscar and Vedral, Vlatko},
  title = {The thermodynamic meaning of negative entropy},
  journal = {Nature},
  volume = {474},
  number = {7349},
  pages = {61--63},
  year = {2011},
  doi = {10.1038/nature10123},
  eprint = {1009.1630},
  archivePrefix = {arXiv},
  primaryClass = {quant-ph},
  url = {https://arxiv.org/abs/1009.1630}
}

@article{ReebWolf2014_Landauer,
  author = {Reeb, David and Wolf, Michael M.},
  title = {An improved {Landauer} principle with finite-size corrections},
  journal = {New Journal of Physics},
  volume = {16},
  number = {10},
  pages = {103011},
  year = {2014},
  doi = {10.1088/1367-2630/16/10/103011},
  eprint = {1306.4352},
  archivePrefix = {arXiv},
  primaryClass = {quant-ph},
  url = {https://arxiv.org/abs/1306.4352}
}

@article{FlammiaLiu2011_DirectFidelity,
  author = {Flammia, Steven T. and Liu, Yi-Kai},
  title = {Direct Fidelity Estimation from Few {Pauli} Measurements},
  journal = {Physical Review Letters},
  volume = {106},
  number = {23},
  pages = {230501},
  year = {2011},
  doi = {10.1103/PhysRevLett.106.230501},
  eprint = {1104.4695},
  archivePrefix = {arXiv},
  primaryClass = {quant-ph},
  url = {https://arxiv.org/abs/1104.4695}
}

@article{KlieschRoth2021_Certification,
  author = {Kliesch, Martin and Roth, Ingo},
  title = {Theory of Quantum System Certification},
  journal = {PRX Quantum},
  volume = {2},
  number = {1},
  pages = {010201},
  year = {2021},
  doi = {10.1103/PRXQuantum.2.010201},
  eprint = {2010.05925},
  archivePrefix = {arXiv},
  primaryClass = {quant-ph},
  url = {https://arxiv.org/abs/2010.05925}
}

@article{TomamichelHayashi2013_Hierarchy,
  author = {Tomamichel, Marco and Hayashi, Masahito},
  title = {A Hierarchy of Information Quantities for Finite Block Length Analysis of Quantum Tasks},
  journal = {IEEE Transactions on Information Theory},
  volume = {59},
  number = {11},
  pages = {7693--7710},
  year = {2013},
  doi = {10.1109/TIT.2013.2276628},
  eprint = {1208.1478},
  archivePrefix = {arXiv},
  primaryClass = {quant-ph},
  url = {https://arxiv.org/abs/1208.1478}
}

@article{HorodeckiEtAl2003_PureMixed,
  author = {Horodecki, Micha{\l} and Horodecki, Pawe{\l} and Oppenheim, Jonathan},
  title = {Reversible transformations from pure to mixed states and the unique measure of information},
  journal = {Physical Review A},
  volume = {67},
  number = {6},
  pages = {062104},
  year = {2003},
  doi = {10.1103/PhysRevA.67.062104},
  eprint = {quant-ph/0212019},
  archivePrefix = {arXiv},
  primaryClass = {quant-ph},
  url = {https://arxiv.org/abs/quant-ph/0212019}
}

@article{SchulmanMorWeinstein2005_HBAC,
  author  = {Schulman, Leonard J. and Mor, Tal and Weinstein, Yossi},
  title   = {Physical Limits of Heat-Bath Algorithmic Cooling},
  journal = {Physical Review Letters},
  volume  = {94},
  number  = {12},
  pages   = {120501},
  year    = {2005},
  doi     = {10.1103/PhysRevLett.94.120501}
}

\end{document}